\documentclass{aa}%
\pdfoutput=1 
\usepackage{graphicx}
\usepackage{natbib}
\usepackage{txfonts}
\usepackage[figuresright]{rotating}
\usepackage{supertabular}
\bibpunct{(}{)}{;}{a}{}{,}%

\font\math = cmmi10
\def\m#1{\hbox{\math \char'#1}} % character codes from p. 430 TeXbook
\def\v{\m{166}}

\def\H2O{H$_2$O}

\begin{document}
   \title{Evolved star water maser cloud size determined by star size}

   \author{A.M.S.~Richards\inst{1}\fnmsep\thanks{\email{\tt amsr@jb.man.ac.uk}}
\and S.~Etoka\inst{1} \and M.D.~Gray\inst{1} \and E.E.~Lekht\inst{2} \and J.E.~Mendoza-Torres\inst{3} \and K.~Murakawa\inst{4} \and G.~Rudnitskij\inst{2} \and J.A.~Yates\inst{5}
          }

   \offprints{}

   \institute{Jodrell Bank Centre for Astrophysics, School of Physics and Astronomy, University of Manchester, M13 9PL, UK 
%              \email{\tt amsr@jb.man.ac.uk}
%             \thanks{}
\and Lomonosov Moscow State University, Sternberg Astronomical Institute,
13 Universitetskij prospekt, Moscow 119234, Russia
\and Instituto Nacional de Astrofísca Optica y Electrónica,
Apartado Postal 51 y 216, Puebla, Pue., Z.P. 72000, Mexico
\and School of Physics and Astronomy, E.C. Stoner Building, University of Leeds,
Leeds, LS2 9JT, UK
\and Department of Physics and Astronomy, University College London, Gower 
Street, London. WC1E 6BT, UK  
%\email{\tt jyates@star.ucl.ac.uk}
             }

   \date{Received September 15, 3000; accepted March 16, 3000}

% \abstract{}{}{}{}{}
% 5 {} token are mandatory

  \abstract {Cool, evolved stars undergo copious mass loss but the
    detailed mechanisms and the form in which the matter is returned
    to the ISM are still under debate.}{We investigated the
    structure and evolution of the wind at 5 to 50 stellar radii from
    Asymptotic Giant Branch and Red Supergiant stars.}  {22-GHz water
    masers around seven evolved stars were imaged using MERLIN, at
    sub-AU resolution.  Each source was observed at between 2 and 7
    epochs, covering several stellar periods. We compared our results
    with long-term single dish monitoring provided by the Pushchino
    radio telescope.}  {The 22-GHz emission is located in
    approximately spherical, thick, unevenly filled shells.  The
    outflow velocity increases twofold or more between the inner and
    outer shell limits. Water maser clumps could be matched at
    successive epochs separated by less than two years for AGB stars,
    or at least 5 years for RSG.  This is much shorter than the
    decades taken for the wind to cross the maser shell, and
    comparison with spectral monitoring shows that some features fade
    and reappear.  In five sources, most of the matched maser features
    brighten or dim in concert from one epoch to the next.  A number
    of individual maser features show idiosyncratic behaviour,
    including one cloud in W Hya caught in the act of passing in front
    of a background cloud leading to 50-fold, transient amplification.
    The masing clouds are one or two orders of magnitude denser than
    the wind average and contain a substantial fraction of the mass
    loss in this region, with a filling factor $<1\%$.  The RSG clouds
    are about ten times bigger than those round the AGB stars.  }{
    Proper motions are dominated by expansion, with no systematic
    rotation.  The maser clouds presumably survive for decades (the
    shell crossing time) but the masers are not always beamed in our
    direction.  Only radiative effects can explain changes in flux
    density throughout the maser shells on short timescales.  The size
    of the clouds is proportional to that of the parent star, being of
    a similar radius to the star once the clumps reach the 22-GHz
    maser shell. Stellar properties such as convection cells must
    determine the clumping scale.}

  \keywords{Masers -- Stars: AGB -- Stars: supergiants --
               circumstellar matter
               }

   \maketitle
%
%________________________________________________________________

\section{Introduction}

Stars of less than eight Solar mass evolve onto the Asymptotic Giant
Branch (AGB) when they have exhausted much of the available H.  They
become very large ($R_{\star} > 1$ AU) and cool ($T_{\star} \approx
2500$ K) and lose mass at the rate of up to an Earth mass per year.
Red Supergiants (RSG) had main sequence masses $>8$ M$_{\odot}$ and
swell to ten times the size of AGB stars, with ten times the mass loss
rate.  Although the internal stellar processes are mass-dependent, the
winds are similar, being cool, dusty and molecular.  O-rich stars
produce SiO, H$_2$O and OH masers with spatially very compact,
spectrally narrow lines which can be imaged at milli-arcsec resolution
using radio interferometry.  A combination of pulsations, radiation
pressure and possibly other e.g. magnetic forces drives material away
from the stellar surface.  The exact mechanisms are still under debate
(e.g. \citealt{Woitke06}) and at a few stellar radii ($R_{\star}$),
SiO masers show both infall and outflow  (e.g. \citealt{Assaf11}).
Dust nucleation is likely to start close to the star
\citep{Wittkowski07} and by 4--6 $R_{\star}$, radiation pressure on
dust is the main force driving the wind.  The H$_2$O masers around
these stars are found in approximately spherical shells extending from
about 5-50 $R_{\star}$.  They provide a very high-resolution probe of
the structure, kinematics and conditions in the circumstellar envelope
(CSE) in the region where the outflow is most strongly accelerated and
passes through the escape velocity, see e.g. \citet{Bowers94},
\citet{Yates94} and also \citet{Habing96} for a general review.

MERLIN\footnote{the UK radio interferometer, operated by the
  University of Manchester on behalf of STFC} has been used to monitor
  H$_2$O and OH masers around a number of evolved stars with no known
  companions.  We analyse 2--7 epochs of images of 22-GHz H$_2$O
  masers around seven objects, listed in Table~\ref{tab:stars}. 
%This  includes supergiants VX Sgr and S Per, estimated masses 10--20
%  M$_{\odot}$. 
The observations of the RSG VX Sgr and S Per (first epoch) were
reported in \citet{Murakawa03} (M03) and \citet{Richards99} (R99),
respectively. The first epochs of the AGB stars IK Tau, U Her, U Ori
and RT Vir were reported in \citet{Bains03} (B03).  W Hya MERLIN
22-GHz observations have not been published previously.

 These stars are close enough and unobscured enough to have AAVSO
 optical monitoring data, \emph{Hipparcos} or other parallax
 measurements and in many cases optical or IR interferometric
 meaurements of the stellar diameter. They are among the objects which
 have been monitored by the Pushchino radio telescope at 22 GHz for
 many decades. Component fitting allows the size of AU-scale clouds to
 be measured from MERLIN data. The H$_2$O maser properties of the
 sources above declination 0\degr\/ were investigated in detail by
 \citet{Richards11} (R11), who found that they are predominantly
 unsaturated.  The maser appearance is mostly characteristic of
 spherical clouds but, in some objects, amplification appears to be
 occuring along the long axis of flattened clouds.

\begin{table*}
\begin{tabular}{lcrrrccccccc}
\hline
Star &Position &Type&\multicolumn{1}{c}{$V_{\star}$}&\multicolumn{1}{c}{Distance}&\multicolumn{1}{c}{$D$} &  $R_{\star}$&$P$&$\dot{M}$& Previous imaging  \\
    &(J2000)& &(km s$^{-1}$) & \multicolumn{1}{c}{(pc)}& \multicolumn{1}{c}{(pc)} &  (AU)&(d) &(M$_{\odot}$ yr$^{-1}$)& \\
\hline
VX Sgr  &18 08 04.05 --22 13 26.6&RSG&--5.3&$1570\pm^{270}_{270}$ &1700  &$7.4\pm0.7$                      &732& $7.2\times10^{-5}$ &C86,B93,M98,M03,V05  \\     
S Per   &02 22 51.71 +58 35 11.4& RSG& --38.5  & $2312\pm^{65}_{32}$ &2300&$8\pm3$             &822& $3.8\times10^{-5}$ &D87,Y94.R99,M98,V01,\\
&&&&&&&&&A10 \\
U Ori   &05 55 49.17 +20 10 30.7&Mira& --39.5 & $260\pm^{50}_{50}$   &266& $1.5\pm0.1$            &368& $2.3\times10^{-7}$ &B88,Y94,B94,B03,V05 \\	
U Her   &16 25 47.47 +18 53 32.9&Mira&--14.5 & $266\pm^{32}_{28}$ &266& $1.3\pm0.1$&406& $3.4\times10^{-7}$&Y94,B94,M98,C00,B03,\\
&&&&&&&&&V02,V05\\ 
IK Tau  &03 53 28.89 +11 24 21.9&Mira& +34.0  & $250\pm^{20}_{20}$ &266& $2.8\pm0.3$                  &470& $2.6\times10^{-6}$&L87,B93,Y94,M98,B03  \\
RT Vir  &13 02 37.98 +05 11 08.4&SRb& +18.2   & $135\pm^{15}_{15}$ &133& $0.8\pm0.04$             &158& $1.3\times10^{-7}$ &Y94,B93,B94,B03,I03\\
W Hya   &13 49 02.00 --28 22 03.4&SRb&+40.6   & $98\pm^{30}_{18}$&98&$2.1\pm0.2$                &375&  $2.3\times10^{-7}$&R90,B93 \\
\multicolumn{6}{l}{References}\\
VX Sgr &vL07&&C86  &\,\,C07 &&\,\,M04  &S11&D10    \\
S Per  &vL07& &D87 & \,\,M08 & & \,\,H94, L05 &S11&vL05\\
U Ori  &vL07& &C91 & \,\,C91  & &\,\,R06  &S11&K98\\
U Her  &vL07& &C94 & \,\,V07   &&\,\,R06  &S11&Y95\\
IK Tau &C06& &K87 & \,\,098  &&M04, R06    &S11&O98, B00\\
RT Vir &vL07& &N86 & \,\,vL07 &  &\,\,M04  &S11&K98, K99\\
W Hya &vL07& &N96  &  \,\,V03  &  & \,\,Z11 &S11&K98\\

\hline
\end{tabular}
\caption{Properties of the sample stars. The stellar velocity
  $V_{\star}$ is given in the Local Standard of Rest (LSR)
  convention. The positions given are for epoch 2000, see
  Section~\ref{sec:align} for more on astrometry. The most recent
  distances are given along with those used in our calculations, $D$,
  in order to remain consistent with M03, R99 and B03.  The values
  of $R_{\star}$ for the AGB stars were measured using IR
  interferometry at H and K bands. The radius of S Per is the average
  of values deduced from spectral fits. The stellar period $P$ is
  complex, changeable and/or uncertain for some objects, see
  Figs.~\ref{VXSgrSPer_AAVSO.png}
  to~\ref{IKTauRTVirWHya_AAVSO.png}. The mass loss rates $\dot{M}$ are
  adjusted to our adopted distances. The final column gives some of
  the previous interferometric 22-GHz H$_2$O maser images which have
  been published. \newline References in the table are:
  A10~\citet{Asaki10}; B88~\citet{Bowers88}; B93~\citet{Bowers93};
  B94~\citet{Bowers94}; B00~\citet{Bieging00};B03~\citet{Bains03};
  C86~\citet{Chapman86}; C91~\citet{Chapman91}; C94~\citet{Chapman94};
  C00~\citet{Colomer00}; C06~\citet{Carlsberg06}; C07~\citet{Chen07};
  D87~\citet{Diamond87}; I03~\citet{Imai03}; K87~\citet{Kirrane87};
  K98~\citet{Knapp98}; K99~\citet{Kerschbaum99}; L87~\citet{Lane87};
  L05~\citet{Levesque05}; M98~\citet{Marvel98}; M03~\citet{Murakawa03}
  M04~\citet{Monnier04}; M08~\citet{Mayne08}; N96~\citet{Neufeld96};
  N86~\citet{Nyman86}; O98~\citet{Olofsson98}; R90~\citet{Reid90};
  R99~\citet{Richards99}; R06~\citet{Ragland06}; S11~\citet{Samus11}
  (GCVS); V01~\citet{Vlemmings01}; V02~\citet{Vlemmings02};
  V03~\citet{Vlemmings03}; V05~\citet{Vlemmings05};
  V07~\citet{Vlemmings07}; vL05~\citet{vanLoon05};
  vL07~\citet{vanLeeuwen07}; Y94~\citet{Yates94};
  Z11~\citet{Zhao-Geisler11}}
\label{tab:stars}
\end{table*}

% Complementary MERLIN+VLBI OH multi-epoch imaging of these sources
% will be presented in a future paper.  
The H$_2$O masers have also
 been imaged using VLBI and the VLA, see Table~\ref{tab:stars}.
 However, a large fraction of the flux (half is not unusual) is
 resolved-out by VLBI, whilst the VLA cannot resolve individual maser
 clumps. Allowing for this, our observational results are consistent
 with other images. Comparisons with previous interpretations are
 included in the relevant sections of this paper.  For completeness,
 we note that we observed several other objects with MERLIN at 22
 GHz. The first epochs of NML Cyg and VY CMa were published by
 \citet{Richards96} and \citet{Richards98v}; further monitoring of
 these two exceptional RSG will be presented by Yates et al., as will
 obervations of the heavily-obscured SRb, R Crt.  We also observed R
 Cas (20000404, 20010511, 20020405) and R Leo (20000429) but these
 were non-detections at a conservative upper limit of 200 mJy.  This
 is consistent with non-detections for several years around this
 period (upper limit 10 Jy) using Pushchino (\citet{Pashchenko04} R
 Cas; \citet{Esipov99} R Leo).

This paper presents a large-scale analysis of multiple observations
(between two and seven epochs per star, spanning up to 7 years). Data
acquisition and analysis is summarised in Section~\ref{sec:data},
including estimating the centre of expansion, aligning epochs and
assessing the astrometric accuracy. We examine proper motions and
cloud survival in Section~\ref{sec:pm} and compare images with
Pushchino single dish monitoring in Section~\ref{sec:Pushchino}, to
investigate the survival of masers with respect to their parent
clouds.  We analyse the kinematics and estimate the mass concentration
in maser clouds in Section~\ref{sec:mass}.  Local and global maser
variability is discussed in Section~\ref{sec:variability}.  The size
of the maser clumps and their relationships with stellar properties is
discussed in Section~\ref{sec:cloudsize} and
Section~\ref{sec:conclusions} presents the summary and future work.

\section{Data acquisition and analysis}
\label{sec:data}

\begin{figure}
   \centering
   \includegraphics[angle=0, width=9cm]{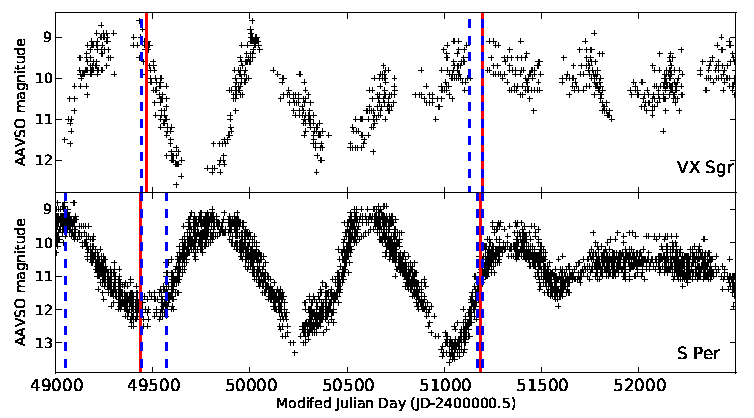}
      \caption{AAVSO light curves for VX Sgr and S Per, with the epochs of observation using MERLIN and Pushchino superimposed in red and blue, respectively.}
         \label{VXSgrSPer_AAVSO.png}
   \end{figure}
\begin{figure}
   \centering
   \includegraphics[angle=0, width=9cm]{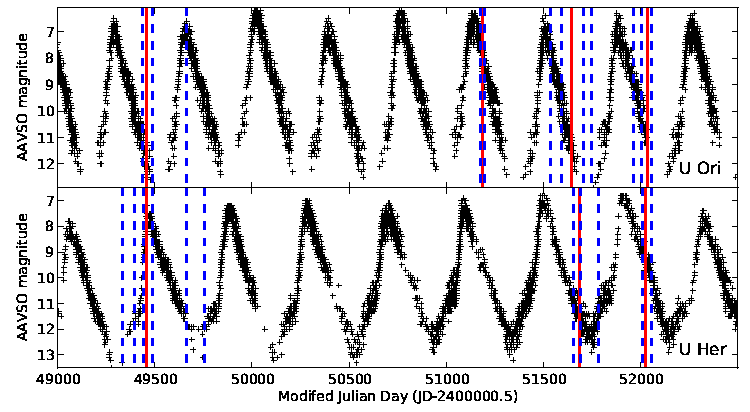}
      \caption{AAVSO light curves for U Ori and U Her, with the epochs of observation using MERLIN and Pushchino superimposed in red and blue, respectively.}
         \label{UOriUHer_AAVSO.png}
   \end{figure}
\begin{figure}
   \centering
   \includegraphics[angle=0, width=9cm]{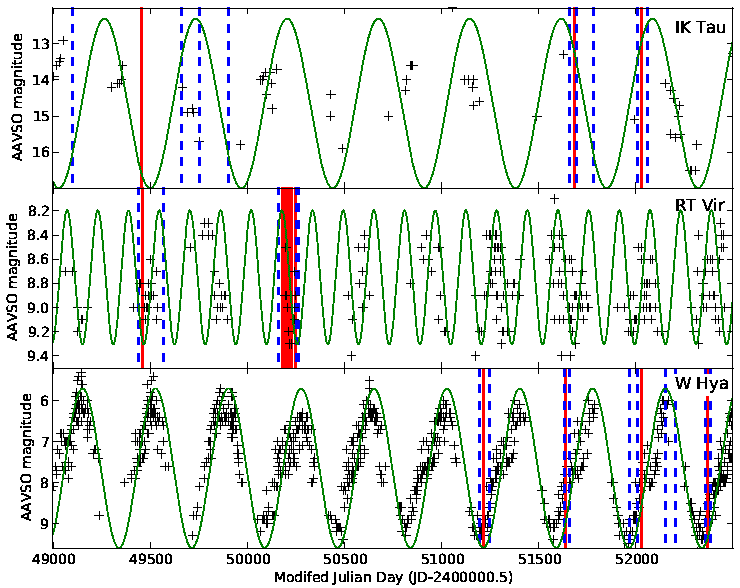}
      \caption{AAVSO light curves for IK Tau, RT Vir and W Hya, with
      the epochs of observation using MERLIN and Pushchino
      superimposed in red and blue, respectively. The optical data for
      these stars is sparse as they are Zodaical and/or heavily
      obscured and tentative, sinusoidal, variability curves
      are overlaid in green. We took the stellar periods from GCVS
      \citep{Samus11} and also the origin and amplitude for RT Vir and
      W Hya, adjusted by eye in the latter case to fit the current
      data. The origin and amplitude for IK Tau were taken
      from \citet{Matsumoto08}.}       
         \label{IKTauRTVirWHya_AAVSO.png}
   \end{figure}

Radio interferometry observations were made using 5 antennas of
MERLIN, longest baseline 217 km, centred on the 22.235079 GHz H$_2$O
maser line, corrected in the correlator to the $V_{\mathrm {LSR}}$ of
the source. 2 GHz of bandwidth, providing 25 km s$^{-1}$ velocity
coverage in 0.105 km s$^{-1}$ channels, was used for most epochs. This
was found to be insufficient for IK Tau and S Per and so 4 GHz
bandwidth, providing 50 km s$^{-1}$ in 0.21 km s$^{-1}$ channels, was
used for these sources in 1999 and later epochs.  VX Sgr was observed
at both epochs in 4 GHz bandwidth and 0.42 km s$^{-1}$ channels were
used for analysis. Data were recorded in two hands of circular
polarization, calibrated independently (apart from amplitude
normalisation). High spatial and spectral resolution is required to
detect the weak polarization of H$_{2}$O (see e.g. \citet{Vlemmings05}
for results using VLBA data), and circular polarization is neglible
for these sources at MERLIN resolution, so we made and analysed images
in total intensity only.

The reduction of the 1994 data (and 1999 data for VX Sgr) is described
in M03, R99 and B03. Similar techniques were used at later epochs,
with the addition of phase referencing in some cases, see
Section~\ref{sec:align}.  All data reduction and initial analysis was
performed using {\sc aips} \citep{Greisen94}. Observational parameters
and measurements are given in Table~\ref{tab:feats}. The epochs of
observation are superimposed (red lines) on AAVSO (American
Association of Variable Star Observers) light curves in
Figs.~\ref{VXSgrSPer_AAVSO.png} to~\ref{IKTauRTVirWHya_AAVSO.png}.
The noise levels achieved depended on the source elevation, the
weather and the total time on target. The weighting used during
imaging was chosen to provide the best resolution without increasing
the noise by more than 20\%.  In all cases, the dynamic range was
limited to a few hundred, or about one hundred for low-declination sources
with poor visibility-plane coverage.

The flux scale was derived from either 3C84 or 3C273, using 22-GHz
monitoring data kindly provided by H. Ter\"{a}sranta from the Mets\"{a}hovi
antenna.  These sources may be slightly resolved by MERLIN
at some epochs and vary during the few weeks or months between the
Metsahovi and MERLIN observations. This dominates the uncertainty in
the flux scales, which is about 10\%.

In Appendix~\ref{apx:kntr} we present contour maps of the maser
emission averaged every few channels.

The position and peak intensity of each patch of maser emission
brighter than 4.5 -- 6 $\sigma_{\mathrm {rms}}$ (depending on
visibility plane coverage) in each channel was measured by fitting a
2-D Gaussian component.  We rejected components fainter than the the
local sidelobe level close to bright peaks.  Their size (with the
restoring beam deconvolved) and integrated intensity was also measured
for all sources at positive declinations i.e. excluding VX Sgr and W
Hya, for which the resolution N-S is too coarse.  The accuracy of
these measurements is proportional to (beam size)/(signal to noise
ratio) (\citealt{Condon97} adapted for sparse arrays by
\citealt{Condon98} and \citealt{Richards99}) so the different
sensitivies are taken account of via the uncertainties used in
analysis. 

\begin{figure}
   \flushleft
\hspace*{-0.5cm}
   \includegraphics[angle=0, width=9.5cm]{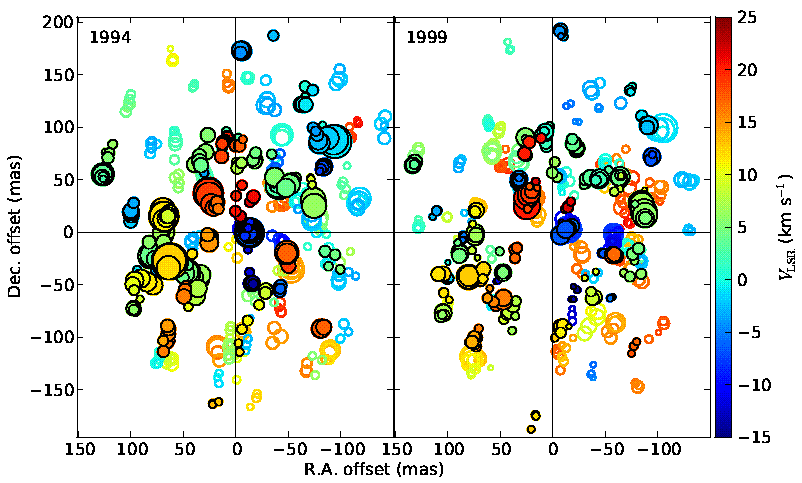}
      \caption{22-GHz H$_{2}$O maser components around VX Sgr imaged
        in 1994 and 1999. Symbol area is proportional to flux
        density. The components with black outlines belong to features
        matched at both epochs. (0,0) is the location of the estimated
        centre of expansion. The colour bar shows the
        $V_{\mathrm{LSR}}$ scale.}
         \label{VXSGR_Match_xyplot.png}
   \end{figure}

\begin{figure}
   \centering
   \includegraphics[angle=0, width=9cm]{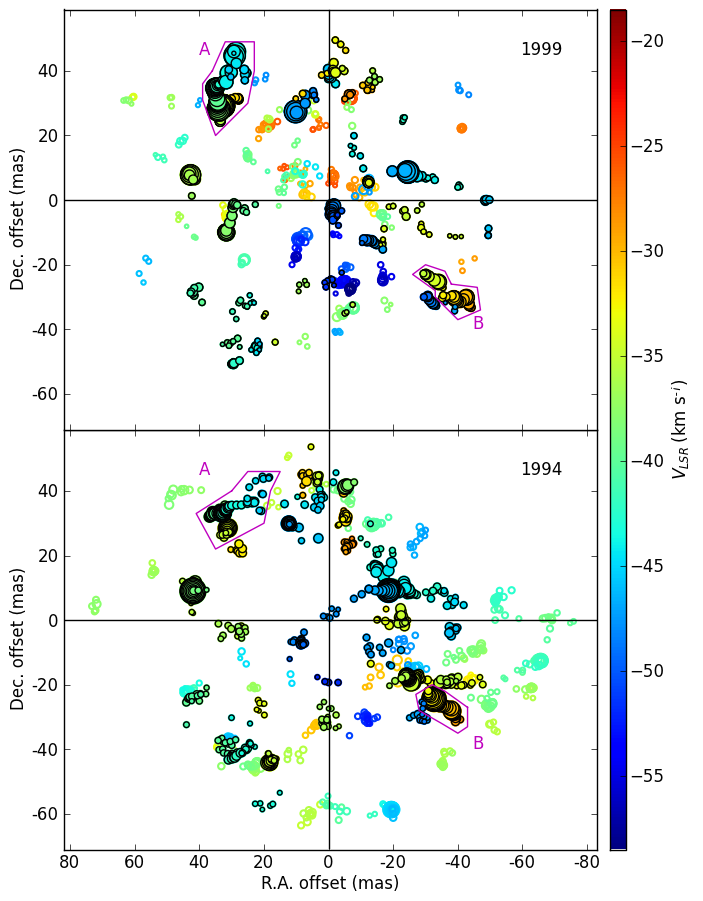}
      \caption{22-GHz H$_{2}$O maser components around S Per imaged in 1994
      and 1999. Symbol
      area is proportional to flux density. The components with black
      outlines belong to features matched at both epochs.  The labelled
      features are enlarged in Fig.~\ref{sper_abfeat.png}.}
         \label{SPER_Match_xyplot.png}
   \end{figure}

\begin{figure}
   \centering
   \includegraphics[angle=0, width=8.5cm]{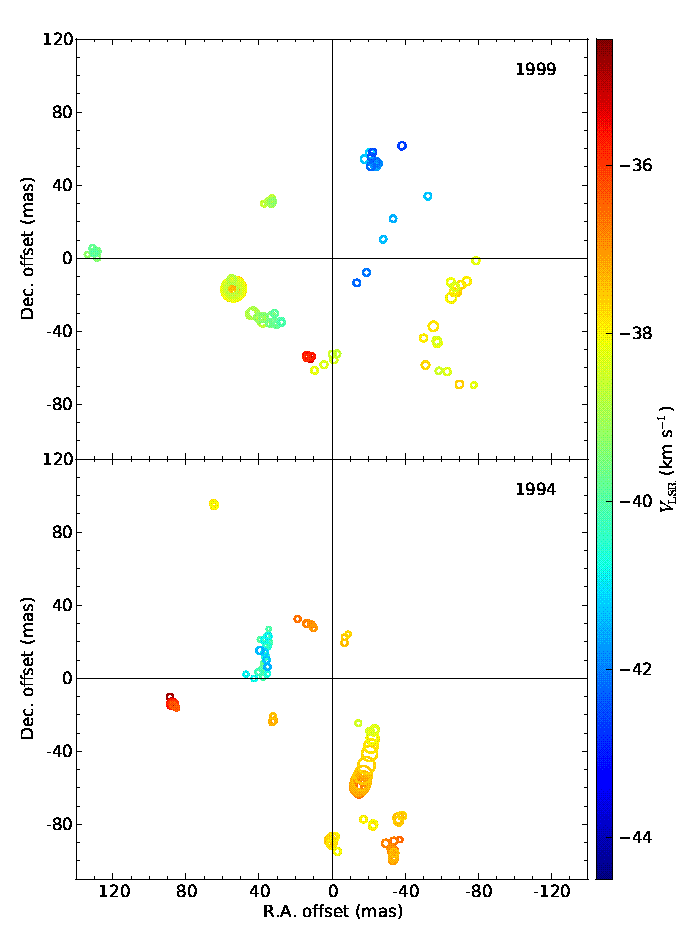}
      \caption{22-GHz H$_{2}$O maser components around U Ori imaged in 1994
      and 1999. Symbol
      area is proportional to flux density.}
         \label{UOri_94-99_xyplot.png}
   \end{figure}

\begin{figure}
   \centering
   \includegraphics[angle=0, width=8.5cm]{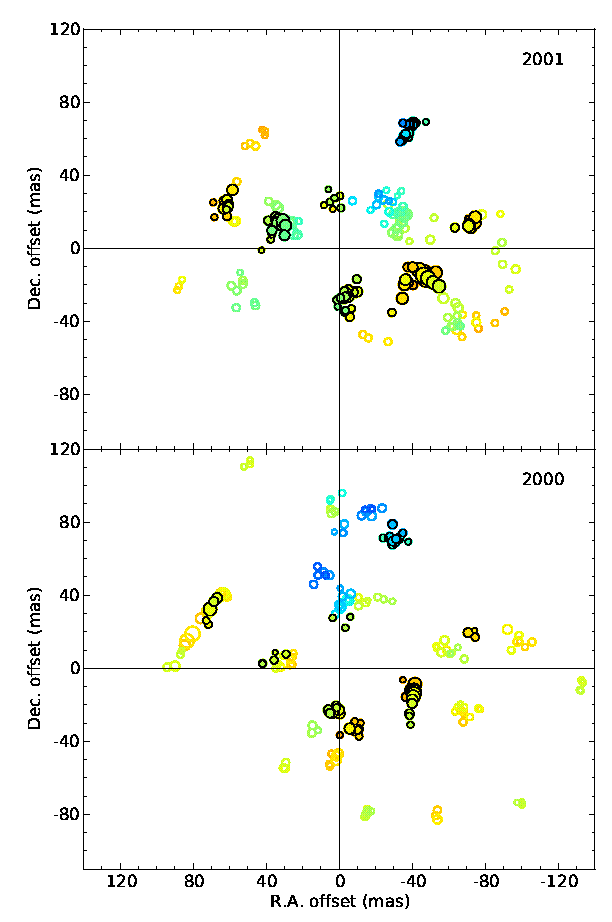}
      \caption{22-GHz H$_{2}$O maser components around U Ori imaged in 2000
      and 2001. Symbol
      area is proportional to flux density. The components with black
      outlines belong to features matched at both epochs. Velocity
      scale as in Fig.~\ref{UOri_94-99_xyplot.png}.}
         \label{UOri_00-01_xyplot.png}
   \end{figure}

\begin{figure}
   \centering
   \includegraphics[angle=0, width=9cm]{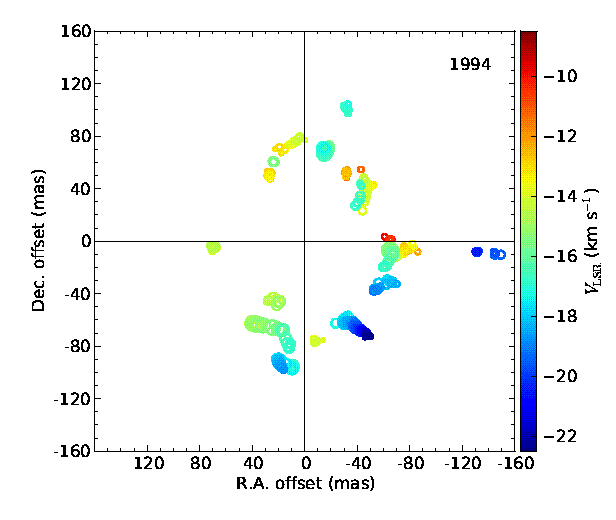}
      \caption{22-GHz H$_{2}$O maser components around U Her imaged in 1994. Symbol
      area is proportional to flux density. }
         \label{UHer_94_xyplot.png}
   \end{figure}

\begin{figure}
   \centering
   \includegraphics[angle=0, width=8.5cm]{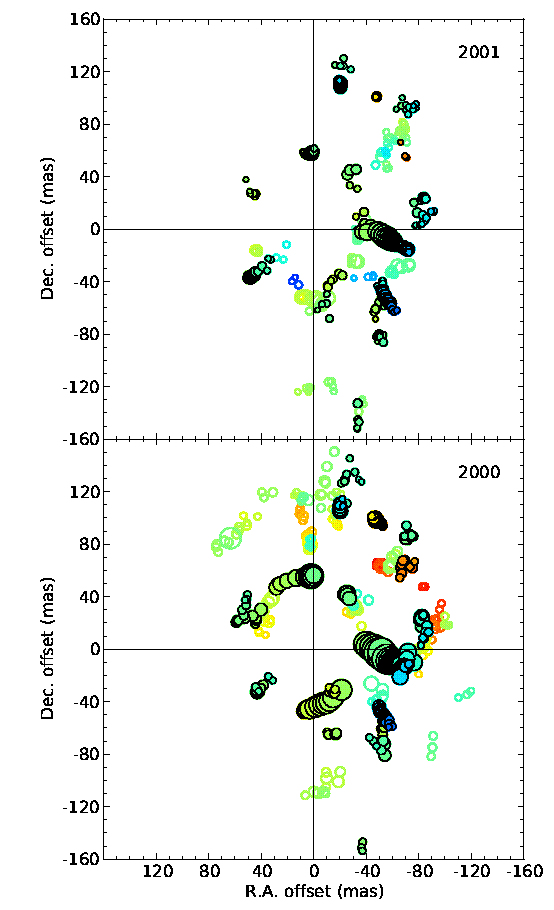}
      \caption{22-GHz H$_{2}$O maser components around U Her imaged in 2000
      and 2001. Symbol
      area is proportional to flux density. The components with black
      outlines belong to features matched at both epochs.  Velocity
      scale as in Fig.~\ref{UHer_94_xyplot.png}. The ring seen in 1994
      (Fig.~\ref{UHer_94_xyplot.png}) may  correspond to the outer
      arcs seen in 2000 and 2001.}
         \label{UHer_00-01_xyplot.png}
   \end{figure}

\begin{figure}
   \centering
   \includegraphics[angle=0, width=9cm]{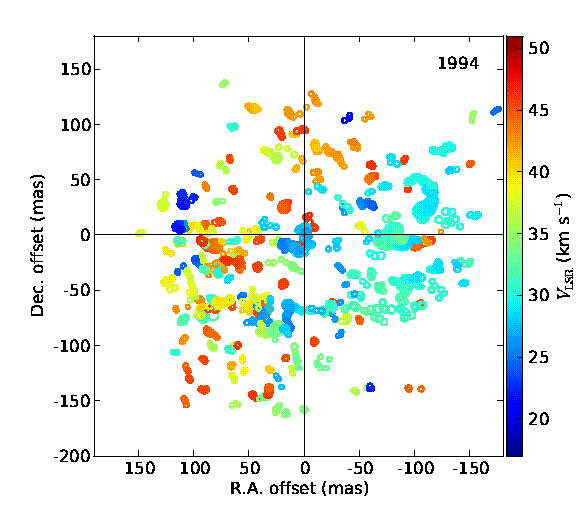}
      \caption{22-GHz H$_{2}$O maser components around IK Tau imaged in 1994. Symbol
      area is proportional to flux density. }
         \label{IKTau_94_xyplot.png}
   \end{figure}

\begin{figure}
   \centering
   \includegraphics[angle=0, width=8.5cm]{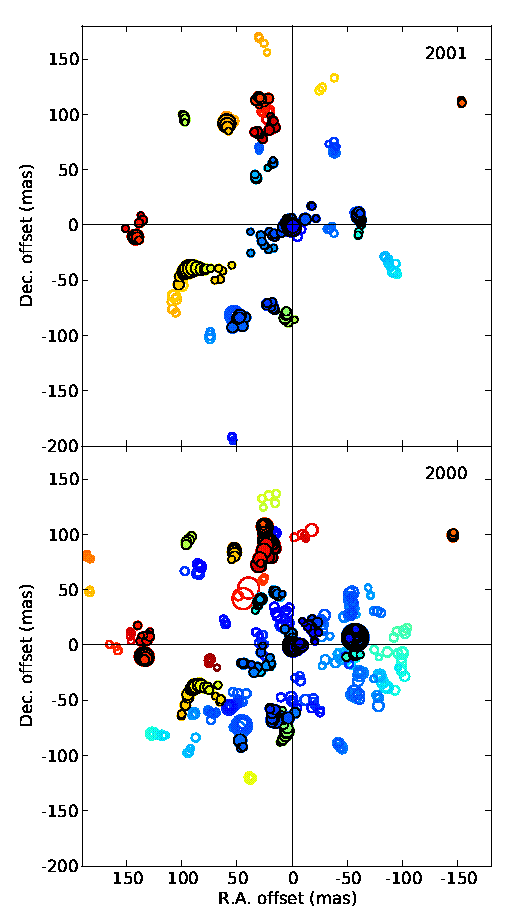}
      \caption{22-GHz H$_{2}$O maser components around IK Tau imaged in 2000
      and 2001. Symbol
      area is proportional to flux density. The components with black
      outlines belong to features matched at both epochs.  Velocity
      scale as in Fig.~\ref{IKTau_94_xyplot.png}.}
         \label{IKTau_00-01_xyplot.png}
   \end{figure}

\begin{figure}
   \centering
   \includegraphics[angle=0, width=9cm]{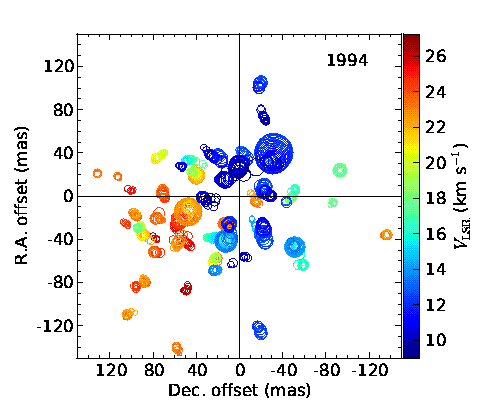}
      \caption{22-GHz H$_{2}$O maser components around RT Vir imaged in 1994. Symbol
      area is proportional to flux density. }
         \label{RTVir_94_xyplot.png}
   \end{figure}

\begin{figure}
   \centering
   \includegraphics[angle=0, width=8.5cm]{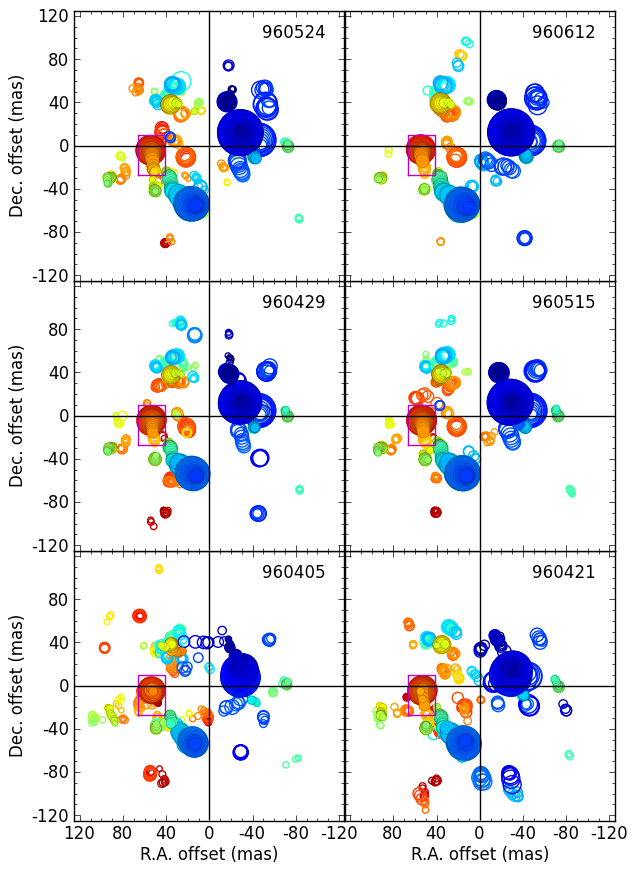}
      \caption{22-GHz H$_{2}$O maser components around RT Vir imaged around
      1996. Symbol
      area is proportional to flux density. The components with black
      outlines belong to features matched at all epochs. The boxed
      feature is enlarged in Fig.~\ref{rtvir_fs.png}.  Velocity
      scale as in Fig.~\ref{RTVir_94_xyplot.png}.}
         \label{RTVir_96_xyplot.png}
   \end{figure}

\begin{figure}
   \centering
   \includegraphics[angle=0, width=9cm]{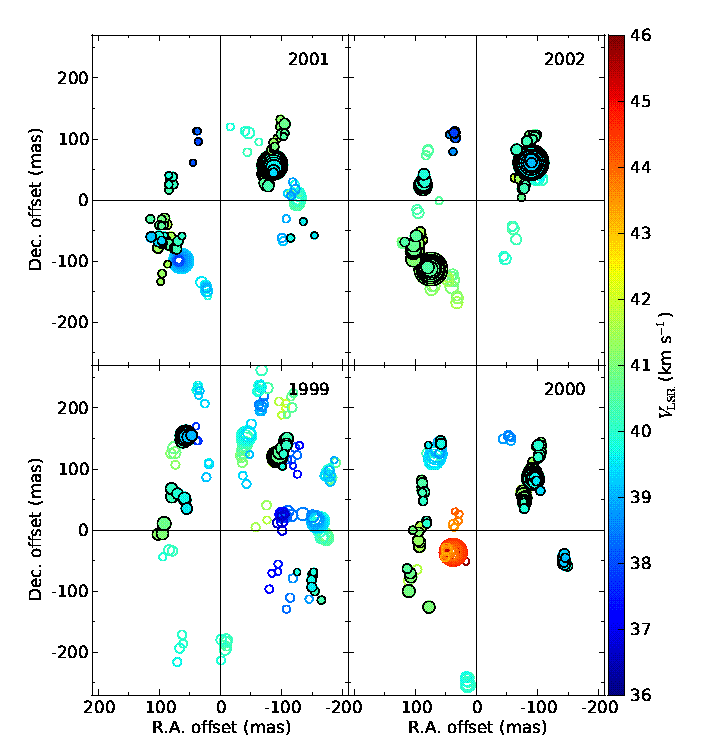}
      \caption{22-GHz H$_{2}$O maser components around W Hya imaged in 1999--2002. Symbol
      area is proportional to flux density. The components with black
      outlines belong to features matched at two or more epochs.}
         \label{WHya_xyplot.png}
   \end{figure}

 The maser component positions are shown in
 Figs.~\ref{VXSGR_Match_xyplot.png} to~\ref{WHya_xyplot.png}.  Maser
 components separated by a few milli-arcsec form series in adjacent channels
 which we term features.  Our selection criteria imply that valid
 features must have a broader minimum total velocity span, if they
 were observed with broader channels.  In the case of S Per and VX
 Sgr, the number of features remains similar despite changes in
 channel width, probably because the broader channels provided greater
 sensitivity.  However, the sensitivity was reduced (due to worse
 weather and shorter observing times) for the 2000--2001 IK Tau
 epochs, despite the doubling of channel width compared with
 1994. This may explain why significantly fewer features were detected
 at the later epochs. 

 The measured parameters for all sources are
 given in Appendix~\ref{apx:feats}, tables~\ref{tab:sp94}
 to~\ref{tab:wh02}. All but the faintest have Gaussian velocity
 profiles and R11 summarise the evidence that these are discrete,
 water vapour clouds.  We rejected isolated components which did not
 form a series of at least 3 components in successive channels
 spanning the average feature size or less.  The width of an
 individual component represents the beamed maser size, but R99
 showed that the largest angular separation of components comprising a
 feature is at least 80\% of the physical size of the cloud (as long
 as the emission is not resolved-out). Feature parameters are given in
Appendix~\ref{apx:feats}, Tables~\ref{tab:sp94} to~\ref{tab:wh02}.

Note that the individual components represent the line of sight
through the cloud where masing is strongest in the velocity interval
sampled by the channel.  The arangement of series of components
represents the velocity gradient with position but it does not
necessarily show the full shape of the cloud.  A linear or curved
series of components can arise from a spherical cloud, especially if
the masers are tightly beamed. This was discussed further in R11,
who selected well-resolved components and features for beaming
analysis whereas here we include all data. In addition, in aligning
the epochs for U Her, we found a few misallocated components and
revised a few feature properties, and we also improved the fitting of
shell limits slightly for some sources/epochs. Hence, there are small
discrepancies between their table 2 and our
Table~\ref{tab:feats}. This does not affect any of the results in
R11 except to reduce the scatter slightly in some of the analysis
involving feature properties.

\begin{table*}
\begin{tabular}{lccccrrrrccrrrcrc}
\hline
Star & Date&$\theta_{\mathrm B}$&$\sigma_{\mathrm{rms}}$ &$r_{\mathrm{i}}$&$r_{\mathrm{o}}$&$\v_{\mathrm{i}}$&$\v_{\mathrm{o}}$&$\epsilon$&$K_{\mathrm{grad}}$&$NC$& $NF$& $N_{\mathrm {prev}}$&$I_{\mathrm{max}}$ &$\overline{l}$ &$\overline{\Delta{V_{1/2}}}$\\
 & (yymmdd)&mas$^2$&mJy b$^{-1}$
&\multicolumn{2}{c}{(AU)}&\multicolumn{2}{c}{(km s$^{-1}$)}&\multicolumn{3}{c}{(km s$^{-1}$ AU$^{-1}$)}& &&(Jy)
&(AU)&(km s$^{-1}$)\\
\hline
VX Sgr &940426   &$40\cdot20$&51&85&350&9.0&20.0&0.57&0.04& 520& 92&- -&294 &$14\pm3  $   &$1.3\pm0.3$ \\
VX Sgr &990116   &$40\cdot20$&42&95&355&10.0&20.5&0.54&0.04& 497& 97& 42&102 &$12\pm4  $  &$1.1\pm0.5$ \\
\\               								                    
S Per$^1$&940324 &$10\cdot10$&17&55&175&9.0&16 &0.51 &0.06& 1040&93&- -& 72&$18\pm9  $    &$0.8\pm0.3$\\
S Per    &990110 &$10\cdot10$&8&30&165&8.5&22 &0.56&0.23 & 689&100& 40& 47&$12\pm6  $     &$0.9\pm0.3$\\
\\               								                   
U Ori  &940417   &$15\cdot15$&12&10& 32&2.5&7.0&0.88&0.20 &  95& 14&- -& 26&$2.7\pm2.6$ &$0.6\pm0.1$\\
U Ori  &990109   &$20\cdot20$&20&12& 36&2.5&5.5&0.72&0.13 &  94& 12&- -& 44&$4.8\pm3.1$ &$0.5\pm0.2$\\
U Ori$^{\phi}$  &000410   &$18\cdot18$&25& 7& 36&2.0&6.0&0.65&0.14 & 187& 33&0&  8&$2.7\pm1.5$ &$0.5\pm0.2$\\
U Ori$^{\phi}$  &010506 &$18\cdot18$&27& 7& 29&2.0&5.5&0.68&0.16 & 212& 25&  9& 10&$4.9\pm2.0$ &$0.7\pm0.3$\\
\\               						                                  
U Her  &940413   &$15\cdot15$&14&13& 47&4.0&9.5&0.69&0.16 & 282& 34&- -& 22&$2.3\pm1.3$        &$0.7\pm0.4$\\
U Her  &000519   &$18\cdot18$&40&10& 41&3.0&8.0&0.72&0.16 & 412& 48&- -&141&$4.4\pm2.0$        &$0.6\pm0.3$\\
U Her$^{\phi}$  &010427 &$18\cdot18$&35&10& 41&3.0&8.0&0.72&0.16 & 324& 36& 20& 38&$3.8\pm1.9$ &$0.6\pm0.3$\\
\\               						                                  
IK Tau$^1$&940415&$15\cdot15$&10&16& 66&5.0&16 &0.82&0.22 &1490&256&- -& 25&$2.0\pm1.3$        &$0.7\pm0.3$\\
IK Tau$^{\phi}$ &000520 &$15\cdot15$&35&16& 72&6.0&18 &0.73&0.21 & 502& 72&- -& 84&$3.4\pm1.8$ &$1.0\pm0.4$\\
IK Tau$^{\phi}$ &010427 &$15\cdot15$&40&20& 72&6.0&18 &0.73 &0.21& 236& 40& 24& 24&$3.2\pm2.0$ &$0.8\pm0.3$\\
\\               						                                  
RT Vir &940416   &$20\cdot20$&12& 6& 25&4.0&10 &0.65&0.32 & 847& 55&- -&394&$1.4\pm1.0$  &$1.0\pm0.4$\\
RT Vir &960405   &$12\cdot12$&25&4.5& 19&3.5&11 &0.80&0.52 & 494& 59&- -&389&$1.2\pm0.9$ &$0.9\pm0.4$\\
RT Vir &960421   &$12\cdot12$&30&4.5& 19&3.5&11 &0.80&0.52 & 512& 52& 23&516&$1.2\pm0.8$ &$0.9\pm0.4$\\
RT Vir &960429   &$12\cdot12$&15&4.5& 19&3.5&11 &0.80&0.52 & 586& 50& 25&526&$0.9\pm0.5$ &$1.0\pm0.4$\\
RT Vir &960515   &$12\cdot12$&25&4.5& 19&3.5&11 &0.80&0.52 & 551& 41& 33&727&$1.0\pm0.7$ &$1.1\pm0.4$\\
RT Vir &960524   &$12\cdot12$&20&4.5& 19&3.5&11 &0.80&0.52 & 483& 51& 36&706&$0.9\pm0.8$ &$0.9\pm0.3$\\
RT Vir &960612   &$12\cdot12$&35&4.5& 19&3.5&11 &0.80&0.52 & 433& 38& 29&791&$0.9\pm0.6$ &$0.8\pm0.3$\\
\\		 												  
W Hya$^{\phi}$  &990209&$50\cdot35$&65& 6& 25&3.5&5.5&0.31&0.11 & 216& 29&- -& 32&$1.5\pm1.0$ &$0.8\pm0.3$ \\
W Hya$^{\phi}$  &000405&$25\cdot25$&55& 6& 25&3.5&5.5&0.31&0.11 & 152& 17&  5& 95&$1.2\pm0.9$ &$0.6\pm0.2$\\
W Hya$^{\phi}$  &010430&$25\cdot25$&40& 7& 16&3.5&5.5&0.55&0.22 & 147& 18&  7&137&$1.5\pm1.1$ &$0.6\pm0.2$\\
W Hya$^{\phi}$  &020405&$28\cdot22$&50& 6& 16&3.5&5.5&0.46&0.20 & 125& 16&  9&268&$1.3\pm1.0$ &$0.7\pm0.1$\\
\hline
\end{tabular}
\caption{Observational parameters and measurements of 22-GHz
  masers. $^1$ denotes epochs where the bandwidth was probably
  slightly less than the total velocity extent of the source. Epochs
  when phase referencing was attempted are marked
  $\phi$. $\theta_{\mathrm B}$ is the restoring beam used (major axis
  N--S). $\sigma_{\mathrm {rms}}$ is the noise in channels which are
  not  dynamic-range limited.
  $r_{\mathrm i}$, $r_{\mathrm o}$, $\v_{\mathrm i}$ and $\v_{\mathrm
  o}$ are the radii and expansion velocities of the inner and outer
  maser shell limits, $\epsilon$ is the logarithmic velocity gradient
  and $K_{\mathrm{grad}}$  is the velocity gradient.  The total number of
  individual maser components ($NC$) and features ($NF$) made up of
  contiguous series of components are given.  $N_{\mathrm{prev}}$ is
  the number of features matched with a previous epoch, see
  Section~\ref{sec:centre}.  $I_{\mathrm{max}}$ is the maximum total
  intensity of the brightest component. $\overline{l}$ is the mean
  feature size, and $\overline{\Delta{V_{1/2}}}$ is the mean
  FWHM of those features to which a Gaussian spectral profile could be
  fitted at $\ge3\sigma$.  }
\label{tab:feats}
\end{table*}



\subsection{Finding the centre of expansion and aligning epochs}
\label{sec:centre}

We initially assume that the winds are dominated by radial expansion
from the central star.  \citet{Yates94} describe finding the centre of
expansion by fitting a thick shell in 3 dimensions to angular radius
and velocity.  This gives well-constrained results for shells which
are symmetric along any axis passing through the centre, although not
necessarily spherical.  This was used for objects with well-filled
CSEs, i.e. all epochs of VX Sgr, S Per and RT Vir and early epochs of
IK Tau and U Her. Many of the individual epochs of AGB stars are
poorly filled and asymmetric, and for these we assumed that emission
at close to $V_{\star}$ should appear within a thick ring centred on
the position which maximises the average angular separation from the
centre of expansion.  \citet{Bains03} found that the two methods gave
similar results for the 1994 observations of RT Vir, IK Tau and U Her.
We used the second method for all epochs of U Ori and W Hya and for
the later epochs of U Her and IK Tau.

We then aligned successive epochs of the same star using the initial
estimates of the centres of expansion and looked for matching
features, allowing for  proper motions consistent with the spectral
extent of the H$_{2}$O maser shell. Matches were found for those
epochs listed in Table~\ref{tab:feats} as having $N_{\mathrm {prev}}$
features in common with the previous epoch. Components belonging to
matched features are outlined in black in
Figs~\ref{VXSGR_Match_xyplot.png} to~\ref{WHya_xyplot.png}.  The
criteria used are described in more detail and the results are
analysed further in Section~\ref{sec:pm}.

Having identified potential matches, we refined the alignment by
minimising the sum of the offsets of matched features
\citep{Richards98v} and if necessary applied additional shifts to the
epoch with the greater uncertainties.  These were in the range 1--5
milli-arcsec for all sources except W Hya, where the uncertainties are greater
in declination and adjustments up to 10 milli-arcsec were needed. In the cases of U
Ori and IK Tau, where phase-referenced positions were available at two
epochs each, the adjustments were within the uncertainties of the
astrometry (see below). The component positions in
Figs~\ref{VXSGR_Match_xyplot.png} to~\ref{WHya_xyplot.png} are given
relative to the matched centres of expansion for each star.

\subsection{Phase referencing results}
\label{sec:align}

All positions are given in J2000 ICRF.
Phase referencing was used at the epochs marked $\phi$ in
Table~\ref{tab:feats}.  For U Ori and W Hya, we also used a more
distant source to provide initial phase-rate and amplitude solutions,
allowing fainter sources within 2\degr\/ to be used for phase
solutions. 
In all cases, the maser peaks were bright
enough to give better signal-to-noise in a single spectral channel,
than were the phase-reference sources observed in the maximum
available 13-MHz bandwiths. There are 4 factors which contribute to
the position uncertainty:
\begin{enumerate}
\item The noise-based error in measuring the position by fitting a
  Gaussian component, negligible ($\la1$ mas) for the brightest masers
  and for the phase reference sources unless otherwise stated.
\item The uncertainty in transfering the phase solutions between
  reference source and maser.  This is estimated from the sky
  separation in minutes of time, e.g. 2\degr \/ = 8 min . We examine
  the phase drift in the equivalent time on the raw data for a bright
  source observed continuously for many minutes, usually the bandpass
  calibrator.  A typical phase drift of half a turn corresponds
  to half a beam position uncertainty. This assumes that the
  atmospheric effects are the same throughout the observations; if the
  weather or the elevation changed significantly we added an allowance
  for this.
\item The phase reference position uncertainty. The target positions
  have been adjusted to the most accurate positions available.
\item Uncertainty in the telescope positions. This is about 1 cm,
  $\sim1$ wavelength at 22 GHz, on the longest MERLIN baseline, giving
  an uncertainty of about one beam (see Table~\ref{tab:feats}).
\end{enumerate}
The naturally-weighted beam size must be used for these estimates.
Only 1. affects data for different components in a single source at a
single epoch. 3. and 4. do not affect data taken using the same array
and the same phase-reference at different epochs. The uncertainties
for each target and phase reference source are given in the paragraphs
following.  Where possible, we initially transfered the phase
reference solutions to the maser and then repeated the process in the
opposite direction. This provides two independent estimates of the
position of maser emission in the reference channel. The discrepancy
is used to refine the estimate of error 2. In other cases, the phase
reference source was too weak to detect until after applying the maser
solutions.

We used \emph{Hipparcos} positions and proper motions
\citep{vanLeeuwen07}, to predict the stellar positions of U Her, U Ori
and W Hya at our epochs of observation.  The formal errors, derived
from the uncertainties in position at epoch 1991.25 and in the proper
motions, are 10--20 mas.  The optical position of IK Tau is less well
known due to its highly obscured location, having an uncertainty of
about 100 milli-arcsec in either coordinate (e.g. \citealt{Zacharias05}). We
compare these positions with our estimates of the centres of expansion derived
from the masers, assumed to be the stellar positions.

\paragraph{\bf U Ori} (Fig.~\ref{UOri_00-01_xyplot.png})
The phase reference source J0552+1913 at 05:52:25.8858 +19:13:40.275,
uncertainty (10, 10) milli-arcsec \citep{Browne98}, just over 1\degr\/ from U
Ori, was used in 2000 and 2001.  The brightest maser channel was at
$-38$ km s$^{-1}$, at both epochs.  This is close to $V_{\star}$ and
thus the maser feature will have a significant proper motion in the
plane of the sky due to the expansion of the CSE, as well as to
stellar proper motion. We therefore compared the positions of the
centre of expansion. These were 05:55:49.1699 +20:10:30.624 and
05:55:49.1697 +20:10:30.619 in 2000 and 2001, respectively, fitting
uncertainty (5, 5) mas, total astrometric uncertainty (23, 32) milli-arcsec per
epoch.  This corresponds to a proper motion of (--3, --5) mas, which,
although close to the fitting uncertainties, is in a direction similar
to the \emph{Hipparcos} proper motion of (--12.12, --6.13),
uncertainty (1.52, 0.73) milli-arcsec yr$^{-1}$.  The positions of U Ori
derived from \emph{Hipparcos} were 05:55:49.1697 +20:10:30.685,
uncertainty (14, 7) mas, and 05:55:49.1688 +10:30.679 (15, 7) mas,
in 2000 and 2001, respectively.  The positions derived from the masers
are offset by (3, --61) and (13, --60) milli-arcsec from the estimated
\emph{Hipparcos} positions. The combined uncertainties are (27, 35)
mas.  The good agreement between the expansion centres estimated
independently from the maser emission at the two epochs suggests that
our alignment is accurate. If the star really was at the location
calculated using the \emph{Hipparcos} data, it would be at the outer
edge of the maser shell (opposite to the maser peak), which suggests
that there is an unrecognised source of astrometric error of order
tens of milli-arcsec in one or both sets of data.

\paragraph{\bf U Her} (Fig.~\ref{UHer_00-01_xyplot.png})
The phase reference source J1635+1831 at 16:35:39.1588 +18:31:03.723,
uncertainty (7, 12) milli-arcsec \citep{Browne98} at 2\degr.4 separation from
U Her, was used in 2001. We obtained a position for the centre of
expansion of U Her of 16:25:47.4710 +18:53:32.887, with a fitting
uncertainty of (4, 4) milli-arcsec and a total astrometric uncertainty of (20,
32) mas.  The \emph{Hipparcos} position at this epoch is 
16:25:47.4702 18:53:32.841, uncertainty (6, 7) mas. The position derived from masers is
offset by (11, 46) mas, close to the combined uncertainties.  The
brightest U Her maser feature observed in 2001 is at --15.88 km
s$^{-1}$, position 16:25:47.4668 +18:53:32.879, total extent 29 mas,
2.1 km s$^{-1}$. \citet{Vlemmings02} carried out MERLIN observations
less than 4 weeks later, optimised for astrometry by using 3 phase
reference sources and a velocity resolution of 0.42 km s$^{-1}$. They
found the brightest feature at --15.7 km s$^{-1}$, position 
16:25:47.468 +18:53:32.849 (10, 10). Our position is offset from this by
(--17, 30) mas, within the combined uncertainties for observations
with the same array but different reference sources

\paragraph{\bf IK Tau} (Fig.~\ref{IKTau_00-01_xyplot.png})
The phase reference source J0409+1217 at 04:09:22.0087 +12:17:39.845,
uncertainty (0.2, 0.2) milli-arcsec \citep{Titov04}, 4\degr\/ separation, was
used in 2000 and 2001. The bright, blue-shifted feature used for
self-calibration is within 1 milli-arcsec of the centre of expansion at
03:53:28.8919 +11:24:21.940, and 03:53:28.8924 +11:24:21.932 in 2000
and 2001, respectively, fitting uncertainty (3, 3) mas, total
astrometric uncertainties (30, 42) milli-arcsec and (33, 43) mas. The 2001
position is offset from the 2000 position by (5, --8) mas. This is
less than expected from the optically measured proper motions, of
(23.5, --26.1) or (8, --80) milli-arcsec yr$^{-1}$ from NOMAD
\citep{Zacharias05} or the PM2000 Bordeaux Proper Motion catalogue
\citep{Ducourant06}, respectively. The 2001 maser-derived position is
(137, --86) milli-arcsec from the Bordeaux position at the same epoch,
03:53:28.8825 +11:24:22.018, uncertainty (27, 27) mas.  The NOMAD
position differs from both by a total of 0\farcs75 (0\farcs15) (with
greater discrepancies at the earlier epoch).

\paragraph{\bf W Hya} (Fig.~\ref{WHya_xyplot.png})
The phase reference source J1342-2900 at 13:42:15.345598
--29:00:41.83163, uncertainty (0.3, 0.8) milli-arcsec \citep{Petrov11}, under
2\degr\/ separation, was used at all epochs.  W Hya has a high proper
motion, (149, --60) milli-arcsec and the phase-referenced maser positions were
offset in the expected direction from the pointing position (by 2--3
arcsec) but the positions at each of the 4 epochs differed by several
tenths of an arcsec or more from the positions predicted from
\emph{Hipparcos} data, in a non-systematic fashion.  Phase transfer at
such low elevation was not accurate enough to give astrometrically
useful results.

In conclusion, the phase referencing results for U Ori and IK Tau
vindicate the methods used to align the epochs. However, the
uncertainties in our measurements and the \emph{Hipparcos} data are too great
for sufficiently accurate comparison to confirm that the star is
located at the centre of expansion.  \citet{Vlemmings02} suggest that
the brightest feature in U Her is in the line of sight to the star
itself.  Our model places it at a separation of about 40--60 milli-arcsec from the centre of expansion.
Unfortunately, the uncertainties are just too great to distinguish
between these models astrometrically.  Such tests will have to wait
for imaging of the radio star and masers simultaneously with
\emph{e}-MERLIN, as was done for W Hya by \citet{Reid90} using the
VLA.

%Bordeaux
%03 53 28.8715 +11 24 22.198
%gives 2000 3 53 28.8810 11 24 22.042 (23,23)
%2001 3 53 28.8825 11 24 22.018 (27,27)

% NOMAD
%2000 3 53 28.8529 11 24 21.419 
%2001 3 53 28.8534 11 24 21.344

\subsection{Maser shells}
\label{sec:shell}

\begin{figure}
   \centering
   \includegraphics[angle=0, width=9cm]{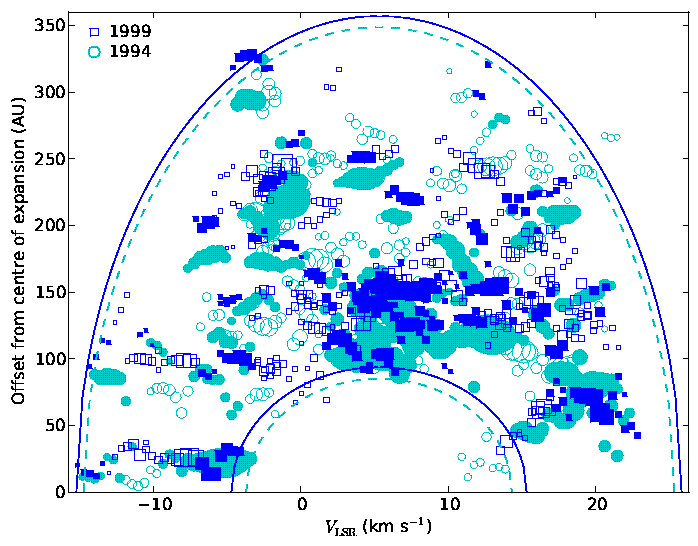}
      \caption{VX Sgr projected distance of masers from the centre of expansion as a
      function of $V_{\mathrm{LSR}}$, centred at $V_{\star}$. Solid
      symbols represent components belonging to features seen at more
      than one epoch. The
      ellipses, fitted by eye, represent the inner and outer limits of
      22-GHz maser emission.}
         \label{VXSgr_ravel.png}
   \end{figure}

\begin{figure}
   \centering
   \includegraphics[angle=0, width=9cm]{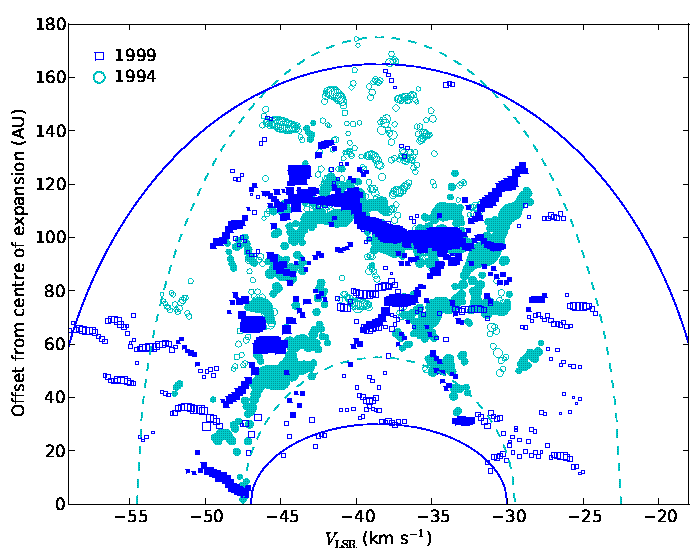}
      \caption{S Per projected distance  of masers from the centre of expansion as a
      function of $V_{\mathrm{LSR}}$, see Fig.~\ref{VXSgr_ravel.png}
      for details. Note that the bandwidth was probably less
      than the full extent of emission in 1994.}
         \label{SPer_ravel.png}
   \end{figure}

\begin{figure}
   \centering
   \includegraphics[angle=0, width=9cm]{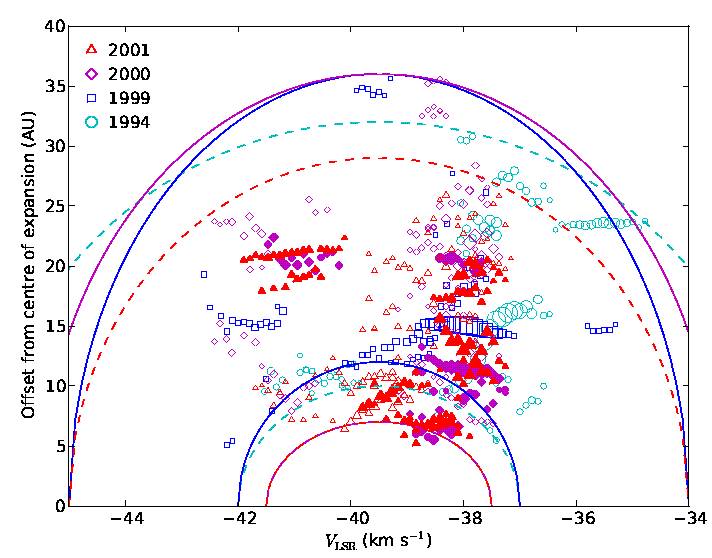}
      \caption{U Ori projected distance  of masers from the centre of expansion as a
      function of $V_{\mathrm{LSR}}$, see Fig.~\ref{VXSgr_ravel.png}
      for details. The inner limits were the same in 2000 and 2001.}
         \label{UOri_ravel.png}
   \end{figure}

\begin{figure}
   \centering
   \includegraphics[angle=0, width=9cm]{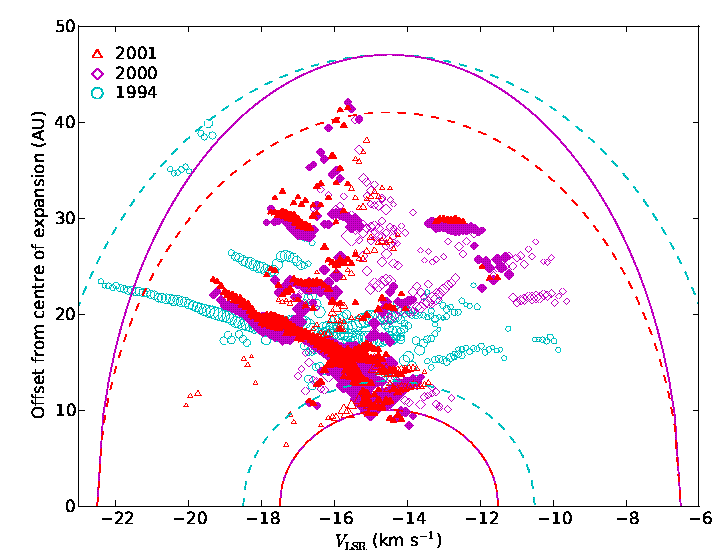}
      \caption{U Her projected distance  of masers from the centre of expansion as a
      function of $V_{\mathrm{LSR}}$, see Fig.~\ref{VXSgr_ravel.png}
      for details. The inner limits were the same in 2000 and 2001.}
         \label{UHer_ravel.png}
   \end{figure}

\begin{figure}
   \centering
   \includegraphics[angle=0, width=9cm]{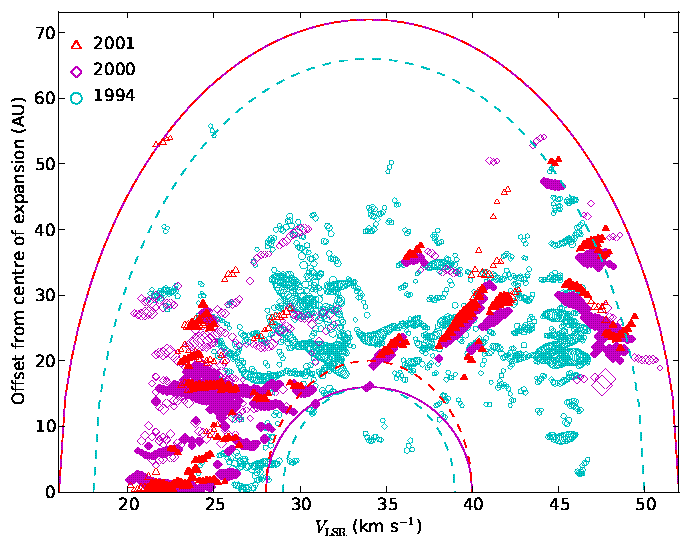}
      \caption{IK Tau projected distance of masers from the centre of
        expansion as a function of $V_{\mathrm{LSR}}$, see
        Fig.~\ref{VXSgr_ravel.png} for details. Note that the
        bandwidth was probably less than the full extent of emission
        in 1994. The outer limits were the same in 2000 and 2001.}
         \label{_ravel.png}
   \end{figure}

\begin{figure}
   \centering
   \includegraphics[angle=0, width=8.5cm]{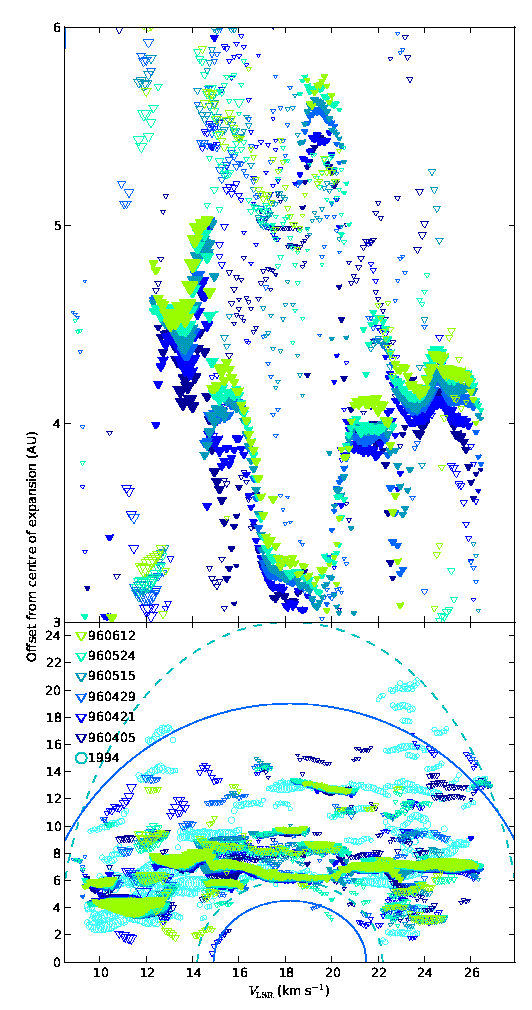}
      \caption{RT Vir projected distance of masers from the centre of
      expansion as a function of $V_{\mathrm{LSR}}$, see
      Fig.~\ref{VXSgr_ravel.png} for details. Features seen at all 6
      1996 epochs are shown by solid symbols. The upper panel shows an enlargement
      of the densest part of the shell. The same shell limits were
      used for all 1996 epochs.}
         \label{RTVir_ravel.png}
   \end{figure}

\begin{figure}
   \centering
   \includegraphics[angle=0, width=9cm]{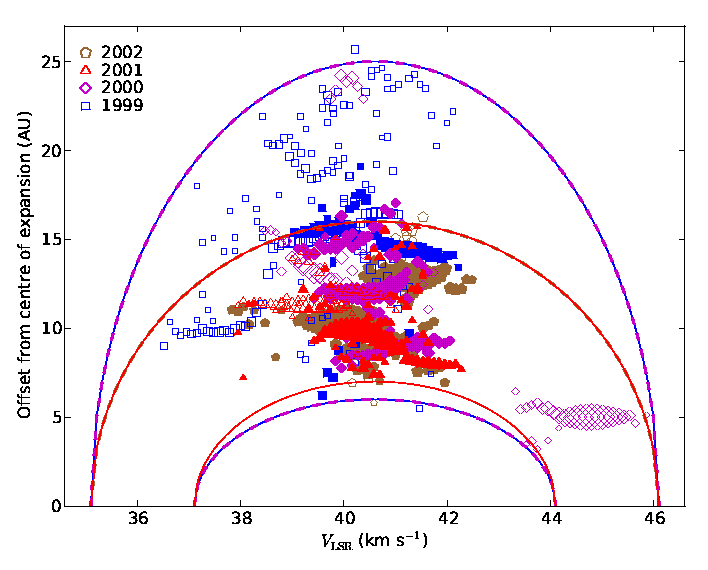}
      \caption{W Hya projected distance  of masers from the centre of expansion as a
      function of $V_{\mathrm{LSR}}$, see Fig.~\ref{VXSgr_ravel.png}
      for details.  The inner limits were the same in 1999, 2000 and
      2002.  The outer limits were the same in 1999 and 2000 and also in
      2001 and 2002. }
         \label{WHya_ravel.png}
   \end{figure}

Figs.~\ref{VXSgr_ravel.png} to~\ref{WHya_ravel.png} show the maser
separations from the centre of expansion projected against the plane
of the sky, $a=\sqrt{x^2 + y^2}$, as a function of $V_{\mathrm {LSR}}$,
centred at $V_{\star}$. We fitted ellipses by eye to the inner and outer
limits of the 22-GHz maser emission, giving the inner and outer
radii and expansion velocities of spherical shells, $r_{\mathrm i}$,
$r_{\mathrm o}$, $\v_{\mathrm i}$ and $\v_{\mathrm o}$. The velocity
gradient is given by
\begin{equation}
K_{\mathrm{grad}} = \frac{\v_{\mathrm o}-\v_{\mathrm i}}{r_{\mathrm
o}-r_{\mathrm i}}
\label{eq:grad}
\end{equation}
and the logarithmic velocity gradient by
\begin{equation}
\epsilon =\frac{\log(\v_{\mathrm o}/\v_{\mathrm i})}{\log (r_{\mathrm o}/r_{\mathrm
i})}
\label{eq:epsilon}
\end{equation}
These parameters are given in Table~\ref{tab:feats} and used to
investigate the mass loss as a whole in Section~\ref{sec:masslossshell}.



\section{Feature persistance and proper motions}
\label{sec:pm}

We matched features at successive epochs as described in
Section~\ref{sec:centre}.  The numbers matched with the previous epoch
are given in Table~\ref{tab:feats}.  We required the mean feature
$V_{\mathrm{LSR}}$ velocities to coincide to within a thermal line
width $\Delta{V}_{\mathrm{th}} \approx$1.4 km s$^{-1}$ for H$_{2}$O in
the maser shell.  We allowed potential feature matches between epochs
if the corresponding proper motions did not exceed the maximum expansion velocity
$|V_{\mathrm{LSR}}-V_{\star}|$, plus an allowance for position
inaccuracy and feature size. For reference, an expansion velocity of 5
km s$^{-1}$ corresponds to a proper motion of $\approx1$ AU yr$^{-1}$
(4--10 milli-arcsec yr$^{-1}$ for the AGB stars at 266--99 pc, and 2--3 milli-arcsec in
5 yr for the RSG at 2.3--1.7 pc, using the stellar distances given in
Table~\ref{tab:stars}).  We tested the the angular and velocity search
regions by decreasing or increasing them by 50\%.  We also tested
whether matches were random coincidences by offsetting one epoch by
the match radius in $x$ and $y$ before looking for matches.  Since one
of the goals of this project was to investigate the CSE kinematics, we
did not want to impose conditions on the directions of proper motions.

We found 5 or more matched pairs of features between the RSG epochs 5
years apart and between most AGB epochs separated by 1 year or less
(up to 3/4 of all features in successive epochs).  The number of
closest matches are listed in Table~\ref{tab:feats}. Changing the
match radius led to fewer matches or more multiple matches but few if
any new unique matches. Applying shifts reduced the number of matches
to 25\% or fewer, suggesting that the majority of the listed
matches are correct.

We only analyse reliable matches, using the above criteria and
requiring the majority of matches to be unique.  In some cases we
found possible matches over longer AGB epoch separations but, as the
potential proper motion and hence search radius increased, there were
more multiple matches and as many matches were found after applying a
shift as before, so these cannot be used.

More details are given for each source:

\paragraph{\bf VX Sgr} We used a maximum allowed separation of 23 milli-arcsec
 in 5 yr, and about half of all features were matched.

\paragraph{\bf S Per }  We used a maximum allowed separation of 13 milli-arcsec
 in 5 yr, and about half of all features were matched.

\paragraph{\bf U Ori} We used a maximum allowed separation of 12 milli-arcsec
 in 1 yr, or 30 milli-arcsec in 5 yr. We only found reliable matches between 2000--2001, for about 1/3 of the features.

\paragraph{\bf U Her} We used a maximum allowed separation of 12 milli-arcsec
 in 1 yr, or 45 milli-arcsec in 6 yr. About 1/2 of the features were matched
 from 2000--2001. Some possible matches were seen between 1994--2000
 but none were unique and a similar number of matches were seen after
 applying the test shifts. 
 After comparing the evolution of the H$_{2}$O
 maser shells as a whole in Section~\ref{sec:time}, we found that the
 ring of emission seen in U Her in 1994 could have expanded to
 the location of the outer ring seen at later epochs, suggesting the
 persistence of a large-scale structure.

\paragraph{\bf IK Tau}  We used a maximum allowed separation of 14 milli-arcsec
 in 1 yr, or 50 milli-arcsec in 6 yr.  About 1/2 of the features were matched
 from 2000--2001. Some possible matches were seen between 1994--2000
 but none were unique and a similar number of matches were seen after
 applying the test shifts. No obvious patterns appear to survive.

\paragraph{\bf RT Vir } We used a maximum allowed separation of 9 milli-arcsec
over 10 weeks in 1996, and 25 milli-arcsec for 1994--1996. Between 1/2 and 3/4
of features were traced from epoch to epoch in 1996, with eleven
(about 1/4) being identified at all six epochs.  No matches were found
between 1994--1996.

\paragraph{\bf W Hya} We used a maximum allowed separation of 20 milli-arcsec
 in 1 yr; 1/3 to 1/2 of the features could be matched from one year to
 the next, including six at three successive epochs. No features were
 matched for all four epochs.

\subsection{Identification of features}
\label{sec:ident}
We are confident that in most of the matched features, we
are seeing masers from the same cloud, for two main reasons.  Firstly,
the 22-GHz H$_{2}$O masers  clouds are much
denser than the average wind density (R99, B03, M03,
Section~\ref{sec:density}).
  Secondly. the more extended features (better-resolved in the
higher-declination sources) often show a comparable internal
structure. This can be seen for some, although not all of the matched
features in Figs.~\ref{SPER_Match_xyplot.png}, \ref{UHer_00-01_xyplot.png}
\ref{IKTau_00-01_xyplot.png} and~\ref{RTVir_96_xyplot.png}.  We zoom
in on two examples from S Per and one from RT Vir.

\begin{figure}
   \hspace*{-0.8cm}
   \includegraphics[angle=0, width=10cm]{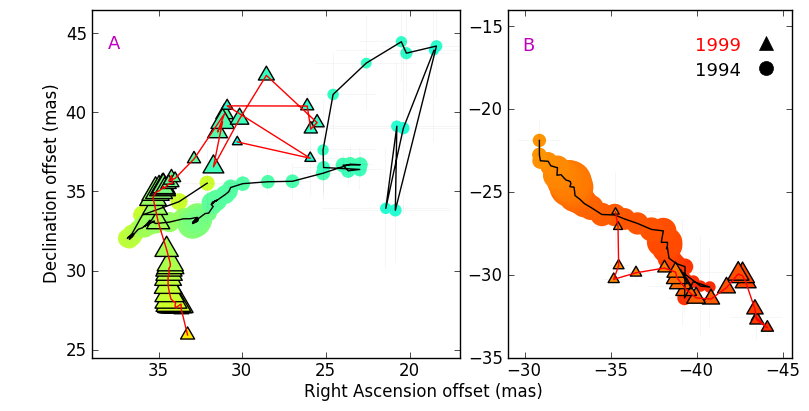}
      \caption{Two of the S Per features matched in 1994 (circles) and
        1999 (triangles), marked on
        Fig.~\ref{SPER_Match_xyplot.png}. The velocity ranges are
        --45.7 to --35.9 km s$^{-1}$ and --32.8 to --28.5 km s$^{-1}$
        respectively, see Fig.~\ref{SPER_Match_xyplot.png} for colour
        scale. The left-hand feature has rotated counterclockwise, in bulk,
        and moved slightly S between 1994 and 1999.  The right-hand
        feature has moved to the SW whilst retaining the elongated
        distribution of the components.}
         \label{sper_abfeat.png}
   \end{figure}

Fig.~\ref{SPER_Match_xyplot.png}  shows a series of features labelled
A, around (30,35) mas, peaking
at --46 -- --47 km s$^{-1}$, which appears to have been rotated
counterclockwise by about 60\degr\/ between epochs (as well as bulk
motion to the S). Another pair of features labelled B, at
around (--37, --25) mas, --30 km s$^{-1}$ are made up of a radial
spoke of components. seen more clearly in the enlargement in
Fig.~\ref{sper_abfeat.png}. 

What effects produce these variations within features?  Bulk rotation
could be responsible for the changes shown in
Fig.~\ref{sper_abfeat.png} (left).  This consists of a series of 4
matched features (separate spectral peaks) whose components form a
spatially and spectrally continuous series, taken to be a single,
intrinsically asymmetric (am$\oe$boid) cloud, with internal density
variations. After removing the mean proper motion of the cloud (3
mas), it is rotating anticlockwise at $\la6\pm4$ km
s$^{-1}$. The feature in Fig.~\ref{sper_abfeat.png} (right) shows an
almost radial gradient of velocity with position, with two slight
wiggles seen at both epochs (with greater position uncertainty in 1999
as it has become fainter). The proper motion corresponds to an
expansion velocity $\approx10$ km s$^{-1}$.

\begin{figure}
   \centering
   \includegraphics[angle=0, width=7cm]{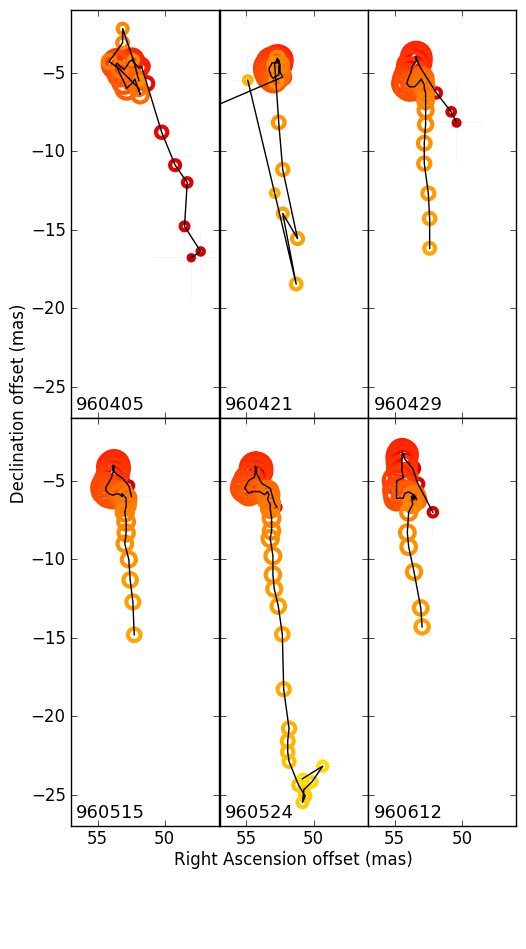}
      \caption{An RT Vir maser feature seen at 6 epochs in 1996,
        marked by the box in Fig.~\ref{RTVir_96_xyplot.png}. The
        velocity range in 21.3 -- 26.4 km s$^{-1}$, see
        Fig.~\ref{RTVir_94_xyplot.png} for colour scale. The
        tadpole-like morphology and orientation is consistent but the
        velocities at which the tail is seen change. }
         \label{rtvir_fs.png}
   \end{figure}

One of the matched RT Vir features is shown in detail in
  Fig.~\ref{rtvir_fs.png}. The bright head of the `tadpole' has a
  similar appearance at all epochs, but it has two faint tails; the
  more redshifted one is seen in 960405, 960429 and (a short piece)
  960612, and the less redshifted (orange-yellow) tail in 960421,
  960429, 960515, 969524 and 960612.  Turbulence inside the feature
  could affect the beaming intensity and/or direction, for example
  causing the higher-velocity (redder) tail to fade and to be replaced
  by a lower-velocity tail. Note that the most sensitive observations
were made in 960429 and 960524 (Table~\ref{tab:feats}), hence the
remnant of the red tail and the longest extension of the new tail seen
at these epochs, respectively. The tails have extents $\ga1$ AU and a
separation of about 0.5 AU.  If the change is due to a single
disturbance, this must be broad enough to affect both regions almost
simultaneously, since a localised, non-dissociative shock could not
traverse the distance in the short time intervals of a few weeks
between observations. Alternatively, the entire cloud has rotated
about an axis close to the plane of the sky such that beaming from
different angles was favoured.

If the clouds are approximately spherical and the masers are
amplification-bounded, the brightest emission at the line centre is
tightly beamed along the line of greatest amplification. Small changes
in the cloud will have little effect on the apparent location of this
emission although they may affect the appearance of weaker, less
strongly beamed emission from the line wings.  On the other hand,
flattened or shocked clouds may possess matter-bounded masers which are
less tightly beamed.  R11 found that U Ori, U Her and to a lesser
extent IK Tau possess some matter-bounded masers, whilst the other
sources studied are mostly amplification-bounded.

Overlapping clouds along the line of sight can produce strong
amplification of a single feature, see Section~\ref{sec:WH-var}.
The survival  of maser features is discussed in
Section~\ref{sec:survive}.
\clearpage

\subsection{Expansion proper motions}
\label{sec:expand}

We aligned the results from successive epochs for each source and
matched features as outlined in Section~\ref{sec:centre}.
Table~\ref{tab:pm} gives the error-weighted proper motions; for
multi-epoch matches, we performed least squares fits to position as a
function of time.
%The average total proper motions correspond to velocities of 19 and 10
%km s$^{-1}$ for VX Sgr and S Per, and up to 12 km s$^{-1}$ for the AGB
%stars (as constrained by the initial matching criteria).
The mean total proper motion (and formal uncertainty) is
$\langle\dot{xy}\rangle$.  We independently estimated mean proper
motions in the radial and tangential directions,
$\langle\dot{a}\rangle$ and $\langle\dot{\theta}\rangle$. 

In a few cases the direction of $\langle\dot{xy}\rangle$ has a large
scatter and $\langle\dot{\theta}\rangle$ and $\langle\dot{a}\rangle$
are consequently small and of low significance. This is most
noticeable at the earlier, noisiest epochs of W Hya, probably due to
measurement errors. Since its values of $\langle\dot{\theta}\rangle$
change sign from epoch to epoch, they are probably not significant.
The U Ori proper motions also show more random behaviour.  Coupled
with the failure to match features in the one-year interval between
1999--2000, this suggests that the U Ori CSE is particularly subject
to turbulence or other disruption.  Many of its H$_2$O masers appear
matter-bounded, hence less tightly beamed, so that the apparent
centroid may be shifted easily by small disturbances (see
Section~\ref{sec:ident} and R11).

\begin{figure}
   \centering
   \includegraphics[angle=0, width=9cm]{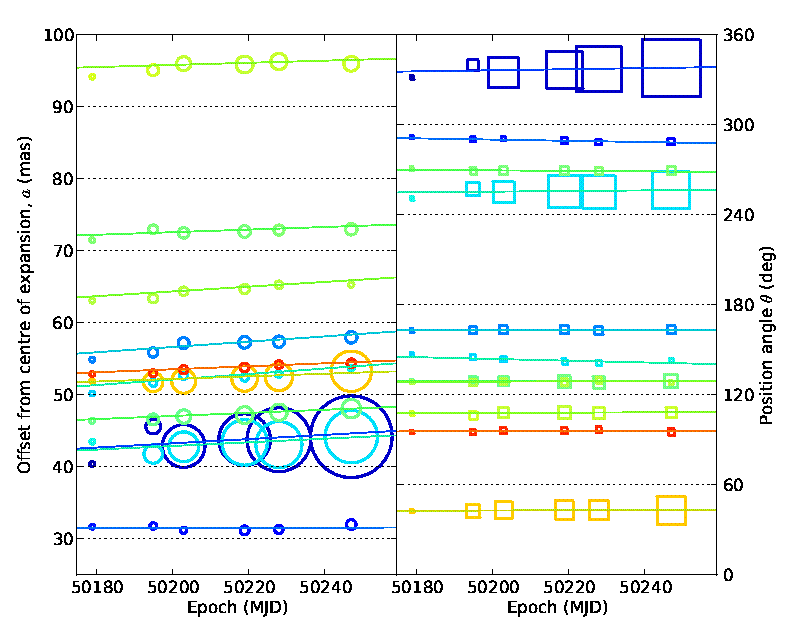}
      \caption{Comparison between the change in position of the 11 RT Vir
      features seen at all epochs, in the radial direction (left) and
      the  tangential direction (right)(see
      Fig.~\ref{RTVir_94_xyplot.png} for $V_{\mathrm {LSR}}$ colour
      scale). Symbol size is proportional to the flux density relative
      to that of the first epoch.}
         \label{RTVir_dadth.png}
   \end{figure}

In all other cases the radial motions $\langle\dot{a}\rangle$ are in
expansion, and exceed $|\langle\dot{\theta}\rangle|$. This is best
illustrated by Fig.~\ref{RTVir_dadth.png} which shows
$\langle\dot{\theta}\rangle \ll \langle\dot{a}\rangle$ for the 11 RT
Vir features matched at all 6 epochs.  
In fact,
$\langle\dot{\theta}\rangle$ is negligible for most sources. IK
Tau maser proper motions are clearly dominated by expansion but a
small rotational component cannot be ruled out, although this is
$<\Delta V_{\mathrm {th}}$.  This is also the case for U Ori although
the majority of the proper motion measurements are in scattered
directions, as mentioned above.
%($\le1\sigma_{\mathrm {rms}}$) in two-epoch measurements
%for VX Sgr, S Per, U Her and W Hya in
%1999--2000. $\langle\dot{\theta}\rangle$ is also negligible for RT Vir
%when all 55 matches (between 2--6 epochs) are considered.  $\langle\dot{\theta}\rangle <
%2\sigma_{\mathrm {rms}}$ for W Hya 2000--2001, $<3\sigma_{\mathrm
%  {rms}}$ for W Hya 2001--2002 and negligible for the two sets of
%3-epoch W Hya data.

\begin{table}
{\tiny
\begin{tabular}{lrrrrrr}
\hline
Source&Epochs&\multicolumn{1}{c}{$\langle\dot{xy}\rangle$}&\multicolumn{1}{c}{$\langle\dot{a}\rangle$}&\multicolumn{1}{c}{$\langle\dot{\theta}\rangle$}&\multicolumn{1}{c}{$\langle{V_{\mathrm a}}\rangle$}\\
      &     &(mas yr$^{-1}$)&(mas yr$^{-1}$)&(mas yr$^{-1}$)&(km s$^{-1}$)   \\
\hline
VX Sgr&94--99 &$2.3\pm0.1$  &$1.4\pm0.1$  &$-0.1\pm0.1$  &$12.0\pm0.2$   \\
S Per &94--99 &$0.9\pm0.1$&$0.5\pm0.1$&$0.0\pm0.1$&$4.9\pm0.1$\\
U Ori &00--01 &$5.0\pm0.1$ &$0.7\pm0.1$&$-0.5\pm0.1$&$0.9\pm0.1$\\
U Her &00--01 &$2.1\pm0.1$ &$1.4\pm0.1$&$-0.1\pm0.1$&$1.7\pm0.1$\\
IK Tau&00--01 &$5.2\pm0.1$ &$4.6\pm0.1$&$-0.5\pm0.1$&$6.0\pm0.1$\\
RT Vir&96(11) &$16.0\pm0.3$ &$11.1\pm0.3$& $0.7\pm0.4$&$6.3\pm0.1$\\
RT Vir&96(55) &$24.3\pm0.2$ &$16.0\pm0.2$&$-0.2\pm0.2$&$8.0\pm0.1$\\
W Hya &99--00 &$10.6\pm1.0$&$1.3\pm1.0$&$ 0.1\pm1.0$&$0.6\pm0.4$\\
W Hya &00--01 &$11.1\pm0.8$&$1.9\pm1.0$&$-1.4\pm1.0$&$0.8\pm0.6$\\
W Hya &01--02 &$ 6.6\pm0.5$&$6.3\pm0.6$&$ 1.7\pm0.6$&$3.2\pm0.3$\\
W Hya&99--00--01&$ 6.3\pm2.1$&$6.3\pm2.0$&$-0.3\pm0.7$&$3.0\pm1.0$\\
W Hya&00--01--02&$ 6.2\pm0.3$&$6.2\pm0.3$&$ 0.1\pm0.1$&$2.9\pm0.2$\\
\hline
\end{tabular}
}
\caption{Proper motions between the epochs indicated as the last two
  digits of the year. The error-weighted means of total proper motion
  ($\langle\dot{xy}\rangle$), its radial and tangential components ($\langle\dot{a}\rangle$ and $\langle\dot{\theta}\rangle$)
  and the radial velocity ($\langle{V_a}\rangle$) are given, with formal uncertainties.
All measurements are relative to the aligned centres
  of expansion, i.e. ignoring the systemic, bulk motion. 
The number of matches between two epochs, $N_{\mathrm{prev}}$, was
given in Table~\ref{tab:feats}.
For W Hya, 2 features were matched at three epochs
from 1999--2000--2001 and 4 from three epochs 2000--2001--2002.
RT Vir was observed 6 times in ten weeks; the first line gives the
 means for the 11 features seen at all epochs and the second line
  gives means for all 55 matches at any epochs. See
  Section~\ref{sec:expand} for more details.}
\label{tab:pm}
\end{table}

\begin{figure}
   \centering
   \includegraphics[angle=0, width=9cm]{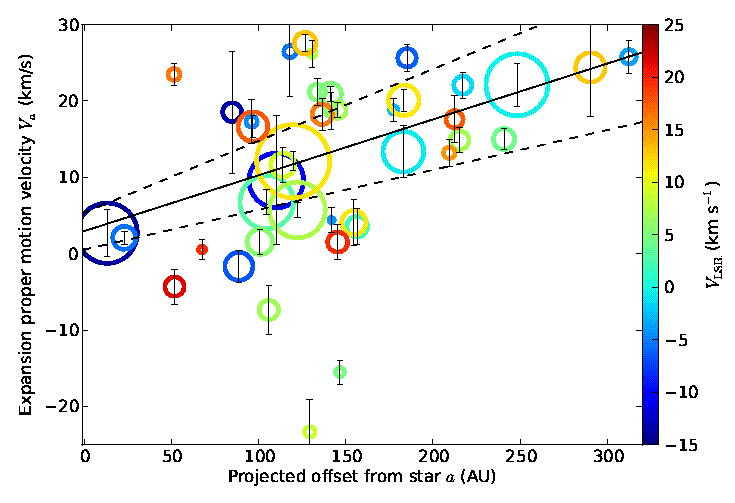}
      \caption{The relationship between $V_{\mathrm
  a}$ and $\overline{a}$ for VX Sgr. Symbol size is proportional to
  feature size. The linear error-weighted least-squares fit and
  dispersion are plotted by the solid and dashed lines.}
         \label{VXSgr_Va-abar.png}
   \end{figure}

\begin{figure}
   \centering
   \includegraphics[angle=0, width=9cm]{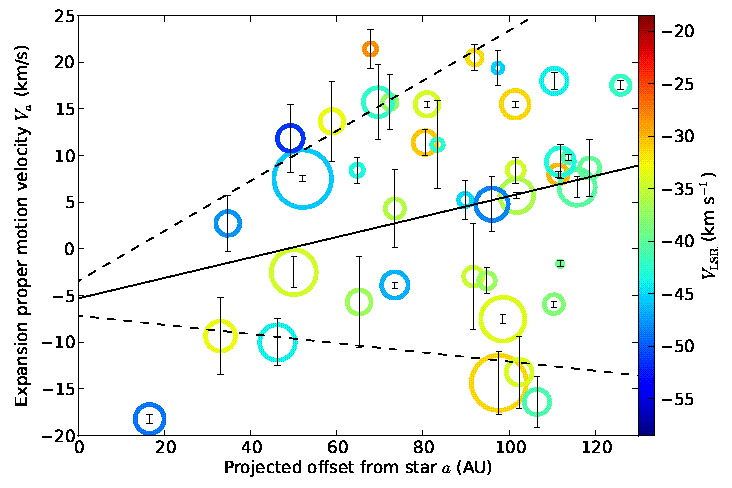}
      \caption{The relationship between $dV_{\mathrm
  a}$ and $\overline{a}$ for S Per, see Fig.~\ref{VXSgr_Va-abar.png}
  for details.}
         \label{SPer_Va-abar.png}
   \end{figure}

\begin{figure}
   \centering
   \includegraphics[angle=0, width=9cm]{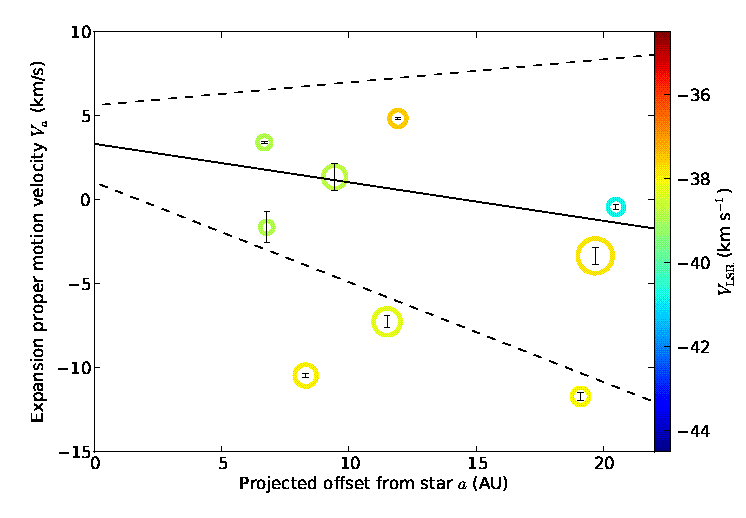}
      \caption{The relationship between $dV_{\mathrm
  a}$ and $\overline{a}$ for U Ori, see Fig.~\ref{VXSgr_Va-abar.png}
  for details.}
         \label{UOri_Va-abar.png}
   \end{figure}

\begin{figure}
   \centering
   \includegraphics[angle=0, width=9cm]{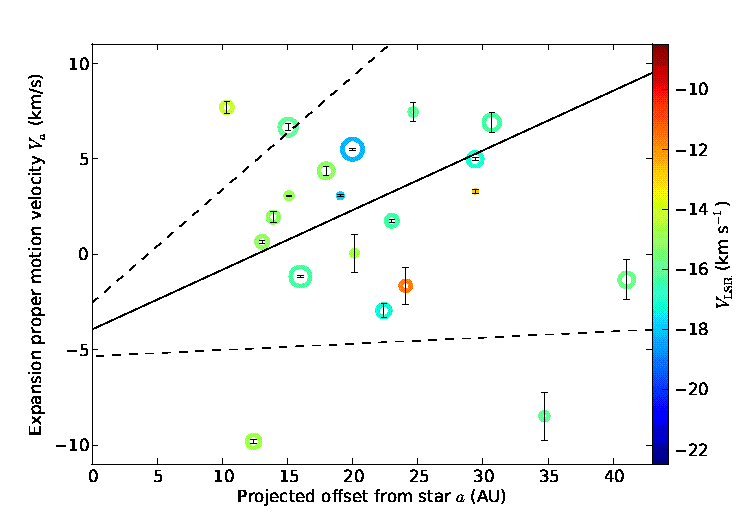}
      \caption{The relationship between $dV_{\mathrm
  a}$ and $\overline{a}$ for U Her, see Fig.~\ref{VXSgr_Va-abar.png}
  for details.}
         \label{UHer_Va-abar.png}
   \end{figure}

\begin{figure}
   \centering
   \includegraphics[angle=0, width=9cm]{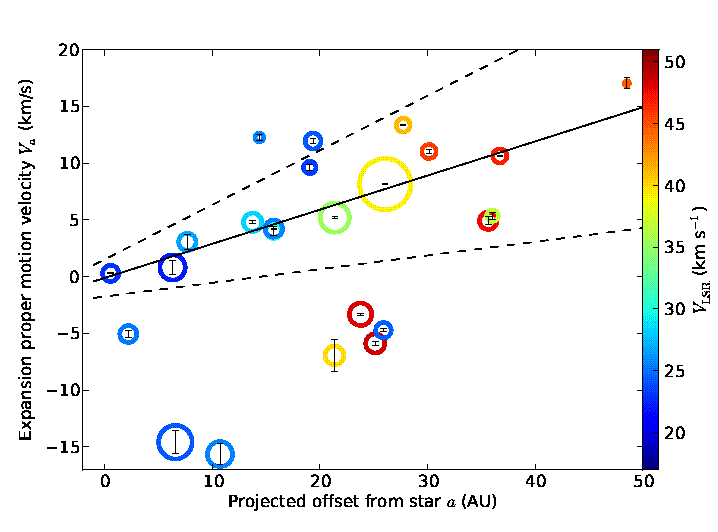}
      \caption{The relationship between $dV_{\mathrm
  a}$ and $\overline{a}$ for IK Tau, see Fig.~\ref{VXSgr_Va-abar.png}
  for details.}
         \label{IKTau_Va-abar.png}
   \end{figure}

\begin{figure}
   \centering
   \includegraphics[angle=0, width=9cm]{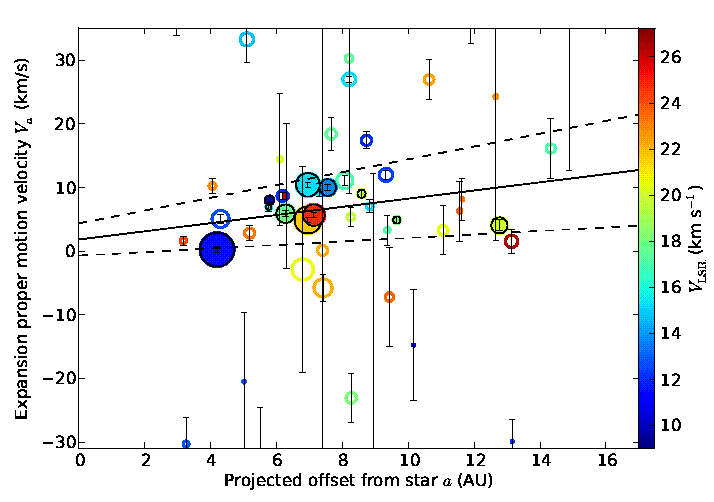}
      \caption{The relationship between $dV_{\mathrm a}$ and
        $\overline{a}$ for RT Vir. Symbol size is proportional to
        feature size. Proper motions derived from features matched at
        all 6 epochs are shown by solid symbols, other
          measurements are represented by hollow symbols (some, with
          very large uncertainties, are not fully shown).  The linear
          error-weighted least-squares fit and dispersion for the
          six-epoch data are plotted by the solid and dashed lines.
      }
         \label{RT Vir_Va-abar.png}
   \end{figure}

\begin{figure}
   \centering
   \includegraphics[angle=0, width=9cm]{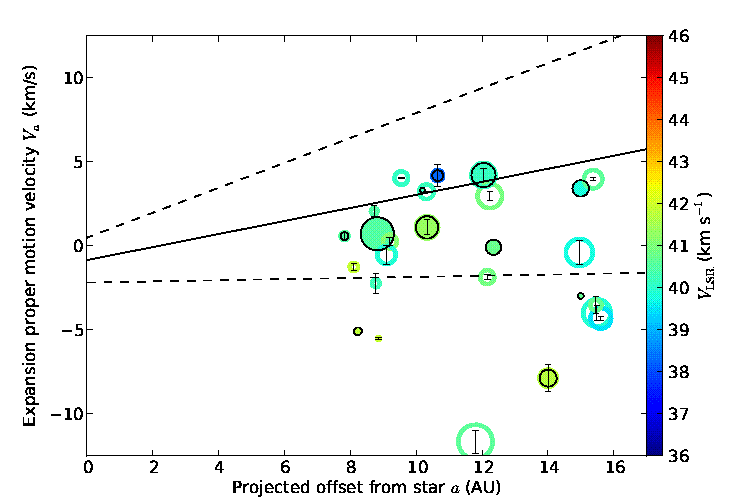}
      \caption{The relationship between $dV_{\mathrm
  a}$ and $\overline{a}$ for W Hya. Symbol size is proportional to
  feature size. Proper motions derived from
  features matched in 2001--2002 or at 3
  epochs are shown by solid symbols. The other two-epoch proper
  motions, with high uncertainties as discussed in the text, are shown
  by hollow symbols. The linear error-weighted least-squares fit and
  dispersion for  the 3-epoch and 2001--2002 data are plotted by the solid and dashed
  lines. }
         \label{WHya_Va-abar.png}
   \end{figure}

Figs.~\ref{VXSgr_Va-abar.png} to~\ref{WHya_Va-abar.png} show expansion
proper motion velocity $V_{\mathrm a}$ as a function of mean angular
separation from the centre of expansion, $\overline{a}$, for the data
in Table~\ref{tab:pm}. We performed a linear error-weighted fit to
find the slope ($dV_{\mathrm a}/d\overline{a}$) and found the
dispersion in the relationship (the formal errors are very
small). This gives an estimate of wind acceleration in the plane of
the sky.
The RSG, VX Sgr and S Per have $dV_{\mathrm a}/d\overline{a}$ of
$0.07\pm0.02$ and $0.11\pm0.16$ km s${-1}$ AU$^{-1}$, respectively.
The Miras U Her and IK Tau have $dV_{\mathrm a}/d\overline{a}$ of
$0.31\pm0.28$ and $0.31\pm0.18$ km s$^{-1}$ AU$^{-1}$,
respectively. The SRb RT Vir has $dV_{\mathrm a}/d\overline{a}
=0.64\pm0.36$ km s$^{-1}$ AU$^{-1}$ for the 11 features matched at all
epochs; if 2--5-epoch matches are considered, the scatter is greater
but the slope is also positive.  W Hya has $dV_{\mathrm
  a}/d\overline{a} =0.39\pm0.35$ km s$^{-1}$ AU$^{-1}$ for the 3-epoch
matches and additional 2-epoch matches in 2001--2002.  The only
exceptions to the positive gradients are U Ori, $dV_{\mathrm
  a}/d\overline{a}=-0.2\pm0.4$ km s$^{-1}$ AU$^{-1}$, and W Hya
1999-2000, $-0.1\pm0.1$ km s$^{-1}$ AU$^{-1}$, probably due to a more turbulent maser shell and to greater uncertainties, respectively, as discussed above.

%There is a stronger, positive correlation between $V_{\mathrm a}$ and
%  $\overline{a}$ if $\langle{\dot{a}}\rangle > 0.5
%  \langle{\dot{xy}}\rangle$. Conversely, U Ori and the earlier W Hya
%  epochs have less ordered proper motions and/or higher measurement
%  errors.  

 For VX Sgr, S Per, IK Tau, U Her, RT Vir and the better W Hya data,
 $dV_{\mathrm a}/d\overline{a}$ is positive.  It is close to
 $K_{\mathrm{grad}}$ (Table~\ref{tab:feats}), within the dispersions,
 for all measurements apart from VX Sgr, where $dV_{\mathrm
   a}/d\overline{a}$ is slightly greater; this may be due to the
 axisymmetry of the outflow, modelled by M03. In the other objects
 the velocity field appears to be approximately spherically symmetric.
%VX Sgr2.94264893475 2.41313695221 0.0731828819496 0.0209562156978
%S Per $V_{\mathrm a} = -5.3\pm1.9 +  0.11\pm0.16 \overline{a}$
%U Her $V_{\mathrm a} = -3.9\pm1.40 + 0.31\pm0.28 \overline{a}$
%U Ori $V_{\mathrm a} = 3.3\pm2.3 + -0.23\pm0.37 \overline{a}$
% W Hya Use 3-epoch matches and 2-epoch matches for components not
%  matched at the third epoch.
%3.93073899054 0.142188981125 -0.0842440778574 0.0138272542294
%Use 3-epoch matches plus 01-02
%$V_{\mathrm a} = -0.87\pm1.33 + 0.39\pm0.35  \overline{a}$
%RT Vir all matches
%-0.712681949793 3.95291842563 1.24977174717 0.135687972523
%11 1.83114151488 2.53481280134 0.641934336743 0.364310998086

These results are consistent with previous proper motion
measurements. \citet{Yates94} identified IK Tau features (at lower
resolution) surviving for 16 months. \citet{Marvel98} observed several
objects including VX Sgr, S Per, U Her and IK Tau for three epochs
during 1995.  They deduced that the
masers are distributed in multiple thin shells or random clumps. They
found that an ellipsoidal model fitted most data better
than a spherically symmetric shell, but comparison between their
figures and our 1994 plots suggests that fainter, extended emission
was resolved out by the VLBA. The proper motions are predominantly
directed radially outward,  consistent with the $LSR$ maser velocities
at distances similar to those given in
Table~\ref{tab:stars}. \citet{Asaki10} performed astrometric
monitoring of S Per with VERA to
obtain the parallax. The maser proper motions were consistent with
a combination of bulk proper motion and a spherically expanding
flow. \citet{Imai03} monitored RT Vir using the VLBA for 5 epochs in
1998. The tendency for the E and W to be dominated by red- and
blue-shifted emission is seen in their results as well as in our
Figs.~\ref{RTVir_94_xyplot.png} and~\ref{RTVir_96_xyplot.png} although
the brightest masers occur in different locations. They also find
expansion (not rotation) proper motions, with at least one feature
showing strong  acceleration.  Note that they use a distance of 220 pc
and so obtain proper motion velocities almost double those resulting
from the distance we adopt (Table~\ref{tab:stars}).

\clearpage

\section{Comparison with single dish data}
\label{sec:Pushchino}
These targets are among a large number of 22-GHz maser sources which
have been monitored by the Pushchino Radio Telescope for several
decades. Some results have already been published in
\citet{Pashchenko99} (VX Sgr); \citet{Lekht05} (S Per);
\citet{Rudnitskij00} (U Ori); \citet{Mendoza-Torres97} and
\citet{Lekht99} (RT Vir).  Earlier epochs of W Hya monitoring were
published by \citet{Rudnitskij99w}.  
\begin{figure}
   \centering
   \includegraphics[angle=0, width=8.5cm]{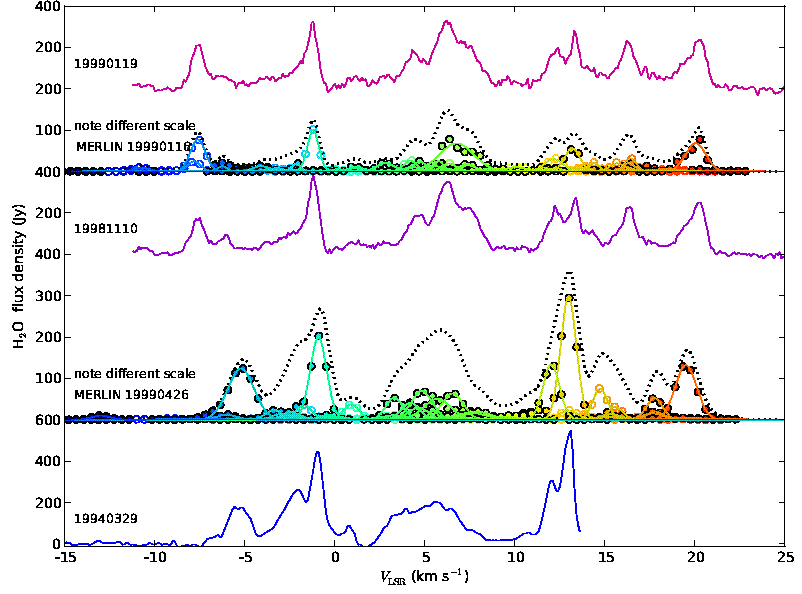}
      \caption{22-GHz H$_{2}$O maser spectra towards VX Sgr at the
        epochs as labelled. Data are from the Pushchino radio
        telescope except where labelled MERLIN.  The coloured MERLIN
        spectra show individual features, the components with black
        outlines were matched at successive epochs.  The dotted black
        lines show the total emission detected per epoch by
        MERLIN. The label of the origin (0) for the vertical scale of
        each spectrum is only shown for the first (lowest) in the
        series.}
         \label{VXSgr_Match94-99_Gauss.png}
   \end{figure}
\begin{figure}
   \centering
   \includegraphics[angle=0, width=8.5cm]{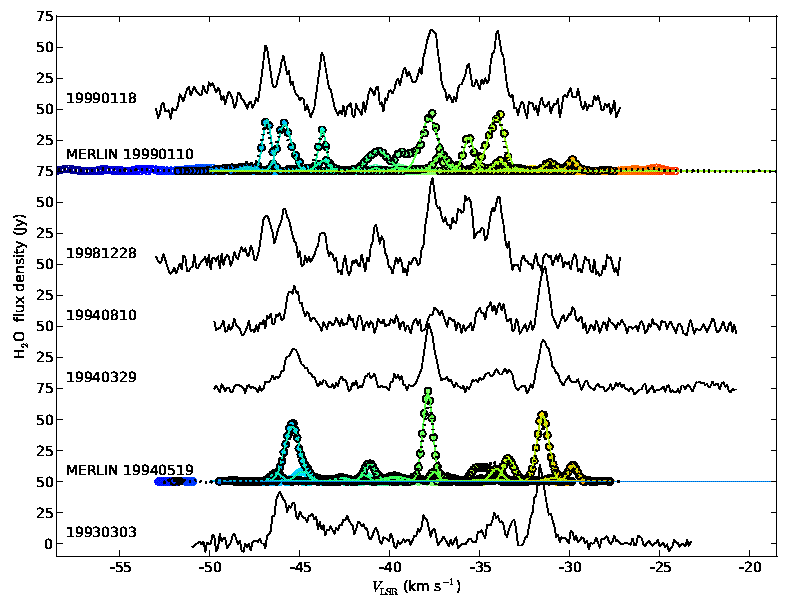}
      \caption{22-GHz H$_{2}$O maser spectra towards S Per at the
      epochs as labelled. See Fig.~\ref{VXSgr_Match94-99_Gauss.png}
      for details. $V_{\mathrm{corr}}$ (1999) = --0.2 km s$^{-1}$.}
         \label{SPer_M94-99_MGauss.png}
   \end{figure}
\begin{figure}
   \centering
   \includegraphics[angle=0, width=8.5cm]{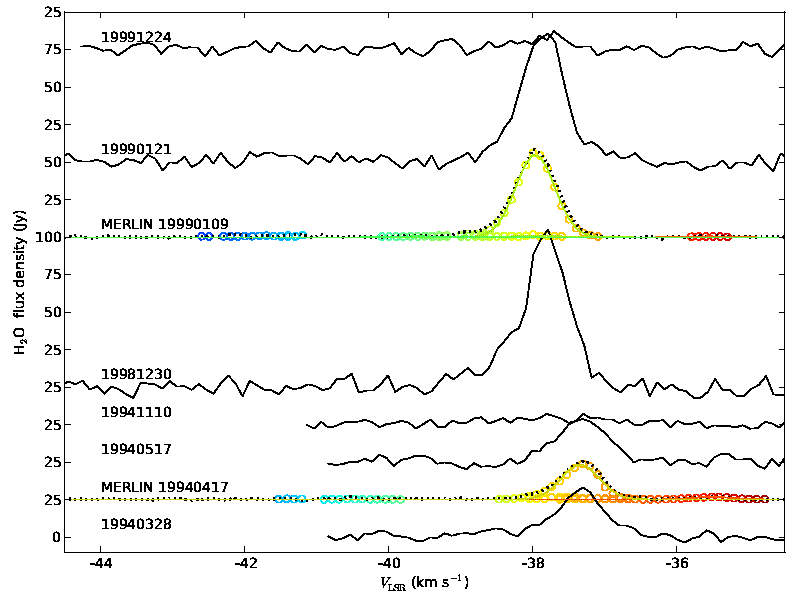}
      \caption{22-GHz H$_{2}$O maser spectra towards U Ori up to
      1999.  See Fig.~\ref{VXSgr_Match94-99_Gauss.png}
      for details. $V_{\mathrm{corr}}$ (1994, 1999) = --0.8, --1.1 km s$^{-1}$.}
         \label{UOri_M94-99_Gauss.png}
   \end{figure}

\begin{table}
\begin{tabular}{lcccc}
\hline
Source & Date  &   $V_{\mathrm{LSR}}$& Peak&  $I_{\mathrm{tot}}$\\
Telescope        & (YYMMDD) &(km s$^{-1}$)   &   (Jy) & (Jy km s$^{-1}$)\\ 
\hline
{\bf VX Sgr} \\
Pushchino   &  940329  &    13.1  &      548&       2696	  \\
MERLIN      &  940426  &    13.0  &      360&       2661$^1$   	  \\
\hline						
Pushchino   &  981110  &     6.3  &       352&  	3208	  \\
MERLIN      &  990116  &     6.3  &       147&          1521	  \\
Pushchino   &  990119  &     6.4  &       321&  	3027	    \\
\hline									
{\bf S Per}     \\							
Pushchino   &   930303 &  --37.83  &      22.9& 	232	     \\
MERLIN      &   940324 &  --37.86  &      63.6& 	222	   \\
Pushchino   &   940329 &  --37.50  &      51.9&         188	     \\
\hline
Pushchino   &   981228 &  --37.72  &      69.3& 	311	     \\
MERLIN      &   990110 &  --37.75  &      45.2&         286	   \\
Pushchino   &   930303 &  --37.74  &      63.5& 	371	     \\
\hline									
{\bf U Ori}     \\							
Pushchino   &   940328 &   --37.3  &      33.0&      30	    \\
MERLIN      &   940417 &   --37.3  &      25.4&      23	   \\
Pushchino   &   940517 &   --37.3  &      28.6&      26	     \\
\hline								
Pushchino   &   981230 &   --37.8  &     105.0&          84	      \\
MERLIN      &   990109 &   --38.0  &      58.7& 	 46	    \\
Pushchino   &   990121 &   --37.8  &      85.8& 	 74	      \\
\hline									
Pushchino   &   000223 &   --37.8  &      29.3& 	 42	         \\
MERLIN      &   000410 &   --38.0  &      18.7& 	 18	   \\
Pushchino   &   000613 &   --37.9  &      28.1&      16	     \\
\hline								
Pushchino   &   010410 &   --37.8  &      25.3&      31	      \\
MERLIN      &   010506 &   --37.9  &      16.6&      19	      \\
Pushchino   &   010530 &   --37.8  &      39.2&      31	     \\
\hline									
{\bf U Her 1994}    \\							
Pushchino   &   940329 &   --14.9  &      21.7& 	 30   	            \\
MERLIN      &   940413 &   --15.0  &      22.1&          44	   \\
Pushchino   &   940517 &   --14.9  &      32.1& 	 56	     \\
\hline									
Pushchino   &   000425 &   --15.6  &      186.3&	406 	        \\
MERLIN      &   000519 &   --15.5  &      209.1&	254	       \\
Pushchino   &   000529 &   --15.6  &      182.3&    267	       \\
\hline									
Pushchino   &   010411 &   --15.8  &       93.5&    118	        \\
MERLIN      &   010427 &   --15.6  &       52.5&     62	      \\
Pushchino   &   010528 &   --15.7  &       73.2&     96	      \\
\hline								
{\bf IK Tau 1994}    \\						
Pushchino   &   930422 &    31.8  &       34.9& 	89	         \\
MERLIN      &   940415 &    31.4  &       24.3& 	48	       \\
Pushchino   &   941109 &    31.7  &       24.7& 	22$^2$	      \\
\hline			   						
Pushchino   &   000425 &    24.9  &      106.4& 	279 	        \\
MERLIN      &   000520 &    24.4  &      117.5& 	360	   \\
Pushchino   &   000530 &    24.6  &      113.0& 	414	     \\
\hline									
Pushchino   &   010409 &    24.2  &       13.4& 	66	     \\
MERLIN      &   010427 &    24.4  &       22.6& 	107	   \\
Pushchino   &   010529 &    24.4  &       23.8& 	118	     \\
\hline									 
{\bf RT Vir}    \\							   
Pushchino   &   940328 &    12.6  &      489.4& 	943	     \\
MERLIN      &   940416 &    12.6  &      394.0& 	590	   \\
Pushchino   &   940808 &    12.6  &      186.0& 	797	     \\
\hline									   
Pushchino   &   960319 &    11.9  &      791.0&    1605	      \\
MERLIN      &   960405 &    11.8  &      389.0&     680	   \\
MERLIN      &   960421 &    11.5  &      515.5&    1190	   \\
MERLIN      &   960429 &    11.5  &      526.3&    1325	   \\
MERLIN      &   960515 &    11.4  &      727.2&    1475	   \\
MERLIN      &   960524 &    11.4  &      705.5&    1415	   \\
MERLIN      &   960612 &    11.3  &      791.6&    1418	   \\
Pushchino   &   960619 &    11.3  &	 359.9&     648          \\
\hline
\end{tabular}
\caption{The velocity and flux density of the peak of the MERLIN
  integrated spectra and the Pushchino spectra taken closest in
  time. The integrated flux density of the whole spectrum is also
  given. In some cases a shift $V_{\mathrm {corr}}$ was applied to the
  Pushchino data, see Figs.~\ref{VXSgr_Match94-99_Gauss.png}
  to~\ref{RTVir_M96all_Gauss.png}. {Table continued.} \newline $^1$Measured over the same
  velocity range as the Pushchino data.
\newline $^2$The Pushchino data only
  cover part of the spectrum.}
\label{tab:P}
\end{table}
\begin{table}
\begin{tabular}{lcccc}
\hline
Source & Date  &   $V_{\mathrm{LSR}}$& Peak& $I_{\mathrm{tot}}$   \\
Telescope        & (YYMMDD) &(km s$^{-1}$)   &   (Jy) & (Jy km s$^{-1}$)    \\ 
\hline
{\bf W Hya}    \\
Pushchino   &   990119 &    39.9  &           128.6&245   	            \\
MERLIN      &   990209 &    39.8  &            91.0&177	     \\
Pushchino   &   990310 &    39.8  &           174.7&323 	      \\
\hline						    								
Pushchino   &   000331 &    40.4  &            81.4&176	       \\
MERLIN &   000405 &    40.6  &            59.2&169$^1$       \\
Pushchino   &   000425 &    40.4  &           169.8&304	      \\
\hline						    								
Pushchino   &   010411 &    40.4  &           377.3&413	      \\
MERLIN      &   010430 &    40.4  &           163.8&204	     \\
Pushchino   &   010904 &    40.3  &           554.8&471	      \\
\hline						    								
Pushchino   &   020326 &    40.2  &           690.8&917	      \\
MERLIN      &   020405 &    40.2  &           296.9&335	     \\
Pushchino   &   020422 &    40.2  &           202.5&255           \\
\hline
\end{tabular}
\addtocounter{table}{-1}
\caption{ Continued. A shift $V_{\mathrm {corr}}$ was applied to the Pushchino
  data, see
Figs.~\ref{WHya_M99-00_Gauss.png}
to~\ref{WHya_M01-02_Gauss.png}. \newline
$^{1}$At this epoch, a peak of 90.8 Jy was seen in the MERLIN
W Hya spectrum at 44.7 km s$^{-1}$. No emission above 10 Jy was detected by
Pushchino at the adjacent epochs around this velocity. 
The integrated MERLIN flux density excluding this feature is 106 Jy km s$^{-1}$
}
\end{table}

\begin{figure}
   \centering
   \includegraphics[angle=0, width=8.5cm]{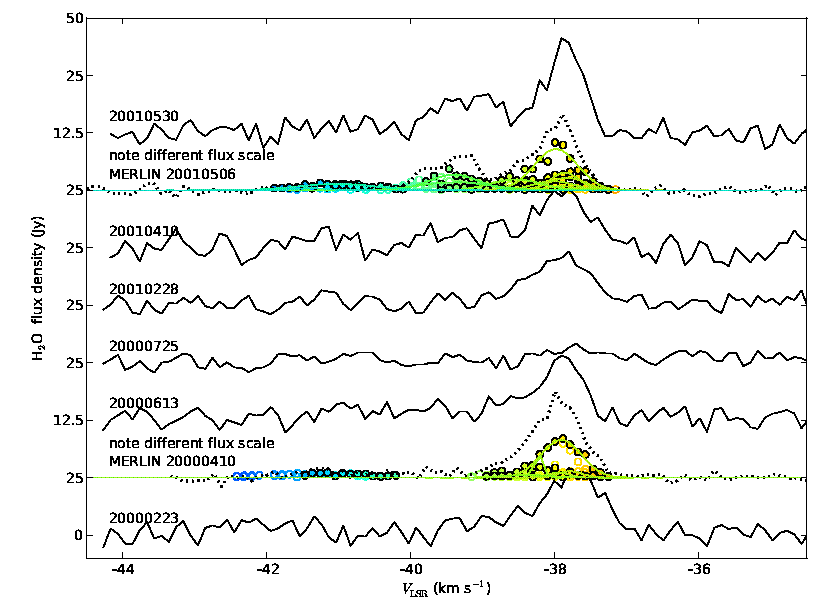}
      \caption{22-GHz H$_{2}$O maser spectra towards U Ori from
      2000.  See Fig.~\ref{VXSgr_Match94-99_Gauss.png}
      for details. $V_{\mathrm{corr}}$ = --1.1 km s$^{-1}$.}
         \label{UOri_M00-01_Gauss.png}
   \end{figure}
\begin{figure}
   \centering
   \includegraphics[angle=0, width=8.5cm]{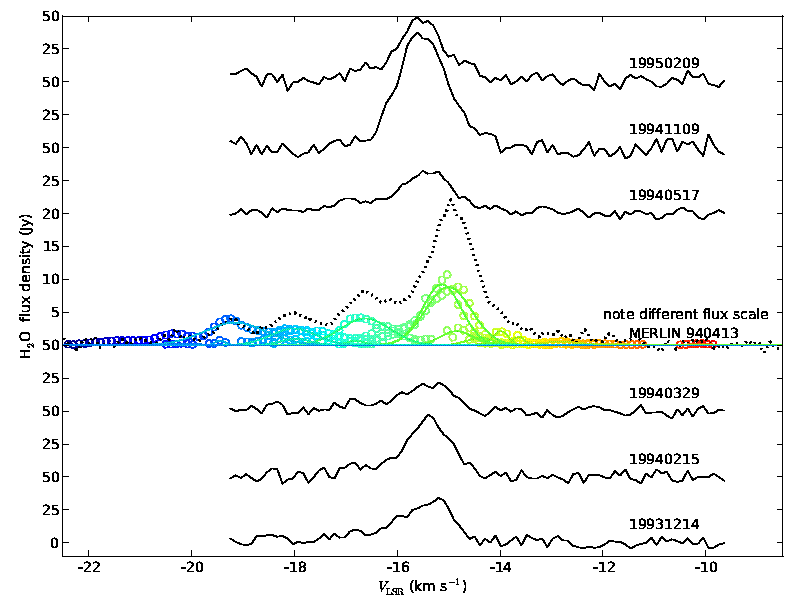}
      \caption{22-GHz H$_{2}$O maser spectra towards U Her around
      1994.  See Fig.~\ref{VXSgr_Match94-99_Gauss.png}
      for details. $V_{\mathrm{corr}}$ = 0.3 km s$^{-1}$.}
         \label{UHer_M94_Gauss.png}
   \end{figure}
\begin{figure}
   \centering
   \includegraphics[angle=0, width=8.5cm]{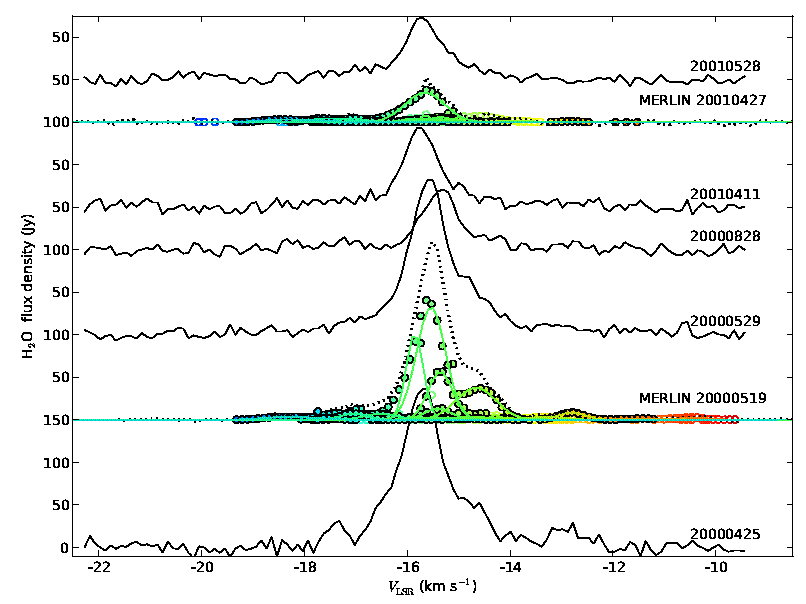}
      \caption{22-GHz H$_{2}$O maser spectra towards U Her from
      2000.  See Fig.~\ref{VXSgr_Match94-99_Gauss.png}
      for details.}
         \label{UHer_M00-01_MGauss.png}
   \end{figure}
\begin{figure}
   \centering
   \includegraphics[angle=0, width=8.5cm]{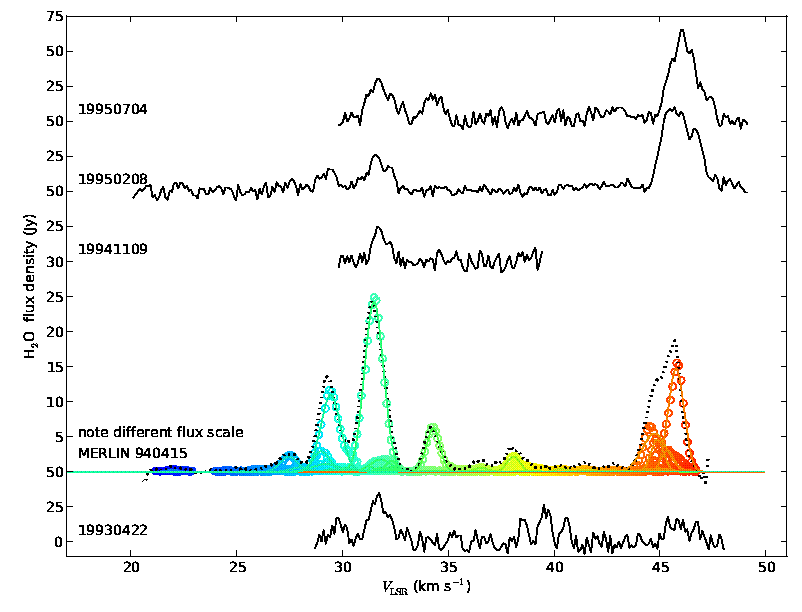}
      \caption{22-GHz H$_{2}$O maser spectra towards IK Tau around
      1994.  See Fig.~\ref{VXSgr_Match94-99_Gauss.png}
      for details. $V_{\mathrm{corr}}$  = --0.6 km s$^{-1}$.}
         \label{IKTau_M94_Gauss.png}
   \end{figure}
\begin{figure}
   \centering
   \includegraphics[angle=0, width=8.5cm]{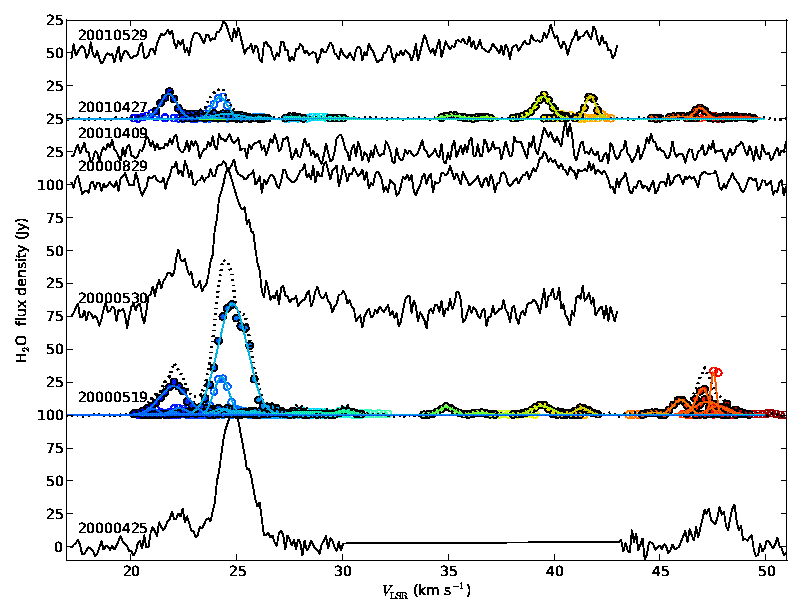}
      \caption{22-GHz H$_{2}$O maser spectra towards IK Tau from
      2000.  See Fig.~\ref{VXSgr_Match94-99_Gauss.png}
      for details. $V_{\mathrm{corr}}$ = --0.3 km s$^{-1}$.}
         \label{IKTau_Match00-01_Gauss.png}
   \end{figure}
\begin{figure}
   \centering
   \includegraphics[angle=0, width=8.5cm]{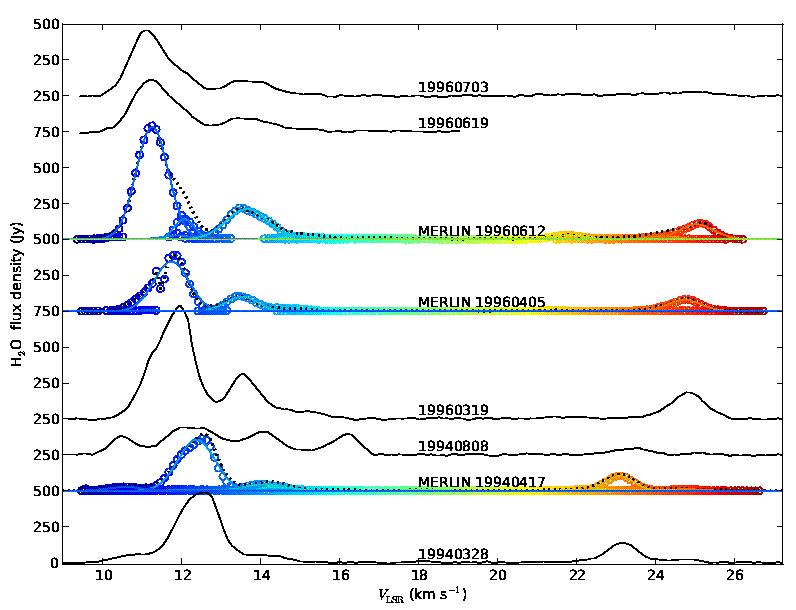}
      \caption{22-GHz H$_{2}$O maser spectra towards RT Vir from
      1994 -- 1996.  See Fig.~\ref{VXSgr_Match94-99_Gauss.png}
      for details. $V_{\mathrm{corr}}$  = --0.3 km s$^{-1}$.}
         \label{RTVir_M94-96_Gauss.png}
   \end{figure}
\begin{figure}
   \centering
   \includegraphics[angle=0, width=8.5cm]{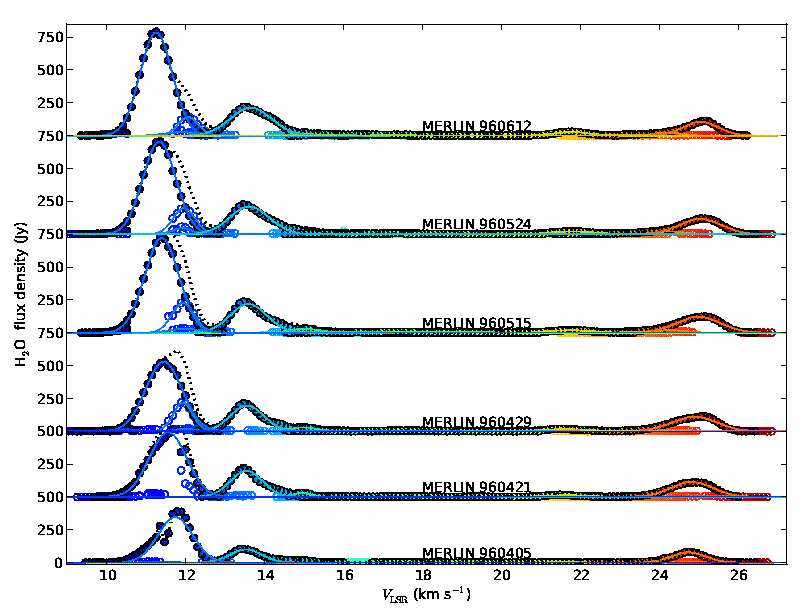}
      \caption{22-GHz H$_{2}$O maser spectra towards RT Vir around
      1996, showing all epochs.  See Fig.~\ref{VXSgr_Match94-99_Gauss.png}
      for details.}
         \label{RTVir_M96all_Gauss.png}
   \end{figure}
\begin{figure}
   \includegraphics[angle=0, width=8.5cm]{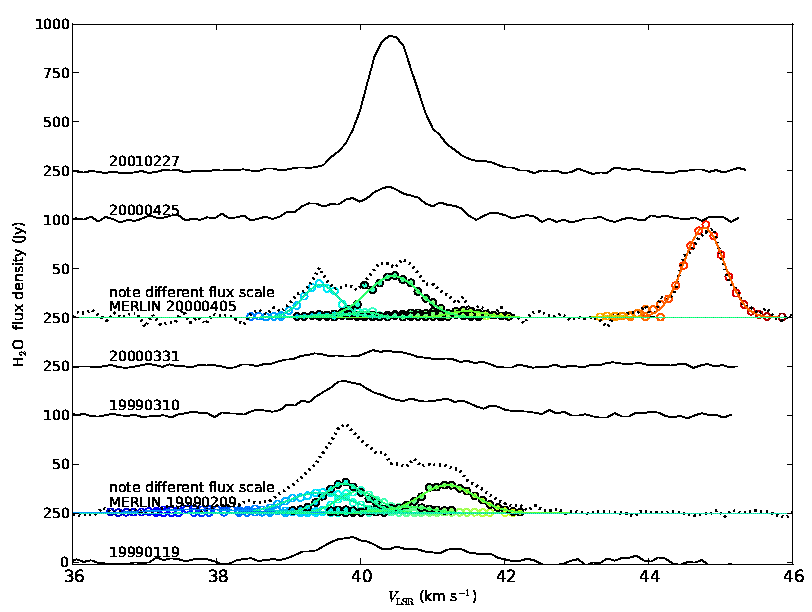}
      \caption{22-GHz H$_{2}$O maser spectra towards W Hya 
      in 1999 -- 2000.  See Fig.~\ref{VXSgr_Match94-99_Gauss.png}
      for details. $V_{\mathrm{corr}}$  = --0.3 km s$^{-1}$.}
         \label{WHya_M99-00_Gauss.png}
   \end{figure}
\begin{figure}
   \centering
   \includegraphics[angle=0, width=8.5cm]{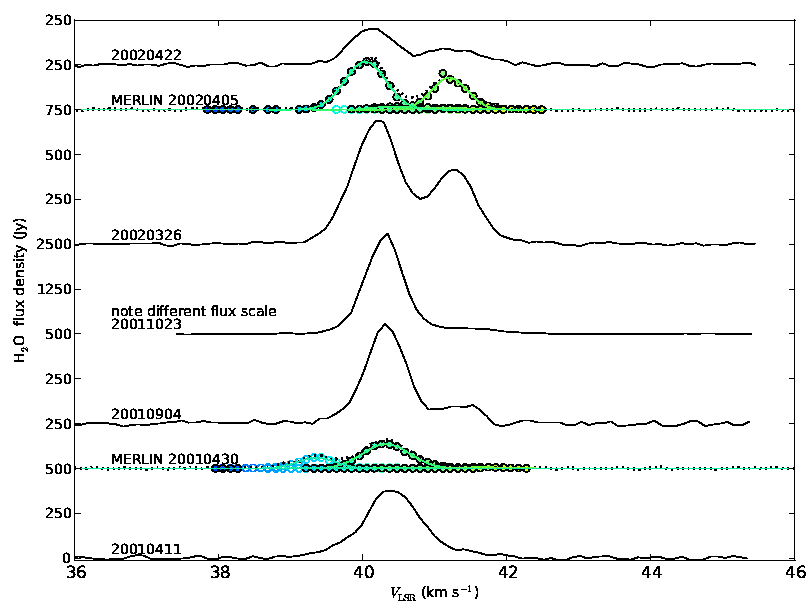}
      \caption{22-GHz H$_{2}$O maser spectra towards W Hya in
      2001 -- 2002.  See Fig.~\ref{VXSgr_Match94-99_Gauss.png}
      for details. $V_{\mathrm{corr}}$  = --0.3 km s$^{-1}$.}
         \label{WHya_M01-02_Gauss.png}
   \end{figure}

The sensitivity limit is $\sim10$
Jy ($\sigma_{\mathrm{rms}}$ 2--4 Jy) and the spectral resolution is
0.101 km s$^{-1}$ (close to the MERLIN resolution of 0.105 km s$^{-1}$
at most epochs). We show the MERLIN velocity profiles and the
Pushchino spectra taken closest in time, along with a few intermediate
epochs, in Figs.~\ref{VXSgr_Match94-99_Gauss.png}
to~\ref{WHya_M01-02_Gauss.png}.  The epochs shown here are
superimposed (blue lines) on AAVSO light curves in
Figs.~\ref{VXSgrSPer_AAVSO.png} to~\ref{IKTauRTVirWHya_AAVSO.png}.
The Pushchino absolute velocities are uncertain at some epochs due to
local oscillator problems so we applied a velocity shift $V_{\mathrm
  {corr}}$, given in the figure captions, to align the Pushchino
spectral peaks with the MERLIN peaks. This correction is only accurate
to within a few tenths km s$^{-1}$ as some the spectral features show
drifts of this magnitude when comparing different epochs taken with
the same instrument.

The dotted black lines in Figs.~\ref{VXSgr_Match94-99_Gauss.png}
to~\ref{WHya_M01-02_Gauss.png} show the integrated MERLIN velocity
profiles. The circles show the individual maser components, outlined
in black for features matched at successive epochs.  The coloured
lines show the Gaussian profiles fitted with $>3\sigma$ accuracy to
individual features.  Comparison with the Puschino data shows that the
maser flux densities vary greatly on timescales of a few weeks. Trends
in spectral variability and velocity drifts are analysed in detail in
the references given at the start of
Section~\ref{sec:Pushchino}. Here, we consider two questions: does
MERLIN recover all the flux, and what does spectral monitoring reveal
about cloud survival when compared with imaging?

\subsection{Flux detected by MERLIN interferometry}
\label{sec:flux}

The shortest MERLIN baseline of 20 km means that emission which is
smooth on scales $>100$ milli-arcsec cannot be imaged reliably.  We tabulate
the peak flux densities at the closest epochs in Table~\ref{tab:P}.
The irregular nature of maser variability means that we cannot
interpolate reliably between Pushchino measurements bracketing a
MERLIN measurement. In some cases, a new feature emerges or
disappears in observations made a few weeks apart.  We can only
investigate whether MERLIN consistently records lower peaks or less
integrated flux ($I_{\mathrm{tot}}$) than Puschino.

{\bf RSG}
For VX Sgr, the MERLIN:Pushchino (M:P) peak ratio is between
0.42 -- 0.66. However, the MERLIN spectral resolution is much coarser,
at 0.42 km s$^{-1}$, which can dilute the flux density of narrow
peaks. The original 0.21 km s$^{-1}$ resolution used in 1994 gave a
peak of 375 Jy \citep{Richards97t} and the integrated flux density at
this epoch is very similar for both instruments.  For S Per, in 1994
M:P is between 2.8--0.65 and the integrated flux densities are very
similar for most adjacent epochs. 

{\bf Miras} For U Ori, M:P is in the range 0.42 -- 0.89. However, the
source is relatively weak and the differences are only significant in 1999.
The MERLIN integrated flux densities are
slightly closer to the Pushchino values, exceeding it at one
epoch.  For U Her, M:P lies in the range 0.56 -- 1.1 and for IK Tau
M:P is between 0.7 -- 1.7, and the integrated flux densities can be
higher or lower at adjacent epochs.

{\bf SR}
For RT Vir, in 1994 M:P 0.81 -- 2.1 (in the latter case the gap is
longer than usual, 15 weeks).  In 1996, six MERLIN observations were
made over 10 weeks during which the peak flux rose from 389 to 792 Jy
but M:P changed from 0.49 to 2.2. The integrated flux densities behave
similarly. For W Hya, M:P is in the range 0.30--1.5, but the Pushchino
peak and integrated flux densities exceed the MERLIN values for all
but the final epoch. 

MERLIN clearly recovers most of the flux (within the flux scale
uncertainty of 10\% for MERLIN and the 10 Jy accuracy limit for
Pushchino) for S Per, U Her, IK Tau and RT Vir.  This also seems
likely for VX Sgr, taking the broader channel width into account, and
for U Ori, which is at a very similar distance as U Her and IK
Tau, so it is unlikely to be more resolved, but it cannot be ruled out
that some of its flux has been missed.  It is possible that more than
10\% of the emission from W Hya has been resolved out, since it is the
closest source, although its low elevation during observations makes
the flux scale even less accurate. Therefore, we regard measurements
of cloud size or flux density to be lower limits for W Hya.

\subsection{Maser features and their parent clouds}
\label{sec:survive}

Section~\ref{sec:centre} describes how we were able to match some
maser features in successive MERLIN epochs (numbers listed in
Table~\ref{tab:feats}). Many are spatially distinct in the MERLIN
images but spectrally blended in Pushchino observations; however, some
of the brightest have unambiguous counterparts.  We investigate the
spectral variability of these features in the more closely
time-sampled Pushchino spectra.

\paragraph{\bf VX Sgr}
 M03 described how about half the VX Sgr maser features detected in
 1994 were also identified in 1999. The brightest spectral peaks are present in
 the intervening epochs \citep{Pashchenko99}.

\paragraph{\bf S Per}
Forty S Per MERLIN features were unambiguously matched at both MERLIN
epochs. Three peaks dominate the 1994 MERLIN spectrum and the
Pushchino spectra at the closest epochs, see 
Fig.~\ref{SPer_M94-99_MGauss.png}. The blue-shifted peak at --46 --
--47 km s$^{-1}$ is seen at all Pushchino epochs (an additional, more
blue-shifted feature appears after 1997). The red-shifted peak becomes
fainter after 1994 and in the 1999 MERLIN spectrum it is under the 10
Jy Pushchino limit. These are seen with a similar internal structure
by MERLIN at both 1994 and 1999 (Fig.~\ref{sper_abfeat.png}).

The central peak around $-37.8$ km s$^{-1}$ exceeds 40 Jy and is
 spatially compact at both MERLIN epochs. However, it has faded
 in Pushchino spectra taken at 19940810, a few months after
 the 1994 MERLIN epoch, and is absent for most of 1995-6 and all of
 1997 \citep{Lekht05}.  It re-appears in the 19981228 and later
 Pushchino spectra as well as in the MERLIN 1999 data.

\paragraph{\bf U Ori}
Nine U Ori MERLIN features were matched at 2000 and 2001.  The brightest
   emission, just under 20 Jy, is close to --38 km s$^{-1}$
   (Fig.~\ref{UOri_M00-01_Gauss.png}). This is dominated by the
   feature at (--40, --15) milli-arcsec (Fig.~\ref{UOri_00-01_xyplot.png}). The
   Puschino spectrum shows a corresponding peak of 20--40 Jy 
   at the closest epochs.  However, no emission above 10 Jy was
   detected from U Ori in intermediate spectra, e.g. 20000725.

\paragraph{\bf U Her}
Twenty U Her MERLIN features were matched at 2000 and 2001. The peak,
around --15.5 km s$^{-1}$, was seen at all Pushchino epochs between these dates.

\paragraph{\bf IK Tau}
Twenty-four IK Tau MERLIN features were matched at 2000 and 2001. In 2000, the
75-Jy peak at 25 km s$^{-1}$ corresponds to the feature at (--50, 10)
milli-arcsec and the 25-Jy peak at 22 km s$^{-1}$ corresponds to the feature at
(0,0) milli-arcsec
(Fig.~\ref{IKTau_00-01_xyplot.png}).
Fig.~\ref{IKTau_Match00-01_Gauss.png} shows that the  25 km s$^{-1}$
peak declines abruptly after 20000530.  The IK Tau
H$_2$O maser shell is well-filled, with several features contributing
to the spectral peaks. It is possible that the 25 km s$^{-1}$ peak was
a flare due to clouds overlapping along the line of sight, similar to
that described for W Hya in Section~\ref{sec:WH-var}. The 22 km
s$^{-1}$ feature is present in the
2001 MERLIN spectrum at a similar flux density, 21 Jy, to that seen in
1994.
The blue-shifted peaks cannot be distinguished in the 200000819 and
20010409 spectra but are present in 20010529.

\paragraph{\bf RT Vir}
Many features were matched between six epochs of RT Vir MERLIN data
taken over 10 weeks in 1996 but there are no intervening Pushchino
spectra.  Fig.~\ref{RTVir_M96all_Gauss.png} shows that the strong
blue- and red-shifted RT Vir peaks show systematic increases in
$|V_{\mathrm{LSR}}-V_{\star}|$ with epoch, totalling 0.43 and 0.17 km
s$^{-1}$ in 10 weeks, respectively, consistent with the strong
acceleration seen in proper motions (Section~\ref{sec:pm}; see also
\citealt{Imai03}).

\paragraph{\bf W Hya} 
Between five and nine W Hya MERLIN features were matched between
successive epochs in 1999, 2000, 2001 and 2002. The brightest part of
the MERLIN spectrum in 1999 is made up of a number of features so
their individual behaviour is hard to trace in the Pushchino spectra.
However, the 30-Jy feature at 41.1 km s$^{-1}$ seen in 1999, located
at (--95, 120) has a much fainter counterpart in 2000
(Figs~\ref{WHya_M99-00_Gauss.png} and~\ref{WHya_xyplot.png}). Faint
emission was also detected at the same velocity in 2001 and rises to
160 Jy in 2002, but this is located on the opposite side of the CSE at
(--70, --110) (Figs~\ref{WHya_M01-02_Gauss.png}
and~\ref{WHya_xyplot.png}). The 2001--2002 flare is discussed in
Section~\ref{sec:WH-var}.

Fig.~\ref{VXSgrSPer_AAVSO.png} to~\ref{IKTauRTVirWHya_AAVSO.png} show
that the intervals over which masers can be matched spans several
stellar periods in the case of VX Sgr, S Per and W Hya, and nearly two
periods for U Her and IK Tau. U Ori features were matched across one
interval close to the stellar period, but not in another, whilst RT
Vir features were only matched over intervals less than one stellar
period. There is no obvious correlation between maser survival and the
number of periods or the phase. 
In other interferometric studies,
\citet{Winnberg11} found that U Her maser components have an average
lifetime of 0.5--1 yr and \citet{Winnberg08} obtained a lifetime
$\sim1$ yr for RX Boo. In general, single dish observations are
affected by spatial blending and may overestimate maser lifetimes, but
we note that \citet{Lekht99} analysed 12 yr of Pushchino monitoring of
RT Vir and found 22 spectral features persisting for 0.66 yr or more,
only 5 of which lasted longer than 2 yr.

The time $t$ taken for the wind to to reach a given distance $r$ in
the 22-GHz maser shell is calculated using the velocity gradient
$K_{\mathrm{grad}}$ (Table~\ref{tab:feats}), assuming linear, radial
acceleration and good coupling of dust and gas velocities for the
H$_{2}$O maser clouds:
\begin{equation}
\frac{t}{\mathrm{yr}} = 4.8
\left(\frac{K_{\mathrm{grad}}}{\mathrm{km\,s}^{-1}/\mathrm{AU}}\right)^{-1}
\ln\left[1+\frac{(r-r_{\mathrm i})K_{\mathrm{grad}}}{\v_{\mathrm
i}}\right]
\label{eq:cross}
\end{equation}
and the total shell crossing time from $r_{\mathrm i}$ to $r_{\mathrm
  o}$, is $t_{\mathrm{cross}}$, given in Table~\ref{tab:mloss}.  This
is about 10--15 yr for SRbs, 20--35 yr for Miras and 35--90 yr for the
RSGs.  This is the minimum survival time for the clouds as discrete,
dense entities, since it is hard to see how they could re-form if
disrupted during passage across the water maser shell.  On the other
hand, most individual maser features become unrecognisable after 2
years or less around AGB stars, and only about half of the RSG
features are matched after 5 years, although M03 found that about one
fifth survive 16 yr; VY CMa shows similar behaviour
(\citealt{Bowers93}, \citealt{Richards98v}).

The comparison with Pushchino data shows that in fact, masers from
individual clouds switch off and on again. This is most obvious for S
Per, U Ori and IK Tau, where spectral peaks can be clearly identified
with spatial features at separate interferometry epochs, but those
peaks are missing in the spectra at some intervening single dish
monitoring epochs. The smaller the clouds (Table~\ref{tab:feats}), the
more short-lived the masers. The sound-crossing times (assuming a
sound speed of 3 km s$^{-1}$) range from 1.5--5 yr for the AGB clouds
and 20--25 yr for the RSG.  The maser lifetimes are comparable to the
time which would be taken for sub-sonic or mildly supersonic shocks to
cross the maser clouds. Fig.~\ref{sper_abfeat.png} shows a localised,
non-radial disturbance propagating at $\la5$ km s$^{-1}$.  These might
temporarily disrupt the maser altogether, or simply cause it to beam
in a direction we cannot detect.

\section{Acceleration and density of maser clouds}
\label{sec:mass}

The inner and outer maser shell limit measurements reveal the
kinematics (Section~\ref{sec:masslossshell}). We then use the
individual cloud sizes to estimate their contribution to the mass loss
(Section~\ref{sec:density}).

\subsection{Acceleration}
\label{sec:masslossshell}

\begin{figure}
   \centering
   \includegraphics[angle=0, width=9cm]{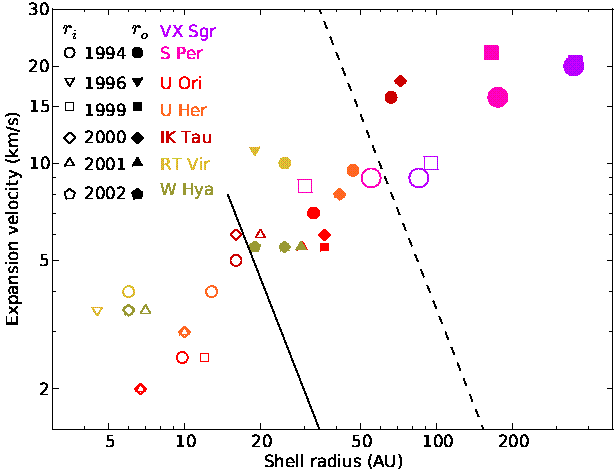}
      \caption{ The inner  and outer 22-GHz
      maser shell radii and velocities are shown by hollow and filled
      symbols, respectively. The
      different stars and epochs are shown by different colours and
      symbols, respectively. The solid and dashed diagonal lines show
      the escape velocities for 1 and 10 M$_{\odot}$ stars, respectively.}
         \label{Allepochs_allrv.png}
   \end{figure}

Fig.~\ref{Allepochs_allrv.png} shows the shell limits for all sources,
taken from Section~\ref{sec:shell}, Figs.~\ref{VXSgr_ravel.png}
to~\ref{WHya_ravel.png} and Table~\ref{tab:feats}. This demonstrates
that the trend for higher velocities at greater distances from the
star is similar when comparing different objects which differ in size
by an order of magnitude.  In all objects, the maser velocity
approximately doubles as the wind traverses the maser shells, passing
through the gravitational escape velocity $V_{\mathrm{esc}}$ in the
process.  This implies ongoing acceleration beyond the dust formation
zone. \citet{Ivezic10} model the CSE velocity field using changes in
optical depth.  However, the consistently strong acceleration over
such a wide range of objects seen here is likely to require additional
force due to changes in the surface properties of the dust as grains
anneal in the wind. This was first suggested by \citet{Chapman86} for
VX Sgr and is discussed for others among our maser sources by
\citet{Richards98d}. \citet{Verhoelst09} observed correlations between
dust composition and outflow velocities for RSG while \citet{Decin10}
modelled \emph{Herschel} observerations of IK Tau using gradual
acceleration.

\begin{figure}
   \centering
   \includegraphics[angle=0, width=9cm]{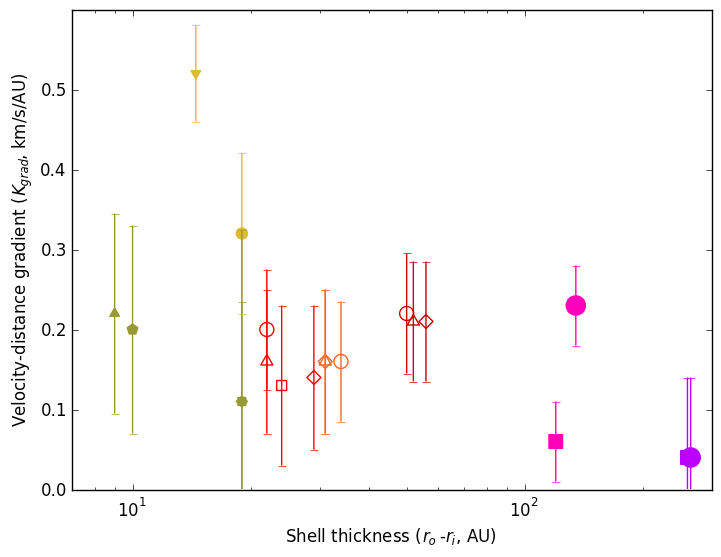}
\caption{The relationship between the velocity gradient $K_{\mathrm {grad}}$
(see Table~\ref{tab:feats}) and total 22-GHz shell thickness.  The
error bars for $K_{\mathrm {grad}}$ contain a rough allowance for
uncertainty in shell fitting. The different stars are indicated as in Fig.~\ref{Allepochs_allrv.png}.}
\label{kgraddr.png}
\end{figure}

Fig.~\ref{kgraddr.png} shows that the SRs, Miras and RSGs form three
distinct groups of increasing 22-GHz shell thickness and this is
weakly correlated with shallower velocity gradients, $K_{\mathrm
  {grad}}$, which measure the increase in expansion velocity with
distance from the star. $K_{\mathrm {grad}}$ is estimated from the
$V_{\mathrm{LSR}}$ measurements along the line of sight. The results
in Section~\ref{sec:expand} demonstrate a similar trend for proper
motions in the plane of the sky. The SRb RT Vir has the highest
gradient of proper motion velocity with position in the shell,
$dV_{\mathrm a}/d\overline{a}$, and the other AGB stars with
significant results have higher values of $dV_{\mathrm
  a}/d\overline{a}$ than the RSG.  The thicker circumstellar envelopes
show more gradual acceleration, but eventually reach higher
velocities.

A theoretical relationship between $\dot{M}$ and terminal velocity was
established by \citet{Cooke85}, who also showed that $\dot{M}$ should
determine the 22-GHz shell $r_{\mathrm i}$. The general forms of these
relationships (and the acceleration through $V_{\mathrm{esc}}$) were
confirmed observationally by \citet{Lane87}, \citet{Yates94},
\citet{Bowers94} and \citet{Colomer00}. However, the observed
$r_{\mathrm i}$ tend to be much larger than predicted. This implies
that the wind is inhomogenous, discussed in Section~\ref{sec:density}.

\subsection{Water maser clump density and the mass in clouds}
\label{sec:density}

We use the relationship between $\dot{M}$ and $r_{\mathrm i}$ to
estimate the density of the H$_2$O maser clouds.  The density of the
stellar wind at radius $r$ is $\rho= \dot{M}/(4 \pi r^2 \v)$, assuming
radial acceleration of a smooth medium. If the average particle mass
is $1.37\times$ the mass of molecular hydrogen, $\rho = 1.37 \times 2
m_{\mathrm H} n$. The overall number density $n$ of the wind is thus
given by:
\begin{equation}
\frac{n}{{\mathrm m}^{-3}} \approx 5\times 10^{22}
\left(\frac{\dot{M}}{\mathrm{M}_{\odot}\,\mathrm{yr}^{-1}}\right)
\left(\frac{r}{\mathrm{AU}}\right)^{-2} \left(\frac{\v}{\mathrm{km\,s}^{-1}}\right)^{-1}
\label{eq:mdot}
\end{equation}

We use this to find the average number density $n_{\mathrm i}$ at
$r_{\mathrm i}$, where $\v = \v_{\mathrm i}$.  The inner edge of the
22-GHz maser shell is determined by the point at which the collision
rate falls below the maser emission rate, at the quenching number
density $n_{\mathrm q}$.  At 1200 K, $n_{\mathrm q}\approx 5 \times
10^{15}$ m$^{-3}$ \citep{Yates97}. We assume that this is the density
of the clouds at $r_{\mathrm i}$.

The values of $\dot{M}$ given in Table~\ref{tab:stars} are taken from
the literature (adjusted to our adopted distances), using high
signal-to-noise CO data, supported by interferometry, where
available. 
%This provides the CSE mass, kinematics and shape and mostly
%consistent results from different teams. 
However, some objects, notably S Per, have faint CO. Other methods of
estimating $\dot{M}$, such as IR observations of dust, are more
model-dependent; the assumed gas:dust ratios vary between about 80 and
several hundred.  The variations in values of $\dot{M}$ in the
literature are a factor of about 5 for most of the objects.

We want to estimate the total number of potentially masing clouds
  in a shell at any one time, $N_{\mathrm{tot}}$.  However, as
  described in Section~\ref{sec:survive}, at any one epoch, some
  clouds may not be producing masers beamed towards Earth. We
  attempted to compensate for this using the multi-epoch observations
  where we should have found reliable matches at neighbouring epochs,
  i.e.  6 epochs for RT Vir, 4 for W Hya, 3 for U Ori and 2 for the
  other stars (Section~\ref{sec:pm}).  This excludes the 1994 data for
  the AGB stars. $N_{\mathrm{tot}}$ is given by the sum of the number
  of features $NF$ at the first epoch ($e=1$), plus the number of
  maser features with no previous matches at subsequent epochs ($NF -
  N_{\mathrm{prev}}$ at $1<e\le$ end) where $e$ = end at the final
  epoch.  We finally subtract the number of clouds likely to have left
  the shell during the time interval for all epochs considered,
  $t_{\mathrm all}$; since $t_{\mathrm all} \ll t_{\mathrm {cross}}$
  this is approximated by the uncorrected value of $N_{\mathrm{tot}}$
  multiplied by $(1- t_{\mathrm all}/ t_{\mathrm {cross}})$.  This
  approach partially compensates for less sensitive epochs (since a
  cloud only has to be detected once to be counted) but
  $N_{\mathrm{tot}}$ is still a lower limit if some clouds are present
  in the CSE but not detected at any epoch.  The degree of
  underestimation may vary from star to star if some have a
  persistently lower fraction of detectably masing clouds.  

%At any one epoch, we are probably only detecting those features
%beaming masers towards Earth. We want to estimate the total number of
%clouds, $N_{\mathrm{tot}}$ in the maser shell at any one time. We do
%this by considering the epochs of observation of each star where
%we should have found reliable matches at neighbouring epochs, i.e.
% 6 epochs for RT Vir, 4 for W Hya, 3 for U Ori and 2 for the other stars
%(Section~\ref{sec:pm}). This excludes the 1994 data for the AGB
%stars. $N_{\mathrm{tot}}$ is given by the sum of the number of
%features $NF$ at the first epoch ($e=1$), plus the number of maser
%features with no previous matches at subsequent epochs ($NF -
%N_{\mathrm{prev}}$ at $e > 1$).  We finally deduct the number of
%clouds likely to have left the shell during the time interval for all
%epochs considered, $t_e$.
%\begin{equation}
%N_{\mathrm{tot}} = \left[\sum^{e}_{e=1} NF_e +
%  NF_{(e+1)}-N_{\mathrm{prev}(e+1)}\right] (1-t_e/t_{\mathrm{cross}})
%\label{eq:Ntot}
%\end{equation}

\begin{equation}
N_{\mathrm{tot}} = \left[NF + \sum^{e = {\mathrm{end}}}_{e=2} \left(NF_{(e)}-N_{\mathrm{prev}(e)}\right)\right](1- t_{\mathrm{all}}/t_{\mathrm{cross}})
\label{eq:Ntot}
\end{equation}

The average number of clouds produced per stellar period is given by $N_{\mathrm{tot}}/P$.

Clouds expanding away from the star under radial acceleration increase
in tangential diameter by $l(r/r_{\mathrm i})$ and in the radial
diameter by $l(r/r_{\mathrm i})^{\epsilon}$ (Equation~\ref{eq:epsilon}). Thus, the number density
in a cloud at $r$ is $n_{\mathrm q} (r/r_{\mathrm i})^{-(2+\epsilon)}$.
We use the brightest part of the maser shell as $r_{\mathrm {typical}}$
(see Section~\ref{sec:cloudsize}) to
estimate the average mass per cloud in the 22-GHz maser shell: 
\begin{equation}
\frac{M_{\mathrm{cloud}}}{{\mathrm M}_{\odot}} = 4\times10^{-24} n_{\mathrm q}
\left(\frac{r_{\mathrm{typical}}}{r_{\mathrm i}}\right)^{-(2+\epsilon)} \left(\frac{l}{\mathrm{AU}}\right)^3
\label{eq:Mtot}
\end{equation}
(making the same assumptions as for Equation~\ref{eq:mdot}).

The mass loss rate in clouds is then given by $\dot{M}_{\mathrm{clouds}}
= M_{\mathrm{cloud}} N_{\mathrm{tot}}/t_{\mathrm{cross}}$.  The filling
  factor of the clouds by volume is given by $f_{\mathrm V} =  N_{\mathrm{tot}}
  (l/2)^3/(r_{\mathrm o}^3 - r_{\mathrm i}^3)$.

 We apply Equations~\ref{eq:mdot} to~\ref{eq:Mtot} to the measurements given in
Table~\ref{tab:feats}, averaged for the epochs considered, to 
 derive the parameters tabulated in Table~\ref{tab:mloss}. The
most striking result is that the H$_{2}$O maser cloud density at
$r_{\mathrm i}$, $n_{\mathrm q}$, is $\ga 45 n_{\mathrm i}$ for all sources.
What are the main causes of uncertainty in this estimate?  The
collision rate, and hence $n_{\mathrm q}$, is proportional to
$\surd$(temperature). The assumed temperature is likely to be an upper
limit, almost double some estimates for the lower-mass stars
\citep{Zubko00}, so our adopted value of
$n_{\mathrm q}$ is probably a lower limit. The main uncertainty in
$r_{\mathrm i}$, 10 milli-arcsec for W Hya and $\le5$ milli-arcsec for the other sources,
arises from determining the stellar position
(Section~\ref{sec:centre}). At most, this could reduce $n_{\mathrm
q}/n_{\mathrm i}$ to 38 for W Hya, and by a smaller proportion for the
other sources. In fact, since one of the main criteria in determining
the centre of expansion was maximising $r_{\mathrm i}$, the errors are
probably lower.  $\v_{\mathrm i}$ is least well-constrained for U Ori
and U Her, but these have lower $\v_{\mathrm i}$ than any other object
considered and any increase  would increase
$n_{\mathrm q}/n_{\mathrm i}$.  

Equation~\ref{eq:mdot} is distance-independent, as long as $r_{\mathrm
i}$ and $\dot{M}$ are estimated consistently and both have a quadratic
dependence on distance. If the highest values of $\dot{M}$ are taken
from the literature, this could reduce $n_{\mathrm
q}/n_{\mathrm i}$ about fivefold, but it seems very unlikely that all the
errors are in the same sense.

 The
mass loss rate in clouds, $\dot{M}_{\mathrm{clouds}}$ is within
$\pm80\%$ of the adopted values of $\dot{M}$. This is a relatively
small discrepancy compared to deviations of $\dot{M}$ in the literature, as
mentioned above.  The space between the clouds cannot be completely
empty so that if accurate values were known,
$\dot{M}_{\mathrm{clouds}}<\dot{M}$ and thus, for S Per, U Her and U Ori,
either $\dot{M}$ must be higher or the mass in clouds must be smaller
(or both). 

In all cases, the cloud filling factor by volume $f_{\mathrm V}$ is
less than 0.01. Using equation (11) of R99, $n_{\mathrm c}/n$ must
be less than $1/f_{\mathrm V}$, where $n_{\mathrm c}/n$ is the cloud
number density at any radius. This constrains $n_{\mathrm
  q}/n_{\mathrm i} < 105$ for U Ori and S Per, or larger values for
the other sources, which is not inconsistent with our estimates of
$n_{\mathrm q}/n_{\mathrm i}$.

There are a number of additional sources of uncertainty in our
estimates of ${M}_{\mathrm{cloud}}$ and derived quantities.
$t_{\mathrm {cross}}$ has a linear dependence on distance and
$M_{\mathrm{cloud}}$ has a cubic dependence, although
$\dot{M}_{\mathrm{cloud}}/\dot{M}$ is independent. $N_{\mathrm{tot}}$
could be an underestimate if there are clouds which have not beamed
masers towards us at any of the epochs used in the analysis, although a
much higher value would produce an unrealistically high
$\dot{M}_{\mathrm{clouds}}$. For the same reason, although the typical
values used for $l$ and $r_{\mathrm{typical}}$ and other assumptions
about cloud behaviour are very simplified, these are unlikely to be
severe overestimates. 
%If the number density at the mid-point of the 22-GHz shell is used,
%this reduces $\dot{M}_{\mathrm{clouds}}/\dot{M}$ to 10--70\%.
 R11 used these estimates of $l$ and $n$ to find
estimates of parameters such as maser brightness temperatures, beaming
and optical depths consistent with existing maser models
e.g. \citet{Elitzur92a}. Altogether, errors of up to factors of a few
are possible.
 
\begin{table*}
\begin{tabular}{lcccccccccccc}
\hline
Star&Epochs&$t_{\mathrm{all}}$&$n_{\mathrm{i}}$&$n_{\mathrm q}/n_{\mathrm i}$&$t_{\mathrm{cross}}$&$N_{\mathrm{tot}}$&$\underline{N_{\mathrm{tot}}}$&$M_{\mathrm{cloud}}$&$\dot{M}_{\mathrm{clouds}}$&
$\underline{\dot{M}_{\mathrm{clouds}}}$ &$f_{\mathrm V}$&$n_{\mathrm
    s}/n$\\
    &      &(yr)           & (m$^{-3}$)	    &      &(yr)&     &  $P$ &(M$_{\odot}$)& (M$_{\odot}$ yr$^{-1}$)&$\dot{M}$&   &   \\
\hline
VX Sgr&  94--99      &4.73 &$4.7\times10^{13}$& 107& 89&   139$^*$& 3&$1.1\times10^{-5}$& $1.7\times10^{-5}$ &  0.2 &   0.0009&0.90    \\
S Per &  94--99      &4.80 &$1.2\times10^{14}$&  43& 38&   134$^*$& 6&$1.4\times10^{-5}$& $4.8\times10^{-5}$ &  1.3 &   0.0095&0.60    \\
U Ori &99--00--01    &2.30 &$6.9\times10^{13}$&  72& 32&    57&    2 &$2.4\times10^{-7}$& $4.2\times10^{-7}$ &  1.8 &   0.0095&0.31    \\
U Her &  00--01      &0.94 &$5.7\times10^{13}$&  88& 29&    62&    2 &$2.9\times10^{-7}$& $6.1\times10^{-7}$ &  1.8 &   0.0079&0.31    \\
IK Tau&  00--01      &0.94 &$6.7\times10^{13}$&  75& 24&    85&    4 &$1.6\times10^{-7}$& $5.4\times10^{-7}$ &  0.2 &   0.0010&0.92    \\
RT Vir&  96(6)       &0.186&$9.2\times10^{13}$&  55& 11&   142&    6 &$3.8\times10^{-9}$& $5.1\times10^{-8}$ &  0.4 &   0.0026&0.86    \\
W Hya &99--00--01--02&3.16 &$9.1\times10^{13}$& 55 & 14&    45&    3 &$1.5\times10^{-8}$& $5.1\times10^{-8}$ &  0.2 &   0.0019&0.90    \\
\hline
\end{tabular}					  
\caption{The quantity and mass of H$_{2}$O maser clouds. $t_{\mathrm
    {all}}$ is the timespan covered by the epochs used in this
    analysis. $n_{\mathrm i}$ is the number density at $r_{\mathrm i}$
    estimated from mass loss rates taken from the literature
    (Table~\ref{tab:stars}) and $n_{\mathrm q}/n_{\mathrm i}$ is the
    ratio of the quenching density to $n_{\mathrm
    i}$. $t_{\mathrm{cross}}$ is the estimated crossing time for the
    22-GHz maser shell. $N_{\mathrm{tot}}$ and $N_{\mathrm{tot}}/P$
    are the total number of maser clouds at any one time and the
    number produced per stellar period, respectively.  The mass per
    H$_{2}$O maser cloud is $M_{\mathrm{cloud}}$,
    $\dot{M}_{\mathrm{clouds}}$ is the mass loss rate in clouds and
    $\dot{M}_{\mathrm{clouds}}/\dot{M}$ is the ratio of this to the
    mass loss rate from the literature.  $f_{\mathrm V}$  is the
     volume filling factor of the clouds in the
    shell. $n_{\mathrm s}/n$ is the fractional number density of the
    wind surrounding the clouds.
\newline $^*$The actual number of clouds may be less since some
RSG    clouds are large enough to contain several features.
}
\label{tab:mloss}
\end{table*}

The inescapable conclusion is that the density of 22-GHz H$_{2}$O
maser clouds is at least an order of magnitude greater than the
average wind density, and probably 50--100 times greater. $\ga20$\% of
the mass loss is concentrated in these clouds, with a filling factor
$<1$\%. \citet{Menut07} modelled IR interferometry of dust emission from
within 300 AU of IRC+10216 using 500 clumps, ten- to a hundred-fold
overdense, of the same radius as the star, which is similar to the
wind structure deduced for our targets.

\section{Time evolution, flares and variability}
\label{sec:variability}

%A number of long-term studies have compared the variability of maser
%spectra with optical periodicity, e.g. ** . 

We model the 3-D changes in maser distribution between epochs in
Section~\ref{sec:time}.  A number of individual maser features
brighten or dim by large factors.  One, in W Hya, is particularly
clearly identified in successive epochs of both MERLIN and Pushchino
data and the reason for its isolated flare is explained in
Section~\ref{sec:WH-var}.  We investigate whether the imaging data
showed any systematic dependence between flux density variations and
location in the maser shell in Section~\ref{sec:shellvary}, and look for
possible causes of variability in Section~\ref{sec:why}.

\subsection{Time evolution of maser distributions}
\label{sec:time}

The positions of masers ($x$, $y$) are measured directly in the plane
of the sky but their velocities ($u = V_{\star}-V_{\mathrm{LSR}}$) are
measured in the perpendicular direction.  We can calculate the full
3-D coordinates using trigonometry, if we assume spherically symmetric
expansion away from the star, obeying $\v = \v_{\mathrm i} + K_{\mathrm{grad}}
(r-r_{\mathrm i})$. $K_{\mathrm{grad}}$ is defined in Equation~\ref{eq:grad}
and $r$ 
and $\v$ are the magnitudes of
the radial  position and velocity vectors for a maser at any point in
the shell, i.e. ${\bf{r} = (x, y, z)}$. Thus, $r^2 = a^2 + z^2$ (where
$a^2 = x^2 + y^2$) and $\v^2 = w^2 + u^2$ (where $w$ is the velocity
component in the plane of the sky, in the direction of $a$).  We use
the acceleration along the $z$-axis to derive an expression which can
be solved for the position and velocity components not directly
measured:
\begin{equation}
u = u_{\mathrm i} + K_{\mathrm{grad}}(z-z_{\mathrm i})
\end{equation}
where $u_{\mathrm i}$ is the line-of-sight velocity component at the
inner maser shell edge at $z_{\mathrm i}$. By simple trigonometry,
$z_{\mathrm i} = r_{\mathrm i}(z/r)$ and $u_{\mathrm i} = \v_{\mathrm
  i}(z/r)$, giving
\begin{equation}
u = \v_{\mathrm i}z/r + K_{\mathrm{grad}}z(1-r_{\mathrm i}/r)
\end{equation}
Substituting $z=\sqrt{r^2 - a^2}$, multiplying by $r$ and then squaring gives
\begin{equation}
u^2 r^2 = [\v_{\mathrm i} + K_{\mathrm{grad}}(r-r_{\mathrm i})]^2 (r^2 - a^2)
\end{equation}
which yields a quartic equation in $r$
\begin{eqnarray}
0 &=& [K_{\mathrm{grad}}^2]\,r^4 \nonumber \\ 
& +& [2K_{\mathrm{grad}}\v_{\mathrm i}-2K_{\mathrm{grad}}^2r_{\mathrm i}]\,r^3 \nonumber \\ 
&+&[-u^2+\v_{\mathrm i}^2 - 2K_{\mathrm{grad}}\v_{\mathrm i}r_{\mathrm i} +
  K_{\mathrm{grad}}^2r_{\mathrm i}^2 - K_{\mathrm{grad}}^2a^2]\,r^2 \nonumber \\ 
& +& [-2\v_{\mathrm i}K_{\mathrm{grad}}a^2 +2K_{\mathrm{grad}}r_{\mathrm i}a^2]\,r \nonumber \\ 
& +&[-\v_{\mathrm i}^2a^2 +2K_{\mathrm{grad}}\v_{\mathrm i}r_{\mathrm i}a^2 -K_{\mathrm{grad}}^2a^2
  r_{\mathrm i}^2]
\label{eq:quartic}
 \end{eqnarray}
where all the coefficients are composed of  measured parameters.
We solved this (using {\sc python numpy}\footnote{{\tt http://numpy.scipy.org/}}) for $r$ and hence can calculate any
other vector component such as $w$.  A similar process was
performed by M03 for VX Sgr.

\begin{figure}
   \centering
   \includegraphics[angle=0, width=9cm]{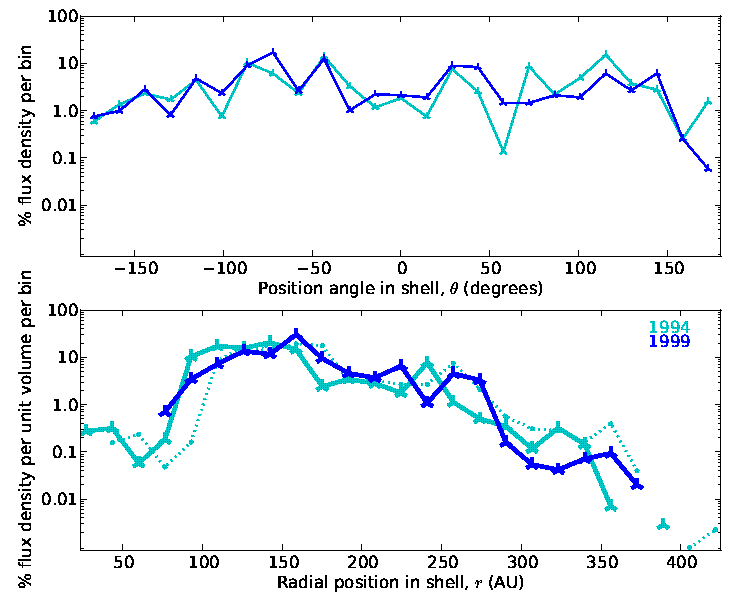}
      \caption{VX Sgr. The lower plot shows the percentage flux
        density per unit volume in concentric spherical shells (heavy
        lines, colour coded by epoch). The fainter dotted lines show
        how the earlier masers would appear if they had flowed out
        unchanged to the time of the latest epoch. The upper plot
        shows percentage flux density per segment of angle in the
        plane of the sky (colour coded by epoch). An isolated marker
        is shown if the bins on either side are empty. }
         \label{VX_r_theta.png}
   \end{figure}

\begin{figure}
   \centering
   \includegraphics[angle=0, width=9cm]{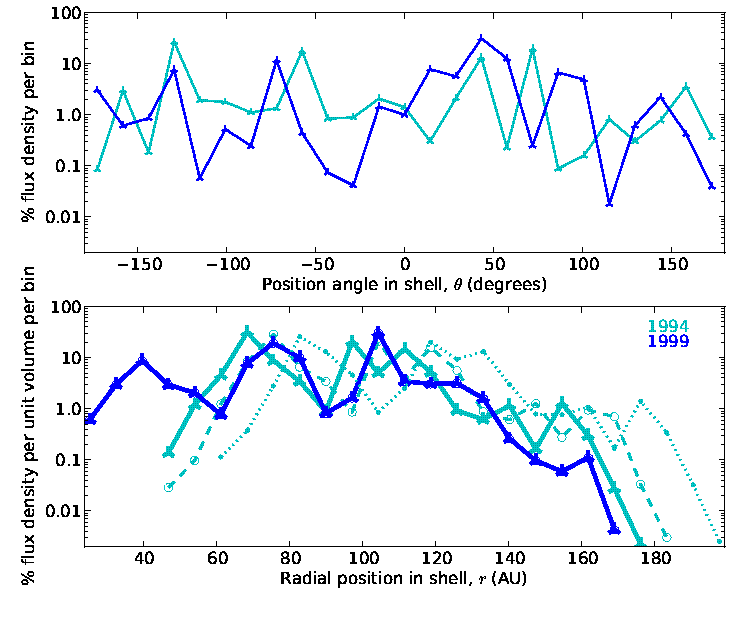}
      \caption{S Per, see Fig.~\ref{VX_r_theta.png} for details. The
      dashed line shows how the 1994 masers would appear if
      they had expanded unchanged at 0.4 of the estimated expansion velocity $\v$, to the time of the latest epoch.}
         \label{SP_r_theta.png}
   \end{figure}

\begin{figure}
   \centering
   \includegraphics[angle=0, width=9cm]{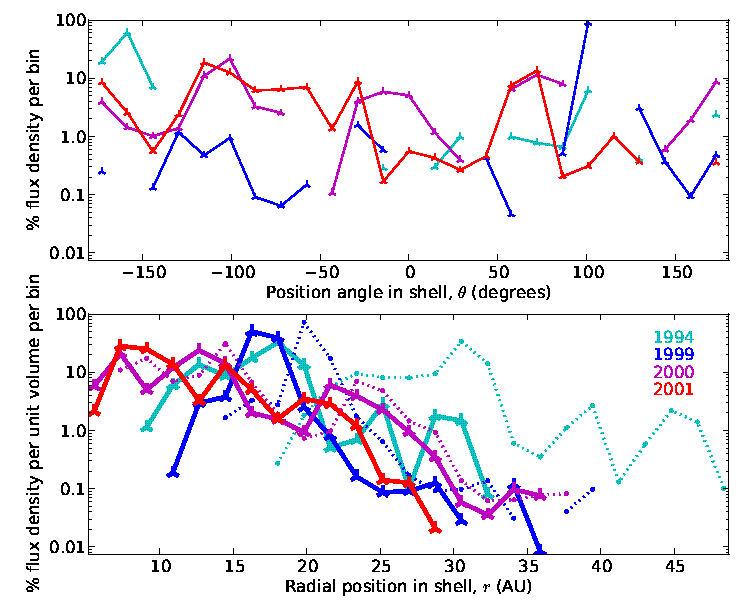}
      \caption{U Ori, see Fig.~\ref{VX_r_theta.png} for details.}
         \label{UO_r_theta.png}
   \end{figure}

\begin{figure}
   \centering
   \includegraphics[angle=0, width=9cm]{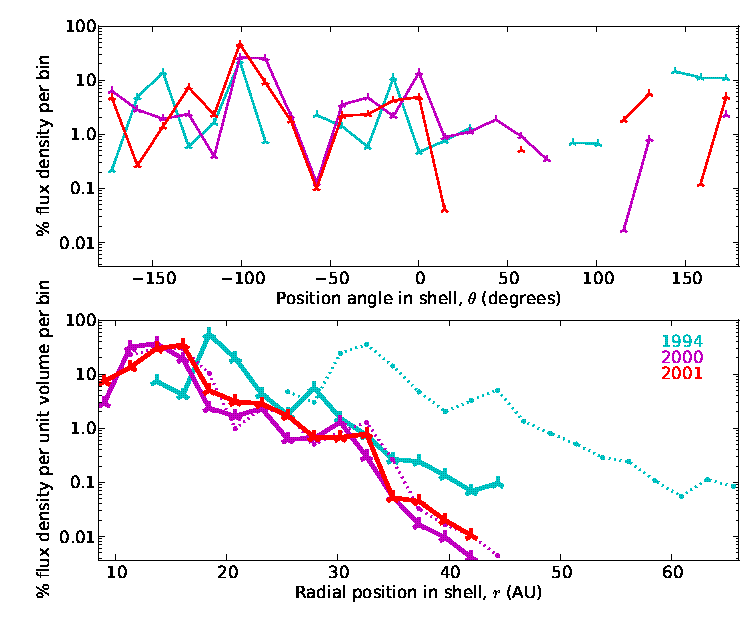}
      \caption{U Her, see Fig.~\ref{VX_r_theta.png} for details.}
         \label{UH_r_theta.png}
   \end{figure}

\begin{figure}
   \centering
   \includegraphics[angle=0, width=9cm]{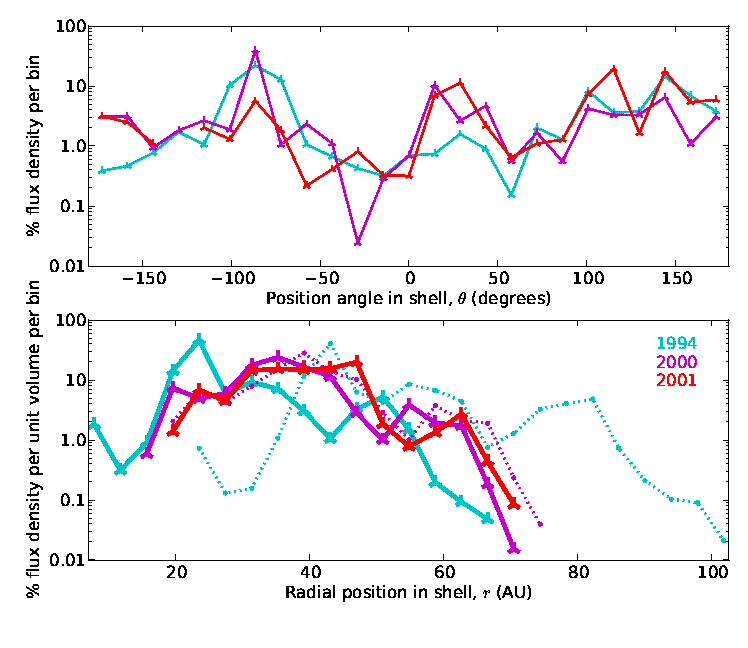}
      \caption{IK Tau, see Fig.~\ref{VX_r_theta.png} for details.}
         \label{IK_r_theta.png}
   \end{figure}

\begin{figure}
   \centering
   \includegraphics[angle=0, width=9cm]{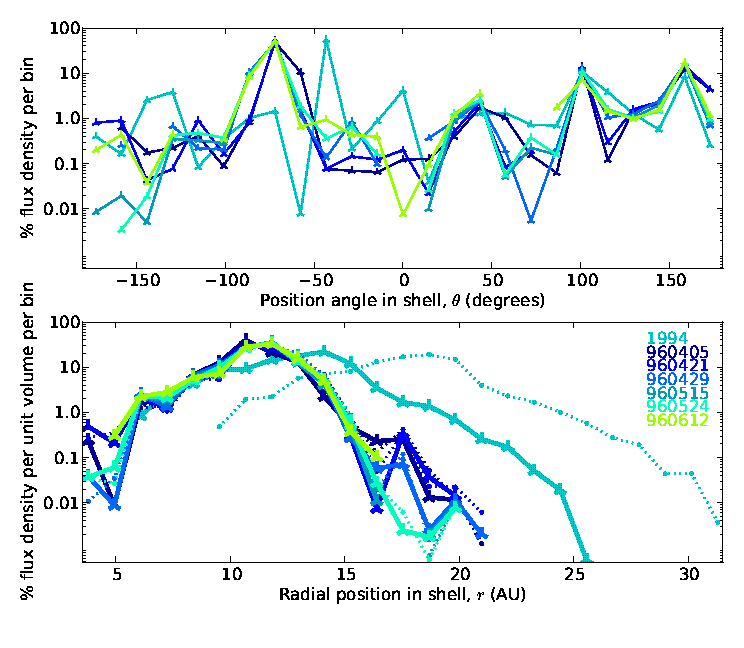}
      \caption{RT Vir, see Fig.~\ref{VX_r_theta.png} for details.}
         \label{RT_r_theta.png}
   \end{figure}

\begin{figure}
   \centering
   \includegraphics[angle=0, width=9cm]{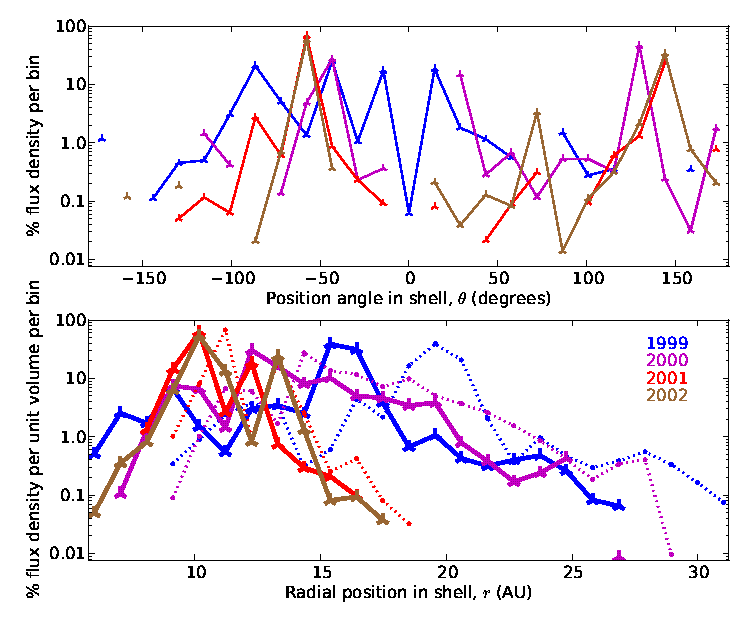}
      \caption{W Hya, see Fig.~\ref{VX_r_theta.png} for details.}
         \label{WH_r_theta.png}
   \end{figure}

We calculated the fractional flux density per unit volume in segments
of arc and in concentric, spherical sub-shells.  In all cases the
ranges shown are divided into 25 bins of equal widths in angle
$\theta$ (as projected against the plane of the sky) or radius $r$
(deduced from 3-D modelling). Since the plots are normalised, the
entire shell waxing or waning in response to stellar luminosity
changes has no effect.  Figs.~\ref{VX_r_theta.png}
to~\ref{WH_r_theta.png} show the angular and radial distributions of
flux as a percentage of the total flux at each epoch.

  We used the total expansion velocity $\v$ to predict the radial
position at the latest epoch, for each component observed at earlier
epochs.  The corresponding extrapolated radial flux distributions are
shown by the faint dotted lines. There are minor discrepancies due to binning different samples into each subshell.
% and since the volume of each sub-shell increases with $r$, the flux
%density per unit volume decreases slightly (the effect is reduced by
%the normalisation).
No compensation has been made for the effects of
changing temperature and density on maser efficiency.



The radial profiles in Figs.~\ref{VX_r_theta.png}
to~\ref{WH_r_theta.png} show that the brightest masers occur within
the inner half of the shell, frequently peaking around the first
quarter (as predicted by e.g. \citealt{Cooke85},
\citealt{Yates97}). However, there are significant differences between
the AGB observations, especially over time intervals of $\ge2$
years. There are most similarities between epochs with the most
matched features, as one would expect. The position angle profiles do
not show any systematic shifts which would indicate rotation, nor any
obvious preferred directions (note that $\theta$ is projected against
the plane of the sky and does not provide a full 3-D model of angular
distribution, such as was made for VX Sgr by M03).

VX Sgr shows fairly good agreement between the radial locations of the
peaks seen in 1999 and the extrapolated 1994 profile.  This is
consistent with the overall similarities in morphology as well as
individual feature matches described by M03. They found
good agreement between  the trigonometric ($w$)
and proper motion expansion ($V_{\mathrm a}$) velocities in the plane
of the sky
if the star is at $1.8\pm0.5$ kpc, within the
uncertainties of the generally adopted value of 1.7 kpc. The emission
at greater $r$ has faded, consistent with \citet{Cooke85}.

When the two brightest peaks in the 1994 S Per radial profile are
extrapolated, they appear to overtake the 1999 emission.  It is
possible that $\v$ has been overestimated for some maser features.  The
mean value of $w$, the component of motion in the plane of the sky
estimated trigonometrically, is 15.2 km s$^{-1}$, compared with 4.9 km
s$^{-1}$ $V_{\mathrm a}$ but Fig.~\ref{SPer_Va-abar.png} shows a large
scatter due to the large uncertainties in the individual measurements
at the large distance of S Per. It is also possible that the velocity
field is axisymmetric, with slower expansion in the plane of the sky,
where the brightest emission is located.  The dashed line in
Fig.~\ref{SP_r_theta.png} shows that if we extrapolate the 1994 data
at $0.4v$, the two highest peaks coincide with the 1999 peaks, with a
trough in between. R99 esimated the time when the material producing
the masers observed in 1994 left the star, and compared this with
AAVSO light curves, finding that the trough corresponded to several
years when the optical period amplitude was only about 1.5 magnitudes
instead of the more usual$\sim3$.

Among the AGB stars, 1996 RT Vir data peak shifts by the predicted
amount (equivalent to a change of about one radial bin from the first
to the last epoch) but the 1994 data are significantly different.  The
mean values of $w$ and $V_{\mathrm a}$ are 10 and $6\pm2$ km/s$^{-1}$,
in reasonable agreement given the model uncertainties.

U Ori and W Hya show only limited similarities between epochs, consistent
with the fewer matches and greater measurement uncertainties,
respectively. IK Tau shows  closer apparent similarities,
including with 1994 data.  However, the IK Tau velocity profiles
(Figs.~\ref{IKTau_M94_Gauss.png} and~\ref{IKTau_Match00-01_Gauss.png})
show that, apart from relatively faint red-shifted emission, the peaks
in 1994 were at very different velocities from later epochs.  
%$w$ is $\sim2 V_{\mathrm a}$, probably due to some spurious matches
%apparently in infall and/or to a non-spherical velocity field. It is
%possible that some maser features seen in all epochs (not just the
%2000--2001 matches identified) but matches between 1994 and epochs
%6--7 years later could not be made with any confidence because the
%large possible proper motions cause too much ambiguity.

The brightest U Her emission is located at a similar velocity at all
epochs, within the thermal line width plus slight acceleration
(Figs.~\ref{UHer_M94_Gauss.png} and~\ref{UHer_M00-01_MGauss.png}). It
is possible that the ring of emission seen in
Fig.~\ref{UHer_94_xyplot.png} corresponds to the fainter, outer ring
in Fig.~\ref{UHer_00-01_xyplot.png}.  The outer ring corresponds to
the small peak or knee in the 2000--2001 emission at $r\sim32$ AU in
Fig.~\ref{UH_r_theta.png}, the location of the extrapolated 1994 peak
(dotted line). If the 1994 peak was scaled down by a factor of about
20, consistent with maser emission fading at larger $r$, this would
produce the flux densities observed in 2000--2001. 

\begin{figure}
   \centering
   \includegraphics[angle=0, width=9cm]{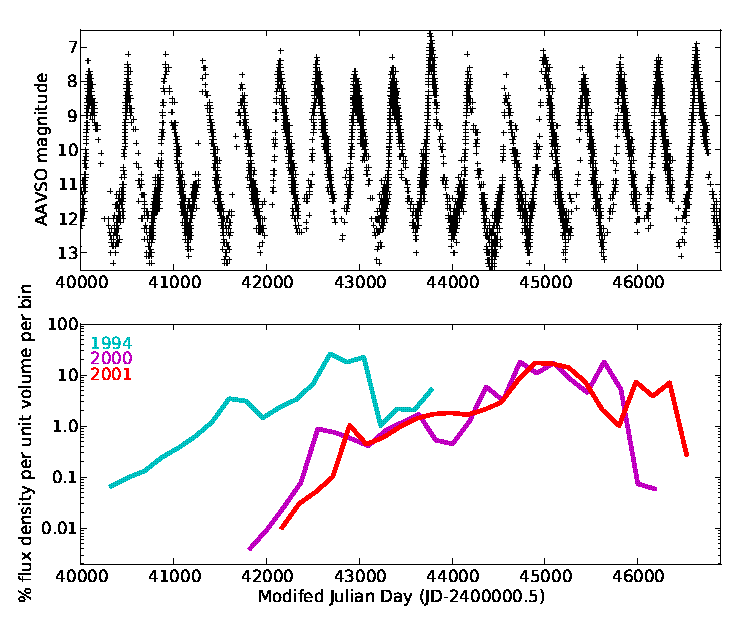}
      \caption{The lower panel shows the flux density profiles for U
      Her as a function of the time  when the masing material left the
      star. The upper panel shows the AAVSO light curve for the same period.}
         \label{UHer_massloss_time.png}
   \end{figure}

We estimated the date
relative to the observation date $t_{\star}$ at which each maser
component at position $r$, velocity $\v$, left the star, based on
Equation~\ref{eq:cross}:
\begin{eqnarray}
\frac{t_{\star}}{\mathrm{day}} &&= -1736 \times \nonumber \\ 
&& \left[ \left(\frac{r_{\mathrm
    i}/\mathrm{AU}}{\v_{\mathrm i} /\mathrm{km s}^{-1}}\right)  + \left(\frac{K_{\mathrm{grad}}}{\mathrm{km\,s}^{-1}/\mathrm{AU}}\right)^{-1} \ln\left(1\!+\!\frac{(r-r_{\mathrm i})K_{\mathrm{grad}}}{\v}\right)\right]
\label{eq:tcomp}
\end{eqnarray}
%\begin{eqnarray}
%t_{\star} &=&  1736\left( \left(\frac{r_{\mathrm i}/\v_{\mathrm i}}{\mathrm{AU} /\mathrm{km s}^{-1}}\right) \nonumber \\
%&+& \left(\frac{K_{\mathrm{grad}}}{\mathrm{km\,s}^{-1}/\mathrm{AU}}\right)^{-1} \ln\left[1+\frac{K_{\mathrm{grad}}(r-r_{\mathrm i})}{\v}\right]\right)
%\label{eq:tcomp}
%\end{eqnarray}
We assume that the material reaches $r_{\mathrm i}$ from the star
travelling at average velocity $\v_{\mathrm i}$ and ignore the light
travel time from U Her to Earth. Fig.~\ref{UHer_massloss_time.png}
shows the flux density summed radially into 25 shells covering this
time interval (converted to MJD), (similar to Fig.~\ref{UH_r_theta.png} but with the density profile reversed with respect to the
abscissa direction), compared with the AAVSO light curve.
The two bright shells of U Her maser emission correspond to MJD around
42800 and 45000, but there is nothing obviously unusual about the
light curved around these times. The intervening maser dip coincides
with a fairly bright U Her maximum but it is not exceptional;
inspection of AAVSO data shows that it has exceeded mag 7 twenty times
in the past century. 
%The mean value of $w$ is greater than the mean $V_{\mathrm a}$. This
%could be due to measurement errors or mismatching but for U Her there
%is also a large uncertainty in estimating $\v_{\mathrm o}$ since we do
%not detect the front and back caps of the shell. An overestimate
%would affect $w$ and also the total crossing time.

%VX Sgr 0.87\pm0.12* S Per 0.12\pm0.13* U Ori 0.47\pm0.03 
%U Her 0.34\pm0.07 IK Tau 0.10\pm0.03 RT Vir 0.01\pm0.42* W Hya 0.19\pm0.04

\subsection{Cloud overlap causes local flare in W Hya}
\label{sec:WH-var}

The 22-GHz peaks from W Hya are generally a few hundred Jy, but
exceeded 1000 Jy on several occasions during Pushchino monitoring
since 1982 \citep{Rudnitskij99w}. Most recorded emission is confined
to the velocity range 36--44 km s$^{-1}$ but the MERLIN data for
20000405 show a 100-Jy peak at 45 km s$^{-1}$
(Fig.~\ref{WHya_M99-00_Gauss.png}).  This was not present in the
Pushchino spectra just a few weeks before or after.

\begin{figure}
   \centering
   \includegraphics[angle=0, width=8.5cm]{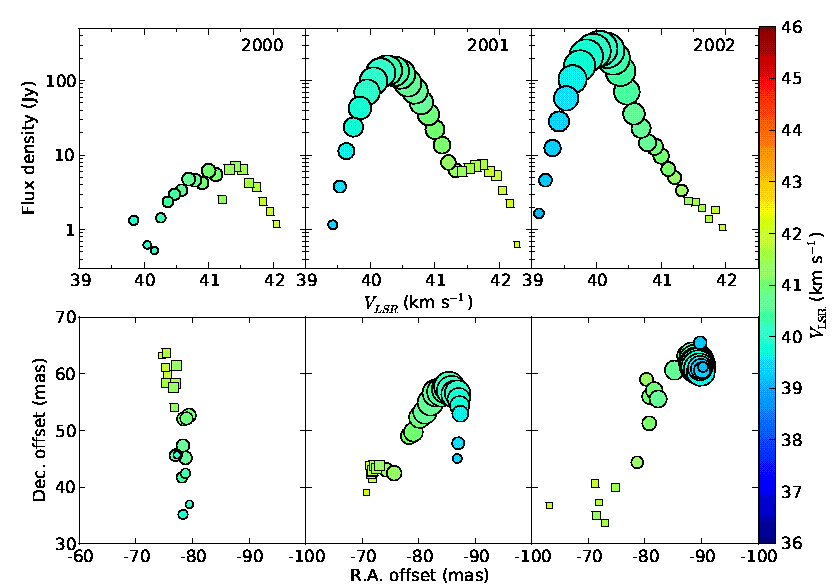}
      \caption{MERLIN observations of the maser feature at the
        velocity of the flare from W Hya seen in Pushchino data at
        20011023 (Fig.~\ref{WHya_M01-02_Gauss.png}). The upper three
        plots show the velocity profiles at the epochs bracketing the
        flare.  The brighter, more blue-shifted feature is shown by
        circles, and an adjacent feature by squares. Symbol size is
        proportional to the square root of flux density. The lower
        three plots show the corresponding maser positions, showing
        that the weaker feature has moved from N to S and brighter
        feature has moved across it in the opposite direction from S
        to NW.}
         \label{WHya_flare.png}
   \end{figure}

An even stronger flare is seen in Fig~\ref{WHya_M01-02_Gauss.png} at
40.4 km s$^{-1}$, reaching nearly 1000 Jy in 20010227 and over 2800 Jy
in 20011023, was detected by both Pushchino and MERLIN.
Fig.~\ref{WHya_flare.png} shows the MERLIN peak in this spectral
region in 2000, 2001 and 2002 (located in the NW of the plots shown in
Fig.~\ref{WHya_xyplot.png}). The peak is shown by circles and a
fainter, nearby feature, with a distinct, Gaussian spectral profile,
is shown by squares.  The brighter peak is 7, 136 and 269 Jy and the
fainter is 6, 7 and 2 Jy, at the three successive epochs. The relative
velocities are $< \Delta V_{\mathrm {th}}$, the fainter peak being
about 1 km s$^{-1}$ more redshifted at all epochs, although both show
a small drift.  Interestingly, the fainter feature is to the N in 2000
but to the S at later epochs, an apparent relative velocity in the
plane of the sky of $\sim5$ km s$^{-1}$.

The MERLIN data were taken close to stellar minimum whilst the
brightest Pushchino peak was seen close to stellar maximum but this
cannot be the main reason for the flare since only one feature is
affected.  It is unlikely that a large-scale shock is involved, since
in that case one would expect both adjacent features to be affected.
The third possible mechanism for an isolated flare is two overlapping
clouds.  The brighter cloud is more blueshifted and therefore in the
foreground, and could dramatically amplify the background cloud
\citep{Kartje99}. Fig.~\ref{WHya_flare.png} shows that the brigher
cloud has a proper motion consistent with outflow; the full extent of
the background cloud may be obscured.

\subsection{Variability of multi-epoch features}
\label{sec:shellvary}

 \begin{figure}
   \centering
   \includegraphics[angle=0, width=9cm]{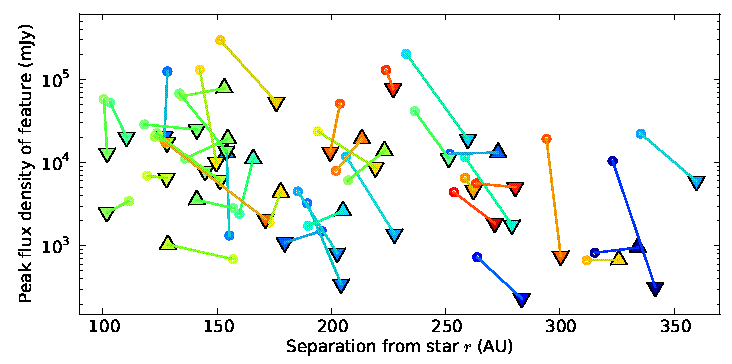}
      \caption{The flux densities of the matched VX Sgr features as a
      function of their radial distance from the star, $r$. Symbol
      colour represents $V_{\mathrm {LSR}}$ as in
      Fig.~\ref{VXSGR_Match_xyplot.png}, and the last epoch is
      outlined with a triangle, pointing in the direction of flux
      change.  }
         \label{VXSgr_Pk_r_match.png}
   \end{figure}

 \begin{figure}
   \centering
   \includegraphics[angle=0, width=9cm]{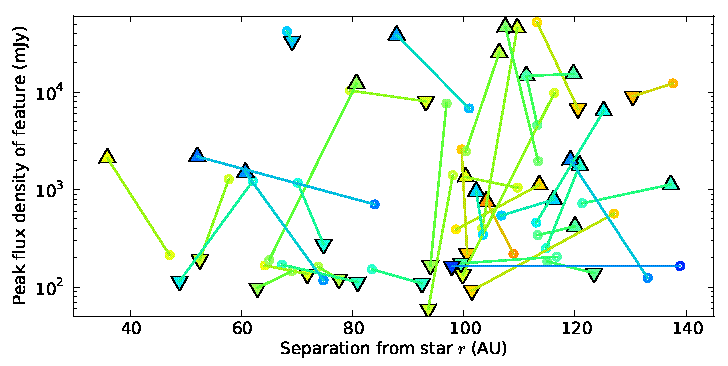}
      \caption{The flux densities of the matched S Per features as a
      function of their radial distance from the star, see
      Fig.~\ref{VXSgr_Pk_r_match.png}. }
         \label{SPer_Pk_r_match.png}
   \end{figure}

 \begin{figure}
   \centering
   \includegraphics[angle=0, width=9cm]{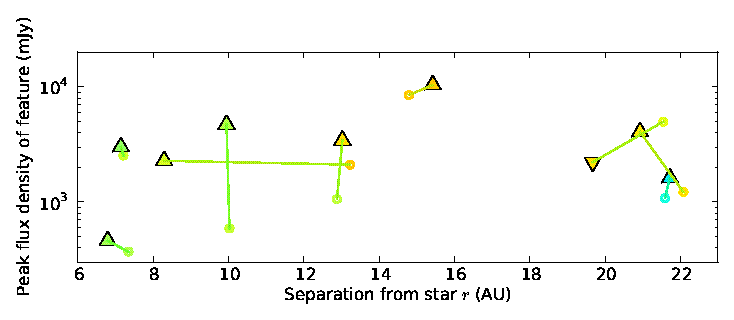}
      \caption{The flux densities of the matched U Ori features as a
      function of their radial distance from the star, see
      Fig.~\ref{VXSgr_Pk_r_match.png}.}
         \label{UOri_Pk_r_match.png}
   \end{figure}

 \begin{figure}
   \centering
   \includegraphics[angle=0, width=9cm]{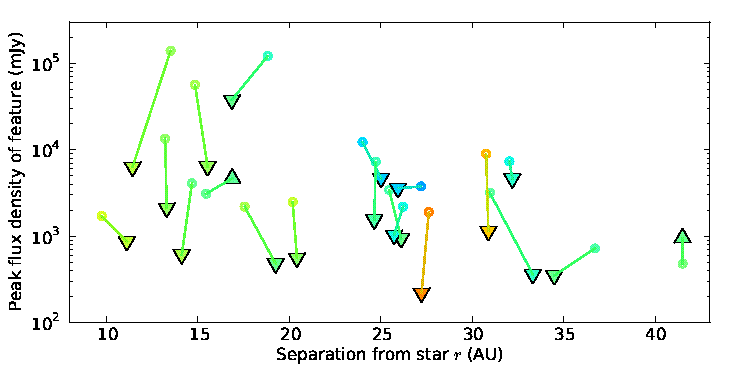}
      \caption{The flux densities of the matched U Her features as a
      function of their radial distance from the star, see
      Fig.~\ref{VXSgr_Pk_r_match.png}. }
         \label{UHer_Pk_r_match.png}
   \end{figure}

 \begin{figure}
   \centering
   \includegraphics[angle=0, width=9cm]{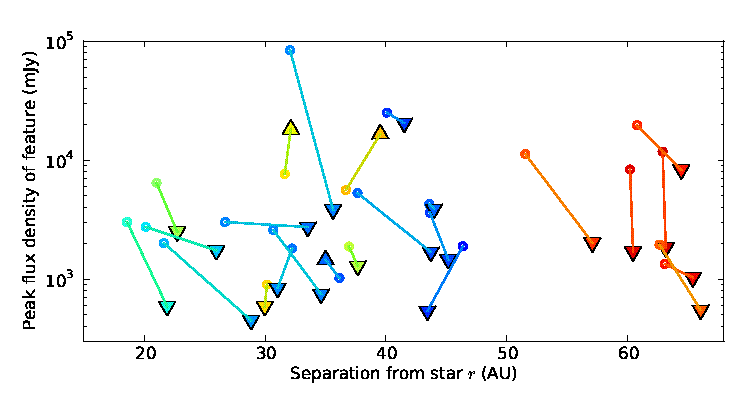}
      \caption{The flux densities of the matched IK Tau features as a
      function of their radial distance from the star,  see
      Fig.~\ref{VXSgr_Pk_r_match.png}. 
}
         \label{IKTau_Pk_r_match.png}
   \end{figure}

\begin{figure}
   \centering
   \includegraphics[angle=0, width=9cm]{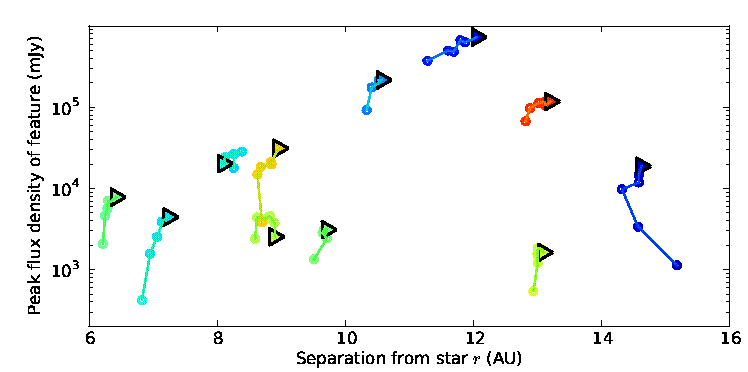}
      \caption{The flux densities of the RT Vir features matched at
      all epochs in 1996  as a
      function of their radial distance from the star,  see
      Fig.~\ref{VXSgr_Pk_r_match.png}}
         \label{RTVir_Pk_r_match.png}
   \end{figure}

\begin{figure}
   \centering
   \includegraphics[angle=0, width=9cm]{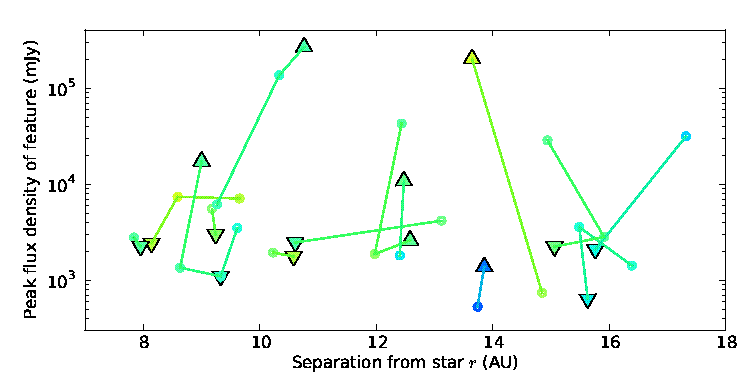}
      \caption{The flux densities of the matched W Hya features as a
      function of their radial distance from the star, see
      Fig.~\ref{VXSgr_Pk_r_match.png}}
         \label{WHya_Pk_r_match.png}
   \end{figure}

We looked for any systematic trends in the brightness evolution of
maser features identified at multiple epochs.  The total distance $r$
of each feature from the assumed stellar position was estimated as
described in Section~\ref{sec:time}.  The 3-D modelling is not only
vulnerable to measurement errors but also to deviations from the
assumptions of spherical symmetry and linear acceleration, but with
only two epochs of data (in most cases) these effects cannot be
separated.  Figs.~\ref{VXSgr_Pk_r_match.png}
to~\ref{WHya_Pk_r_match.png} show the change in flux density as a
function of $r$ for matched features.

S Per and W Hya do not show a clear trend, although this may be
obscured due to the larger position uncertainties owing to the greater
distance and low elevation, respectively.  There may be some spurious
matches such as the blue-shifted S Per features with very large
changes in $r$, since we did not want to impose an a-priori 3D model
but used the same position matching critera at all
velocities. However, if spherical expansion dominates, the apparent
large proper motions are unlikely for highly blue-shifted features. A
possible sign of the effects of asymmetric expansion is described in
Section~\ref{sec:time}.

Features throughout the maser shell behave similarly in the remaining
sources.  The majority of VX Sgr, U Her and IK Tau matched
features get fainter, generally consistent with
Figs.~\ref{VXSgr_Match94-99_Gauss.png},
~\ref{UHer_M00-01_MGauss.png} and  \ref{IKTau_Match00-01_Gauss.png}.  Most of the U Ori matched features
get slightly brighter, consistent with
Fig.~\ref{UOri_M00-01_Gauss.png}.  
The tendency for $r$ to decrease
in U Ori may be due to a large random component to its proper motions
(Section~\ref{sec:expand}). 

RT Vir has the best sampled data.  Fig.~\ref{RTVir_Pk_r_match.png}
shows that all 11 matched features increase in brightness relative to
the first interferometry epoch of 1996.  Nine of the features continue
to increase, although a few level off or slightly decrease at the
final epochs. The other two features reach maximum at intermediate epochs.
Fig.~\ref{RTVir_dadth.png} indicates the fractional increase.
Figs.~\ref{RTVir_M94-96_Gauss.png} and~\ref{RTVir_M96all_Gauss.png}
show that the main features of the RT Vir spectrum dropped in flux
density between the Pushchino observations in 19960319 and the MERLIN
observations in 19960405. The brightness then increased steadily
during the interferometric monitoring but dropped sharply again
between the last MERLIN epoch in 19960612 and Pushchino observations
in 19960619. Fig.~\ref{IKTauRTVirWHya_AAVSO.png} shows that the MERLIN
observations were made during the decline of optical phase.

\subsection{Possible causes of variability}
\label{sec:why}

 In all objects, individual maser features can show idiosyncratic
behaviour. Some of this can be attributed to local turbulence
(Section~\ref{sec:survive}).  Extreme flares are confined to small
spectral and spatial regions in our data, such as the W Hya event
(Section~\ref{sec:WH-var}). There are several possible types of more
general behaviour.  Masers switch on, brighten relatively rapidly and
then decline slowly as the parent gas enters and eventually leaves the
region where maser excitation is possible, over several decades, as
predicted by \citet{Cooke85} and seen in Figs.~\ref{VX_r_theta.png}
to~\ref{WH_r_theta.png}. 

There is no obvious correlation between flux density changes and
radial position in the shell of any CSE studied. In those objects
which do show systematic behaviour, the whole shell is affected,
suggesting a response to stellar illumination (the light crossing time
for an entire 22-GHz H$_2$O shell is a couple of days or less).
Figs.~\ref{VXSgrSPer_AAVSO.png} to~\ref{IKTauRTVirWHya_AAVSO.png} show
that there is no systematic relationship between maser
brighness and the optical phase at the times of our observations. 
However, this is not surprising, since IR is expected to have more
effect on 22-GHz masers.  \citet{Smith06} found $0.1-0.2 P$ lags of IR
variability with respect to the optical light curves for Miras, where
$P$ is the stellar period. There was no clear relationship for other
sources such as SRbs.  

Long-term, well sampled single dish data have shown some correlations
between H$_{2}$O maser variability and stellar periods.
\citet{Shintani08} monitored a large number of AGB and RSG, including
all those in our sample, for $3-4$ yr using the VERA Iriki single
dish. They fitted periods to the integrated 22-GHz emission and
compared these with AAVSO data. Excluding fits marked `bad', they
deduced lags, as fractions of $P$, for: U Ori $0.47\pm0.03$; U Her
$0.34\pm0.07$; IK Tau $0.10\pm0.03$; W Hya $0.19\pm0.04$.  $\sim20$ yr
of Pushchino monitoring showed similar lags of 0.2--0.4 $P$ for U Ori
\citep{Rudnitskij00} but W Hya produces semi-regular flares with lags
of $0-2.1 P$ \citep{Rudnitskij99w}. No clear relationship was found
for RT Vir. As for the RSG, \citet{Pashchenko99} and \citet{Lekht05}
found phase delays of $\le1$ and $0.01-0.5 P$ for VX Sgr and S Per,
respectively. In summary, the clearest relationship is seen for Miras,
and tends to align 22-GHz maser and IR periodicity.  A relationship
between stellar IR and maser brightness is not excluded for the SRbs
and RSGs but would not be expected to follow the optical light
curve. In fact, OH mainline masers around SRbs show a mixture of
cyclic and erruptive behaviour with variability on timescales a short
as a month \citep{Etoka01}, so H$_2$O masers  closer to the star
would be expected to be even more variable.

Changing stellar IR will affect the 22-GHz maser in at least 3 ways.
Firstly, it is collisionally pumped, so the intensity is sensitive to
heating \citep{Cooke85}.  Secondly the pump cycle may be suppressed by
a strong IR radiation field \citep{Yates97}.  These phenomena would
affect the whole shell at close to light speed, i.e. within a few
days. This could explain the behaviour of VX Sgr, U Ori, U Her, IK Tau
and RT Vir, where the majority of matched maser features change flux
density in concert between our observational epochs.  Thirdly, the
changing radiation pressure on dust will affect the velocity field
and, for example, if acceleration decreases, radial beaming will be
enhanced at the expense of tangential beaming \citep{Engels97},
producing anti-correlated behaviour in the extremes versus the middle
part of the spectrum.  Our data are not well enough sampled over more
than one stellar period, which would be required to test whether there
is a correlation between the maser structure and the phase cycle.

\citet{Rudnitskij90} suggested that the stellar pulsations shock the
maser shell, producing brightness variations which would lag the
optical phase by an interval equivalent to the shock crossing time
from the stellar surface to the 22-GHz $r_{\mathrm{i}}$.
\citet{Shintani08} found evidence for lags of $>0.5 P$ in many
objects, taken to support a similar model.  \citet{Reid97} showed that
the pulsation velocity at or above the radio photosphere is unlikely
to exceed 5 km s$^{-1}$ in AGB stars.  The maximum outflow velocities
of SiO masers provide another indication of the velocity of possible
impacts on the H$_{2}$O maser shell. The SiO $J=1-0$, $\v=1$ and $2$
transitions in all our targets were surveyed by \citet{Kim10}.
Although SiO masers are even more variable than H$_{2}$O, results from
other published observations are generally within a thermal linewidth.
The SiO velocities are 9--10 km s$^{-1}$ for VX Sgr and S Per, 6 km
s$^{-1}$ for U Ori, IK Tau and W Hya, 4 km s$^{-1}$ for U Her and 5 km
s$^{-1}$ for RT Vir.  This is comparable to $\v_{\mathrm{i}}$ and $\ll
\v_{\mathrm{o}}$ for all objects except U Ori and W Hya where it is
comparable to $\v_{\mathrm{o}}$.

A shock velocity of 5--10 km s$^{-1}$ corresponds to a proper motion
of $\approx1-2$ AU yr$^{-1}$ so a single shock would not affect an
entire shell; even the smallest (the SRbs) are $\ga10$ AU thick.
It is possible that shocks affect the inner edge of the 22-GHz shell,
and possibly penetrate deeper in objects where $\v_{\mathrm{i}}$ is
smallest, such as U Ori and U Her, which show distinctive `matter
bounded' beaming properties for some masers (R11).

\citet{Hofner95} developed a model in which C-rich dust production is
intrinsically episodic. The formation of dust during one time interval
leads to back-warming and suppression of further grain growth nearer
the star, until radiation pressure has driven the initial dust shell
away from the star.  The models predict enhanced dust formation
episodes at intervals similar to, or longer than the stellar period,
but not all episodes are equally efficient. The dust:gas ratio, and
thus acceleration, is therefore a function of distance from the star.
\citet{Jeong03} applied this model to an O-rich Mira.  The amplitude
of velocity variations depend on the extinction coefficients used,
giving $\sim1-7$ km s$^{-1}$ at 15 $R_{\star}$.  Our data could be
used to test more precise models for our stars.

\section{Relationship between maser cloud size and parent star}
\label{sec:cloudsize}

\begin{figure}
   \centering
   \includegraphics[angle=0, width=9cm]{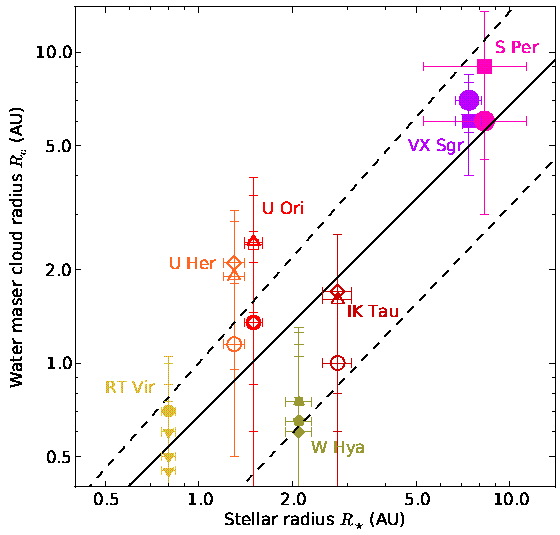}
      \caption{Water maser cloud radius $R_{\mathrm c}$ as a function
      of $R_{\star}$. The different epochs are shown by different
      symbol shapes as in Fig.~\ref{Allepochs_allrv.png}.  RSG, Miras
      and SRb are shown by large, hollow and small symbols,
      respectively. The solid and dashed lines show the slope of an
      error-weighted fit to the relationship between $R_{\mathrm c}$
      and  $R_{\star}$, and the dispersion in the relationship.}
         \label{CloudsizeRadius.png}
   \end{figure}

The average cloud radius for each epoch $R_{\mathrm c} =
\overline{l}/2$. Fig.~\ref{CloudsizeRadius.png} shows that this is an
almost linear
function of stellar radius $R_{\star}$.  The error-weighted fit and
dispersion are shown by the solid and dashed lines, where the
relationship is
\begin{equation}
\log R_{\mathrm c} = (-0.17\pm0.17) + (1.00\pm0.13)\log R_{\star}
\label{eq:csize}
\end{equation}
Note that W Hya cloud sizes
 are most likely to be underestimated, see
Section~\ref{sec:flux}. 

%The IK Tau average cloud size in 1994 also seems small; at this epoch
%an exceptionally large number of clouds (256) were identified;
%Fig~\ref{IKTau_94_xyplot.png} suggests that conditions were
%exceptionally favourable for milli-arcsec

 The brightest part of the
shell, where the highest number of clouds are detected, is close to
the inner radius (see Section~\ref{sec:time}), so we use $r_{\mathrm {typical}} =r_{\mathrm i} +
(r_{\mathrm o}-r_{\mathrm i})/4$ as the typical location in the shell
where clouds have radius $R_{\mathrm c}$.
If the clouds are formed at the surface of the star and expand
uniformly, radially in the wind, the average cloud radius,
extrapolated back to the surface of the star, is approximately
(0.04--0.09)$R_{\star}$ for the RSG, (0.05--0.12)$R_{\star}$ for the
Miras and (0.06--0.07)$R_{\star}$ for the SRbs, i.e. $\sim 5-10\%
R_{\star}$ for all stars.

This implies that the size of the clouds is determined by the size of
the star. The 22-GHz H$_{2}$O masers arise from clouds which are much
denser than the average wind density (R99, B03, M03,
Section~\ref{sec:density}) and which must survive for decades (even if
individual masers do not). If these clouds were to be formed in the
outflow after it had left the star, the cause would be some local wind
phenomenon such as cooling due to dust formation.  The length scale of
this process would be determined by basic physical properties, implying that all
clouds would form on the same size scales. This is not consistent with
our data.

%\citet{Hartquist97} suggested that Parker instabilities in the stellar
%atmosphere could produce magnetised, SiO-masing clumps.  More recent
%stellar, Zeeman splitting and SiO maser models (\citealt{Reid97};
%\citealt{Elitzur98}; \citealt{Gray09}) provide values for the magnetic
%field, density and so on which differ by orders of magnitude from
%those used by \citet{Hartquist97}, so the conditions for fragmentation
%are probably not met.

Other observations have revealed inhomogeneities and clumps closer to
the star.  \citet{Chiavassa09} compared VLTI image reconstruction of
VX Sgr at $\lambda$ 1.6 $\mu$m with three-dimensional simulations of
surface convection. The best-fitting synthetic image possessed two
convective cells of 4--5 milli-arcsec (comparable to R$_{\star}$ at this
wavelength), with many 0.5--1 milli-arcsec granules, two being particularly
prominent.  A third spot appears at $\sim10$ milli-arcsec separation from the
centre, i.e. outside the photosphere. Although the fine details of the
models are not uniquely constrained, the stellar surface clearly
possesses structure on scales including the extrapolated birth size of
the clouds, $\sim 5-10\%$ $R_{\star}$.  Similar models for AGB stars
have so far only been published for C-rich objects \citep{Freytag08},
which also produce a few large convection cells. These have a
piston effect comparable to traditional pulsations, and thus are
potentially capable of enhancing mass loss on sub-surface scales. U
Ori was imaged by \citet{Pluzhnik09} using IOTA. The H$_2$O band image
at 1.78 $\mu$m shows three bright clumps just outside the photospheric
radius.\footnote{\citet{Pluzhnik09} identify the (non-astrometric) IR
  clumps with H$_2$O maser features imaged by \citet{Vlemmings05} but
  even if the alignment were correct, it ignores projection effects;
  22-GHz masers cannot exist so close to the star.}  Recent VLTI
observations of four O-righ Miras (not in our sample) by
\citet{Wittkowski11} show deviations from symmetry in the H$_2$O band
indicating the presence of clumpiness on scales approaching 10\% of
the stellar size.

At much greater distances, around carbon stars, detatched CO shells
have been observed and modelled by \citet{Bergman93} and
\citet{Olofsson96}.  The clumps were not individually resolved, but
the inhmogeneity was inferred from modelling larger-scale images. The
objects studied have mass loss rates in the range
$(0.4-4)\times10^{-5}$ M$_{\odot}$ yr$^{-1}$ concentrated in clumps of
mass $\sim2 \times 10^{-5}$ M$_{\odot}$, similar to the values for our
RSG (Table~\ref{tab:mloss}).  The model of \citet{Bergman93} is based
on S Sct, using a clump radius $R_\mathrm{c} \approx 1350$ AU at a
distance $r\approx36000$ AU from the star. \citet{Olofsson96} model
the cloud expansion as $R_\mathrm{c} \propto r^{0.8}$ for steady
outflow and limited cooling.  Extrapolated back to $r=60$ AU gives
$R_\mathrm{c} \approx 8$ AU, in agreement with the estimated H$_2$O
cloud radii around S Per and VX Sgr.  Recent measurements of dust
clumps in two detatched CO shells \citep{Olofsson10} yield similar
sizes.

\section{Summary and further work}
\label{sec:conclusions}

We have resolved the detailed structure of the approximately
spherical, thick H$_{2}$O maser shells around two RSG (VX Sgr, S Per),
three Miras (U Ori, U Her, IK Tau) and two SRb stars (RT Vir, W Hya),
at multiple epochs. MERLIN detects all the 22-GHz maser emission (or
possibly almost all, in the case of the closest object), at high enough
resolution to measure the sizes and proper motions of individual
clouds. We compare these images with single dish monitoring by the
Pushchino radio telescope. Our results provide new insights into the
development of the stellar wind in the region where the escape
velocity is exceeded.

\paragraph{\bf Survival}

The H$_{2}$O maser shells are 10--250 AU thick with crossing times
from a few tens of years for AGB stars to nearly a century for RSG, but
most individual maser features survive less than 1--2 yr and 1--2
decades, in the AGB stars and RSG, respectively. The disappearance and
reappearance of individual masers identified by comparing imaging with
single dish monitoring shows that this contradiction is resolved if
the masers emanate from long-lived clouds. These undergo subsonic or
mildly supersonic turbulence of a few km s$^{-1}$, which causes
detectable masers to be alternately boosted and supressed without
destroying the clouds themselves.

\paragraph{\bf Expansion proper motions}

The proper motions of VX Sgr, S Per, U Her, IK Tau, RT Vir and most
epochs of W Hya are unambiguously dominated by spheroidal expansion.
There is a hint of rotation in the IK Tau maser proper motions.  The
small tangential component could be caused by random motions, which is
also likely to be a significant factor affecting U Ori. The radial
proper motion velocities (with respect to the assumed stellar
position) are generally consistent with $V_{\mathrm{LSR}}$
measurements, but the former have larger measurement errors and
scatter, especially for the low declination source W Hya. Discrepancies
between velocities derived from proper motion and $V_{\mathrm{LSR}}$
measurements could also be due to asymmetry, as seen for VX Sgr
(M03). 

\paragraph{\bf Acceleration}
The 22-GHz masers lie outside the dust formation zone, where the wind
is driven by radiation pressure on grains.  The outflow velocity
increases twofold or more across the maser shells of all
sources. Moreover, comparison of different sources shows that the
larger the inner and outer shell radii, the greater the expansion
velocities at the outer rims, although the acceleration is more
gradual. This is likely to be due to changes in the dust optical
properties so a higher proportion of radiation is absorbed, and/or
increasing optical depth so that the stellar photons are re-emitted
making radiation pressure more efficient.  This is consistent with
\emph{Herschel} and other results which show that terminal velocity is
not reached until hundreds of $R_{\star}$.

\paragraph{\bf Dense clumps}
The wind at $\sim5-50 R_{\star}$ is concentrated in clumps which are
40--110 times denser than the wind average. They comprise between a
fifth and almost all of the mass in the stellar wind in the maser
shell, although accurate comparisons are difficult due to
uncertainties in measuring the total mass loss rate. The volume
filling factor is less than 1\%. At least two to six clouds
are formed per stellar period.

\paragraph{\bf Flare due to cloud overlap}
Single-dish monitoring of H$_2$O masers reveals various forms of
variability; in some instances a single spectral feature brightens by
more than one order of magnitude for between a few weeks and a year or
two.  One such flare, in W Hya, was imaged at 3 epochs from
2000--2002, bracketing the 2500-Jy maximum captured by Pushchino. The
images show that the variable cloud appears to pass in front of
another. Two sets of components are spatially differentiated, at
slightly different $V_{\mathrm{LSR}}$, both with distinct Gaussian
spectral profiles. The flaring feature is initially to the S of the
other cloud, both $<10$ Jy. Two years later it has exceeded 400 Jy and
moved to the NW of the fainter feature (in a direction consistent with
outflow).  This is a dramatic confirmation of the predictions of
\citet{Kartje99} for maser amplification by overlapping clouds.

\paragraph{\bf Variability throughout the maser shell}
The majority of matched features in each of VX Sgr, U Her and IK Tau
decrease in flux density between the two imaging epochs regardless of
their position in the maser shell; the majority of U Ori matched
features increase.  The intervals between epochs are of order 1--2
stellar periods, but there is no obvious connection between optical
phase and brightening or dimming.  Features throughout the RT Vir
maser shell, imaged six times during the decline of optical
brightness, rise and then slightly decline again in 22-GHz flux
density. S Per and W Hya matched features show an apparently random
mixture of brightening and dimming. The coordinated behaviour of
features in 5 CSEs suggests that radiative effects play a major role
in determining maser brightness, probably linked to the IR stellar
phase (which lags the optical). The 22-GHz maser shells are 10--200 AU
thick, so the light travel time is less than a few days but a shock at
5--10 km S$^{-1}$ would take many years to cross even the thinnest
shell.  Nonetheless, the H$_2$O masers may be affected by shocks close
to the inner rim of the shell.

\paragraph{\bf Cloud size depends on star size}

The average radius of water maser clouds is 0.5--2 AU for the AGB
stars and 6--9 AU for the RSG, showing a close dependence on
$R_{\star}$.  Assuming the clouds expand steadily in the outflow, this
corresponds to birth radii $\sim5-10\% R_{\star}$.   Previous CO and
dust imaging have hinted at similar clumping scales and overdensity,
but only MERLIN H$_{2}$O maser observations have well-resolved all the
emission from such clumps, close to the star. The dependence on
stellar size suggests that stellar surface phenomena such as
convection cells determine the scale of clumps.

\paragraph{\bf Future work}
We are in the process of analysing multi-epoch MERLIN and VLBI
observations of OH masers which probe different conditions in these
objects. A future paper will compare the H$_2$O morphology with the
distribution of OH masers to investigate asymmetry and inhomogeneity.
There are very few estimates published for the sizes of SiO or OH
maser clouds in our sources, but we will seek available data where the
clouds are likely to be resolved but not resolved-out, to test whether
the sizes do increase with distance from the star.  

The advent of ALMA, \emph{e}-MERLIN, upgraded VLBI and other
interferometers will allow the unanswered questions to be tackled.
The current baselines of ALMA, up to 1 km, will locate sub-mm H$_2$O
masers with respect to the 22 GHz transition and the dust formation
zone, mapping changes in excitation conditions.  Eventual longer
baselines and full sensitivity will resolve both dust and molecular
species in clumps, showing whether the 22-GHz clouds are indeed
dustier and denser.  ALMA will also resolve the stars, as will
\emph{e}-MERLIN and the EVLA at its highest frequencies.  This will
reveal any convective disurbances and timely VLBA SiO maser monitoring
will show whether these lead directly to clumpy mass loss, and if the
SiO clumps in turn become dusty water maser clouds, resolved by ALMA
and \emph{e}-MERLIN in multiple transitions.  Simultaneous imaging of
the star and masers at 22 GHz (only previously achieved for a very few
objects e.g. \citealt{Reid90}; \citealt{Reid97}) will solve the
current astrometric ambiguities.

\begin{acknowledgements}
We recall the tremendous contributions of Jim Cohen (1948--2006) in
the early years of this project. We warmly thank the referee, Dieter
Engels, for a very thorough reading of the paper which led to great
improvements in clarity, Indra Bains for many useful contributions and
Robert Laing for trigonometric discussions.  We acknowledge, with
thanks, our use of data from the AAVSO (American Association of
Variable Star Observers), the GCVS (General Catalogue of Variable
Stars) and the Vizier and ADS services.

\end{acknowledgements}

\bibliographystyle{aa}

\bibliography{cse}

\begin{thebibliography}{92}
\expandafter\ifx\csname natexlab\endcsname\relax\def\natexlab#1{#1}\fi

\bibitem[{{Asaki} {et~al.}(2010){Asaki}, {Deguchi}, {Imai}, {Hachisuka},
  {Miyoshi}, \& {Honma}}]{Asaki10}
{Asaki}, Y., {Deguchi}, S., {Imai}, H., {et~al.} 2010, ApJ, 721, 267

\bibitem[{{Assaf} {et~al.}(2011){Assaf}, {Diamond}, {Richards}, \&
  {Gray}}]{Assaf11}
{Assaf}, K.~A., {Diamond}, P.~J., {Richards}, A.~M.~S., \& {Gray}, M.~D. 2011,
  MNRAS, 415, 1083

\bibitem[{Bains {et~al.}(2003)Bains, Cohen, Louridas, Richards,
  Rosa-Gonzal\'{e}z, \& Yates}]{Bains03}
Bains, I., Cohen, R.~J., Louridas, A., {et~al.} 2003, MNRAS, 342, 8, (B03)

\bibitem[{Bergman {et~al.}(1993)Bergman, Carlstr{\protect\"{o}}m, \&
  Olofsson}]{Bergman93}
Bergman, P., Carlstr{\protect\"{o}}m, U., \& Olofsson, H. 1993, A\&A, 268, 685

\bibitem[{{Bieging} {et~al.}(2000){Bieging}, {Shaked}, \&
  {Gensheimer}}]{Bieging00}
{Bieging}, J.~H., {Shaked}, S., \& {Gensheimer}, P.~D. 2000, ApJ, 543, 897

\bibitem[{Bowers {et~al.}(1993)Bowers, Claussen, \& Johnston}]{Bowers93}
Bowers, P.~F., Claussen, M.~J., \& Johnston, K.~J. 1993, AJ, 105, 284

\bibitem[{Bowers \& Johnston(1988)}]{Bowers88}
Bowers, P.~F. \& Johnston, K.~J. 1988, ApJ, 330, 339

\bibitem[{Bowers \& Johnston(1994)}]{Bowers94}
Bowers, P.~F. \& Johnston, K.~J. 1994, ApJS, 92, 189

\bibitem[{{Browne} {et~al.}(1998){Browne}, {Wilkinson}, {Patnaik}, \&
  {Wrobel}}]{Browne98}
{Browne}, I.~W.~A., {Wilkinson}, P.~N., {Patnaik}, A.~R., \& {Wrobel}, J.~M.
  1998, \mnras, 293, 257

\bibitem[{{Carlsberg Meridian Catalog 14 }(2006)}]{Carlsberg06}
{Carlsberg Meridian Catalog 14 }. 2006, {Copenhagen University Obs., IoA,
  Cambridge, UK and Real Instituto y Observatorio de La Armada en San
  Fernando}, 1304, 0

\bibitem[{Chapman \& Cohen(1986)}]{Chapman86}
Chapman, J.~M. \& Cohen, R.~J. 1986, MNRAS, 220, 513

\bibitem[{Chapman {et~al.}(1991)Chapman, Cohen, \& Saikia}]{Chapman91}
Chapman, J.~S., Cohen, R.~J., \& Saikia, D.~J. 1991, MNRAS, 249, 227

\bibitem[{Chapman {et~al.}(1994)Chapman, Sivagnanam, Cohen, \&
  LeSqueren}]{Chapman94}
Chapman, J.~S., Sivagnanam, P., Cohen, R.~J., \& LeSqueren, A.~M. 1994, MNRAS,
  268, 475

\bibitem[{{Chen} {et~al.}(2007){Chen}, {Shen}, \& {Xu}}]{Chen07}
{Chen}, X., {Shen}, Z.-Q., \& {Xu}, Y. 2007, CJAA, 7, 531

\bibitem[{{Chiavassa} {et~al.}(2009){Chiavassa}, {Plez}, {Josselin}, \&
  {Freytag}}]{Chiavassa09}
{Chiavassa}, A., {Plez}, B., {Josselin}, E., \& {Freytag}, B. 2009, A\&A, 506,
  1351

\bibitem[{Colomer {et~al.}(2000)Colomer, Reid, Menten, \&
  Bujarrabal}]{Colomer00}
Colomer, F., Reid, M.~J., Menten, K.~M., \& Bujarrabal, V. 2000, A\&A, 355, 979

\bibitem[{Condon(1997)}]{Condon97}
Condon, J.~J. 1997, PASP, 109, 166

\bibitem[{Condon {et~al.}(1998)Condon, Cotton, Greisen, Yin, Perley, Taylor, \&
  Broderick}]{Condon98}
Condon, J.~J., Cotton, W.~D., Greisen, E.~W., {et~al.} 1998, AJ, 115, 1693

\bibitem[{Cooke \& Elitzur(1985)}]{Cooke85}
Cooke, B. \& Elitzur, M. 1985, ApJ, 295, 175

\bibitem[{{Decin} {et~al.}(2010){Decin}, {Justtanont}, {De Beck}, {Lombaert},
  {de Koter}, {Waters}, {Marston}, {Teyssier}, {Sch{\"o}ier}, {Bujarrabal},
  {Alcolea}, {Cernicharo}, {Dominik}, {Melnick}, {Menten}, {Neufeld},
  {Olofsson}, {Planesas}, {Schmidt}, {Szczerba}, {de Graauw}, {Helmich},
  {Roelfsema}, {Dieleman}, {Morris}, {Gallego}, {D{\'{\i}}ez-Gonz{\'a}lez}, \&
  {Caux}}]{Decin10}
{Decin}, L., {Justtanont}, K., {De Beck}, E., {et~al.} 2010, A\&A, 521, L4

\bibitem[{Diamond {et~al.}(1987)Diamond, Johnston, Chapman, Lane, Bowers,
  Spencer, \& Booth}]{Diamond87}
Diamond, P.~J., Johnston, K.~J., Chapman, J.~M., {et~al.} 1987, A\&A, 174, 95

\bibitem[{{Ducourant} {et~al.}(2006){Ducourant}, {Le Campion}, {Rapaport},
  {Camargo}, {Soubiran}, {P{\'e}rie}, {Teixeira}, {Daigne}, {Triaud},
  {R{\'e}qui{\`e}me}, {Fresneau}, \& {Colin}}]{Ducourant06}
{Ducourant}, C., {Le Campion}, J.~F., {Rapaport}, M., {et~al.} 2006, A\&A, 448,
  1235

\bibitem[{Elitzur {et~al.}(1992)Elitzur, Hollenbach, \& McKee}]{Elitzur92a}
Elitzur, M., Hollenbach, D.~J., \& McKee, C.~F. 1992, ApJ, 394, 221, (EHM92)

\bibitem[{Engels {et~al.}(1997)Engels, Winnberg, Walmsley, \& Brand}]{Engels97}
Engels, D., Winnberg, A., Walmsley, C.~M., \& Brand, J. 1997, A\&A, 322, 291

\bibitem[{{Esipov} {et~al.}(1999){Esipov}, {Pashchenko}, {Rudnitskii}, \&
  {Fomin}}]{Esipov99}
{Esipov}, V.~F., {Pashchenko}, M.~I., {Rudnitskii}, G.~M., \& {Fomin}, S.~V.
  1999, Astronomy Letters, 25, 672

\bibitem[{Etoka {et~al.}(2001)Etoka, Blaskiewicz, Szymczak, \&
  Le~Squeren}]{Etoka01}
Etoka, S., Blaskiewicz, L., Szymczak, M., \& Le~Squeren, A.~M. 2001, A\&A, 378,
  522

\bibitem[{{Freytag} \& {H{\"o}fner}(2008)}]{Freytag08}
{Freytag}, B. \& {H{\"o}fner}, S. 2008, A\&A, 483, 571

\bibitem[{Greisen(1994)}]{Greisen94}
Greisen, E., ed. 1994, AIPS Cookbook (NRAO, Charlottesville, VA 22903-2475,
  USA)

\bibitem[{Habing(1996)}]{Habing96}
Habing, H.~J. 1996, A\&A Rev., 7, 97

\bibitem[{H{\protect\"{o}}fner {et~al.}(1995)H{\protect\"{o}}fner, Feuchtinger,
  \& Dorfi}]{Hofner95}
H{\protect\"{o}}fner, S., Feuchtinger, M.~U., \& Dorfi, E.~A. 1995, A\&A, 297,
  815

\bibitem[{{Imai} {et~al.}(2003){Imai}, {Shibata}, {Marvel}, {Diamond}, {Sasao},
  {Miyoshi}, {Inoue}, {Migenes}, \& {Murata}}]{Imai03}
{Imai}, H., {Shibata}, K.~M., {Marvel}, K.~B., {et~al.} 2003, ApJ, 590, 460

\bibitem[{{Ivezi{\'c}} \& {Elitzur}(2010)}]{Ivezic10}
{Ivezi{\'c}}, Z. \& {Elitzur}, M. 2010, MNRAS, 404, 1415

\bibitem[{{Jeong} {et~al.}(2003){Jeong}, {Winters}, {Le Bertre}, \&
  {Sedlmayr}}]{Jeong03}
{Jeong}, K.~S., {Winters}, J.~M., {Le Bertre}, T., \& {Sedlmayr}, E. 2003,
  A\&A, 407, 191

\bibitem[{Kartje {et~al.}(1999)Kartje, K\protect{\"o}nigl, \&
  Elitzur}]{Kartje99}
Kartje, J.~F., K\protect{\"o}nigl, A., \& Elitzur, M. 1999, ApJ, 513, 180

\bibitem[{{Kerschbaum} \& {Olofsson}(1999)}]{Kerschbaum99}
{Kerschbaum}, F. \& {Olofsson}, H. 1999, A\&AS, 138, 299

\bibitem[{{Kim} {et~al.}(2010){Kim}, {Cho}, {Oh}, \& {Byun}}]{Kim10}
{Kim}, J., {Cho}, S.-H., {Oh}, C.~S., \& {Byun}, D.-Y. 2010, ApJS, 188, 209

\bibitem[{Kirrane(1987)}]{Kirrane87}
Kirrane, T.-M. 1987, PhD thesis, University of Manchester

\bibitem[{{Knapp} {et~al.}(1998){Knapp}, {Young}, {Lee}, \&
  {Jorissen}}]{Knapp98}
{Knapp}, G.~R., {Young}, K., {Lee}, E., \& {Jorissen}, A. 1998, ApJS, 117, 209

\bibitem[{Lane {et~al.}(1987)Lane, Johnston, Bowers, Spencer, \&
  Diamond}]{Lane87}
Lane, A.~P., Johnston, K.~J., Bowers, P.~F., Spencer, J.~H., \& Diamond, P.~J.
  1987, ApJ, 323, 756

\bibitem[{Lekht {et~al.}(1999)Lekht, Mendoza-Torres, Pashchenko, \&
  Berulis}]{Lekht99}
Lekht, E.~E., Mendoza-Torres, J.~E., Pashchenko, M.~I., \& Berulis, I.~I. 1999,
  A\&A, 343, 241

\bibitem[{{Lekht} {et~al.}(2005){Lekht}, {Rudnitskij}, {Mendoza-Torres}, \&
  {Tolmachev}}]{Lekht05}
{Lekht}, E.~E., {Rudnitskij}, G.~M., {Mendoza-Torres}, J.~E., \& {Tolmachev},
  A.~M. 2005, A\&A, 437, 127

\bibitem[{{Levesque} {et~al.}(2005){Levesque}, {Massey}, {Olsen}, {Plez},
  {Josselin}, {Maeder}, \& {Meynet}}]{Levesque05}
{Levesque}, E.~M., {Massey}, P., {Olsen}, K.~A.~G., {et~al.} 2005, ApJ, 628,
  973

\bibitem[{{Marvel} {et~al.}(1998){Marvel}, {Diamond}, \& {Kemball}}]{Marvel98}
{Marvel}, K.~B., {Diamond}, P.~J., \& {Kemball}, A.~J. 1998, in ASP Conference
  Series, Vol. 154, Cool Stars, Stellar Systems, and the Sun, ed.
  {R.~A.~Donahue \& J.~A.~Bookbinder}, 1621

\bibitem[{{Matsumoto} {et~al.}(2008){Matsumoto}, {Omodaka}, {Imai}, {Shimizu},
  {Bushimata}, {Choi}, {Hirota}, {Honma}, {Inomata}, {Iwadate}, {Jike},
  {Kameno}, {Kameya}, {Kamohara}, {Kan-Ya}, {Kawaguchi}, {Kobayashi}, {Kuji},
  {Kurayama}, {Maeda}, {Manabe}, {Miyaji}, {Nakagawa}, {Nagayama}, {Nakashima},
  {Oh}, {Oyama}, {Sakai}, {Sakakibara}, {Sasao}, {Sato}, {Shibata}, {Shintani},
  {Sofue}, {Sora}, {Suda}, {Tamura}, {Tsushima}, \& {Yamashita}}]{Matsumoto08}
{Matsumoto}, N., {Omodaka}, T., {Imai}, H., {et~al.} 2008, PASJ, 60, 1039

\bibitem[{{Mayne} \& {Naylor}(2008)}]{Mayne08}
{Mayne}, N.~J. \& {Naylor}, T. 2008, MNRAS, 386, 261

\bibitem[{Mendoza-Torres {et~al.}(1997)Mendoza-Torres, Lekht, Pashchenko, \&
  Berulis}]{Mendoza-Torres97}
Mendoza-Torres, J.~E., Lekht, E.~E., Pashchenko, M.~I., \& Berulis, I.~I. 1997,
  A\&AS, 126, 257

\bibitem[{{Menut} {et~al.}(2007){Menut}, {Gendron}, {Schartmann}, {Tuthill},
  {Lopez}, {Danchi}, {Wolf}, {Lagrange}, {Flament}, {Rouan}, {Cl{\'e}net}, \&
  {Berruyer}}]{Menut07}
{Menut}, J.-L., {Gendron}, E., {Schartmann}, M., {et~al.} 2007, MNRAS, 376, L6

\bibitem[{Monnier {et~al.}(2004)Monnier, Millan-Gabet, Tuthill, T, T, T, T, T,
  T, T, T, \& T}]{Monnier04}
Monnier, J.~D., Millan-Gabet, R., Tuthill, P.~G., {et~al.} 2004, ApJ, 605, 436

\bibitem[{Murakawa {et~al.}(2003)Murakawa, Yates, Richards, \&
  Cohen}]{Murakawa03}
Murakawa, K., Yates, J.~A., Richards, A. M.~S., \& Cohen, R.~J. 2003, MNRAS,
  344, 1, (M03)

\bibitem[{{Neufeld} {et~al.}(1996){Neufeld}, {Chen}, {Melnick}, {de Graauw},
  {Feuchtgruber}, {Haser}, {Lutz}, \& {Harwit}}]{Neufeld96}
{Neufeld}, D.~A., {Chen}, W., {Melnick}, G.~J., {et~al.} 1996, A\&A, 315, L237

\bibitem[{Nyman {et~al.}(1986)Nyman, Johansson, \& Booth}]{Nyman86}
Nyman, L.-A., Johansson, L. E.~B., \& Booth, R.~S. 1986, A\&A, 160, 352

\bibitem[{Olofsson {et~al.}(1996)Olofsson, Bergmann, Eriksson, \&
  Gustafsson}]{Olofsson96}
Olofsson, H., Bergmann, P., Eriksson, K., \& Gustafsson, B. 1996, A\&A, 311,
  587

\bibitem[{{Olofsson} {et~al.}(1998){Olofsson}, {Lindqvist}, {Nyman}, \&
  {Winnberg}}]{Olofsson98}
{Olofsson}, H., {Lindqvist}, M., {Nyman}, L.-A., \& {Winnberg}, A. 1998, A\&A,
  329, 1059

\bibitem[{{Olofsson} {et~al.}(2010){Olofsson}, {Maercker}, {Eriksson},
  {Gustafsson}, \& {Sch{\"o}ier}}]{Olofsson10}
{Olofsson}, H., {Maercker}, M., {Eriksson}, K., {Gustafsson}, B., \&
  {Sch{\"o}ier}, F. 2010, \aap, 515, A27

\bibitem[{{Pashchenko} \& {Rudnitskii}(1999)}]{Pashchenko99}
{Pashchenko}, M.~I. \& {Rudnitskii}, G.~M. 1999, Astronomy Reports, 43, 311

\bibitem[{{Pashchenko} \& {Rudnitskii}(2004)}]{Pashchenko04}
{Pashchenko}, M.~I. \& {Rudnitskii}, G.~M. 2004, Astronomy Reports, 48, 380

\bibitem[{{Petrov} {et~al.}(2011){Petrov}, {Kovalev}, {Fomalont}, \&
  {Gordon}}]{Petrov11}
{Petrov}, L., {Kovalev}, Y.~Y., {Fomalont}, E.~B., \& {Gordon}, D. 2011, AJ,
  142, 35

\bibitem[{{Pluzhnik} {et~al.}(2009){Pluzhnik}, {Ragland}, {LeCoroller},
  {Cotton}, {Danchi}, {Traub}, \& {Willson}}]{Pluzhnik09}
{Pluzhnik}, E.~A., {Ragland}, S., {LeCoroller}, H., {et~al.} 2009, ApJ, 700,
  114

\bibitem[{Ragland {et~al.}(2006)Ragland, Traub, Berger, A, A, A, A, A, A, A, \&
  A}]{Ragland06}
Ragland, S., Traub, W.~A., Berger, J.-P., {et~al.} 2006, ApJ, 652, 650

\bibitem[{Reid \& Menten(1990)}]{Reid90}
Reid, M.~J. \& Menten, K.~M. 1990, ApJ, 360, L51

\bibitem[{Reid \& Menten(1997)}]{Reid97}
Reid, M.~J. \& Menten, K.~M. 1997, ApJ, 476, 327

\bibitem[{Richards(1997)}]{Richards97t}
Richards, A. M.~S. 1997, PhD thesis, University of Manchester

\bibitem[{{Richards} {et~al.}(2011){Richards}, {Elitzur}, \&
  {Yates}}]{Richards11}
{Richards}, A.~M.~S., {Elitzur}, M., \& {Yates}, J.~A. 2011, A\&A, 525, A56,
  (R11)

\bibitem[{Richards \& Yates(1998)}]{Richards98d}
Richards, A. M.~S. \& Yates, J.~A. 1998, Ir. AJ, 25, 7

\bibitem[{Richards {et~al.}(1996)Richards, Yates, \& Cohen}]{Richards96}
Richards, A. M.~S., Yates, J.~A., \& Cohen, R.~J. 1996, MNRAS, 282, 665

\bibitem[{Richards {et~al.}(1998)Richards, Yates, \& Cohen}]{Richards98v}
Richards, A. M.~S., Yates, J.~A., \& Cohen, R.~J. 1998, MNRAS, 299, 319

\bibitem[{Richards {et~al.}(1999)Richards, Yates, \& Cohen}]{Richards99}
Richards, A. M.~S., Yates, J.~A., \& Cohen, R.~J. 1999, MNRAS, 306, 954, (R99)

\bibitem[{{Rudnitskii} \& {Chuprikov}(1990)}]{Rudnitskij90}
{Rudnitskii}, G.~M. \& {Chuprikov}, A.~A. 1990, Soviet Astronomy, 34, 147

\bibitem[{{Rudnitskii} {et~al.}(1999){Rudnitskii}, {Lekht}, \&
  {Berulis}}]{Rudnitskij99w}
{Rudnitskii}, G.~M., {Lekht}, E.~E., \& {Berulis}, I.~I. 1999, Astronomy
  Letters, 25, 398

\bibitem[{Rudnitskij {et~al.}(2000)Rudnitskij, Mendoza-Torres, Pashchenko, \&
  Berulis}]{Rudnitskij00}
Rudnitskij, G.~M., Mendoza-Torres, J.~E., Pashchenko, M.~I., \& Berulis, I.~I.
  2000, A\&AS, 146, 385

\bibitem[{{Samus} {et~al.}(2011){Samus}, {Durlevich}, \& {et al.}}]{Samus11}
{Samus}, N.~N., {Durlevich}, O.~V., \& {et al.} 2011, VizieR Online Data
  Catalog, 1, 2025

\bibitem[{{Shintani} {et~al.}(2008){Shintani}, {Imai}, {Ando}, {Nakashima},
  {Hirota}, {Inomata}, {Kai}, {Kameno}, {Kijima}, {Kobayashi}, {Kuroki},
  {Maeda}, {Maruyama}, {Matsumoto}, {Miyaji}, {Nagayama}, {Nagayoshi},
  {Nakamura}, {Nakagawa}, {Namikawa}, {Omodaka}, {Oyama}, {Sakakibara},
  {Shimizu}, {Sora}, {Tsushima}, {Ueda}, {Ueda}, \& {Yamashita}}]{Shintani08}
{Shintani}, M., {Imai}, H., {Ando}, K., {et~al.} 2008, PASJ, 60, 1077

\bibitem[{{Smith} {et~al.}(2006){Smith}, {Price}, \& {Moffett}}]{Smith06}
{Smith}, B.~J., {Price}, S.~D., \& {Moffett}, A.~J. 2006, AJ, 131, 612

\bibitem[{{Titov}(2004)}]{Titov04}
{Titov}, O.~A. 2004, Astronomy Reports, 48, 941

\bibitem[{van Leeuwen(2007)}]{vanLeeuwen07}
van Leeuwen, F. 2007, A\&A, 474, 653

\bibitem[{{van Loon} {et~al.}(2005){van Loon}, {Cioni}, {Zijlstra}, \&
  {Loup}}]{vanLoon05}
{van Loon}, J.~T., {Cioni}, M.-R.~L., {Zijlstra}, A.~A., \& {Loup}, C. 2005,
  A\&A, 438, 273

\bibitem[{{Verhoelst} {et~al.}(2009){Verhoelst}, {van der Zypen}, {Hony},
  {Decin}, {Cami}, \& {Eriksson}}]{Verhoelst09}
{Verhoelst}, T., {van der Zypen}, N., {Hony}, S., {et~al.} 2009, A\&A, 498, 127

\bibitem[{{Vlemmings} {et~al.}(2001){Vlemmings}, {Diamond}, \& {van
  Langevelde}}]{Vlemmings01}
{Vlemmings}, W., {Diamond}, P.~J., \& {van Langevelde}, H.~J. 2001, A\&A, 375,
  L1

\bibitem[{{Vlemmings} {et~al.}(2002){Vlemmings}, {Diamond}, \& {van
  Langevelde}}]{Vlemmings02}
{Vlemmings}, W.~H.~T., {Diamond}, P.~J., \& {van Langevelde}, H.~J. 2002, A\&A,
  394, 589

\bibitem[{{Vlemmings} \& {van Langevelde}(2007)}]{Vlemmings07}
{Vlemmings}, W.~H.~T. \& {van Langevelde}, H.~J. 2007, A\&A, 472, 547

\bibitem[{{Vlemmings} {et~al.}(2005){Vlemmings}, {van Langevelde}, \&
  {Diamond}}]{Vlemmings05}
{Vlemmings}, W.~H.~T., {van Langevelde}, H.~J., \& {Diamond}, P.~J. 2005, A\&A,
  434, 1029

\bibitem[{{Vlemmings} {et~al.}(2003){Vlemmings}, {van Langevelde}, {Diamond},
  {Habing}, \& {Schilizzi}}]{Vlemmings03}
{Vlemmings}, W.~H.~T., {van Langevelde}, H.~J., {Diamond}, P.~J., {Habing},
  H.~J., \& {Schilizzi}, R.~T. 2003, A\&A, 407, 213

\bibitem[{{Winnberg} {et~al.}(2011){Winnberg}, {Brand}, \&
  {Engels}}]{Winnberg11}
{Winnberg}, A., {Brand}, J., \& {Engels}, D. 2011, in ASP Conference Series,
  Vol. 445, Why Galaxies Care about AGB Stars II, ed. {F.~Kerschbaum,
  T.~Lebzelter, \& R.~F.~Wing}, 375

\bibitem[{{Winnberg} {et~al.}(2008){Winnberg}, {Engels}, {Brand}, {Baldacci},
  \& {Walmsley}}]{Winnberg08}
{Winnberg}, A., {Engels}, D., {Brand}, J., {Baldacci}, L., \& {Walmsley}, C.~M.
  2008, A\&A, 482, 831

\bibitem[{{Wittkowski} {et~al.}(2011){Wittkowski}, {Boboltz}, {Ireland},
  {Karovicova}, {Ohnaka}, {Scholz}, {van Wyk}, {Whitelock}, {Wood}, \&
  {Zijlstra}}]{Wittkowski11}
{Wittkowski}, M., {Boboltz}, D.~A., {Ireland}, M., {et~al.} 2011, A\&A, 532, L7

\bibitem[{{Wittkowski} {et~al.}(2007){Wittkowski}, {Boboltz}, {Ohnaka},
  {Driebe}, \& {Scholz}}]{Wittkowski07}
{Wittkowski}, M., {Boboltz}, D.~A., {Ohnaka}, K., {Driebe}, T., \& {Scholz}, M.
  2007, A\&A, 470, 191

\bibitem[{{Woitke}(2006)}]{Woitke06}
{Woitke}, P. 2006, A\&A, 460, L9

\bibitem[{Yates \& Cohen(1994)}]{Yates94}
Yates, J.~A. \& Cohen, R.~J. 1994, MNRAS, 270, 958

\bibitem[{Yates {et~al.}(1997)Yates, Field, \& Gray}]{Yates97}
Yates, J.~A., Field, D., \& Gray, M.~D. 1997, MNRAS, 285, 383

\bibitem[{{Zacharias} {et~al.}(2005){Zacharias}, {Monet}, {Levine}, {Urban},
  {Gaume}, \& {Wycoff}}]{Zacharias05}
{Zacharias}, N., {Monet}, D.~G., {Levine}, S.~E., {et~al.} 2005, VizieR Online
  Data: NOMAD Catalog, 1297

\bibitem[{{Zhao-Geisler} {et~al.}(2011){Zhao-Geisler}, {Quirrenbach},
  {K{\"o}hler}, {Lopez}, \& {Leinert}}]{Zhao-Geisler11}
{Zhao-Geisler}, R., {Quirrenbach}, A., {K{\"o}hler}, R., {Lopez}, B., \&
  {Leinert}, C. 2011, A\&A, 530, A120

\bibitem[{Zubko \& Elitzur(2000)}]{Zubko00}
Zubko, V. \& Elitzur, M. 2000, ApJ, 544, L137

\end{thebibliography}

%\end{document}

\appendix

\section{Contour plots of H$_{2}$O maser emission}
\label{apx:kntr}

These plots  show channel-averaged emission (see Section~\ref{sec:data}
for the original channel widths used). The restoring beams are shown in
the first or last panel; in most cases these are slightly larger than
the beams used in the full-resolution images for analysis
(Table~\ref{tab:feats}).  The lowest contour is at the average
$3\sigma_{\mathrm {rms}}$ for the cube but bright channels have been
blanked where they are dynamic-range limited. 
Plots  for VX Sgr were
published by M03.

The image cubes for these plots were made by averaging 5 to 10
channels ($\sim1$ km s$^{-1}$) and thus, due to frequency dilution,
the peaks are often lower than those measured at full resolution.  The
same contour levels were used for all epochs of each source, at
$\sim3-5\sigma_{\mathrm{rms}}$, to aid comparison.   Negative contours
are not shown to avoid crowding the plots; these generally do not
exceed the first positive but blanking has been applied to
dynamic-range limited channels where indicated (identified by broken contours).

\begin{figure*}
\centering
\includegraphics[angle=0, width=16cm]{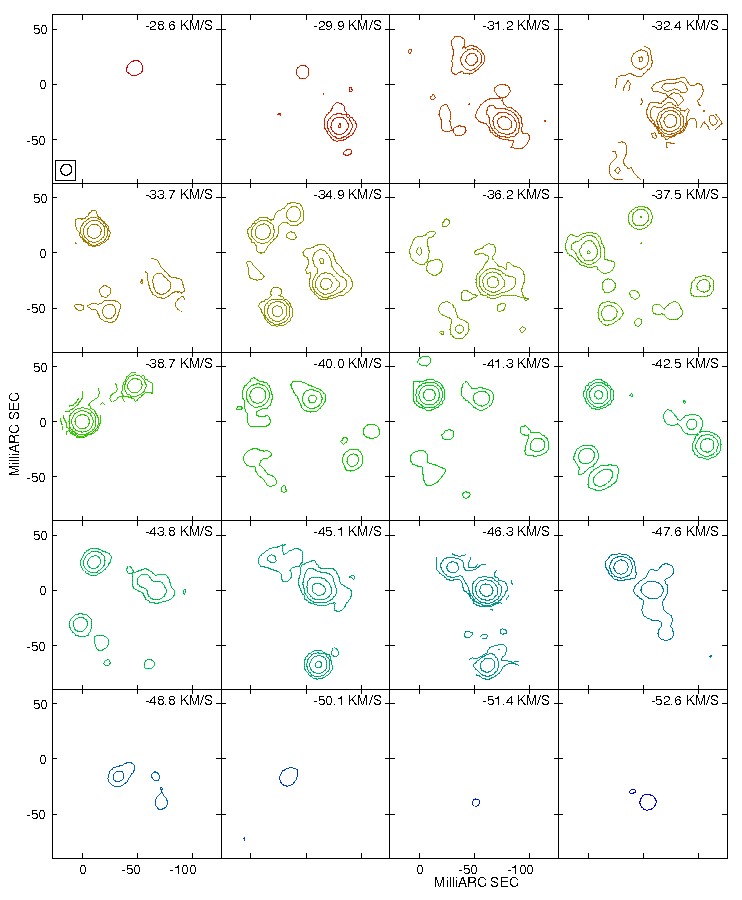}
\caption{S Per 1994, contours at (1, 4, 16, 64, 256) $\times$ 20 mJy
  beam$^{-1}$, blanking applied to bright channels.}
\label{SPER22_94.png}
\end{figure*}

\begin{figure*}
\centering
\includegraphics[angle=0, width=16cm]{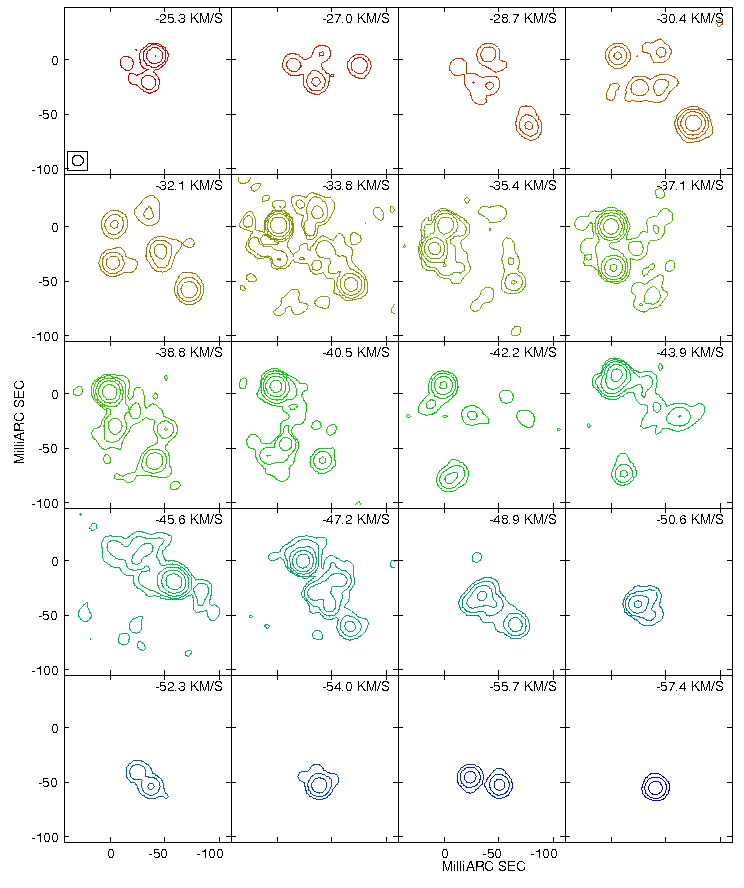}
\caption{ S Per 1999, contours at (1, 4, 16, 64, 256) $\times$ 20 mJy
  beam$^{-1}$.}
\label{SPER22_99.png}
\end{figure*}
\begin{figure*}
\centering
\includegraphics[angle=0, width=16cm]{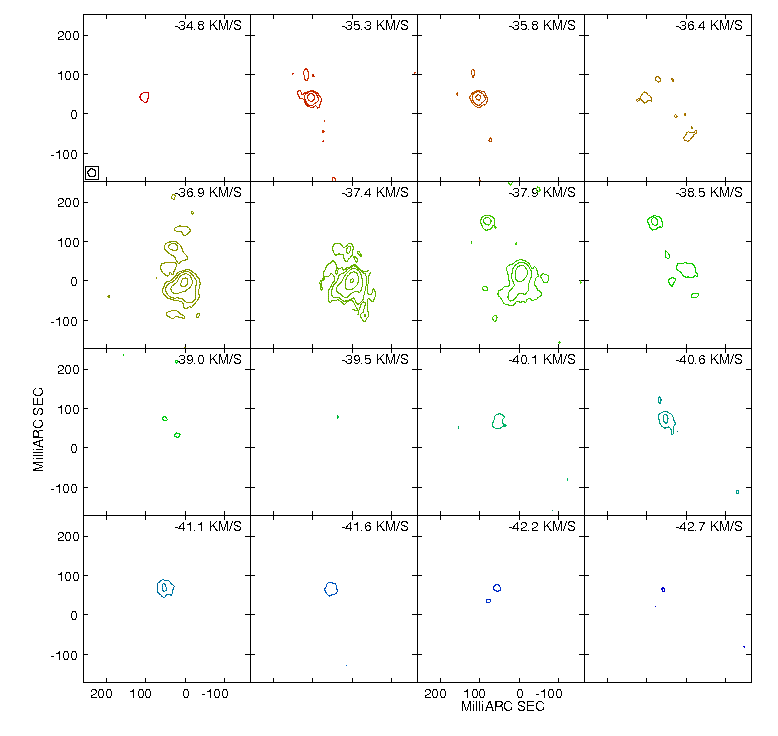}
\caption{ U Ori 1994, contours at (1, 4, 16, 64, 256) $\times$ 50 mJy
  beam$^{-1}$. }
\label{UORI_22_94.png}
\end{figure*}
\begin{figure*}
\centering
\includegraphics[angle=0, width=16cm]{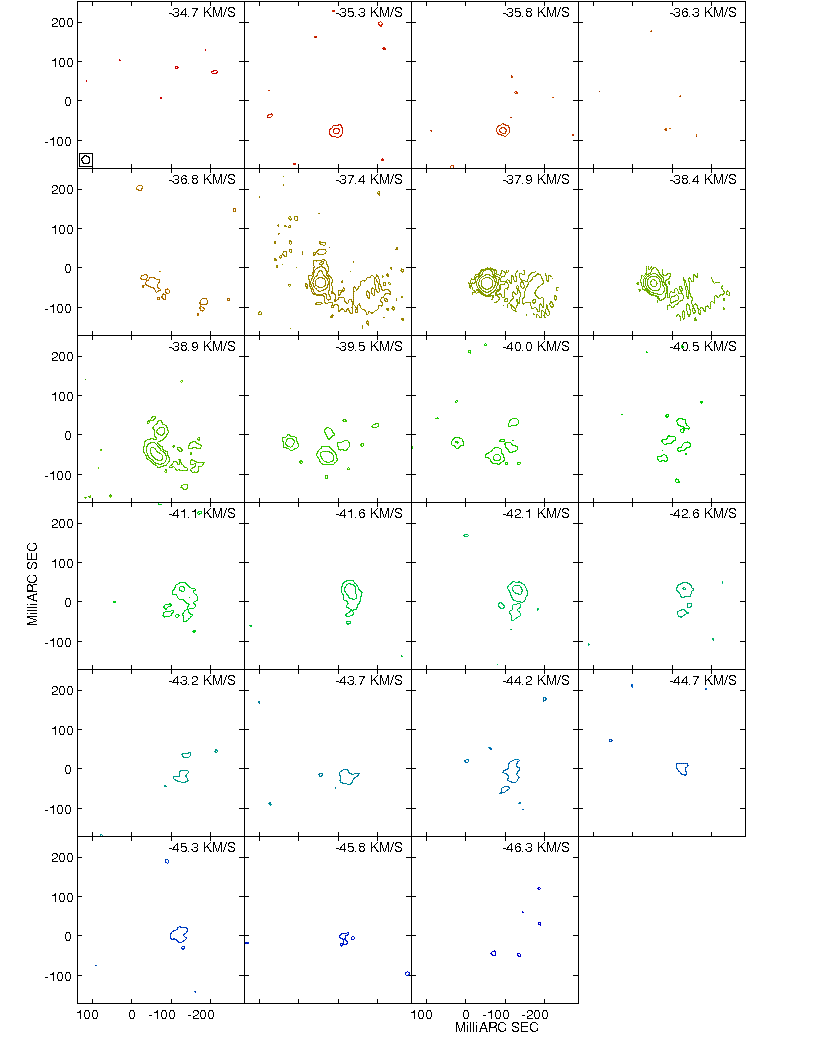}
\caption{ U Ori 1999, contours at (1, 4, 16, 64, 256) $\times$ 50 mJy
  beam$^{-1}$, blanking applied to bright channels. }
\label{UORI_22_99.png}
\end{figure*}
\begin{figure*}
\centering
\includegraphics[angle=0, width=16cm]{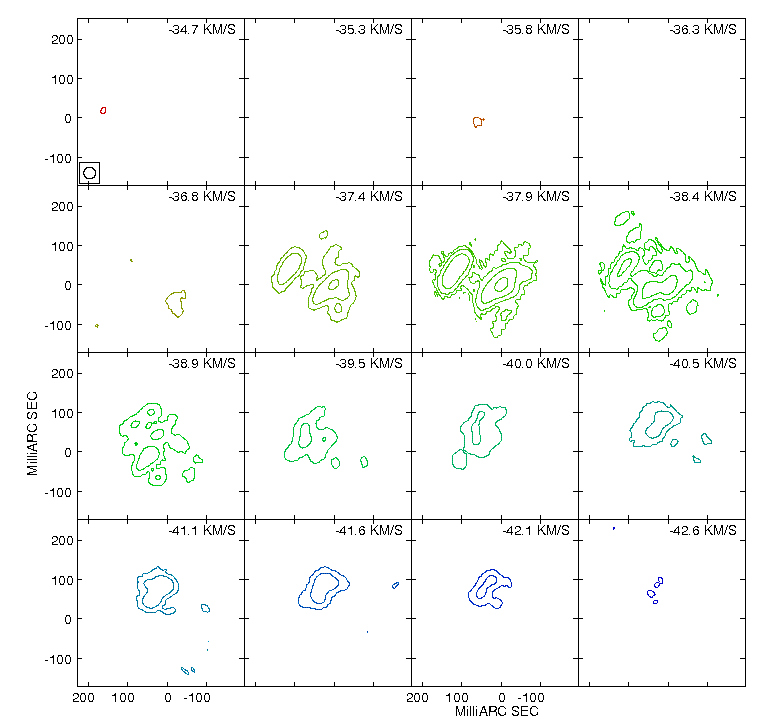}
\caption{ U Ori 2000, contours at (1, 4, 16, 64, 256) $\times$ 50 mJy
  beam$^{-1}$. }
\label{UORI_22_00_5.png}
\end{figure*}
\begin{figure*}
\centering
\includegraphics[angle=0, width=16cm]{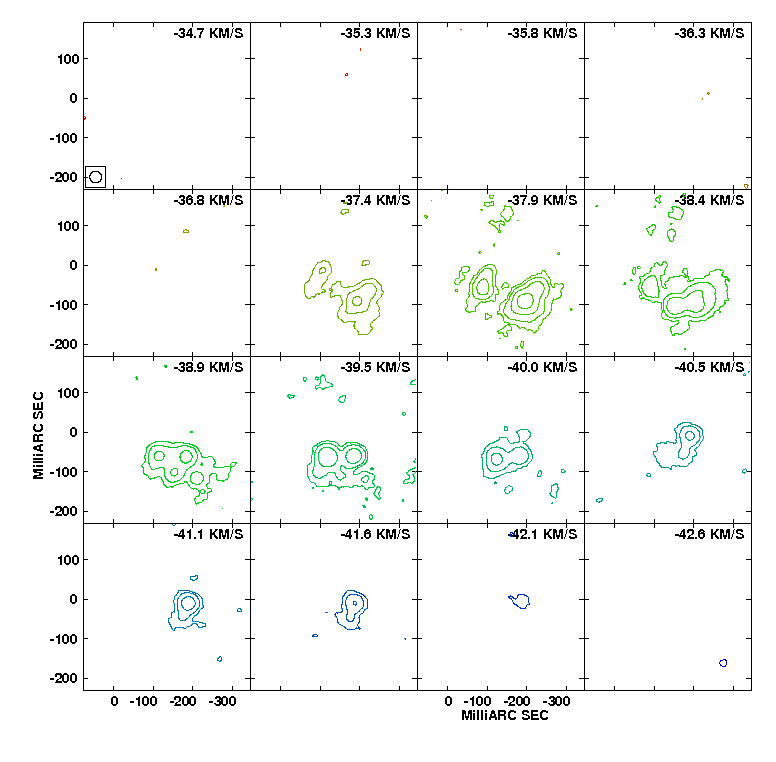}
\caption{ U Ori 2001, contours at (1, 4, 16, 64, 256) $\times$ 50 mJy
  beam$^{-1}$. }
\label{UORI22_01_5.png}
\end{figure*}
\begin{figure*}
\centering
\includegraphics[angle=0, width=16cm]{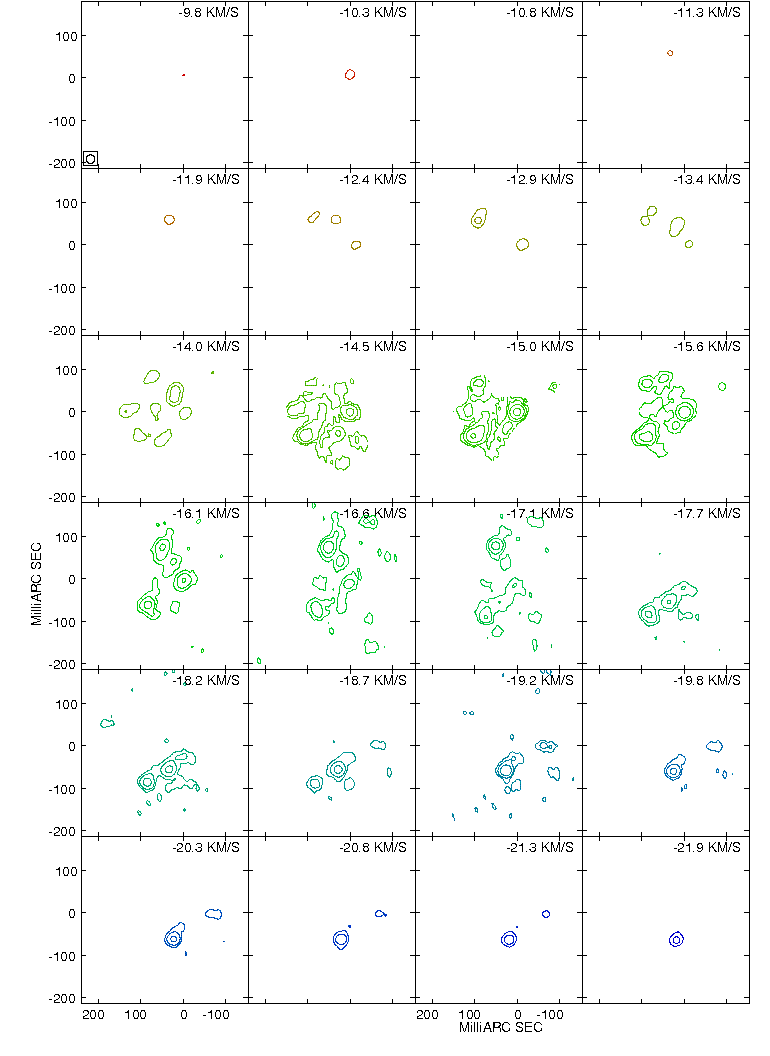}
\caption{ U Her 1994, contours at (1, 4, 16, 64, 256) $\times$ 60 mJy
  beam$^{-1}$, blanking applied to bright channels.  }
\label{UHER_22_94_5.png}
\end{figure*}
\begin{figure*}
\centering
\includegraphics[angle=0, width=16cm]{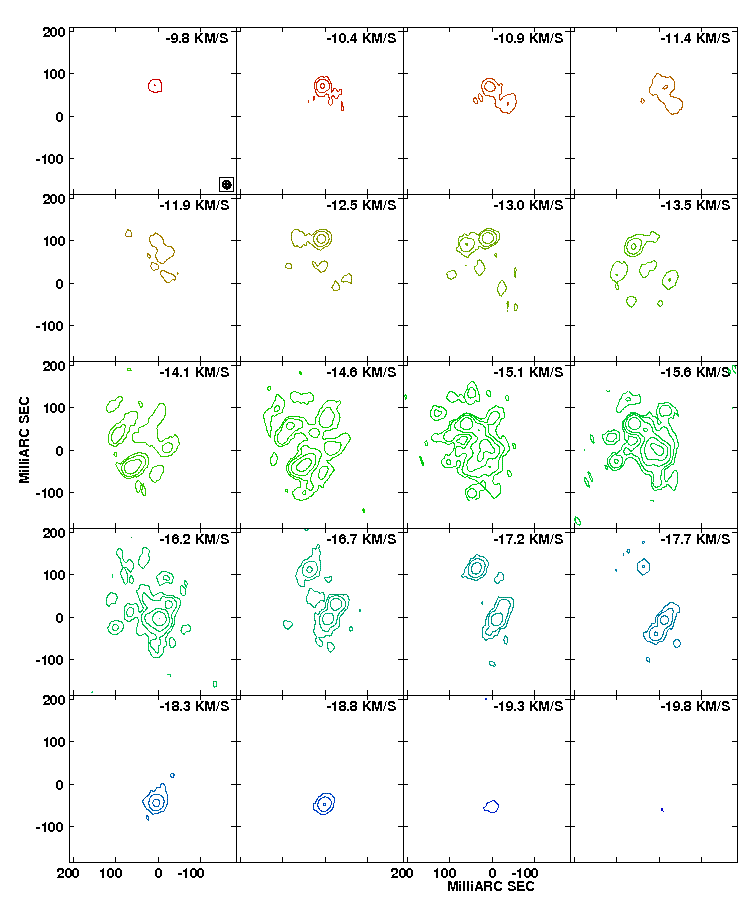}
\caption{ U Her 2001, contours at (1,  4, 16, 64, 256) $\times$ 60 mJy
  beam$^{-1}$. }
\label{UHER22_00.png}
\end{figure*}
\begin{figure*}
\centering
\includegraphics[angle=0, width=16cm]{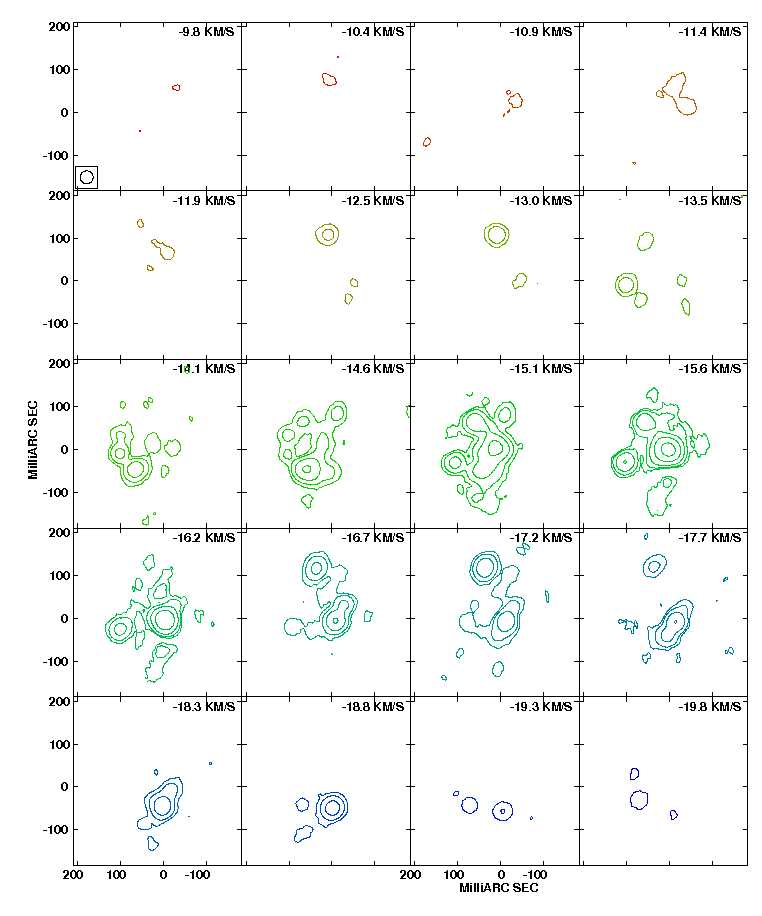}
\caption{ U Her 2011, contours at (1,  4, 16, 64, 256) $\times$ 60 mJy
  beam$^{-1}$. }
\label{UHER22_01_5.png}
\end{figure*}
\begin{figure*}
\centering
\includegraphics[angle=0, width=16cm]{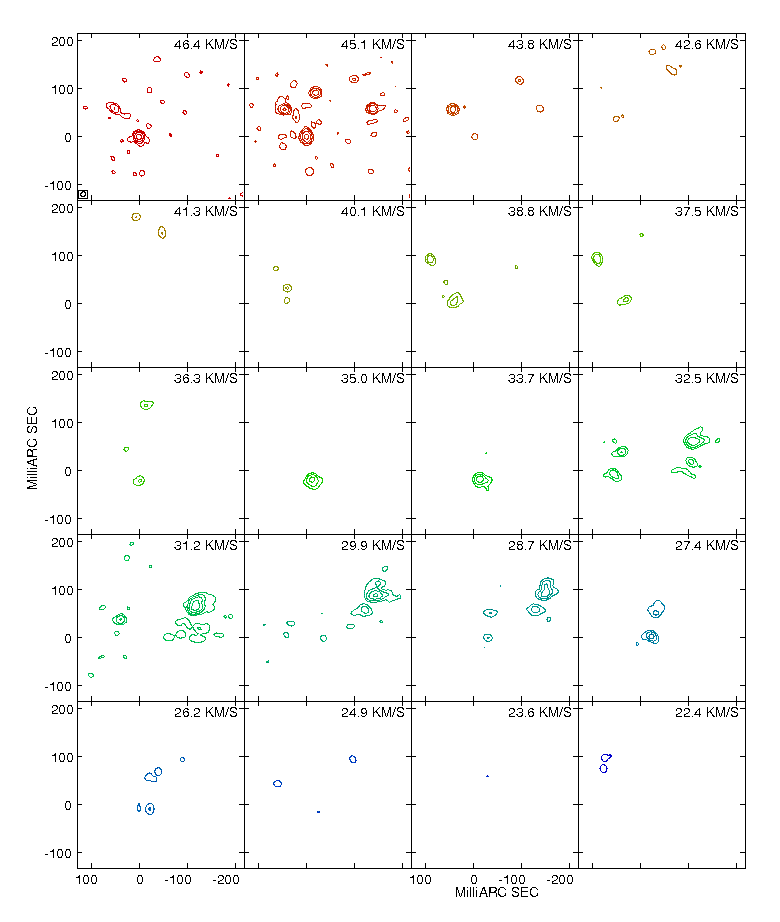}
\caption{ IK Tau 1994, contours at (1,  4, 16, 64, 256) $\times$ 40 mJy
  beam$^{-1}$. }
\label{IKTAU22_94.png}
\end{figure*}
\begin{figure*}
\centering
\includegraphics[angle=0, width=16cm]{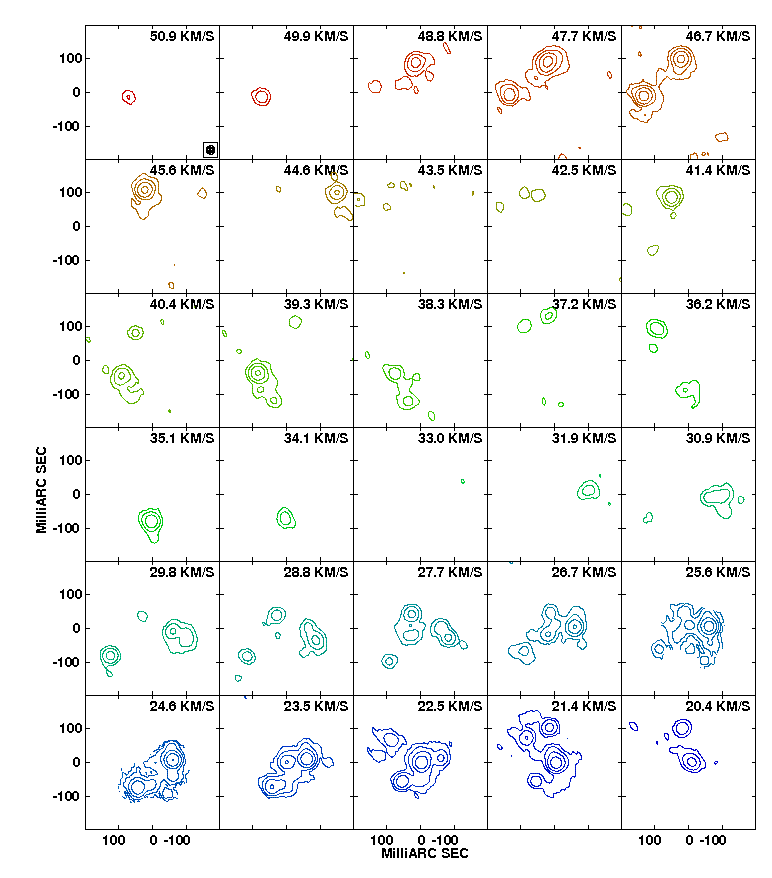}
\caption{ IK Tau 2000, contours at (1,  4, 16, 64, 256) $\times$ 40 mJy
  beam$^{-1}$. }
\label{IKTAU_0022_5.png}
\end{figure*}
\begin{figure*}
\centering
\includegraphics[angle=0, width=16cm]{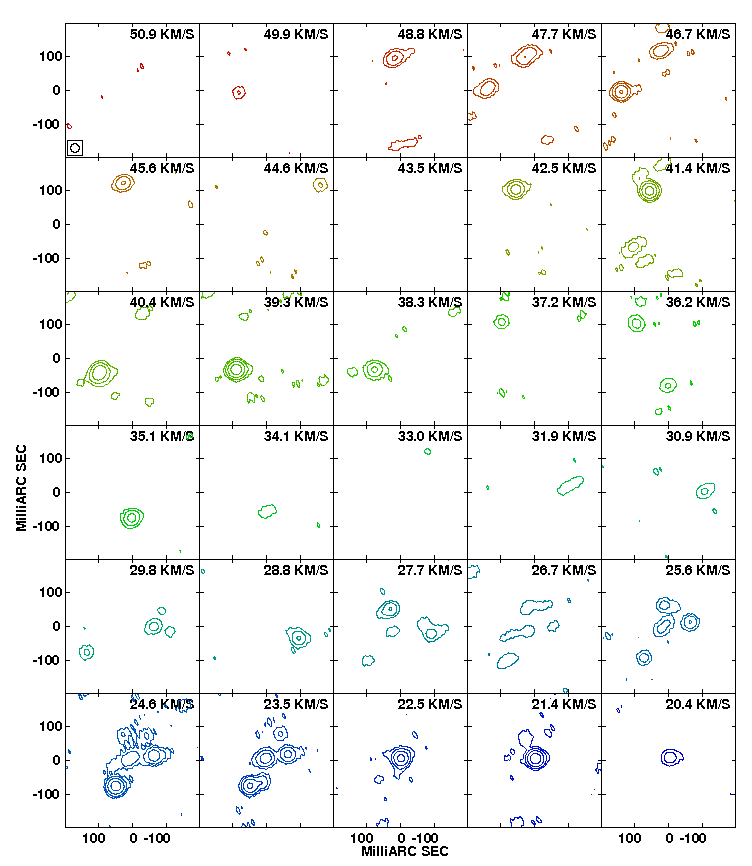}
\caption{ IK Tau 2001, contours at (1,  4, 16, 64, 256) $\times$ 40 mJy
  beam$^{-1}$.}
\label{IK22_01_5.png}
\end{figure*}
\clearpage
\begin{figure*}
\centering
\includegraphics[angle=0, width=16cm]{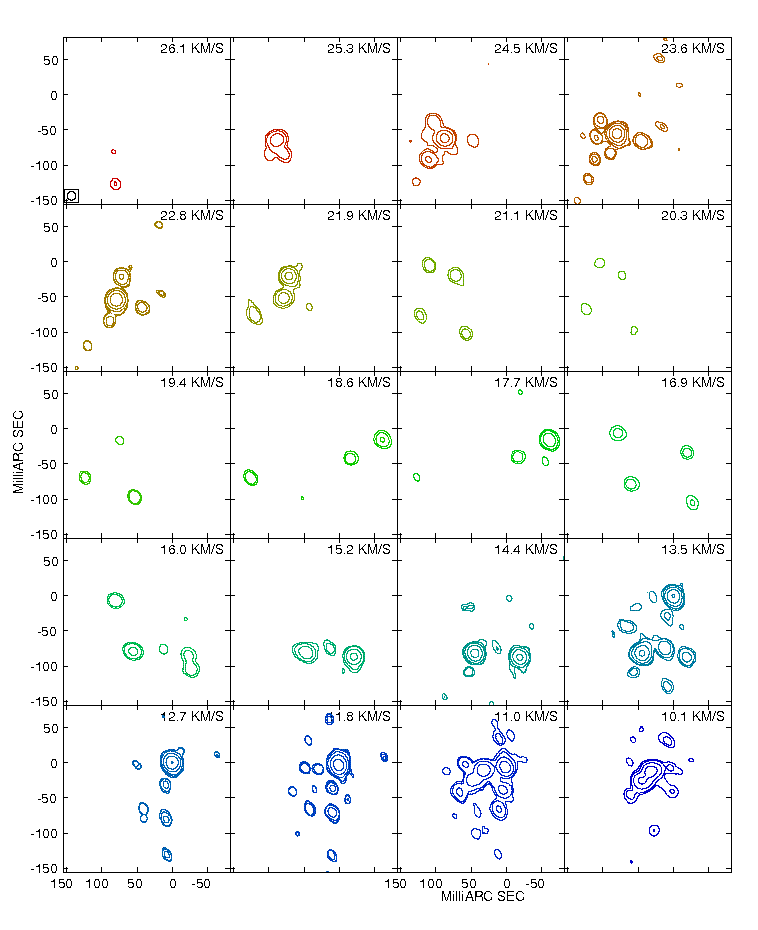}
\caption{ RT Vir 1994, contours at (1,  4, 16, 64, 256) $\times$ 90 mJy
  beam$^{-1}$. }
\label{RTVIR22_94.png}
\end{figure*}
\begin{figure*}
\centering
\includegraphics[angle=0, width=16cm]{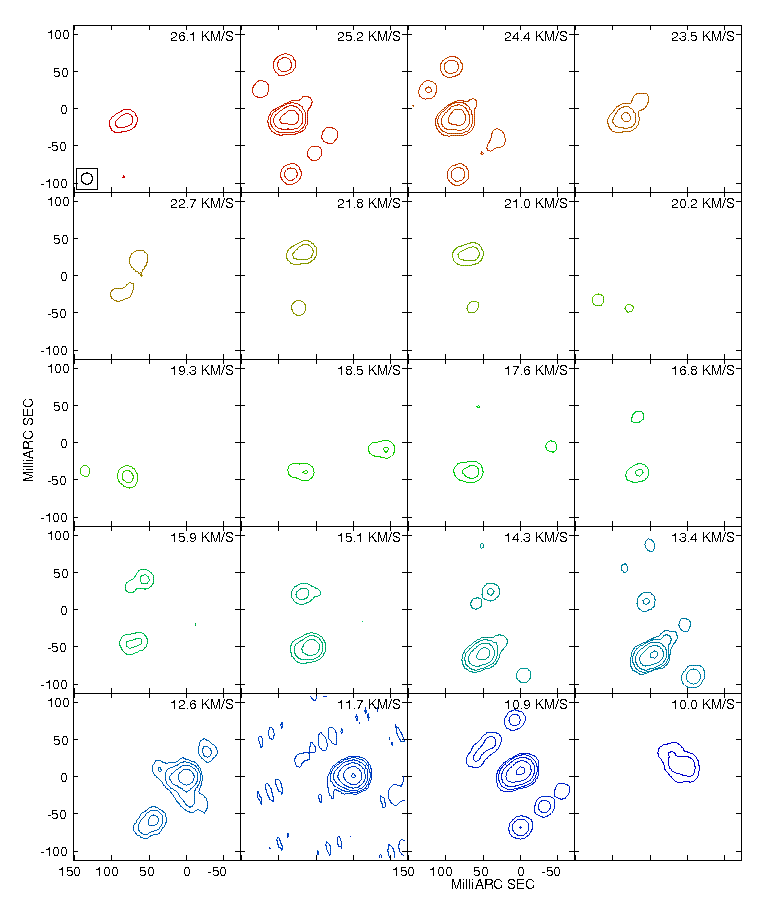}
\caption{ RT Vir 960405, contours at (1,  4, 16, 64, 256) $\times$ 90 mJy
  beam$^{-1}$. }
\label{RTVIR22_96_1.png}
\end{figure*}
\begin{figure*}
\centering
\includegraphics[angle=0, width=16cm]{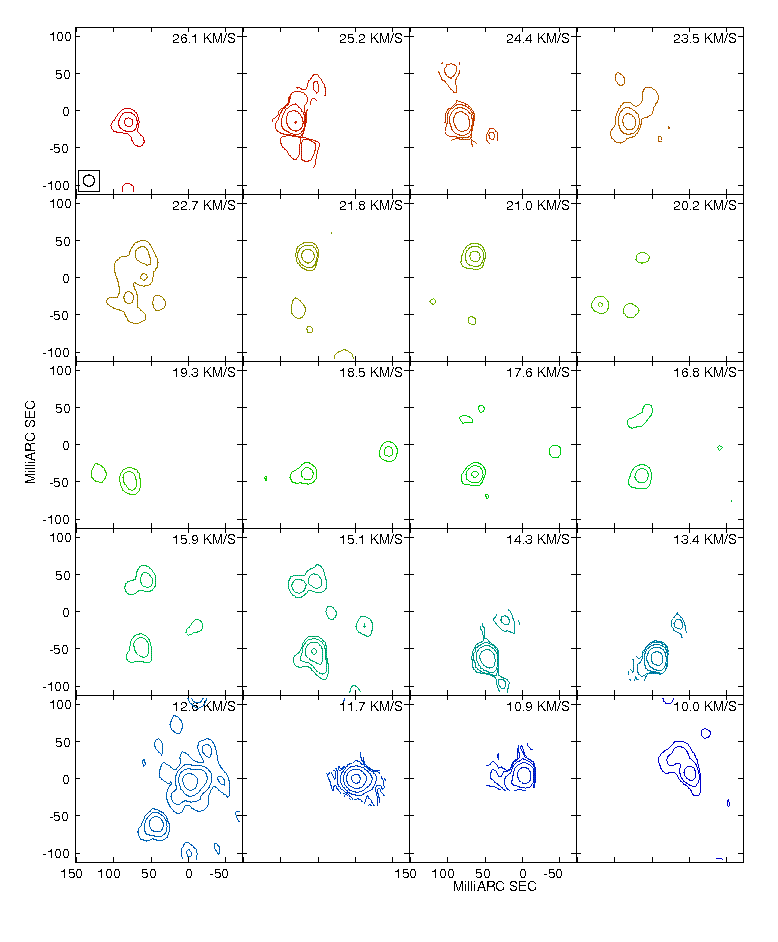}
\caption{ RT Vir 960421, contours at (1,  4, 16, 64, 256) $\times$ 90 mJy
  beam$^{-1}$, blanking applied to bright channels.}
\label{RTVIR22_96_2.png}
\end{figure*}
\begin{figure*}
\centering
\includegraphics[angle=0, width=16cm]{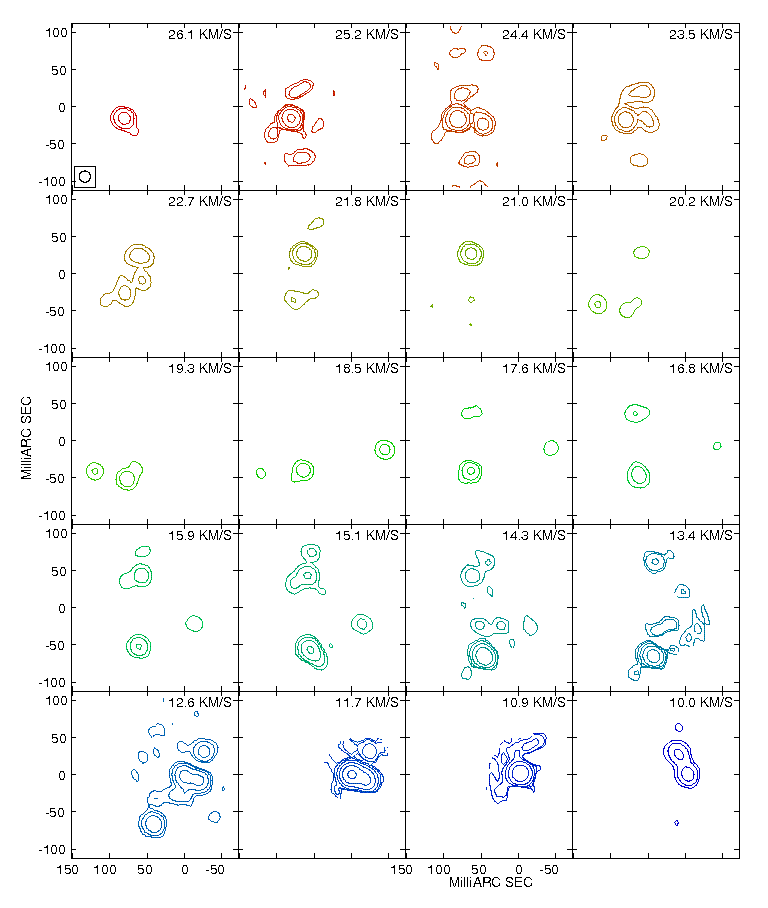}
\caption{ RT Vir 960429, contours at (1,  4, 16, 64, 256) $\times$ 90 mJy
  beam$^{-1}$, blanking applied to bright channels. }
\label{RTVIR22_96_3.png}
\end{figure*}
\begin{figure*}
\centering
\includegraphics[angle=0, width=16cm]{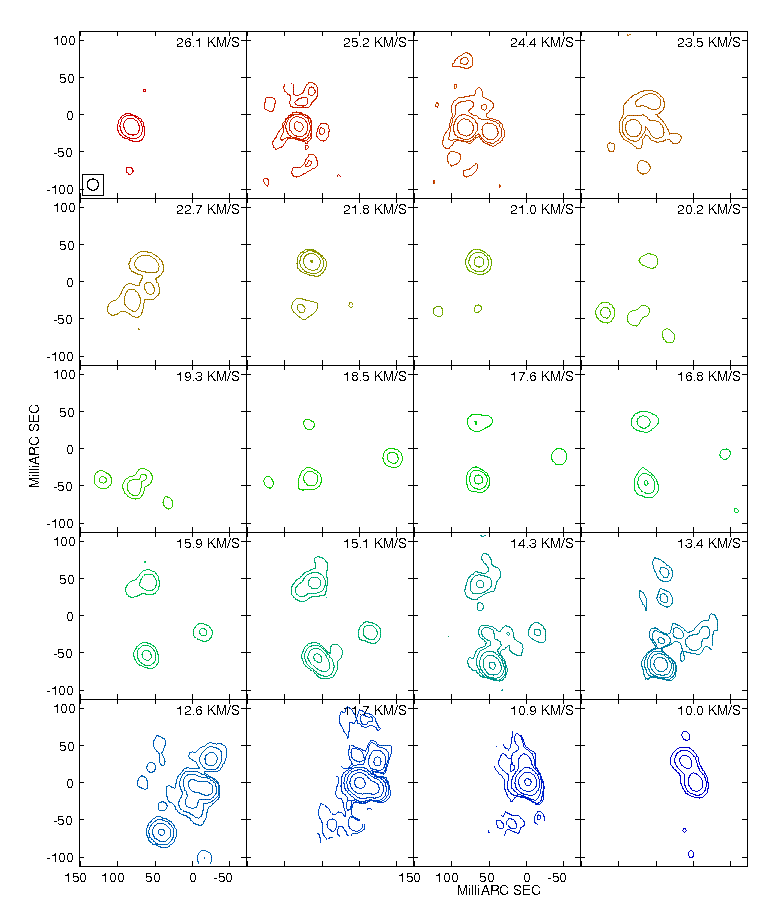}
\caption{ RT Vir 960515, contours at (1,  4, 16, 64, 256) $\times$ 90 mJy
  beam$^{-1}$, blanking applied to bright channels. }
\label{RTVIR22_96_4.png}
\end{figure*}
\begin{figure*}
\centering
\includegraphics[angle=0, width=16cm]{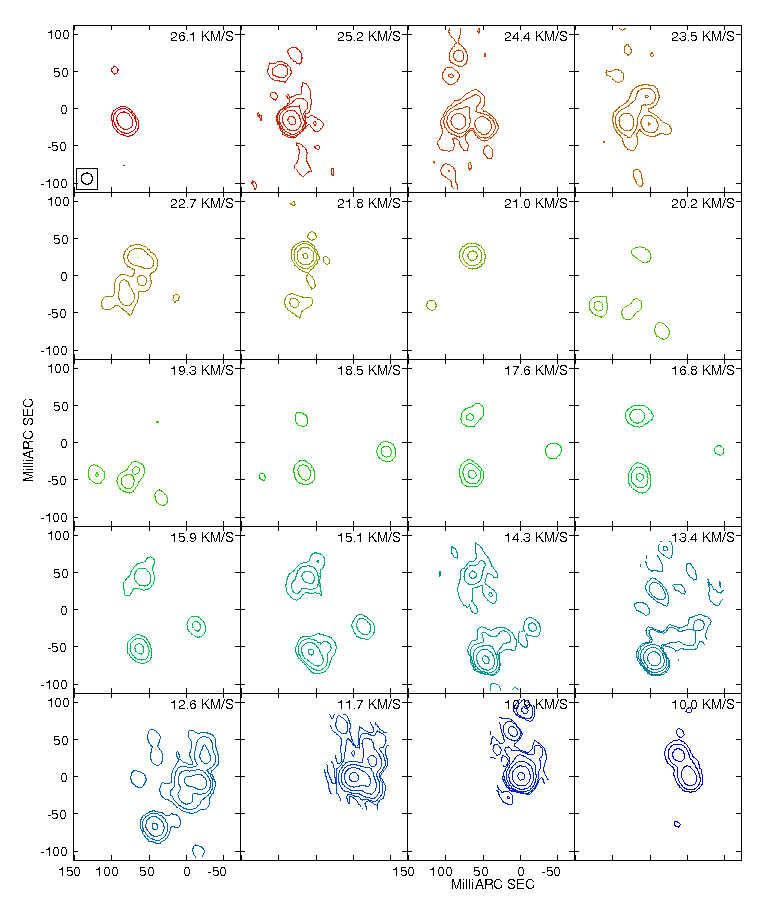}
\caption{ RT Vir 960524, contours at (1,  4, 16, 64, 256) $\times$ 90 mJy
  beam$^{-1}$, blanking applied to bright channels.}
\label{RTVIR22_96_5.png}
\end{figure*}
\begin{figure*}
\centering
\includegraphics[angle=0, width=16cm]{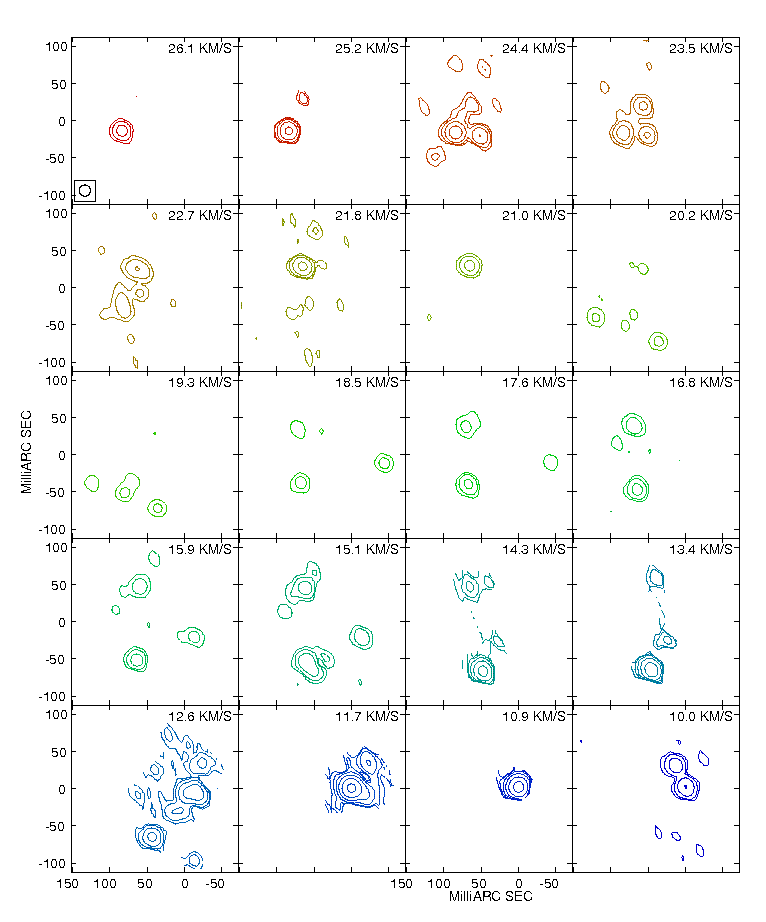}
\caption{ RT Vir 960612, contours at (1,  4, 16, 64, 256) $\times$ 90 mJy
  beam$^{-1}$, blanking applied to bright channels. }
\label{RTVIR22_96_6.png}
\end{figure*}
\clearpage
\begin{figure*}
\centering
\includegraphics[angle=0, width=16cm]{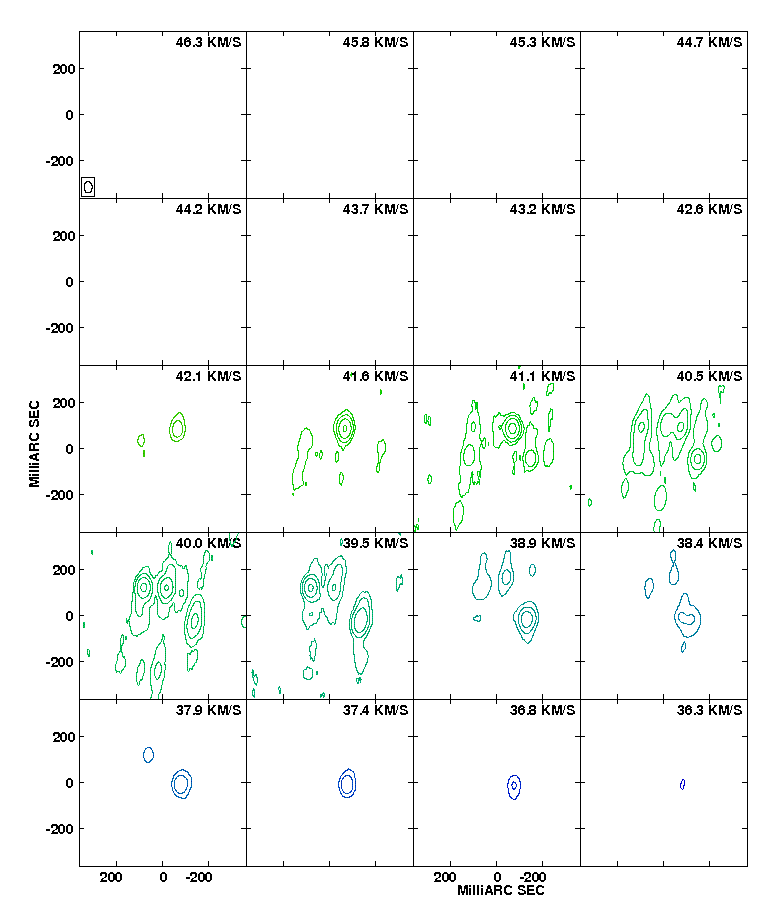}
\caption{ W Hya 1999, contours at (1,  4, 16, 64, 256) $\times$ 200 mJy
  beam$^{-1}$. }
\label{WHYA_22_99_5.png}
\end{figure*}
\begin{figure*}
\centering
\includegraphics[angle=0, width=16cm]{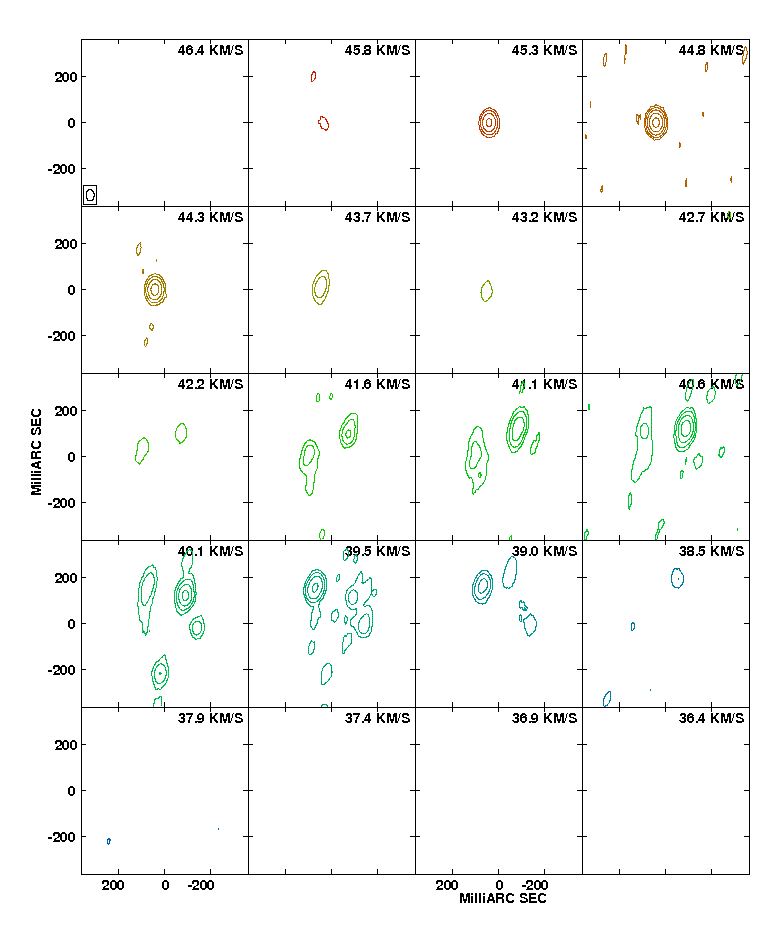}
\caption{ W Hya 2000, contours at (1,  4, 16, 64, 256) $\times$ 200 mJy
  beam$^{-1}$. }
\label{WHYA00_5.png}
\end{figure*}
\begin{figure*}
\centering
\includegraphics[angle=0, width=16cm]{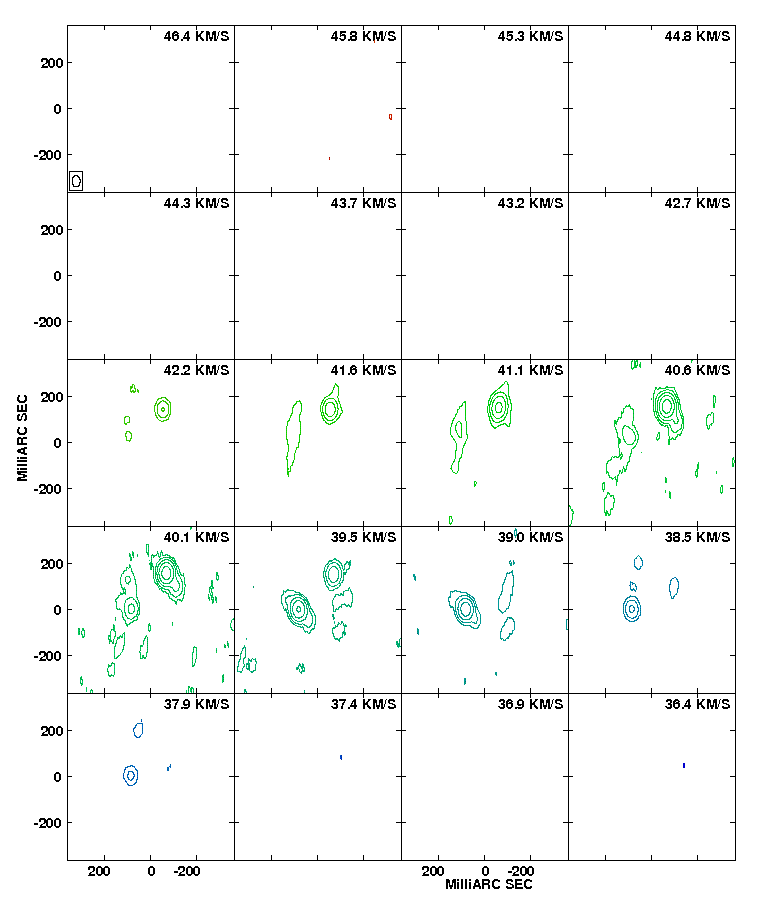}
\caption{ W Hya 2001, contours at (1,  4, 16, 64, 256) $\times$ 200 mJy
  beam$^{-1}$. }
\label{WHYA01_5.png}
\end{figure*}
\begin{figure*}
\centering
\includegraphics[angle=0, width=16cm]{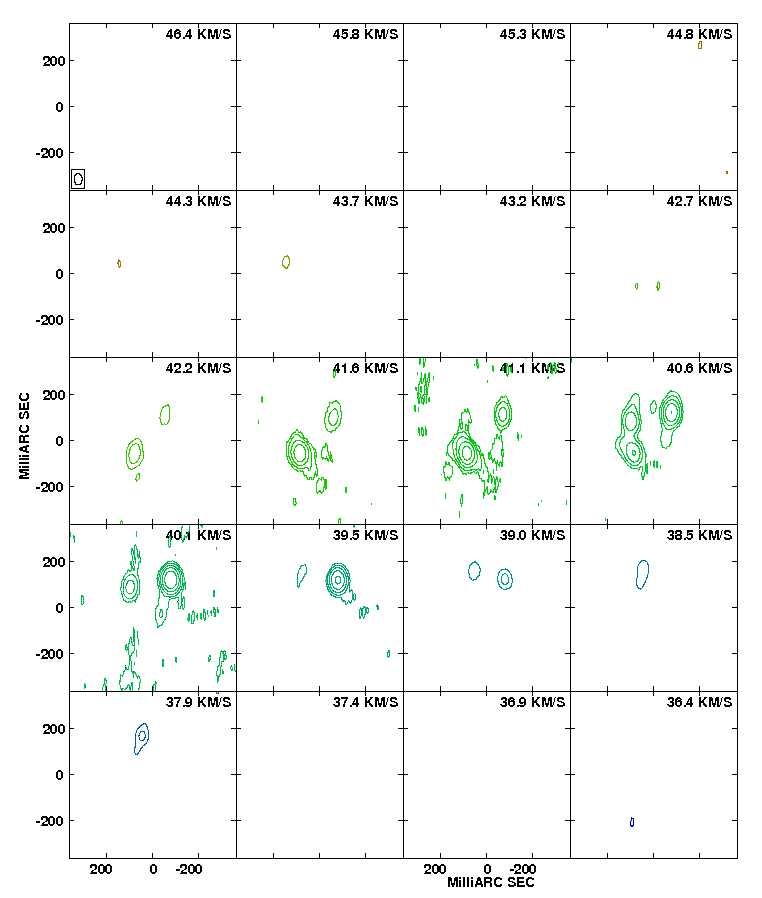}
\caption{ W Hya 2002, contours at (1,  4, 16, 64, 256) $\times$ 200 mJy
  beam$^{-1}$. }
\label{WHYA02_5.png}
\end{figure*}
\clearpage
\section{Measured maser feature parameters}
\label{apx:feats}

 These tables list the maser feature parameters.
Lists for VX Sgr were
published by M03.
% S Per 1994
\begin{table*}
\begin{tabular}{ccrrrrrr}
\hline
$V_{\mathrm{LSR}}$&$\Delta{V}$&\multicolumn{1}{c}{$\Delta{V_{1/2}}$}&\multicolumn{1}{c}{$x$}&\multicolumn{1}{c}{$y$}&\multicolumn{1}{c}{$a$}&\multicolumn{1}{c}{$l$}&\multicolumn{1}{c}{$I$}\\ 
(km s$^{-1}$)&(km s$^{-1}$)&\multicolumn{1}{c}{(km s$^{-1}$)}&\multicolumn{1}{c}{(mas)}&\multicolumn{1}{c}{(mas)}&\multicolumn{1}{c}{(mas)}&\multicolumn{1}{c}{(mas)}&\multicolumn{1}{c}{(Jy)}\\ 
\hline
 -52.5 &0.6&$ 0.00\pm0.00$&$ -11.6\pm 1.3$&$ -29.7\pm 1.3$&$  31.8\pm 1.3$&$   2.8\pm 7.2$&  0.15\\
 -51.8 &0.4&$ 0.39\pm0.13$&$   0.5\pm 1.5$&$ -18.7\pm 1.5$&$  18.8\pm 1.5$&$   6.6\pm 7.1$&  0.22\\
 -51.6 &1.4&$ 1.09\pm0.17$&$ -11.4\pm 0.7$&$ -29.7\pm 0.7$&$  31.8\pm 0.7$&$  11.3\pm 8.0$&  0.35\\
 -48.7 &1.8&$ 1.03\pm0.04$&$   8.8\pm 0.2$&$  -6.8\pm 0.2$&$  11.2\pm 0.2$&$   9.5\pm 6.0$&  0.80\\
 -48.2 &0.8&$ 0.00\pm0.00$&$ -28.8\pm 1.3$&$ -28.7\pm 1.3$&$  40.7\pm 1.3$&$   6.3\pm 6.8$&  0.19\\
 -47.6 &1.0&$ 1.56\pm1.21$&$ -23.1\pm 1.2$&$  -7.6\pm 1.2$&$  24.4\pm 1.2$&$  11.4\pm 7.0$&  0.13\\
\hline
\end{tabular}
\caption{S Per 1994 22-GHz H$_2$O maser feature
  parameters. $\Delta{V}$ is the full velocity extent of the feature
  and $\Delta{V}_{1/2}$ is the FWHM, set to zero if a Gaussian
  spectral profile could not be fitted at $\ge3\sigma$. $x$, $y$ and
  $a$ are the Cartesian and radial (in the plane of the sky) offsets
  from the centre of expansion. $l$ is the angular extent of the
  feature and $I$ is its peak.   (\emph{see next page}).}
\end{table*}

\begin{table*}
\begin{tabular}{ccrrrrrr}
\hline
$V_{\mathrm{LSR}}$&$\Delta{V}$&\multicolumn{1}{c}{$\Delta{V_{1/2}}$}&\multicolumn{1}{c}{$x$}&\multicolumn{1}{c}{$y$}&\multicolumn{1}{c}{$a$}&\multicolumn{1}{c}{$l$}&\multicolumn{1}{c}{$I$}\\ 
(km s$^{-1}$)&(km s$^{-1}$)&\multicolumn{1}{c}{(km s$^{-1}$)}&\multicolumn{1}{c}{(mas)}&\multicolumn{1}{c}{(mas)}&\multicolumn{1}{c}{(mas)}&\multicolumn{1}{c}{(mas)}&\multicolumn{1}{c}{(Jy)}\\ 
\hline
 -35.4 &1.0&$ 0.00\pm0.00$&$  27.7\pm 0.9$&$  -2.9\pm 0.9$&$  27.8\pm 0.9$&$   3.3\pm 4.9$&  1.44\\
 -35.4 &1.3&$ 1.19\pm0.32$&$   6.0\pm 1.3$&$ -59.6\pm 1.3$&$  60.0\pm 1.3$&$   8.7\pm11.2$&  0.44\\
-35.3 &1.4&$ 0.84\pm0.32$&$ -23.9\pm 0.1$&$ -17.1\pm 0.1$&$  29.4\pm 0.1$&$   6.8\pm 4.1$& 11.24\\
 -35.2 &0.7&$ 0.49\pm0.04$&$  14.0\pm 1.6$&$  41.4\pm 1.6$&$  44.2\pm 1.6$&$  21.4\pm 7.2$&  0.29\\
 -35.2 &0.6&$ 0.00\pm0.00$&$ -20.4\pm 1.0$&$  10.4\pm 1.0$&$  22.9\pm 1.0$&$   4.2\pm 5.3$&  0.50\\

 -34.9 &1.9&$ 0.97\pm0.01$&$ -35.8\pm 0.1$&$ -19.0\pm 0.1$&$  40.5\pm 0.1$&$   6.3\pm 1.5$&  1.44\\
\hline
\end{tabular}
\addtocounter{table}{-1}
\caption{S Per 1994 22-GHz H$_2$O maser feature parameters (\emph{continued}).}
\label{tab:sp94}
\end{table*}

\begin{table*}
\begin{tabular}{ccrrrrrr}
\hline
$V_{\mathrm{LSR}}$&$\Delta{V}$&\multicolumn{1}{c}{$\Delta{V_{1/2}}$}&\multicolumn{1}{c}{$x$}&\multicolumn{1}{c}{$y$}&\multicolumn{1}{c}{$a$}&\multicolumn{1}{c}{$l$}&\multicolumn{1}{c}{$I$}\\ 
(km s$^{-1}$)&(km s$^{-1}$)&\multicolumn{1}{c}{(km s$^{-1}$)}&\multicolumn{1}{c}{(mas)}&\multicolumn{1}{c}{(mas)}&\multicolumn{1}{c}{(mas)}&\multicolumn{1}{c}{(mas)}&\multicolumn{1}{c}{(Jy)}\\ 
\hline
 -29.0 &0.7&$ 0.42\pm0.00$&$  -5.1\pm 1.4$&$  22.5\pm 1.4$&$  23.1\pm 1.4$&$   2.6\pm 7.2$&  0.32\\
 -28.1 &0.8&$ 0.53\pm0.04$&$  -6.2\pm 0.9$&$  24.0\pm 0.9$&$  24.8\pm 0.9$&$   3.4\pm 4.2$&  0.48\\
 -57.9 &2.5&$ 0.91\pm0.06$&$  -4.6\pm 0.1$&$ -26.3\pm 0.1$&$  26.7\pm 0.1$&$   4.0\pm 1.8$&  1.83\\
 -55.8 &1.9&$ 0.75\pm0.03$&$ -14.6\pm 0.1$&$ -23.5\pm 0.1$&$  27.7\pm 0.1$&$   6.0\pm 3.1$&  1.29\\
 -55.5 &2.3&$ 1.08\pm0.00$&$  12.1\pm 0.1$&$ -16.3\pm 0.1$&$  20.3\pm 0.1$&$   3.3\pm 4.7$&  1.08\\
 -53.9 &0.8&$ 1.18\pm0.93$&$  -0.2\pm 1.8$&$  -9.5\pm 1.3$&$   9.5\pm 1.3$&$   2.2\pm 5.9$&  0.08\\
\hline
\end{tabular}
\caption{S Per 1999 22-GHz H$_2$O maser feature parameters (\emph{see next page}).}
\end{table*}

\begin{table*}
\begin{tabular}{ccrrrrrr}
\hline
$V_{\mathrm{LSR}}$&$\Delta{V}$&\multicolumn{1}{c}{$\Delta{V_{1/2}}$}&\multicolumn{1}{c}{$x$}&\multicolumn{1}{c}{$y$}&\multicolumn{1}{c}{$a$}&\multicolumn{1}{c}{$l$}&\multicolumn{1}{c}{$I$}\\ 
(km s$^{-1}$)&(km s$^{-1}$)&\multicolumn{1}{c}{(km s$^{-1}$)}&\multicolumn{1}{c}{(mas)}&\multicolumn{1}{c}{(mas)}&\multicolumn{1}{c}{(mas)}&\multicolumn{1}{c}{(mas)}&\multicolumn{1}{c}{(Jy)}\\ 
\hline
 -43.1 &1.3&$ 0.57\pm0.01$&$  24.4\pm 0.1$&$ -44.4\pm 0.1$&$  50.7\pm 0.1$&$   1.5\pm 3.2$&  1.89\\
 -42.5 &0.6&$ 0.32\pm0.21$&$ -38.4\pm 1.6$&$   5.6\pm 1.4$&$  38.7\pm 1.6$&$   1.0\pm 4.9$&  0.17\\
-42.5 &1.0&$ 0.00\pm0.00$&$  10.4\pm 0.6$&$   9.1\pm 0.6$&$  13.8\pm 0.6$&$   4.9\pm 2.2$&  0.28\\ 
  -42.4 &0.8&$ 0.95\pm0.39$&$ -21.0\pm 1.3$&$  26.4\pm 1.3$&$  33.7\pm 1.3$&$   1.6\pm 5.2$&  0.14\\
 -42.1 &1.3&$ 0.68\pm0.03$&$  31.6\pm 0.1$&$ -49.4\pm 0.1$&$  58.6\pm 0.1$&$   2.7\pm 0.9$&  1.24\\
 -41.9 &1.7&$ 1.01\pm0.14$&$  48.7\pm 0.6$&$  18.4\pm 0.5$&$  52.2\pm 0.6$&$  11.5\pm 4.4$&  0.26\\
\hline
\end{tabular}
\addtocounter{table}{-1}
\caption{S Per 1999 22-GHz H$_2$O maser feature parameters (\emph{continued}).}
\label{tab:sp99}
\end{table*}

% U Ori 94
\begin{table*}
\begin{tabular}{ccrrrrrr}
\hline
$V_{\mathrm{LSR}}$&$\Delta{V}$&\multicolumn{1}{c}{$\Delta{V_{1/2}}$}&\multicolumn{1}{c}{$x$}&\multicolumn{1}{c}{$y$}&\multicolumn{1}{c}{$a$}&\multicolumn{1}{c}{$l$}&\multicolumn{1}{c}{$I$}\\ 
(km s$^{-1}$)&(km s$^{-1}$)&\multicolumn{1}{c}{(km s$^{-1}$)}&\multicolumn{1}{c}{(mas)}&\multicolumn{1}{c}{(mas)}&\multicolumn{1}{c}{(mas)}&\multicolumn{1}{c}{(mas)}&\multicolumn{1}{c}{(Jy)}\\ 
\hline
 -40.8 &0.3&$ 0.15\pm0.10$&$  39.0\pm 1.3$&$   2.1\pm 0.1$&$  39.1\pm 2.4$&$  12.1\pm 5.8$&  0.04\\
 -40.7 &1.5&$ 0.00\pm0.00$&$  36.1\pm 0.6$&$  18.5\pm 0.1$&$  40.6\pm 0.4$&$  21.4\pm 5.6$&  0.37\\
 -40.1 &0.5&$ 0.00\pm0.00$&$  37.6\pm 1.3$&$   3.8\pm 0.1$&$  37.8\pm 1.1$&$   8.5\pm 7.4$&  0.07\\
 -38.0 &0.3&$ 0.15\pm0.10$&$  64.5\pm 0.4$&$  94.9\pm 0.1$&$ 114.7\pm 0.4$&$   1.5\pm 1.1$&  0.10\\
 -37.9 &0.3&$ 0.15\pm0.10$&$ -20.3\pm 0.7$&$ -79.3\pm 0.1$&$  81.9\pm 0.7$&$   6.5\pm 3.5$&  0.15\\
 -37.6 &0.7&$ 0.72\pm0.42$&$   0.4\pm 0.2$&$ -88.7\pm 0.1$&$  88.7\pm 0.2$&$   9.0\pm 4.0$&  0.92\\
 -37.5 &0.3&$ 0.15\pm0.10$&$ -36.9\pm 0.5$&$ -77.1\pm 0.1$&$  85.5\pm 0.5$&$   3.5\pm 0.9$&  0.34\\
 -37.5 &0.3&$ 0.15\pm0.10$&$  -7.4\pm 1.1$&$  21.7\pm 0.1$&$  22.9\pm 1.1$&$   5.2\pm 3.1$&  0.06\\
 -37.4 &2.0&$ 0.62\pm0.01$&$ -17.6\pm 0.1$&$ -52.1\pm 0.1$&$  55.0\pm 0.1$&$  40.2\pm 5.1$& 15.76\\
 -37.3 &0.4&$ 0.66\pm0.54$&$ -33.7\pm 0.3$&$ -98.0\pm 0.1$&$ 103.6\pm 0.3$&$   3.9\pm 1.0$&  0.35\\
 -37.3 &0.3&$ 0.15\pm0.10$&$  32.4\pm 0.6$&$ -23.2\pm 0.1$&$  39.8\pm 0.5$&$   3.3\pm 1.8$&  0.08\\
 -36.8 &0.7&$ 0.58\pm0.24$&$ -32.7\pm 0.6$&$ -93.4\pm 0.1$&$  99.0\pm 0.6$&$  10.0\pm 3.9$&  0.24\\
 -36.8 &0.6&$ 0.56\pm0.21$&$  11.8\pm 0.5$&$  29.3\pm 0.1$&$  31.6\pm 0.5$&$   9.8\pm 3.1$&  0.15\\
 -35.5 &1.8&$ 0.67\pm0.02$&$  87.3\pm 0.1$&$ -14.0\pm 0.1$&$  88.4\pm 0.1$&$   7.1\pm10.9$&  1.22\\
\hline
\end{tabular}
\caption{U Ori 1994 22-GHz H$_2$O maser feature parameters.}
\label{tab:uo94}
\end{table*}

% U Ori 1999
\begin{table*}
\begin{tabular}{ccrrrrrr}
\hline
$V_{\mathrm{LSR}}$&$\Delta{V}$&\multicolumn{1}{c}{$\Delta{V_{1/2}}$}&\multicolumn{1}{c}{$x$}&\multicolumn{1}{c}{$y$}&\multicolumn{1}{c}{$a$}&\multicolumn{1}{c}{$l$}&\multicolumn{1}{c}{$I$}\\ 
(km s$^{-1}$)&(km s$^{-1}$)&\multicolumn{1}{c}{(km s$^{-1}$)}&\multicolumn{1}{c}{(mas)}&\multicolumn{1}{c}{(mas)}&\multicolumn{1}{c}{(mas)}&\multicolumn{1}{c}{(mas)}&\multicolumn{1}{c}{(Jy)}\\ 
\hline
-42.1 &0.2&$ 0.15\pm0.10$&$ -16.6\pm 5.3$&$  -9.0\pm 0.1$&$  19.9\pm 5.7$&$   7.9\pm11.9$&  0.38\\
 -41.8 &1.5&$ 0.00\pm0.00$&$ -23.0\pm 1.0$&$  52.9\pm 0.1$&$  57.7\pm 1.0$&$  23.4\pm10.7$&  0.83\\
 -41.4 &0.3&$ 0.15\pm0.10$&$ -32.7\pm 5.1$&$  21.3\pm 0.1$&$  37.5\pm 4.3$&$  34.0\pm11.2$&  0.27\\
 -39.6 &0.7&$ 0.47\pm0.01$&$ 129.8\pm 0.8$&$   3.1\pm 0.1$&$ 129.9\pm 0.9$&$   7.4\pm 5.3$&  0.42\\
 -39.1 &1.6&$ 0.00\pm0.00$&$  38.8\pm 0.3$&$ -33.6\pm 0.1$&$  51.3\pm 0.3$&$  17.8\pm 5.2$&  1.68\\
 -39.0 &0.7&$ 0.68\pm0.22$&$  33.5\pm 1.2$&$  31.1\pm 0.1$&$  45.7\pm 1.2$&$   5.4\pm 8.1$&  0.30\\
 -38.4 &0.5&$ 0.00\pm0.00$&$   3.2\pm 1.6$&$ -57.1\pm 0.1$&$  57.3\pm 1.6$&$  15.2\pm 5.9$&  0.36\\
 -38.3 &0.5&$ 0.00\pm0.00$&$ -63.4\pm 4.6$&$ -61.2\pm 0.1$&$  83.2\pm 4.3$&$  31.7\pm14.1$&  0.85\\
 -38.0 &1.9&$ 0.63\pm0.04$&$  53.6\pm 0.1$&$ -16.2\pm 0.1$&$  56.1\pm 0.1$&$   6.7\pm 1.6$& 55.93\\
 -37.9 &0.8&$ 0.64\pm0.06$&$ -70.7\pm 2.3$&$ -14.9\pm 0.1$&$  72.4\pm 2.0$&$  24.5\pm15.7$&  0.77\\
 -37.8 &0.4&$ 0.21\pm0.02$&$ -52.6\pm 3.8$&$ -49.8\pm 0.1$&$  72.1\pm 4.0$&$  37.5\pm18.7$&  0.38\\
 -35.6 &0.6&$ 0.42\pm0.02$&$  12.7\pm 0.9$&$ -53.8\pm 0.1$&$  55.3\pm 0.9$&$   3.5\pm 3.5$&  0.39\\
\hline
\end{tabular}
\caption{U Ori 1999 22-GHz H$_2$O maser feature parameters.}
\label{tab:uo99}
\end{table*}

% U Ori 2000
\begin{table*}
\begin{tabular}{ccrrrrrr}
\hline
$V_{\mathrm{LSR}}$&$\Delta{V}$&\multicolumn{1}{c}{$\Delta{V_{1/2}}$}&\multicolumn{1}{c}{$x$}&\multicolumn{1}{c}{$y$}&\multicolumn{1}{c}{$a$}&\multicolumn{1}{c}{$l$}&\multicolumn{1}{c}{$I$}\\ 
(km s$^{-1}$)&(km s$^{-1}$)&\multicolumn{1}{c}{(km s$^{-1}$)}&\multicolumn{1}{c}{(mas)}&\multicolumn{1}{c}{(mas)}&\multicolumn{1}{c}{(mas)}&\multicolumn{1}{c}{(mas)}&\multicolumn{1}{c}{(Jy)}\\ 
\hline
-42.3 &0.5&$ 0.00\pm0.00$&$  11.0\pm 0.8$&$  49.9\pm 1.1$&$  52.1\pm 0.1$&$  11.9\pm 3.3$&  0.51\\
 -42.1 &1.1&$ 0.00\pm0.00$&$ -16.4\pm 0.5$&$  86.3\pm 0.5$&$  87.6\pm 0.1$&$  11.9\pm 2.5$&  0.93\\
 -41.6 &0.5&$ 0.15\pm0.10$&$  -0.2\pm 0.8$&$  75.2\pm 0.8$&$  74.9\pm 0.0$&$   7.2\pm 2.7$&  0.67\\
 -41.3 &1.5&$ 1.50\pm0.76$&$  -1.0\pm 0.3$&$  38.1\pm 0.3$&$  35.9\pm 0.0$&$  27.4\pm 1.2$&  0.85\\
 -40.7 &1.5&$ 1.41\pm0.61$&$ -30.4\pm 0.3$&$  70.7\pm 0.2$&$  77.2\pm 0.1$&$  17.7\pm 4.6$&  1.08\\
 -40.6 &0.5&$ 0.15\pm0.10$&$   3.9\pm 1.0$&$  92.7\pm 1.0$&$  94.5\pm 0.0$&$   7.8\pm 3.2$&  0.30\\
\hline
\end{tabular}
\caption{U Ori 2000 22-GHz H$_2$O maser feature parameters (\emph{see next page}).}
\end{table*}

\begin{table*}
\begin{tabular}{ccrrrrrr}
\hline
$V_{\mathrm{LSR}}$&$\Delta{V}$&\multicolumn{1}{c}{$\Delta{V_{1/2}}$}&\multicolumn{1}{c}{$x$}&\multicolumn{1}{c}{$y$}&\multicolumn{1}{c}{$a$}&\multicolumn{1}{c}{$l$}&\multicolumn{1}{c}{$I$}\\ 
(km s$^{-1}$)&(km s$^{-1}$)&\multicolumn{1}{c}{(km s$^{-1}$)}&\multicolumn{1}{c}{(mas)}&\multicolumn{1}{c}{(mas)}&\multicolumn{1}{c}{(mas)}&\multicolumn{1}{c}{(mas)}&\multicolumn{1}{c}{(Jy)}\\ 
\hline
 -37.7 &0.5&$ 0.15\pm0.10$&$ -53.7\pm 1.2$&$ -78.6\pm 1.1$&$  95.8\pm 0.7$&$   5.2\pm 4.0$&  0.86\\
 -37.5 &0.9&$ 0.00\pm0.00$&$  -9.2\pm 0.1$&$ -33.2\pm 0.3$&$  35.7\pm 0.0$&$  13.7\pm 2.8$&  2.11\\
 -37.5 &1.4&$ 0.65\pm0.02$&$ -40.5\pm 0.1$&$ -13.3\pm 0.1$&$  42.8\pm 0.1$&$  14.5\pm 1.8$&  8.49\\
 -37.5 &0.5&$ 0.15\pm0.10$&$  72.2\pm 0.3$&$  30.9\pm 0.3$&$  78.6\pm 0.3$&$   8.9\pm 1.0$&  3.43\\
 -37.5 &0.7&$ 0.38\pm0.05$&$   1.9\pm 0.2$&$ -49.4\pm 0.3$&$  48.5\pm 0.0$&$   8.4\pm 2.1$&  1.37\\

\hline
\end{tabular}
\addtocounter{table}{-1}
\caption{U Ori 2000 22-GHz H$_2$O maser feature parameters (\emph{continued}).}
\label{tab:uo00}
\end{table*}

% U Ori 2001
\begin{table*}
\begin{tabular}{ccrrrrrr}
\hline
$V_{\mathrm{LSR}}$&$\Delta{V}$&\multicolumn{1}{c}{$\Delta{V_{1/2}}$}&\multicolumn{1}{c}{$x$}&\multicolumn{1}{c}{$y$}&\multicolumn{1}{c}{$a$}&\multicolumn{1}{c}{$l$}&\multicolumn{1}{c}{$I$}\\ 
(km s$^{-1}$)&(km s$^{-1}$)&\multicolumn{1}{c}{(km s$^{-1}$)}&\multicolumn{1}{c}{(mas)}&\multicolumn{1}{c}{(mas)}&\multicolumn{1}{c}{(mas)}&\multicolumn{1}{c}{(mas)}&\multicolumn{1}{c}{(Jy)}\\ 
\hline
-41.4 &0.7&$ 0.52\pm0.09$&$ -24.5\pm 0.8$&$  26.8\pm 0.6$&$  36.6\pm 0.7$&$  10.8\pm 3.1$&  0.52\\
 -41.0 &2.0&$ 1.26\pm0.20$&$ -40.0\pm 0.1$&$  68.3\pm 0.1$&$  79.2\pm 0.1$&$  12.8\pm 2.0$&  1.61\\
 -41.0 &1.3&$ 1.18\pm0.69$&$ -36.1\pm 0.2$&$  61.6\pm 0.3$&$  71.4\pm 0.3$&$   7.0\pm 1.9$&  1.14\\
 -40.8 &0.6&$ 0.00\pm0.00$&$ -20.5\pm 1.0$&$  22.6\pm 0.7$&$  28.4\pm 0.8$&$  23.7\pm 2.5$&  0.53\\
 -40.4 &0.5&$ 0.15\pm0.10$&$ -30.5\pm 0.6$&$  30.7\pm 0.6$&$  42.5\pm 0.6$&$   8.6\pm 1.7$&  0.20\\
 -39.9 &1.4&$ 0.92\pm0.31$&$  29.0\pm 0.1$&$  16.4\pm 0.1$&$  33.2\pm 0.1$&$  24.6\pm 1.1$&  2.86\\
\hline
\end{tabular}
\caption{U Ori 2001 22-GHz H$_2$O maser feature parameters.}
\label{tab:uo01}
\end{table*}

% U Her 1994
\begin{table*}
\begin{tabular}{ccrrrrrr}
\hline
$V_{\mathrm{LSR}}$&$\Delta{V}$&\multicolumn{1}{c}{$\Delta{V_{1/2}}$}&\multicolumn{1}{c}{$x$}&\multicolumn{1}{c}{$y$}&\multicolumn{1}{c}{$a$}&\multicolumn{1}{c}{$l$}&\multicolumn{1}{c}{$I$}\\ 
(km s$^{-1}$)&(km s$^{-1}$)&\multicolumn{1}{c}{(km s$^{-1}$)}&\multicolumn{1}{c}{(mas)}&\multicolumn{1}{c}{(mas)}&\multicolumn{1}{c}{(mas)}&\multicolumn{1}{c}{(mas)}&\multicolumn{1}{c}{(Jy)}\\ 
\hline
-21.6 &1.3&$ 1.32\pm0.76$&$ -46.4\pm 0.1$&$ -70.2\pm 0.1$&$  84.2\pm 0.1$&$   5.2\pm 2.5$&  0.75\\
 -20.4 &1.2&$ 1.23\pm0.41$&$ -42.9\pm 0.1$&$ -68.1\pm 0.1$&$  80.5\pm 0.1$&$   4.9\pm 0.4$&  1.72\\
 -20.3 &0.6&$ 0.54\pm0.10$&$-130.9\pm 0.5$&$  -8.1\pm 0.4$&$ 131.2\pm 0.5$&$   3.2\pm 2.2$&  0.22\\
 -19.5 &0.4&$ 0.23\pm0.00$&$-146.0\pm 0.4$&$ -10.1\pm 0.6$&$ 146.4\pm 0.4$&$   5.6\pm 1.4$&  0.78\\
 -19.1 &1.7&$ 1.43\pm0.36$&$ -37.5\pm 0.1$&$ -63.9\pm 0.1$&$  74.1\pm 0.1$&$   8.9\pm 0.3$&  3.96\\
 -18.9 &0.3&$ 0.15\pm0.10$&$ -53.9\pm 0.8$&$ -37.1\pm 0.6$&$  65.4\pm 0.7$&$   3.3\pm 2.0$&  0.51\\
\hline
\end{tabular}
\caption{U Her 1994 22-GHz H$_2$O maser feature parameters (\emph{see next page}).}
\end{table*}

\begin{table*}
\begin{tabular}{ccrrrrrr}
\hline
$V_{\mathrm{LSR}}$&$\Delta{V}$&\multicolumn{1}{c}{$\Delta{V_{1/2}}$}&\multicolumn{1}{c}{$x$}&\multicolumn{1}{c}{$y$}&\multicolumn{1}{c}{$a$}&\multicolumn{1}{c}{$l$}&\multicolumn{1}{c}{$I$}\\ 
(km s$^{-1}$)&(km s$^{-1}$)&\multicolumn{1}{c}{(km s$^{-1}$)}&\multicolumn{1}{c}{(mas)}&\multicolumn{1}{c}{(mas)}&\multicolumn{1}{c}{(mas)}&\multicolumn{1}{c}{(mas)}&\multicolumn{1}{c}{(Jy)}\\ 
\hline
 -13.7 &0.7&$ 0.98\pm0.64$&$   9.5\pm 0.6$&$  74.1\pm 0.7$&$  74.7\pm 0.7$&$   4.7\pm 3.4$&  0.32\\
 -13.6 &0.6&$ 0.24\pm0.03$&$ -48.2\pm 0.5$&$  43.1\pm 0.6$&$  64.5\pm 0.6$&$   7.1\pm 3.9$&  0.55\\
 -13.4 &0.4&$ 0.00\pm0.00$&$ -77.0\pm 2.0$&$  -6.0\pm 2.3$&$  77.4\pm 2.0$&$  13.1\pm 6.6$&  0.27\\

 -13.0 &0.7&$ 0.00\pm0.00$&$  18.4\pm 0.9$&$  71.3\pm 0.8$&$  73.8\pm 0.8$&$  11.6\pm 6.2$&  0.25\\
 -12.9 &1.1&$ 0.89\pm0.20$&$  27.2\pm 0.3$&$  50.8\pm 0.3$&$  57.6\pm 0.3$&$   5.4\pm 2.3$&  0.44\\
 -12.8 &0.8&$ 0.00\pm0.00$&$ -77.6\pm 0.7$&$  -6.0\pm 0.7$&$  77.8\pm 0.7$&$   5.3\pm 3.6$&  0.38\\
 -12.2 &1.1&$ 1.06\pm0.34$&$ -31.7\pm 0.6$&$  52.8\pm 0.6$&$  61.6\pm 0.6$&$   4.9\pm 7.7$&  0.23\\
 -12.2 &0.4&$ 0.00\pm0.00$&$ -80.3\pm 3.4$&$  -8.0\pm 2.7$&$  80.7\pm 3.3$&$   8.3\pm11.6$&  0.14\\
 -11.5 &0.5&$ 0.00\pm0.00$&$ -32.2\pm 1.8$&$  51.6\pm 1.3$&$  60.8\pm 1.5$&$  13.6\pm18.4$&  0.12\\
 -10.2 &0.8&$ 0.52\pm0.06$&$ -65.2\pm 0.8$&$   1.4\pm 0.8$&$  65.2\pm 0.8$&$   5.7\pm11.2$&  0.27\\

\hline
\end{tabular}
\addtocounter{table}{-1}
\caption{U Her 1994 22-GHz H$_2$O maser feature parameters (\emph{continued}).}
\label{tab:uh94}
\end{table*}

%U Her 2000
\begin{table*}
\begin{tabular}{ccrrrrrr}
\hline
$V_{\mathrm{LSR}}$&$\Delta{V}$&\multicolumn{1}{c}{$\Delta{V_{1/2}}$}&\multicolumn{1}{c}{$x$}&\multicolumn{1}{c}{$y$}&\multicolumn{1}{c}{$a$}&\multicolumn{1}{c}{$l$}&\multicolumn{1}{c}{$I$}\\ 
(km s$^{-1}$)&(km s$^{-1}$)&\multicolumn{1}{c}{(km s$^{-1}$)}&\multicolumn{1}{c}{(mas)}&\multicolumn{1}{c}{(mas)}&\multicolumn{1}{c}{(mas)}&\multicolumn{1}{c}{(mas)}&\multicolumn{1}{c}{(Jy)}\\ 
\hline
-18.5 &2.3&$ 0.00\pm0.00$&$ -53.1\pm 0.0$&$ -50.2\pm 0.0$&$  73.1\pm 0.0$&$  21.2\pm 2.3$&  3.77\\
 -17.8 &1.3&$ 0.82\pm0.00$&$ -69.3\pm 0.0$&$ -13.1\pm 0.0$&$  70.5\pm 0.0$&$  20.5\pm 0.7$& 12.30\\
 -17.3 &0.8&$ 0.63\pm0.00$&$ -84.5\pm 0.1$&$   8.3\pm 0.2$&$  85.3\pm 0.1$&$  15.5\pm 0.7$&  2.19\\
 -17.3 &1.6&$ 0.00\pm0.00$&$ -20.1\pm 0.0$&$ 106.8\pm 0.0$&$ 108.7\pm 0.0$&$  12.1\pm 1.5$&  7.36\\
 -16.8 &1.7&$ 0.15\pm0.10$&$ -59.8\pm 0.0$&$  -9.1\pm 0.0$&$  60.5\pm 0.0$&$  21.8\pm 0.2$&122.30\\
 -16.7 &0.9&$ 0.00\pm0.00$&$ -28.9\pm 0.3$&$  37.4\pm 0.5$&$  47.5\pm 0.5$&$  20.8\pm 3.0$&  1.16\\
\hline
\end{tabular}
%\addtocounter{table}{-1}
\caption{U Her 2000 22-GHz H$_2$O maser feature parameters.}
\label{tab:uh00}
\end{table*}

%U Her 2001
\begin{table*}
\begin{tabular}{ccrrrrrr}
\hline
$V_{\mathrm{LSR}}$&$\Delta{V}$&\multicolumn{1}{c}{$\Delta{V_{1/2}}$}&\multicolumn{1}{c}{$x$}&\multicolumn{1}{c}{$y$}&\multicolumn{1}{c}{$a$}&\multicolumn{1}{c}{$l$}&\multicolumn{1}{c}{$I$}\\ 
(km s$^{-1}$)&(km s$^{-1}$)&\multicolumn{1}{c}{(km s$^{-1}$)}&\multicolumn{1}{c}{(mas)}&\multicolumn{1}{c}{(mas)}&\multicolumn{1}{c}{(mas)}&\multicolumn{1}{c}{(mas)}&\multicolumn{1}{c}{(Jy)}\\ 
\hline
-19.9 &0.5&$ 0.15\pm0.10$&$  15.0\pm 0.7$&$ -37.8\pm 0.7$&$  40.7\pm 0.7$&$   7.7\pm 4.7$&  0.48\\
 -18.3 &0.5&$ 0.45\pm0.28$&$ -42.9\pm 0.5$&$ -36.5\pm 0.3$&$  56.3\pm 0.5$&$  15.1\pm 1.9$&  0.32\\
 -17.9 &1.4&$ 0.00\pm0.00$&$ -71.2\pm 0.0$&$ -15.2\pm 0.0$&$  72.8\pm 0.0$&$   6.2\pm 0.3$&  4.69\\
 -17.9 &2.6&$ 1.16\pm0.02$&$ -55.8\pm 0.0$&$ -53.4\pm 0.0$&$  77.2\pm 0.0$&$  28.1\pm 1.6$&  3.58\\
 -17.3 &1.5&$ 0.49\pm0.04$&$ -82.7\pm 0.2$&$   5.9\pm 0.2$&$  83.1\pm 0.2$&$  18.6\pm 0.9$&  1.03\\
 -17.3 &0.6&$ 0.00\pm0.00$&$ -55.0\pm 0.4$&$  58.6\pm 0.4$&$  80.3\pm 0.4$&$   3.2\pm 1.3$&  0.28\\
\hline
\end{tabular}
%\addtocounter{table}{-1}
\caption{U Her 2001 22-GHz H$_2$O maser feature parameters.}
\label{tab:uh01}
\end{table*}

%IK Tau 1994
\begin{table*}
\begin{tabular}{ccrrrrrr}
\hline
$V_{\mathrm{LSR}}$&$\Delta{V}$&\multicolumn{1}{c}{$\Delta{V_{1/2}}$}&\multicolumn{1}{c}{$x$}&\multicolumn{1}{c}{$y$}&\multicolumn{1}{c}{$a$}&\multicolumn{1}{c}{$l$}&\multicolumn{1}{c}{$I$}\\ 
(km s$^{-1}$)&(km s$^{-1}$)&\multicolumn{1}{c}{(km s$^{-1}$)}&\multicolumn{1}{c}{(mas)}&\multicolumn{1}{c}{(mas)}&\multicolumn{1}{c}{(mas)}&\multicolumn{1}{c}{(mas)}&\multicolumn{1}{c}{(Jy)}\\ 
\hline
21.3 &0.3&$ 0.15\pm0.10$&$ -40.0\pm 2.7$&$ 106.0\pm 0.1$&$ 113.4\pm 2.7$&$   6.6\pm 7.1$&  0.03\\
  21.8 &0.7&$ 0.70\pm0.19$&$ 112.6\pm 1.2$&$   5.0\pm 0.1$&$ 112.7\pm 1.2$&$   5.2\pm 4.0$&  0.13\\
  21.9 &0.7&$ 0.56\pm0.07$&$ 110.5\pm 0.8$&$  28.1\pm 0.1$&$ 113.9\pm 0.8$&$   6.9\pm 4.8$&  0.19\\
  22.2 &0.3&$ 0.15\pm0.10$&$ -60.0\pm 2.9$&$-137.9\pm 0.1$&$ 150.4\pm 2.9$&$   3.6\pm 7.8$&  0.02\\
  22.3 &0.3&$ 0.15\pm0.10$&$ 100.2\pm 2.4$&$  33.5\pm 0.1$&$ 105.7\pm 2.4$&$   4.2\pm 6.1$&  0.04\\
  22.5 &0.8&$ 1.25\pm1.06$&$ 114.6\pm 0.8$&$   8.4\pm 0.1$&$ 114.9\pm 0.8$&$   5.2\pm 6.2$&  0.16\\
\hline
\end{tabular}
%\addtocounter{table}{-1}
\caption{IK Tau 1994 22-GHz H$_2$O maser feature parameters (\emph{see next page}).}
\end{table*}
\clearpage
\begin{table*}
\begin{tabular}{ccrrrrrr}
\hline
$V_{\mathrm{LSR}}$&$\Delta{V}$&\multicolumn{1}{c}{$\Delta{V_{1/2}}$}&\multicolumn{1}{c}{$x$}&\multicolumn{1}{c}{$y$}&\multicolumn{1}{c}{$a$}&\multicolumn{1}{c}{$l$}&\multicolumn{1}{c}{$I$}\\ 
(km s$^{-1}$)&(km s$^{-1}$)&\multicolumn{1}{c}{(km s$^{-1}$)}&\multicolumn{1}{c}{(mas)}&\multicolumn{1}{c}{(mas)}&\multicolumn{1}{c}{(mas)}&\multicolumn{1}{c}{(mas)}&\multicolumn{1}{c}{(Jy)}\\ 
\hline
  25.9 &0.3&$ 0.15\pm0.10$&$ 107.9\pm 3.6$&$   8.8\pm 0.1$&$ 108.3\pm 3.6$&$   5.3\pm10.1$&  0.02\\
  26.0 &0.5&$ 0.00\pm0.00$&$  39.7\pm 2.2$&$ -67.3\pm 0.1$&$  78.1\pm 2.2$&$   7.7\pm 7.9$&  0.06\\
  26.1 &0.3&$ 0.15\pm0.10$&$ -32.8\pm 4.0$&$ -29.9\pm 0.1$&$  44.4\pm 4.0$&$   5.9\pm10.3$&  0.02\\
  26.2 &0.3&$ 0.15\pm0.10$&$ -53.0\pm 3.1$&$  25.9\pm 0.1$&$  59.0\pm 3.1$&$   1.7\pm 9.7$&  0.04\\
  26.4 &0.4&$ 0.00\pm0.00$&$  35.0\pm 2.7$&$ -75.2\pm 0.1$&$  83.0\pm 2.7$&$  10.0\pm 8.1$&  0.07\\
  26.5 &0.3&$ 0.15\pm0.10$&$ -35.4\pm 3.4$&$  23.2\pm 0.1$&$  42.4\pm 3.4$&$   5.9\pm 8.4$&  0.02\\
\hline
\end{tabular}
\addtocounter{table}{-1}
\caption{IK Tau 1994 22-GHz H$_2$O maser feature parameters (\emph{continued; see next page}).}
\end{table*}

% 23 + 3*66 + remainder

\begin{table*}
\begin{tabular}{ccrrrrrr}
\hline
$V_{\mathrm{LSR}}$&$\Delta{V}$&\multicolumn{1}{c}{$\Delta{V_{1/2}}$}&\multicolumn{1}{c}{$x$}&\multicolumn{1}{c}{$y$}&\multicolumn{1}{c}{$a$}&\multicolumn{1}{c}{$l$}&\multicolumn{1}{c}{$I$}\\ 
(km s$^{-1}$)&(km s$^{-1}$)&\multicolumn{1}{c}{(km s$^{-1}$)}&\multicolumn{1}{c}{(mas)}&\multicolumn{1}{c}{(mas)}&\multicolumn{1}{c}{(mas)}&\multicolumn{1}{c}{(mas)}&\multicolumn{1}{c}{(Jy)}\\ 
\hline

  31.4 &0.4&$ 0.25\pm0.00$&$ -76.6\pm 2.1$&$ -66.6\pm 0.1$&$ 101.4\pm 2.1$&$  10.6\pm 7.2$&  0.21\\
  31.5 &0.5&$ 0.73\pm0.46$&$ -61.3\pm 2.4$&$ -35.3\pm 0.1$&$  70.7\pm 2.4$&$   5.8\pm 9.8$&  0.13\\
  31.5 &0.3&$ 0.15\pm0.10$&$ -91.9\pm 2.7$&$ -49.9\pm 0.1$&$ 104.5\pm 2.7$&$   4.5\pm 6.9$&  0.15\\
  31.6 &2.1&$ 1.05\pm0.01$&$ -78.1\pm 0.1$&$  -1.2\pm 0.1$&$  78.2\pm 0.1$&$  20.1\pm 1.0$& 26.74\\
  31.6 &0.8&$ 0.76\pm0.15$&$ -44.7\pm 2.5$&$ -35.7\pm 0.1$&$  57.2\pm 2.5$&$   5.3\pm11.2$&  0.21\\
  31.6 &0.5&$ 0.00\pm0.00$&$-145.5\pm 3.5$&$ -25.0\pm 0.1$&$ 147.6\pm 3.5$&$   9.4\pm13.0$&  0.11\\
\hline
\end{tabular}
\addtocounter{table}{-1}
\caption{IK Tau 1994 22-GHz H$_2$O maser feature parameters (\emph{continued; see next page}).}
\end{table*}

\begin{table*}
\begin{tabular}{ccrrrrrr}
\hline
$V_{\mathrm{LSR}}$&$\Delta{V}$&\multicolumn{1}{c}{$\Delta{V_{1/2}}$}&\multicolumn{1}{c}{$x$}&\multicolumn{1}{c}{$y$}&\multicolumn{1}{c}{$a$}&\multicolumn{1}{c}{$l$}&\multicolumn{1}{c}{$I$}\\ 
(km s$^{-1}$)&(km s$^{-1}$)&\multicolumn{1}{c}{(km s$^{-1}$)}&\multicolumn{1}{c}{(mas)}&\multicolumn{1}{c}{(mas)}&\multicolumn{1}{c}{(mas)}&\multicolumn{1}{c}{(mas)}&\multicolumn{1}{c}{(Jy)}\\ 
\hline
  38.5 &0.3&$ 0.15\pm0.10$&$  94.5\pm 3.1$&$   6.0\pm 0.1$&$  94.7\pm 3.1$&$   4.7\pm 8.0$&  0.03\\
  38.5 &0.4&$ 0.00\pm0.00$&$ -52.5\pm 2.3$&$   7.4\pm 0.1$&$  53.1\pm 2.3$&$   7.9\pm 6.9$&  0.04\\
  38.6 &0.8&$ 0.77\pm0.24$&$  94.2\pm 1.3$&$ -21.6\pm 0.1$&$  96.7\pm 1.3$&$   5.8\pm 7.2$&  0.09\\
  38.6 &0.4&$ 0.00\pm0.00$&$  76.4\pm 0.5$&$ -62.4\pm 0.1$&$  98.6\pm 0.5$&$   2.9\pm 1.6$&  0.40\\
  38.7 &0.5&$ 0.00\pm0.00$&$  35.0\pm 3.2$&$ -56.2\pm 0.1$&$  66.3\pm 3.2$&$   7.4\pm 9.9$&  0.04\\
  38.8 &0.3&$ 0.15\pm0.10$&$  93.2\pm 2.7$&$ -70.9\pm 0.1$&$ 117.1\pm 2.7$&$  14.4\pm 7.4$&  0.05\\
\hline
\end{tabular}
\addtocounter{table}{-1}
\caption{IK Tau 1994 22-GHz H$_2$O maser feature parameters (\emph{continued; see next page}).}
\end{table*}

\clearpage
\begin{table*}
\begin{tabular}{ccrrrrrr}
\hline
$V_{\mathrm{LSR}}$&$\Delta{V}$&\multicolumn{1}{c}{$\Delta{V_{1/2}}$}&\multicolumn{1}{c}{$x$}&\multicolumn{1}{c}{$y$}&\multicolumn{1}{c}{$a$}&\multicolumn{1}{c}{$l$}&\multicolumn{1}{c}{$I$}\\ 
(km s$^{-1}$)&(km s$^{-1}$)&\multicolumn{1}{c}{(km s$^{-1}$)}&\multicolumn{1}{c}{(mas)}&\multicolumn{1}{c}{(mas)}&\multicolumn{1}{c}{(mas)}&\multicolumn{1}{c}{(mas)}&\multicolumn{1}{c}{(Jy)}\\ 
\hline
  45.2 &1.4&$ 0.62\pm0.14$&$  76.6\pm 1.8$&$  12.1\pm 0.1$&$  77.6\pm 1.8$&$   7.3\pm 9.0$&  0.11\\
  45.2 &0.5&$ 0.00\pm0.00$&$  65.0\pm 3.2$&$ -63.1\pm 0.1$&$  90.6\pm 3.2$&$   5.5\pm11.3$&  0.07\\
  45.2 &0.3&$ 0.15\pm0.10$&$  58.3\pm 2.0$&$ -23.9\pm 0.1$&$  63.0\pm 2.0$&$   3.6\pm 5.0$&  0.08\\
  45.3 &1.6&$ 0.69\pm0.01$&$  18.1\pm 0.1$&$  25.0\pm 0.1$&$  30.9\pm 0.1$&$   7.9\pm 8.8$&  2.30\\
  45.3 &0.3&$ 0.15\pm0.10$&$  67.2\pm 3.0$&$  68.3\pm 0.1$&$  95.8\pm 3.0$&$   2.9\pm 8.4$&  0.03\\
  45.3 &0.4&$ 0.50\pm0.23$&$  84.2\pm 3.4$&$ -66.9\pm 0.1$&$ 107.6\pm 3.4$&$   8.3\pm 9.6$&  0.07\\
\hline
\end{tabular}
\addtocounter{table}{-1}
\caption{IK Tau 1994 22-GHz H$_2$O maser feature parameters (\emph{continued}).}
\label{tab:ik94}
\end{table*}

%IK Tau 2000
\begin{table*}
\begin{tabular}{ccrrrrrr}
\hline
$V_{\mathrm{LSR}}$&$\Delta{V}$&\multicolumn{1}{c}{$\Delta{V_{1/2}}$}&\multicolumn{1}{c}{$x$}&\multicolumn{1}{c}{$y$}&\multicolumn{1}{c}{$a$}&\multicolumn{1}{c}{$l$}&\multicolumn{1}{c}{$I$}\\ 
(km s$^{-1}$)&(km s$^{-1}$)&\multicolumn{1}{c}{(km s$^{-1}$)}&\multicolumn{1}{c}{(mas)}&\multicolumn{1}{c}{(mas)}&\multicolumn{1}{c}{(mas)}&\multicolumn{1}{c}{(mas)}&\multicolumn{1}{c}{(Jy)}\\ 
\hline
  21.0 &1.5&$ 0.65\pm0.01$&$  18.3\pm 0.0$&$ 100.8\pm 0.0$&$ 102.4\pm 0.0$&$   9.8\pm 1.2$&  4.92\\
  21.1 &1.1&$ 0.00\pm0.00$&$   7.5\pm 0.3$&$  21.8\pm 0.2$&$  22.5\pm 0.2$&$  18.7\pm 1.0$&  1.13\\
  21.5 &1.1&$ 0.00\pm0.00$&$   4.6\pm 0.2$&$  17.5\pm 0.1$&$  18.2\pm 0.1$&$   7.7\pm 0.9$&  1.30\\
  21.6 &3.4&$ 2.17\pm0.80$&$ -16.9\pm 0.1$&$  16.6\pm 0.1$&$  23.5\pm 0.1$&$  25.1\pm 1.9$&  1.89\\
  21.8 &2.3&$ 2.22\pm0.65$&$  84.0\pm 0.1$&$  69.0\pm 0.1$&$ 108.8\pm 0.1$&$  17.4\pm 1.6$&  1.67\\
  21.8 &1.1&$ 0.00\pm0.00$&$  61.0\pm 0.7$&$  20.7\pm 0.8$&$  64.4\pm 0.7$&$   5.5\pm 2.1$&  0.49\\
\hline
\end{tabular}
%\addtocounter{table}{-1}
\caption{IK Tau 2000 22-GHz H$_2$O maser feature parameters (\emph{see next page}).}
\end{table*}

\begin{table*}
\begin{tabular}{ccrrrrrr}
\hline
$V_{\mathrm{LSR}}$&$\Delta{V}$&\multicolumn{1}{c}{$\Delta{V_{1/2}}$}&\multicolumn{1}{c}{$x$}&\multicolumn{1}{c}{$y$}&\multicolumn{1}{c}{$a$}&\multicolumn{1}{c}{$l$}&\multicolumn{1}{c}{$I$}\\ 
(km s$^{-1}$)&(km s$^{-1}$)&\multicolumn{1}{c}{(km s$^{-1}$)}&\multicolumn{1}{c}{(mas)}&\multicolumn{1}{c}{(mas)}&\multicolumn{1}{c}{(mas)}&\multicolumn{1}{c}{(mas)}&\multicolumn{1}{c}{(Jy)}\\ 
\hline
  24.8 &1.9&$ 0.82\pm0.00$&$  45.7\pm 0.0$&$ -73.0\pm 0.0$&$  86.1\pm 0.0$&$   8.6\pm 1.4$& 27.63\\
  25.0 &1.1&$ 0.30\pm0.20$&$ -77.3\pm 0.8$&$  30.2\pm 0.7$&$  82.9\pm 0.8$&$   6.6\pm 1.9$&  0.86\\
  25.2 &1.1&$ 1.01\pm0.34$&$  40.9\pm 0.5$&$ -17.6\pm 0.5$&$  46.2\pm 0.5$&$  16.9\pm 2.2$&  1.81\\
  25.4 &0.8&$ 0.30\pm0.20$&$ -56.0\pm 0.1$&$ -45.8\pm 0.1$&$  72.4\pm 0.1$&$   4.3\pm 0.3$&  3.07\\
  25.4 &1.3&$ 0.64\pm0.02$&$ -62.3\pm 0.2$&$ -25.5\pm 0.1$&$  67.5\pm 0.2$&$  15.8\pm 0.8$&  3.36\\
  25.6 &1.3&$ 0.00\pm0.00$&$ -77.8\pm 0.3$&$ -48.0\pm 0.3$&$  91.5\pm 0.3$&$   9.8\pm 3.0$&  2.11\\
\hline
\end{tabular}
\addtocounter{table}{-1}
\caption{IK Tau 2000 22-GHz H$_2$O maser feature parameters (\emph{continued}).}
\label{tab:ik00}
\end{table*}

%IK Tau 2001
\begin{table*}
\begin{tabular}{ccrrrrrr}
\hline
$V_{\mathrm{LSR}}$&$\Delta{V}$&\multicolumn{1}{c}{$\Delta{V_{1/2}}$}&\multicolumn{1}{c}{$x$}&\multicolumn{1}{c}{$y$}&\multicolumn{1}{c}{$a$}&\multicolumn{1}{c}{$l$}&\multicolumn{1}{c}{$I$}\\ 
(km s$^{-1}$)&(km s$^{-1}$)&\multicolumn{1}{c}{(km s$^{-1}$)}&\multicolumn{1}{c}{(mas)}&\multicolumn{1}{c}{(mas)}&\multicolumn{1}{c}{(mas)}&\multicolumn{1}{c}{(mas)}&\multicolumn{1}{c}{(Jy)}\\ 
\hline

 20.6 &1.7&$ 0.50\pm0.07$&$  -1.8\pm 0.1$&$  -0.2\pm 0.1$&$   1.9\pm 0.1$&$  16.6\pm 0.9$&  3.89\\
  22.1 &1.1&$ 0.86\pm0.65$&$  53.8\pm 0.5$&$-193.9\pm 0.5$&$ 201.3\pm 0.5$&$   4.6\pm 1.7$&  0.51\\
  22.7 &0.6&$ 0.15\pm0.10$&$ -18.2\pm 0.5$&$  15.3\pm 0.5$&$  24.1\pm 0.5$&$  12.3\pm 1.3$&  0.54\\
  23.0 &2.7&$ 0.92\pm0.11$&$   0.5\pm 0.0$&$  -2.0\pm 0.0$&$   2.3\pm 0.0$&$   9.5\pm 0.3$& 20.66\\
  23.4 &0.8&$ 0.95\pm0.50$&$ -59.8\pm 0.2$&$  11.0\pm 0.1$&$  60.8\pm 0.2$&$   4.3\pm 0.4$&  1.64\\
  23.4 &0.6&$ 0.15\pm0.10$&$  22.8\pm 0.2$&$ -71.7\pm 0.2$&$  75.2\pm 0.2$&$   5.6\pm 1.6$&  1.46\\
  23.4 &1.3&$ 0.00\pm0.00$&$ -38.3\pm 0.4$&$  71.2\pm 0.4$&$  80.7\pm 0.4$&$  10.6\pm 1.8$&  0.54\\
  23.7 &1.7&$ 0.83\pm0.02$&$  52.5\pm 0.0$&$ -81.6\pm 0.0$&$  97.1\pm 0.0$&$   4.3\pm 0.2$& 16.61\\
  23.9 &0.8&$ 0.63\pm0.07$&$  17.3\pm 0.1$&$ -75.3\pm 0.1$&$  77.3\pm 0.1$&$   4.3\pm 0.4$&  1.69\\
  24.2 &0.8&$ 0.15\pm0.10$&$  29.9\pm 0.4$&$  69.2\pm 0.4$&$  75.3\pm 0.4$&$   3.3\pm 0.9$&  0.49\\
  24.6 &1.9&$ 0.00\pm0.00$&$  17.2\pm 0.2$&$  -8.7\pm 0.1$&$  19.4\pm 0.2$&$  29.5\pm 2.2$&  1.35\\
  24.6 &1.5&$ 1.44\pm0.61$&$  47.7\pm 0.1$&$ -83.0\pm 0.1$&$  95.7\pm 0.1$&$  15.7\pm 1.0$&  3.89\\
  24.9 &1.9&$ 0.00\pm0.00$&$   2.4\pm 0.2$&$   5.5\pm 0.2$&$   6.5\pm 0.2$&$  17.8\pm 3.2$&  2.74\\

\hline
\end{tabular}
%\addtocounter{table}{-1}
\caption{IK Tau 2001 22-GHz H$_2$O maser feature parameters (\emph{see next page}).}
\end{table*}

\begin{table*}
\begin{tabular}{ccrrrrrr}
\hline
$V_{\mathrm{LSR}}$&$\Delta{V}$&\multicolumn{1}{c}{$\Delta{V_{1/2}}$}&\multicolumn{1}{c}{$x$}&\multicolumn{1}{c}{$y$}&\multicolumn{1}{c}{$a$}&\multicolumn{1}{c}{$l$}&\multicolumn{1}{c}{$I$}\\ 
(km s$^{-1}$)&(km s$^{-1}$)&\multicolumn{1}{c}{(km s$^{-1}$)}&\multicolumn{1}{c}{(mas)}&\multicolumn{1}{c}{(mas)}&\multicolumn{1}{c}{(mas)}&\multicolumn{1}{c}{(mas)}&\multicolumn{1}{c}{(Jy)}\\ 
\hline
  24.9 &0.8&$ 0.00\pm0.00$&$ -35.6\pm 0.3$&$  -3.7\pm 0.3$&$  35.9\pm 0.3$&$   8.9\pm 1.3$&  0.49\\
  25.1 &1.3&$ 0.00\pm0.00$&$ -37.7\pm 0.3$&$  66.0\pm 0.3$&$  76.0\pm 0.3$&$   6.3\pm 0.9$&  0.65\\
  25.4 &2.9&$ 1.48\pm0.07$&$ -60.3\pm 0.0$&$   7.3\pm 0.0$&$  60.7\pm 0.0$&$  10.3\pm 1.0$&  3.84\\
  25.6 &1.3&$ 0.00\pm0.00$&$  17.4\pm 0.2$&$  56.0\pm 0.2$&$  58.6\pm 0.2$&$   3.5\pm 0.8$&  0.75\\
  25.8 &0.6&$ 0.15\pm0.10$&$  28.7\pm 0.4$&$ -19.2\pm 0.4$&$  34.6\pm 0.4$&$  20.1\pm 1.6$&  0.84\\
  25.9 &0.8&$ 0.75\pm0.20$&$  74.2\pm 0.1$&$-100.4\pm 0.1$&$ 124.9\pm 0.1$&$   6.3\pm 0.8$&  1.16\\
\hline
\end{tabular}
\addtocounter{table}{-1}
\caption{IK Tau 2001 22-GHz H$_2$O maser feature parameters (\emph{continued}).}
\label{tab:ik01}
\end{table*}

\begin{table*}
\begin{tabular}{ccrrrrrr}
\hline
$V_{\mathrm{LSR}}$&$\Delta{V}$&\multicolumn{1}{c}{$\Delta{V_{1/2}}$}&\multicolumn{1}{c}{$x$}&\multicolumn{1}{c}{$y$}&\multicolumn{1}{c}{$a$}&\multicolumn{1}{c}{$l$}&\multicolumn{1}{c}{$I$}\\ 
(km s$^{-1}$)&(km s$^{-1}$)&\multicolumn{1}{c}{(km s$^{-1}$)}&\multicolumn{1}{c}{(mas)}&\multicolumn{1}{c}{(mas)}&\multicolumn{1}{c}{(mas)}&\multicolumn{1}{c}{(mas)}&\multicolumn{1}{c}{(Jy)}\\ 
\hline
 10.4 &1.9&$ 0.00\pm0.00$&$   3.3\pm 0.1$&$  19.0\pm 0.1$&$  23.7\pm 0.1$&$  36.5\pm 1.9$& 11.37\\
  10.5 &1.8&$ 1.23\pm0.23$&$ -22.7\pm 0.8$&$  72.8\pm 0.8$&$  76.3\pm 0.8$&$  14.0\pm 6.1$&  0.80\\
  10.5 &0.9&$ 1.25\pm0.95$&$  -3.5\pm 1.4$&$ -56.8\pm 1.4$&$  57.0\pm 1.4$&$  17.8\pm 7.7$&  0.66\\
  10.7 &2.1&$ 1.05\pm0.01$&$   0.6\pm 0.0$&$  27.8\pm 0.0$&$  27.9\pm 0.0$&$  12.8\pm 1.6$& 26.71\\
  10.9 &0.9&$ 1.16\pm0.87$&$  53.3\pm 1.6$&$  27.5\pm 1.6$&$  60.0\pm 1.6$&$   4.2\pm 7.8$&  0.56\\
  11.0 &1.3&$ 0.81\pm0.11$&$  25.6\pm 0.4$&$  37.3\pm 0.4$&$  45.3\pm 0.4$&$  16.8\pm 8.6$&  2.24\\
\hline
\end{tabular}
%\addtocounter{table}{-1}
\caption{RT Vir 1994 22-GHz H$_2$O maser feature parameters (\emph{see next page}).}
\end{table*}

\begin{table*}
\begin{tabular}{ccrrrrrr}
\hline
$V_{\mathrm{LSR}}$&$\Delta{V}$&\multicolumn{1}{c}{$\Delta{V_{1/2}}$}&\multicolumn{1}{c}{$x$}&\multicolumn{1}{c}{$y$}&\multicolumn{1}{c}{$a$}&\multicolumn{1}{c}{$l$}&\multicolumn{1}{c}{$I$}\\ 
(km s$^{-1}$)&(km s$^{-1}$)&\multicolumn{1}{c}{(km s$^{-1}$)}&\multicolumn{1}{c}{(mas)}&\multicolumn{1}{c}{(mas)}&\multicolumn{1}{c}{(mas)}&\multicolumn{1}{c}{(mas)}&\multicolumn{1}{c}{(Jy)}\\ 
\hline
   19.4 &1.4&$ 0.96\pm0.11$&$  21.4\pm 0.5$&$ -57.4\pm 0.5$&$  61.2\pm 0.5$&$   3.4\pm 5.5$&  0.97\\
  20.9 &1.7&$ 1.35\pm0.36$&$  76.4\pm 0.6$&$  35.3\pm 0.6$&$  84.1\pm 0.6$&$  10.6\pm 5.4$&  1.23\\
  21.1 &1.2&$ 0.58\pm0.01$&$  25.0\pm 0.8$&$ -62.2\pm 0.8$&$  67.0\pm 0.8$&$   2.2\pm 6.1$&  0.86\\
  21.7 &1.7&$ 1.10\pm0.07$&$  89.1\pm 0.4$&$ -35.6\pm 0.4$&$  96.0\pm 0.4$&$  12.4\pm 6.4$&  1.29\\
  22.2 &2.9&$ 0.96\pm0.03$&$  39.6\pm 0.0$&$  19.4\pm 0.0$&$  44.0\pm 0.0$&$   7.6\pm12.0$& 12.43\\
  23.0 &1.4&$ 0.76\pm0.02$&$  56.5\pm 0.7$&$ -43.1\pm 0.7$&$  71.1\pm 0.7$&$  12.9\pm 8.0$&  4.20\\
\hline
\end{tabular}
\addtocounter{table}{-1}
\caption{RT Vir 1994 22-GHz H$_2$O maser feature parameters (\emph{continued}).}
\label{tab:rt94}
\end{table*}

\begin{table*}
\begin{tabular}{ccrrrrrr}
\hline
$V_{\mathrm{LSR}}$&$\Delta{V}$&\multicolumn{1}{c}{$\Delta{V_{1/2}}$}&\multicolumn{1}{c}{$x$}&\multicolumn{1}{c}{$y$}&\multicolumn{1}{c}{$a$}&\multicolumn{1}{c}{$l$}&\multicolumn{1}{c}{$I$}\\ 
(km s$^{-1}$)&(km s$^{-1}$)&\multicolumn{1}{c}{(km s$^{-1}$)}&\multicolumn{1}{c}{(mas)}&\multicolumn{1}{c}{(mas)}&\multicolumn{1}{c}{(mas)}&\multicolumn{1}{c}{(mas)}&\multicolumn{1}{c}{(Jy)}\\ 
\hline
 9.8 &0.5&$ 0.00\pm0.00$&$ -19.1\pm 0.2$&$  35.6\pm 0.2$&$  40.4\pm 0.2$&$   8.4\pm 1.5$&  1.50\\
10.3 &0.3&$ 0.36\pm0.09$&$  -5.6\pm 0.2$&$  41.1\pm 0.2$&$  41.5\pm 1.5$&$  18.2\pm 0.7$&  0.70\\
10.6 &0.4&$ 0.47\pm0.19$&$   5.1\pm 0.2$&$  39.9\pm 0.2$&$  40.3\pm 0.2$&$  21.4\pm 1.0$&  1.00\\
11.1 &0.6&$ 0.59\pm0.15$&$ -29.1\pm 0.3$&$ -61.6\pm 0.3$&$  68.1\pm 1.1$&$   1.1\pm 0.8$&  0.80\\
11.7 &2.9&$ 1.04\pm0.02$&$ -29.4\pm 0.1$&$  11.9\pm 0.1$&$  31.7\pm 0.1$&$  26.6\pm 0.4$&  0.40\\
12.5 &0.5&$ 0.42\pm0.19$&$ -50.2\pm 0.3$&$ -29.9\pm 0.3$&$  58.4\pm 1.0$&$   6.9\pm 1.7$&  1.70\\
\hline
\end{tabular}
%\addtocounter{table}{-1}
\caption{RT Vir 19960405 22-GHz H$_2$O maser feature parameters (\emph{see next page}).}
\end{table*}

\begin{table*}
\begin{tabular}{ccrrrrrr}
\hline
$V_{\mathrm{LSR}}$&$\Delta{V}$&\multicolumn{1}{c}{$\Delta{V_{1/2}}$}&\multicolumn{1}{c}{$x$}&\multicolumn{1}{c}{$y$}&\multicolumn{1}{c}{$a$}&\multicolumn{1}{c}{$l$}&\multicolumn{1}{c}{$I$}\\ 
(km s$^{-1}$)&(km s$^{-1}$)&\multicolumn{1}{c}{(km s$^{-1}$)}&\multicolumn{1}{c}{(mas)}&\multicolumn{1}{c}{(mas)}&\multicolumn{1}{c}{(mas)}&\multicolumn{1}{c}{(mas)}&\multicolumn{1}{c}{(Jy)}\\ 
\hline
20.7 &0.4&$ 0.00\pm0.00$&$  82.2\pm 1.0$&$ -11.4\pm 1.0$&$  83.0\pm 1.0$&$   4.5\pm 3.5$&  3.50\\
20.8 &0.6&$ 0.65\pm0.25$&$  35.8\pm 0.4$&$ -36.5\pm 0.4$&$  51.1\pm 0.4$&$  13.1\pm 3.5$&  3.50\\
20.9 &0.3&$ 0.15\pm0.10$&$  18.6\pm 1.2$&$ -25.8\pm 1.3$&$  31.8\pm 2.0$&$   0.9\pm 3.4$&  3.40\\
21.1 &0.3&$ 0.15\pm0.10$&$  92.4\pm 1.6$&$  63.7\pm 1.7$&$ 112.3\pm 2.2$&$   7.2\pm 5.2$&  5.20\\
21.4 &0.7&$ 0.73\pm0.24$&$  46.5\pm 0.4$&$ 107.7\pm 0.4$&$ 117.4\pm 1.2$&$   2.2\pm 2.3$&  2.30\\
21.4 &1.7&$ 1.18\pm0.12$&$  49.1\pm 0.1$&$  33.2\pm 0.1$&$  59.3\pm 0.7$&$   9.5\pm 1.4$&  1.40\\
\hline
\end{tabular}
\addtocounter{table}{-1}
\caption{RT Vir 19960405 22-GHz H$_2$O maser feature parameters (\emph{continued}).}
\label{tab:rt96_1}
\end{table*}

\begin{table*}
\begin{tabular}{ccrrrrrr}
\hline
$V_{\mathrm{LSR}}$&$\Delta{V}$&\multicolumn{1}{c}{$\Delta{V_{1/2}}$}&\multicolumn{1}{c}{$x$}&\multicolumn{1}{c}{$y$}&\multicolumn{1}{c}{$a$}&\multicolumn{1}{c}{$l$}&\multicolumn{1}{c}{$I$}\\ 
(km s$^{-1}$)&(km s$^{-1}$)&\multicolumn{1}{c}{(km s$^{-1}$)}&\multicolumn{1}{c}{(mas)}&\multicolumn{1}{c}{(mas)}&\multicolumn{1}{c}{(mas)}&\multicolumn{1}{c}{(mas)}&\multicolumn{1}{c}{(Jy)}\\ 
\hline

  9.8 &0.8&$ 0.42\pm0.03$&$ -29.2\pm 0.2$&$  40.7\pm 0.3$&$  50.1\pm 0.4$&$  12.2\pm 4.3$&  4.30\\
 10.3 &0.4&$ 0.00\pm0.00$&$ -15.7\pm 0.2$&$  42.8\pm 0.2$&$  45.6\pm 0.2$&$   5.2\pm 0.5$&  0.50\\
 10.4 &0.8&$ 0.94\pm0.77$&$  -0.9\pm 0.2$&$  34.0\pm 0.2$&$  34.0\pm 0.2$&$   6.7\pm 1.3$&  1.30\\
 10.4 &0.4&$ 0.00\pm0.00$&$ -79.4\pm 0.4$&$ -21.6\pm 0.6$&$  82.3\pm 1.1$&$   7.1\pm 1.6$&  1.60\\
 11.1 &0.8&$ 0.88\pm0.82$&$ -28.3\pm 0.3$&$ -84.0\pm 0.4$&$  88.6\pm 1.1$&$  11.8\pm 1.5$&  1.50\\
 11.2 &0.6&$ 0.65\pm0.41$&$ -13.6\pm 0.1$&$   3.3\pm 0.1$&$  14.0\pm 1.0$&$   4.0\pm 0.5$&  0.50\\
\hline
\end{tabular}
%\addtocounter{table}{-1}
\caption{RT Vir 19960421 22-GHz H$_2$O maser feature parameters (\emph{see next page}).}
\end{table*}

\begin{table*}
\begin{tabular}{ccrrrrrr}
\hline
$V_{\mathrm{LSR}}$&$\Delta{V}$&\multicolumn{1}{c}{$\Delta{V_{1/2}}$}&\multicolumn{1}{c}{$x$}&\multicolumn{1}{c}{$y$}&\multicolumn{1}{c}{$a$}&\multicolumn{1}{c}{$l$}&\multicolumn{1}{c}{$I$}\\ 
(km s$^{-1}$)&(km s$^{-1}$)&\multicolumn{1}{c}{(km s$^{-1}$)}&\multicolumn{1}{c}{(mas)}&\multicolumn{1}{c}{(mas)}&\multicolumn{1}{c}{(mas)}&\multicolumn{1}{c}{(mas)}&\multicolumn{1}{c}{(Jy)}\\ 
\hline
 22.3 &0.3&$ 0.15\pm0.10$&$  61.6\pm 0.9$&$  37.9\pm 1.0$&$  72.3\pm 1.7$&$   1.5\pm 2.4$&  2.40\\
 22.5 &0.3&$ 0.15\pm0.10$&$  77.1\pm 0.9$&$ -24.2\pm 0.9$&$  80.8\pm 0.9$&$   6.7\pm 2.3$&  2.30\\
 22.8 &0.3&$ 0.15\pm0.10$&$  22.6\pm 0.5$&$  28.8\pm 0.5$&$  36.6\pm 1.8$&$   0.6\pm 1.4$&  1.40\\
 22.8 &0.6&$ 0.00\pm0.00$&$  13.9\pm 0.7$&$ -22.5\pm 1.1$&$  26.4\pm 1.7$&$  13.9\pm 4.5$&  4.50\\
 22.9 &1.7&$ 1.42\pm0.27$&$  32.3\pm 0.2$&$  13.9\pm 0.2$&$  35.2\pm 0.2$&$  19.1\pm 1.8$&  1.80\\
 23.3 &0.8&$ 0.70\pm0.13$&$  24.5\pm 0.2$&$  29.2\pm 0.2$&$  38.1\pm 0.2$&$   5.2\pm 1.2$&  1.20\\
\hline
\end{tabular}
\addtocounter{table}{-1}
\caption{RT Vir 19960421 22-GHz H$_2$O maser feature parameters (\emph{continued}).}
\label{tab:rt96_2}
\end{table*}

\begin{table*}
\begin{tabular}{ccrrrrrr}
\hline
$V_{\mathrm{LSR}}$&$\Delta{V}$&\multicolumn{1}{c}{$\Delta{V_{1/2}}$}&\multicolumn{1}{c}{$x$}&\multicolumn{1}{c}{$y$}&\multicolumn{1}{c}{$a$}&\multicolumn{1}{c}{$l$}&\multicolumn{1}{c}{$I$}\\ 
(km s$^{-1}$)&(km s$^{-1}$)&\multicolumn{1}{c}{(km s$^{-1}$)}&\multicolumn{1}{c}{(mas)}&\multicolumn{1}{c}{(mas)}&\multicolumn{1}{c}{(mas)}&\multicolumn{1}{c}{(mas)}&\multicolumn{1}{c}{(Jy)}\\ 
\hline

  9.1 &0.6&$ 0.47\pm0.07$&$ -18.9\pm 0.8$&$  52.7\pm 0.9$&$  56.0\pm 0.9$&$   6.9\pm 3.1$&  3.10\\
 10.2 &0.3&$ 0.15\pm0.10$&$ -18.2\pm 0.8$&$  75.3\pm 0.8$&$  77.5\pm 0.8$&$   2.8\pm 2.3$&  2.30\\
 10.3 &1.6&$ 0.70\pm0.07$&$ -18.1\pm 0.1$&$  38.9\pm 0.1$&$  42.9\pm 0.1$&$  11.8\pm 2.9$&  2.90\\
 10.7 &0.3&$ 0.15\pm0.10$&$ -27.4\pm 0.3$&$  32.8\pm 0.4$&$  42.7\pm 1.5$&$   1.8\pm 0.9$&  0.90\\
 11.1 &0.4&$ 0.00\pm0.00$&$ -46.9\pm 0.5$&$ -39.5\pm 0.4$&$  61.3\pm 1.2$&$   1.1\pm 1.4$&  1.40\\
 11.4 &3.3&$ 1.08\pm0.01$&$ -29.2\pm 0.1$&$  11.1\pm 0.1$&$  31.2\pm 0.1$&$  13.1\pm 0.6$&  0.60\\
\hline
\end{tabular}
%\addtocounter{table}{-1}
\caption{RT Vir 19960429 22-GHz H$_2$O maser feature parameters (\emph{see next page}).}
\end{table*}

\begin{table*}
\begin{tabular}{ccrrrrrr}
\hline
$V_{\mathrm{LSR}}$&$\Delta{V}$&\multicolumn{1}{c}{$\Delta{V_{1/2}}$}&\multicolumn{1}{c}{$x$}&\multicolumn{1}{c}{$y$}&\multicolumn{1}{c}{$a$}&\multicolumn{1}{c}{$l$}&\multicolumn{1}{c}{$I$}\\ 
(km s$^{-1}$)&(km s$^{-1}$)&\multicolumn{1}{c}{(km s$^{-1}$)}&\multicolumn{1}{c}{(mas)}&\multicolumn{1}{c}{(mas)}&\multicolumn{1}{c}{(mas)}&\multicolumn{1}{c}{(mas)}&\multicolumn{1}{c}{(Jy)}\\ 
\hline
 24.3 &0.6&$ 0.92\pm0.75$&$  45.8\pm 0.3$&$  28.3\pm 0.2$&$  53.9\pm 1.1$&$  14.0\pm 1.3$&  1.30\\
 24.3 &1.6&$ 0.82\pm0.04$&$  20.9\pm 0.1$&$ -10.8\pm 0.1$&$  23.6\pm 0.1$&$   6.5\pm 0.7$&  0.70\\
 24.7 &0.6&$ 0.00\pm0.00$&$  36.3\pm 0.4$&$ -59.2\pm 0.3$&$  69.5\pm 1.2$&$   3.3\pm 1.2$&  1.20\\
 24.8 &3.9&$ 1.33\pm0.01$&$  53.2\pm 0.1$&$  -5.7\pm 0.1$&$  53.5\pm 0.1$&$  12.7\pm 0.7$&  0.70\\
 26.1 &0.3&$ 0.15\pm0.10$&$  43.0\pm 0.4$&$ -19.4\pm 0.4$&$  47.2\pm 0.4$&$   2.7\pm 1.4$&  1.40\\
 26.1 &0.3&$ 0.15\pm0.10$&$  53.2\pm 0.9$&$ -99.4\pm 1.0$&$ 112.8\pm 1.0$&$   6.4\pm 2.8$&  2.80\\
 26.4 &0.8&$ 0.48\pm0.02$&$  40.6\pm 0.2$&$ -90.1\pm 0.2$&$  98.8\pm 0.2$&$   5.3\pm 3.2$&  3.20\\
\hline
\end{tabular}
\addtocounter{table}{-1}
\caption{RT Vir 19960429 22-GHz H$_2$O maser feature parameters (\emph{continues}).}
\label{tab:rt96_3}
\end{table*}

\begin{table*}
\begin{tabular}{ccrrrrrr}
\hline
$V_{\mathrm{LSR}}$&$\Delta{V}$&\multicolumn{1}{c}{$\Delta{V_{1/2}}$}&\multicolumn{1}{c}{$x$}&\multicolumn{1}{c}{$y$}&\multicolumn{1}{c}{$a$}&\multicolumn{1}{c}{$l$}&\multicolumn{1}{c}{$I$}\\ 
(km s$^{-1}$)&(km s$^{-1}$)&\multicolumn{1}{c}{(km s$^{-1}$)}&\multicolumn{1}{c}{(mas)}&\multicolumn{1}{c}{(mas)}&\multicolumn{1}{c}{(mas)}&\multicolumn{1}{c}{(mas)}&\multicolumn{1}{c}{(Jy)}\\ 
\hline

 10.2 &1.3&$ 0.00\pm0.00$&$ -17.3\pm 0.1$&$  40.2\pm 0.1$&$  43.8\pm 0.1$&$   4.3\pm 1.2$&  1.20\\
 11.4 &3.5&$ 1.01\pm0.00$&$ -29.4\pm 0.1$&$  10.5\pm 0.1$&$  31.2\pm 0.1$&$  22.6\pm 0.6$&  0.60\\
 12.0 &1.4&$ 0.66\pm0.05$&$ -46.5\pm 0.1$&$   4.4\pm 0.1$&$  46.7\pm 0.1$&$   5.5\pm 0.4$&  0.40\\
 12.1 &1.3&$ 0.86\pm0.18$&$ -53.0\pm 0.1$&$  42.9\pm 0.1$&$  68.2\pm 0.1$&$   6.5\pm 0.3$&  0.30\\
 12.7 &0.9&$ 1.25\pm0.78$&$ -27.8\pm 0.1$&$ -18.1\pm 0.1$&$  33.2\pm 0.1$&$  14.9\pm 0.4$&  0.40\\
 12.7 &0.3&$ 0.15\pm0.10$&$  37.0\pm 0.3$&$   9.3\pm 0.3$&$  38.1\pm 0.3$&$   0.4\pm 1.1$&  1.10\\
\hline
\end{tabular}
%\addtocounter{table}{-1}
\caption{RT Vir 19960515 22-GHz H$_2$O maser feature parameters.}
\label{tab:rt96_4}
\end{table*}

\begin{table*}
\begin{tabular}{ccrrrrrr}
\hline
$V_{\mathrm{LSR}}$&$\Delta{V}$&\multicolumn{1}{c}{$\Delta{V_{1/2}}$}&\multicolumn{1}{c}{$x$}&\multicolumn{1}{c}{$y$}&\multicolumn{1}{c}{$a$}&\multicolumn{1}{c}{$l$}&\multicolumn{1}{c}{$I$}\\ 
(km s$^{-1}$)&(km s$^{-1}$)&\multicolumn{1}{c}{(km s$^{-1}$)}&\multicolumn{1}{c}{(mas)}&\multicolumn{1}{c}{(mas)}&\multicolumn{1}{c}{(mas)}&\multicolumn{1}{c}{(mas)}&\multicolumn{1}{c}{(Jy)}\\ 
\hline
  9.1 &0.4&$ 0.00\pm0.00$&$ -20.4\pm 1.1$&$  51.3\pm 1.3$&$  55.2\pm 1.3$&$   5.5\pm 3.9$&  3.90\\
 10.2 &1.3&$ 0.00\pm0.00$&$ -16.8\pm 0.1$&$  40.4\pm 0.1$&$  43.8\pm 0.1$&$   3.1\pm 1.1$&  1.10\\
 10.5 &0.3&$ 0.15\pm0.10$&$ -17.8\pm 0.4$&$  73.8\pm 0.5$&$  75.9\pm 0.5$&$   0.9\pm 1.1$&  1.10\\
 11.3 &3.4&$ 0.98\pm0.01$&$ -29.7\pm 0.1$&$  10.0\pm 0.1$&$  31.3\pm 0.1$&$  18.4\pm 3.0$&  3.00\\
 12.0 &0.9&$ 0.54\pm0.07$&$ -53.2\pm 0.1$&$  35.5\pm 0.1$&$  63.9\pm 0.1$&$   5.9\pm 0.5$&  0.50\\
 12.1 &1.4&$ 0.65\pm0.07$&$ -46.6\pm 0.1$&$   4.8\pm 0.1$&$  46.8\pm 0.1$&$  14.2\pm 0.7$&  0.70\\
\hline
\end{tabular}
%\addtocounter{table}{-1}
\caption{RT Vir 19960524 22-GHz H$_2$O maser feature parameters (\emph{see next page}).}
\end{table*}

\begin{table*}
\begin{tabular}{ccrrrrrr}
\hline
$V_{\mathrm{LSR}}$&$\Delta{V}$&\multicolumn{1}{c}{$\Delta{V_{1/2}}$}&\multicolumn{1}{c}{$x$}&\multicolumn{1}{c}{$y$}&\multicolumn{1}{c}{$a$}&\multicolumn{1}{c}{$l$}&\multicolumn{1}{c}{$I$}\\ 
(km s$^{-1}$)&(km s$^{-1}$)&\multicolumn{1}{c}{(km s$^{-1}$)}&\multicolumn{1}{c}{(mas)}&\multicolumn{1}{c}{(mas)}&\multicolumn{1}{c}{(mas)}&\multicolumn{1}{c}{(mas)}&\multicolumn{1}{c}{(Jy)}\\ 
\hline
 12.2 &1.2&$ 0.00\pm0.00$&$ -54.0\pm 0.1$&$  49.3\pm 0.1$&$  73.1\pm 0.1$&$  15.8\pm 0.6$&  0.60\\
 12.6 &0.3&$ 0.15\pm0.10$&$  36.4\pm 0.3$&$   7.9\pm 0.3$&$  37.3\pm 0.3$&$   2.2\pm 0.8$&  0.80\\
 12.7 &1.2&$ 1.38\pm0.76$&$ -27.0\pm 0.1$&$ -17.8\pm 0.1$&$  32.3\pm 0.1$&$  15.7\pm 0.6$&  0.60\\
 13.7 &2.5&$ 1.16\pm0.00$&$  16.6\pm 0.1$&$ -55.0\pm 0.1$&$  57.4\pm 0.1$&$  11.6\pm 0.2$&  0.20\\
 14.7 &0.6&$ 0.00\pm0.00$&$   8.3\pm 0.2$&$ -38.2\pm 0.1$&$  39.1\pm 0.1$&$   6.4\pm 0.9$&  0.90\\
 15.0 &0.3&$ 0.15\pm0.10$&$  19.0\pm 0.1$&$ -58.1\pm 0.1$&$  61.1\pm 2.0$&$   6.5\pm 5.9$&  5.90\\
\hline
\end{tabular}
\addtocounter{table}{-1}
\caption{RT Vir 19960524 22-GHz H$_2$O maser feature parameters (\emph{continued}).}
\label{tab:rt96_5}
\end{table*}

\begin{table*}
\begin{tabular}{ccrrrrrr}
\hline
$V_{\mathrm{LSR}}$&$\Delta{V}$&\multicolumn{1}{c}{$\Delta{V_{1/2}}$}&\multicolumn{1}{c}{$x$}&\multicolumn{1}{c}{$y$}&\multicolumn{1}{c}{$a$}&\multicolumn{1}{c}{$l$}&\multicolumn{1}{c}{$I$}\\ 
(km s$^{-1}$)&(km s$^{-1}$)&\multicolumn{1}{c}{(km s$^{-1}$)}&\multicolumn{1}{c}{(mas)}&\multicolumn{1}{c}{(mas)}&\multicolumn{1}{c}{(mas)}&\multicolumn{1}{c}{(mas)}&\multicolumn{1}{c}{(Jy)}\\ 
\hline

 10.2 &1.3&$ 0.89\pm0.40$&$ -16.6\pm 0.1$&$  41.0\pm 0.2$&$  44.2\pm 0.2$&$   5.1\pm 3.1$&  3.10\\
 11.3 &3.0&$ 0.96\pm0.01$&$ -30.3\pm 0.1$&$  10.4\pm 0.1$&$  32.0\pm 0.1$&$  12.2\pm 2.3$&  2.30\\
 12.1 &1.2&$ 0.60\pm0.05$&$ -47.2\pm 0.1$&$   4.3\pm 0.1$&$  47.4\pm 0.1$&$   2.7\pm 0.4$&  0.40\\
 12.1 &1.3&$ 0.92\pm0.15$&$ -53.4\pm 0.1$&$  43.7\pm 0.2$&$  69.0\pm 0.1$&$  14.2\pm 1.3$&  1.30\\
 12.2 &0.5&$ 0.47\pm0.21$&$ -41.6\pm 0.5$&$ -85.8\pm 0.5$&$  95.4\pm 0.5$&$   0.3\pm 1.8$&  1.80\\
 12.6 &0.7&$ 0.68\pm0.17$&$ -25.7\pm 0.2$&$ -20.3\pm 0.2$&$  32.8\pm 0.2$&$  10.9\pm 1.3$&  1.30\\
 13.1 &0.6&$ 0.91\pm0.66$&$  -7.8\pm 0.3$&$ -15.1\pm 0.3$&$  17.0\pm 1.0$&$   8.0\pm 1.4$&  1.40\\
 13.7 &2.5&$ 1.23\pm0.03$&$  16.6\pm 0.1$&$ -55.6\pm 0.1$&$  58.0\pm 0.1$&$  13.8\pm 1.3$&  1.30\\
 14.4 &0.8&$ 0.00\pm0.00$&$  -2.4\pm 0.3$&$ -16.3\pm 0.3$&$  16.5\pm 0.9$&$   4.8\pm 1.1$&  1.10\\
 14.6 &0.3&$ 0.15\pm0.10$&$ -15.1\pm 1.1$&$ -32.7\pm 1.0$&$  36.0\pm 1.8$&$   2.5\pm 2.7$&  2.70\\
 14.7 &0.9&$ 0.90\pm0.26$&$  33.5\pm 0.1$&$  55.5\pm 0.1$&$  64.8\pm 0.1$&$   5.4\pm 0.8$&  0.80\\
 14.8 &0.3&$ 0.15\pm0.10$&$  20.4\pm 0.8$&$  75.0\pm 1.3$&$  77.7\pm 2.1$&$   3.5\pm 2.8$&  2.80\\
 14.9 &0.4&$ 0.00\pm0.00$&$   7.1\pm 0.4$&$ -37.9\pm 0.4$&$  38.6\pm 0.4$&$   2.4\pm 1.2$&  1.20\\
 15.3 &0.4&$ 0.15\pm0.10$&$  26.3\pm 0.1$&$ -48.0\pm 0.3$&$  54.7\pm 0.3$&$   9.7\pm 1.3$&  1.30\\
 15.3 &1.6&$ 1.07\pm0.10$&$ -43.0\pm 0.1$&$ -10.2\pm 0.1$&$  44.2\pm 0.1$&$   4.1\pm 2.3$&  2.30\\
\hline
\end{tabular}
%\addtocounter{table}{-1}
\caption{RT Vir 19960612 22-GHz H$_2$O maser feature parameters (\emph{see next page}).}
\end{table*}
\clearpage
\begin{table*}
\begin{tabular}{ccrrrrrr}
\hline
$V_{\mathrm{LSR}}$&$\Delta{V}$&\multicolumn{1}{c}{$\Delta{V_{1/2}}$}&\multicolumn{1}{c}{$x$}&\multicolumn{1}{c}{$y$}&\multicolumn{1}{c}{$a$}&\multicolumn{1}{c}{$l$}&\multicolumn{1}{c}{$I$}\\ 
(km s$^{-1}$)&(km s$^{-1}$)&\multicolumn{1}{c}{(km s$^{-1}$)}&\multicolumn{1}{c}{(mas)}&\multicolumn{1}{c}{(mas)}&\multicolumn{1}{c}{(mas)}&\multicolumn{1}{c}{(mas)}&\multicolumn{1}{c}{(Jy)}\\ 
\hline
 15.6 &1.1&$ 0.77\pm0.26$&$  30.4\pm 0.1$&$  56.1\pm 0.1$&$  63.8\pm 0.9$&$  19.3\pm 2.6$&  2.60\\
 15.7 &2.2&$ 0.00\pm0.00$&$  32.4\pm 0.1$&$ -42.8\pm 0.1$&$  53.7\pm 0.1$&$  14.4\pm 0.2$&  0.20\\
 15.8 &0.4&$ 0.00\pm0.00$&$  11.9\pm 1.0$&$  96.0\pm 1.1$&$  96.7\pm 1.9$&$   4.1\pm 3.3$&  3.30\\
 17.1 &1.5&$ 0.78\pm0.01$&$  39.4\pm 0.1$&$  47.1\pm 0.1$&$  61.4\pm 0.1$&$   6.1\pm 3.2$&  3.20\\
 17.9 &3.3&$ 1.84\pm0.87$&$  37.2\pm 0.1$&$ -30.5\pm 0.1$&$  48.1\pm 0.1$&$   8.5\pm 1.4$&  1.40\\
 18.3 &1.3&$ 0.66\pm0.01$&$ -73.0\pm 0.4$&$  -0.7\pm 0.2$&$  73.0\pm 0.4$&$   4.4\pm 4.4$&  4.40\\
\hline
\end{tabular}
\addtocounter{table}{-1}
\caption{RT Vir 19960612 22-GHz H$_2$O maser feature parameters (\emph{continued}.}
\label{tab:rt96_6}
\end{table*}

\begin{table*}
\begin{tabular}{ccrrrrrr}
\hline
$V_{\mathrm{LSR}}$&$\Delta{V}$&\multicolumn{1}{c}{$\Delta{V_{1/2}}$}&\multicolumn{1}{c}{$x$}&\multicolumn{1}{c}{$y$}&\multicolumn{1}{c}{$a$}&\multicolumn{1}{c}{$l$}&\multicolumn{1}{c}{$I$}\\ 
(km s$^{-1}$)&(km s$^{-1}$)&\multicolumn{1}{c}{(km s$^{-1}$)}&\multicolumn{1}{c}{(mas)}&\multicolumn{1}{c}{(mas)}&\multicolumn{1}{c}{(mas)}&\multicolumn{1}{c}{(mas)}&\multicolumn{1}{c}{(Jy)}\\ 
\hline
 37.0 &1.9&$ 1.40\pm0.35$&$-100.8\pm 0.1$&$  25.0\pm 0.2$&$ 103.9\pm 0.1$&$  30.9\pm 5.8$& 12.85\\
  37.3 &0.5&$ 0.15\pm0.10$&$-125.8\pm 2.1$&$ 107.6\pm 3.2$&$ 167.1\pm 2.8$&$  48.0\pm 7.7$& 10.20\\
  37.3 &0.5&$ 0.00\pm0.00$&$ -88.1\pm 1.9$&$ -66.7\pm 2.9$&$ 113.5\pm 2.4$&$  43.5\pm 7.4$& 14.43\\
  37.8 &0.4&$ 0.15\pm0.10$&$  37.9\pm 1.4$&$ 148.1\pm 2.6$&$ 152.8\pm 2.6$&$  24.5\pm 9.7$&  2.95\\
  38.2 &0.7&$ 0.00\pm0.00$&$-119.8\pm 1.9$&$ 123.1\pm 3.5$&$ 172.1\pm 2.8$&$  25.8\pm 8.6$& 16.34\\
  38.5 &0.6&$ 0.43\pm0.27$&$-117.8\pm 1.9$&$-108.3\pm 3.6$&$ 159.8\pm 2.8$&$  62.5\pm 7.9$& 37.15\\
\hline
\end{tabular}
%\addtocounter{table}{-1}
\caption{W Hya 1999 22-GHz H$_2$O maser feature parameters.}
\label{tab:wh99}
\end{table*}

\begin{table*}
\begin{tabular}{ccrrrrrr}
\hline
$V_{\mathrm{LSR}}$&$\Delta{V}$&\multicolumn{1}{c}{$\Delta{V_{1/2}}$}&\multicolumn{1}{c}{$x$}&\multicolumn{1}{c}{$y$}&\multicolumn{1}{c}{$a$}&\multicolumn{1}{c}{$l$}&\multicolumn{1}{c}{$I$}\\ 
(km s$^{-1}$)&(km s$^{-1}$)&\multicolumn{1}{c}{(km s$^{-1}$)}&\multicolumn{1}{c}{(mas)}&\multicolumn{1}{c}{(mas)}&\multicolumn{1}{c}{(mas)}&\multicolumn{1}{c}{(mas)}&\multicolumn{1}{c}{(Jy)}\\ 
\hline
  38.7 &0.6&$ 0.00\pm0.00$&$ -53.5\pm 0.8$&$ 152.7\pm 0.9$&$ 161.8\pm 0.9$&$  16.4\pm 4.3$& 14.04\\
  39.7 &1.8&$ 0.15\pm0.10$&$  67.2\pm 0.1$&$ 121.3\pm 0.1$&$ 138.8\pm 0.1$&$  38.9\pm 2.3$& 20.24\\
  39.7 &0.4&$ 0.83\pm0.06$&$  58.3\pm 0.3$&$ 142.3\pm 0.3$&$ 153.9\pm 0.3$&$  21.9\pm 3.1$& 20.87\\
  39.8 &2.0&$ 1.82\pm1.70$&$ -92.6\pm 0.0$&$  85.7\pm 0.0$&$ 126.2\pm 0.0$&$  37.9\pm 5.0$& 18.03\\
  39.9 &1.7&$ 0.45\pm0.03$&$-143.8\pm 0.2$&$ -52.6\pm 0.2$&$ 153.1\pm 0.2$&$  21.6\pm 3.8$&  6.78\\
  40.0 &0.4&$ 0.58\pm0.02$&$  84.8\pm 1.1$&$  37.8\pm 1.4$&$  92.2\pm 1.1$&$  37.3\pm 5.6$&  6.09\\
\hline
\end{tabular}
%\addtocounter{table}{-1}
\caption{W Hya 2000 22-GHz H$_2$O maser feature parameters.}
\label{tab:wh00}
\end{table*}

\begin{table*}
\begin{tabular}{ccrrrrrr}
\hline
$V_{\mathrm{LSR}}$&$\Delta{V}$&\multicolumn{1}{c}{$\Delta{V_{1/2}}$}&\multicolumn{1}{c}{$x$}&\multicolumn{1}{c}{$y$}&\multicolumn{1}{c}{$a$}&\multicolumn{1}{c}{$l$}&\multicolumn{1}{c}{$I$}\\ 
(km s$^{-1}$)&(km s$^{-1}$)&\multicolumn{1}{c}{(km s$^{-1}$)}&\multicolumn{1}{c}{(mas)}&\multicolumn{1}{c}{(mas)}&\multicolumn{1}{c}{(mas)}&\multicolumn{1}{c}{(mas)}&\multicolumn{1}{c}{(Jy)}\\ 
\hline
 38.1 &0.5&$ 0.00\pm0.00$&$  38.5\pm 1.0$&$  97.8\pm 1.3$&$ 104.6\pm 1.2$&$  52.6\pm 4.8$&  8.67\\
  38.5 &2.3&$ 0.61\pm0.03$&$  64.2\pm 0.0$&$ -99.4\pm 0.0$&$ 118.3\pm 0.0$&$   6.1\pm 0.9$&  4.95\\
  39.0 &0.5&$ 0.15\pm0.10$&$-103.3\pm 1.1$&$ -56.9\pm 1.6$&$ 117.7\pm 1.3$&$  37.4\pm 4.8$&  6.91\\
  39.0 &0.8&$ 0.00\pm0.00$&$-117.8\pm 1.0$&$  18.3\pm 1.3$&$ 119.3\pm 1.0$&$  25.6\pm 4.1$& 11.64\\
  39.5 &1.1&$ 0.60\pm0.09$&$  23.6\pm 0.3$&$-143.4\pm 0.3$&$ 145.4\pm 0.3$&$  24.8\pm 3.2$&  9.96\\
  39.5 &0.8&$ 0.00\pm0.00$&$  97.5\pm 0.7$&$ -67.1\pm 0.6$&$ 118.7\pm 0.7$&$  25.6\pm 3.2$& 21.46\\
\hline
\end{tabular}
%\addtocounter{table}{-1}
\caption{W Hya 2001 22-GHz H$_2$O maser feature parameters.}
\label{tab:wh01}
\end{table*}

\begin{table*}
\begin{tabular}{ccrrrrrr}
\hline
$V_{\mathrm{LSR}}$&$\Delta{V}$&\multicolumn{1}{c}{$\Delta{V_{1/2}}$}&\multicolumn{1}{c}{$x$}&\multicolumn{1}{c}{$y$}&\multicolumn{1}{c}{$a$}&\multicolumn{1}{c}{$l$}&\multicolumn{1}{c}{$I$}\\ 
(km s$^{-1}$)&(km s$^{-1}$)&\multicolumn{1}{c}{(km s$^{-1}$)}&\multicolumn{1}{c}{(mas)}&\multicolumn{1}{c}{(mas)}&\multicolumn{1}{c}{(mas)}&\multicolumn{1}{c}{(mas)}&\multicolumn{1}{c}{(Jy)}\\ 
\hline
 38.4 &1.2&$ 0.00\pm0.00$&$  35.3\pm 0.4$&$ 107.3\pm 0.5$&$ 112.9\pm 0.5$&$  34.5\pm 3.6$&  9.71\\
  39.9 &0.7&$ 0.00\pm0.00$&$ -98.0\pm 0.2$&$  34.2\pm 0.2$&$ 103.8\pm 0.2$&$  10.9\pm 3.0$&  9.14\\
  40.2 &0.7&$ 0.00\pm0.00$&$ -50.2\pm 0.7$&$ -80.8\pm 1.2$&$  95.4\pm 1.0$&$  56.9\pm 3.0$& 20.18\\
  40.4 &2.4&$ 0.00\pm0.00$&$ -89.2\pm 0.0$&$  61.6\pm 0.0$&$ 108.4\pm 0.0$&$  24.1\pm 1.7$& 11.63\\
  40.5 &0.4&$ 0.00\pm0.00$&$  85.7\pm 0.7$&$ -17.0\pm 0.9$&$  87.0\pm 0.7$&$  41.3\pm 2.2$& 36.24\\
  40.5 &1.4&$ 0.00\pm0.00$&$  87.7\pm 0.0$&$  24.6\pm 0.1$&$  91.1\pm 0.0$&$  23.6\pm 3.1$&  4.38\\
\hline
\end{tabular}
%\addtocounter{table}{-1}
\caption{W Hya 2002 22-GHz H$_2$O maser feature parameters.}
\label{tab:wh02}
\end{table*}

\end{document}